\documentclass[12pt]{report}
\usepackage{epsfig,cite,feynarts,amsmath,psfrag,axodraw}

\input paperdef

\evensidemargin \oddsidemargin
\allowdisplaybreaks

\def\_{\rule{.3em}{.15ex}} 
\def\slash#1{\setbox0=\hbox{$#1$}#1\hskip-\wd0\dimen0=5pt\advance
       \dimen0 by-\ht0\advance\dimen0 by\dp0\lower0.5\dimen0\hbox
         to\wd0{\hss\sl/\/\hss}}

\def\ra{\rightarrow}

\begin{document}

\thispagestyle{empty}
\setcounter{page}{0}
\def\thefootnote{\fnsymbol{footnote}}

\begin{flushright}
CERN--PH--TH/2004--224\\
DCPT/04/162\\ 
IPPP/04/81\\
MPP--2004--145\\
hep-ph/0412214 \\
\end{flushright}

\vspace{1mm}

\begin{center}

{\large\sc {\bf Electroweak Precision Observables
             in the Minimal Supersymmetric Standard Model}}

\vspace{1cm}

{\sc 
S.~Heinemeyer$^{1}$%
\footnote{email: Sven.Heinemeyer@cern.ch}%
, W. Hollik$^{2}$%
\footnote{email: hollik@mppmu.mpg.de}%
~and~G.~Weiglein$^{3}$%
\footnote{email: Georg.Weiglein@durham.ac.uk}
}

\vspace*{1cm}

{\sl
$^1$CERN, TH Division, Dept.\ of Physics, CH-1211 Geneva 23, Switzerland 

\vspace*{0.4cm} 

$^2$Max-Planck-Institut f\"ur Physik (Werner-Heisenberg-Institut),\\ 
F\"ohringer Ring 6, D--80805 Munich, Germany

\vspace*{0.4cm} 

$^3$Institute for Particle Physics Phenomenology, University of Durham,\\ 
Durham DH1~3LE, UK 

} 

\end{center}

\vspace*{0.2cm}

\BC
{\bf Abstract}
\EC
The current status of electroweak precision observables in the Minimal 
Supersymmetric Standard Model (MSSM) is reviewed.
We focus in particular on the $W$~boson mass, $\MW$, the effective leptonic 
weak mixing angle, $\sweff$, the anomalous magnetic moment of the muon, 
$(g-2)_\mu$, and the lightest $\cp$-even MSSM Higgs boson mass, $\mh$. 
We summarize the current experimental situation and the status of the
theoretical evaluations. An estimate of the current theoretical
uncertainties from unknown higher-order corrections and from the
experimental errors of the input parameters is given. We discuss future 
prospects for both the experimental accuracies and the precision of the
theoretical predictions. Confronting the precision data with the theory
predictions within the unconstrained MSSM and within specific
SUSY-breaking scenarios, we analyse how well the data are described by
the theory. The mSUGRA scenario with cosmological constraints yields a
very good fit to the data, showing a clear preference for a relatively 
light mass scale of the SUSY particles. The constraints on the parameter
space from the precision data is discussed, and it is shown that the
prospective accuracy at the next generation of colliders will enhance
the sensitivity of the precision tests very significantly.

\def\thefootnote{\arabic{footnote}}
\setcounter{page}{0}
\setcounter{footnote}{0}


\newpage
\thispagestyle{empty}
\mbox{}
\setcounter{page}{0}
\newpage
\tableofcontents
\newpage
\mbox{}
\newpage

\newpage

\chapter{Introduction}
\label{chap:intro}


\section{Motivation}
\label{sec:MSSMmotivation}

Theories based on Supersymmetry (SUSY)~\cite{mssm} are widely
considered as the theoretically most appealing extension of the
Standard Model (SM)~\cite{sm}. They are consistent with the approximate
unification of the gauge coupling constants at the GUT scale and
provide a way to cancel the quadratic divergences in the Higgs sector
hence stabilizing the huge hierarchy between the GUT and the Fermi
scales. Furthermore, in SUSY theories the breaking of the electroweak
symmetry is naturally induced at the Fermi scale, and the lightest
supersymmetric particle can be neutral, weakly interacting and
absolutely stable, providing therefore a natural solution for the dark
matter problem.

SUSY predicts the existence of scalar partners $\tilde{f}_L,
\tilde{f}_R$ to each SM chiral fermion, and spin--1/2 partners to the
gauge bosons and to the scalar Higgs bosons. So far, the direct search
for SUSY particles has not been successful.
One can only set lower bounds of ${\cal O}(100)$~GeV on
their masses~\cite{pdg}. The search reach will be extended in various
ways in
the ongoing Run~II at the upgraded Fermilab Tevatron~\cite{susytev}.
The LHC~\cite{atlastdr,cms} and the $e^+e^-$ International Linear Collider
(ILC)~\cite{teslatdr,orangebook,acfarep} have very good prospects for
exploring SUSY at the TeV scale, which is favoured from naturalness
arguments. From the interplay of both machines detailed
information on the SUSY spectrum can be expected in this
case~\cite{lhclc}.

In the Minimal Supersymmetric extension of the Standard Model (MSSM)
two Higgs doublets are
required resulting in five physical Higgs bosons~\cite{hhg}. The
direct search resulted in lower limits of about $90 \gev$ for the
neutral Higgs bosons and about $80 \gev$ for the charged
ones~\cite{LEPHiggsearly,LEPHiggsSM}. The Higgs search at the Tevatron
will be able to   
probe significant parts of the MSSM parameter space at the 95\% C.L.\
even with rather moderate luminosity~\cite{higgstev}. The LHC will
discover at least one MSSM Higgs boson over most of the MSSM parameter
space~\cite{atlastdr,cms,howiemarcela,ehow,asbs2}. The ILC will be able
to detect any Higgs boson that couples to the $Z$~boson in a decay-mode
independent way. The properties of all Higgs-bosons which are within the
kinematic reach of the ILC will be determined with high
precision~\cite{teslatdr,orangebook,acfarep}.

Contrary to the SM case, where the mass of the Higgs boson is a
free parameter, within the MSSM the quartic
couplings of the Higgs potential are fixed in terms of the gauge
couplings as a consequence of SUSY~\cite{hhg}. Thus, at the
tree-level, the Higgs sector
is determined by just two independent parameters besides the SM
electroweak gauge couplings $g$ and $g'$, conventionally chosen as
$\tb = v_2/v_1$, the ratio of the vacuum expectation values of the two
Higgs doublets, and $\MA$, the mass of the $\cp$-odd $A$~boson.
As a consequence, the mass of the lightest $\cp$-even MSSM Higgs boson
can be predicted in terms of the other model parameters.

\bigskip

Besides the direct detection of SUSY particles and Higgs bosons, 
SUSY can also be probed via the virtual effects of the
additional particles to precision observables.
This requires a very high precision
of the experimental results as well as of the theoretical predictions.
The wealth of high-precision measurements carried out at LEP, SLC and
the Tevatron~\cite{LEPEWWG} as well as the ``Muon $g-2$ Experiment''
(E821)~\cite{g-2exp} and further low-energy experiments provide a
powerful tool for testing the electroweak theory and probing indirect
effects of SUSY particles.
The most relevant electroweak precision observables (EWPO) in this
context are the $W$~boson mass, $\MW$, the effective leptonic weak
mixing angle, $\sweff$, the anomalous magnetic moment of the muon, 
$\amu \equiv (g-2)_\mu/2$, and
the mass of the lightest $\cp$-even MSSM Higgs
boson, $\mh$.
While the current exclusion bounds on $\mh$ already allow to constrain
the MSSM parameter space, 
the prospective accuracy for the measurement of the mass of a light 
Higgs boson at the
LHC of about $200 \mev$~\cite{atlastdr,cms} or at the ILC
of even $50 \mev$~\cite{teslatdr,orangebook,acfarep} could promote $\mh$ to a
precision observable. Owing to the sensitive dependence of $\mh$ on
especially the scalar top sector, the measured value of $\mh$ will allow
to set stringent constraints on the parameters in this sector.

Since the experimental data --- with few exceptions --- are well
described by the SM~\cite{LEPEWWG}, the electroweak precision tests at
present mainly yield constraints on possible extensions of the SM, e.g.\ 
lower limits on SUSY particle masses. Nevertheless, one can use the
available data to investigate whether small deviations from the SM
predictions could be caused by
quantum effects of the SUSY particles: sleptons, squarks, gluinos,
charginos/neutralinos and additional Higgs bosons, and what regions of
the SUSY parameter space might be favoured.


\section{The structure of the MSSM}
\label{sec:MSSMstructure}

The MSSM constitutes the minimal supersymmetric extension of the
SM. The number of SUSY generators is $N=1$, the smallest possible value.
In order to keep anomaly cancellation, contrary to the SM a second
Higgs doublet is needed~\cite{glawei}. One Higgs doublet, $\cHe$,
gives mass to the $d$-type fermions (with weak isospin -1/2), the
other doublet, $\cHz$, gives mass to the $u$-type fermions (with weak
isospin +1/2). 
All SM multiplets, including the two Higgs doublets (2HDM), are extended to
supersymmetric multiplets, resulting in scalar partners for quarks and
leptons (``squarks'' and ``sleptons'') and fermionic partners for the
SM gauge boson and the Higgs bosons (``gauginos'' and ``gluinos'').   
In \refta{tab:MSSMfields} the spectrum of the MSSM fields is
summarized (family indices are suppressed).
In this report we do not consider effects of complex phases, i.e.\
we treat all MSSM parameters as real.

\begin{table}[htb]
\begin{center}
\renewcommand{\arraystretch}{1.8}
\begin{tabular}{|cc|cc|cc|} \hline
\rule[-2ex]{0mm}{5ex}%
\bf{superfield} & $(SU(3),SU(2),U(1))$ & 
\bf{2HDM particle} & spin & \bf{SUSY partner} & spin \\
\hline \hline 
$\hat{Q}$ & (3, 2, $\frac{1}{3}$) & $(u,d)_L$ & $\frac{1}{2}$ & 
$(\tilde{u},\tilde{d})_L$ & 0 \\  
$\hat{U}$ & ($3^*$, 1, $-\frac{4}{3}$) & 
$\bar{u}_R$ & $\frac{1}{2}$ & $\tilde{u}^*_R$ & 0 \\ 
$\hat{D}$ & ($3^*$, 1, $\frac{2}{3}$) & 
$\bar{d}_R$ & $\frac{1}{2}$ & $\tilde{d}^*_R$ & 0 \\
\hline 
$\hat{L}$ & (1, 2, $-1$) & $(\nu,e)_L$ & $\frac{1}{2}$ & 
$\tilde{L}_L=(\tilde{\nu},\tilde{e})_L$ & 0 \\
$\hat{E}$ & (1, 1, 2) 
& $\bar{e}_R$ & $\frac{1}{2}$ & $\tilde{e}^*_R$ & 0 \\ 
\hline 
$\hat{H}_1$ & (1, 2, $-1$) & $(H^0_1,H^-_1)_L$ & 0 & 
$(\tilde{H}^0_1,\tilde{H}^-_1)_L$ & $\frac{1}{2}$ \\
$\hat{H}_2$ & (1, 2, 1) & $(H^+_2,H^0_2)_L$ & 0 & 
$(\tilde{H}^+_2,\tilde{H}^0_2)_L$ & $\frac{1}{2}$ \\ 
\hline 
$\hat{W}$ & (1, 3, 0) & 
$W^i$ & 1 & $\tilde{W}^i$ & $\frac{1}{2}$ \\
$\hat{B}$ & (1, 1, 0) & 
$B^0$ & 1 & $\tilde{B}^0$ & $\frac{1}{2}$ \\
\hline
$\hat{G}_a$ & (8, 1, 0) & 
$g_a$ & 1 & $\tilde{g}_a$ & $\frac{1}{2}$ \\ 
\hline
\end{tabular}
\end{center}
\caption{Superfields and particle content of the MSSM.}
\label{tab:MSSMfields}
\end{table}

The mass eigenstates of the gauginos are linear combinations of these
fields, denoted as ``neutralinos'' and ``charginos''. Also the left- and
right-handed squarks (and sleptons) can mix, yielding the mass
eigenstates (denoted by the indices
$1,2$ instead of $L,R$). All physical particles
of the MSSM are given in \refta{tab:MSSMparticles}.

\begin{table}[htb]
\begin{center}
\renewcommand{\arraystretch}{1.8}
\begin{tabular}{|ll|c||ll|c|} \hline
\rule[-2ex]{0mm}{5ex}%
\bf{2HDM particle} & & \bf{spin} & \bf{SUSY particle} & & \bf{spin} \\
\hline \hline  
quarks:  & $q$ & $\frac{1}{2}$ & 
squarks: & $\tilde{q}_1,\tilde{q}_2$ & $0$ \\
leptons: & $l$ & $\frac{1}{2}$ & 
sleptons: & $\tilde{l}_1,\tilde{l}_2$ & $0$ \\ 
gluons: & $g_a$ & $1$ & Gluinos: & $\tilde{g}_a$ & $\frac{1}{2}$ \\ 
gauge bosons: & $W^\pm, Z^0, \gamma$ & $1$ & Neutralinos: & 
$\tilde{\chi}_1^0$, $\tilde{\chi}_2^0$, $\tilde{\chi}_3^0$,
$\tilde{\chi}_4^0$ & $\frac{1}{2}$ \\ 
Higgs bosons: & $h^0, H^0, A^0, H^\pm$ & $0$ & Charginos: & 
$\tilde{\chi}_1^{\pm}$, $\tilde{\chi}_2^{\pm}$ & $\frac{1}{2}$ \\ 
\hline
\end{tabular}
\end{center}
\caption{The particle content of the MSSM.}
\label{tab:MSSMparticles}
\end{table}


\subsection{The Higgs sector of the MSSM}
\label{subsec:higgssector}

The two Higgs doublets form the Higgs potential~\cite{hhg}
\BEA
V &=& (m_1^2 + |\mu|^2) |\cHe|^2 + (m_2^2 + |\mu|^2) |\cHz|^2 
      - m_{12}^2 (\epsilon_{ab} \cHe^a\cHz^b + \hc)  \non \\
  & & + \frac{1}{8}({g_1}^2+{g_2}^2) \left[ |\cHe|^2 - |\cHz|^2 \right]^2
        + \frac{1}{2} {g_2}^2|\cHe^{\dag} \cHz|^2~,
\label{higgspot}
\EEA
which contains $m_1, m_2, m_{12}$ as soft SUSY breaking parameters and
$\mu$ as the Higgsino mass parameter; 
$g, g'$ are the $SU(2)$ and $U(1)$ gauge couplings, and 
$\epsilon_{12} = -1$.

The doublet fields $\cHe$ and $\cHz$ are decomposed  in the following way:
\BEA
\cHe &=& \VL \cHe^0 \\[0.5ex] \cHe^- \VR \; = \; \VL v_1 
        + \frac{1}{\sqrt2}(\phi_1^0 - i\chi_1^0) \\[0.5ex] -\phi_1^- \VR  
        \non \\
\cHz &=& \VL \cHz^+ \\[0.5ex] \cHz^0 \VR \; = \; \VL \phi_2^+ \\[0.5ex] 
        v_2 + \frac{1}{\sqrt2}(\phi_2^0 + i\chi_2^0) \VR~.
\label{higgsfeldunrot}
\EEA
The potential (\ref{higgspot}) can be described with the help of two  
independent parameters (besides $g$ and $g'$): 
$\Tb = v_2/v_1$ and $M_A^2 = -m_{12}^2(\Tb+\CTb)$,
where $M_A$ is the mass of the $\cp$-odd $A$ boson.

The diagonalization of the bilinear part of the Higgs potential,
i.e.\ the Higgs mass matrices, is performed via the orthogonal
transformations 
\BEA
\label{hHdiag}
\VL H^0 \\[0.5ex] h^0 \VR &=& \ML \Ca & \Sa \\[0.5ex] -\Sa & \Ca \MR 
\VL \phi_1^0 \\[0.5ex] \phi_2^0 \VR  \\
\label{AGdiag}
\VL G^0 \\[0.5ex] A^0 \VR &=& \ML \Cb & \sbe \\[0.5ex] -\sbe & \Cb \MR 
\VL \chi_1^0 \\[0.5ex] \chi_2^0 \VR  \\
\label{Hpmdiag}
\VL G^{\pm} \\[0.5ex] H^{\pm} \VR &=& \ML \Cb & \sbe \\[0.5ex] -\sbe & 
\Cb \MR \VL \phi_1^{\pm} \\[0.5ex] \phi_2^{\pm} \VR~.
\EEA
The mixing angle $\al$ is determined through
\BE
\tan 2\al = \tan 2\be \; \frac{\MA^2 + M_Z^2}{\MA^2 - M_Z^2} ;
\qquad  -\frac{\pi}{2} < \al < 0~.
\label{alphaborn}
\end{equation}

One gets the following Higgs spectrum:
\BEA
\mbox{2 neutral bosons},\, {\cal CP} = +1 &:& h^0, H^0 \non \\
\mbox{1 neutral boson},\, {\cal CP} = -1  &:& A^0 \non \\
\mbox{2 charged bosons}                   &:& H^+, H^- \non \\
\mbox{3 unphysical Goldstone bosons}      &:& G^0, G^+, G^- .
\EEA

The masses of the gauge bosons are given in analogy to the SM:
\BE
M_W^2 = \frac{1}{2} g_2^2 (v_1^2+v_2^2) ;\qquad
M_Z^2 = \frac{1}{2}(g_1^2+g_2^2)(v_1^2+v_2^2) ;\qquad M_\gamma=0.
\end{equation}

\bigskip
At tree level the mass matrix of the neutral $\cp$-even Higgs bosons
is given in the $\Pe$-$\Pz$-basis 
in terms of $\MZ$, $\MA$, and $\Tb$ by
\BEA
M_{\rm Higgs}^{2, {\rm tree}} &=& \ML \mpe^2 & \mpez^2 \\ 
                           \mpez^2 & \mpz^2 \MR \non\\
&=& \ML \MA^2 \SQb + \MZ^2 \CQb & -(\MA^2 + \MZ^2) \sbe \Cb \\
    -(\MA^2 + \MZ^2) \sbe \Cb & \MA^2 \CQb + \MZ^2 \SQb \MR,
\label{higgsmassmatrixtree}
\EEA
which by diagonalization according to \refeq{hHdiag} yields the
tree-level Higgs boson masses
\BE
M_{\rm Higgs}^{2, {\rm tree}} 
   \stackrel{\al}{\longrightarrow}
   \ML m_{H,{\rm tree}}^2 & 0 \\ 0 &  m_{h,{\rm tree}}^2 \MR~.
\end{equation}
The mixing angle $\alpha$ satisfies
\BE
\tan 2\alpha = \tan 2\beta \frac{\MA^2 + \MZ^2}{\MA^2 - \MZ^2},
\quad - \frac{\pi}{2} < \alpha < 0 .
\label{alpha}
\end{equation}
Since we treat all MSSM parameters as real there is no mixing between
$\cp$-even and $\cp$-odd Higgs bosons.

The tree-level results for the neutral 
$\cp$-even Higgs-boson masses of the MSSM read
\BE
m^2_{H, h} =
 \frac{1}{2} \KKL \MA^2 + \MZ^2
         \pm \sqrt{(\MA^2 + \MZ^2)^2 - 4 \MZ^2 \MA^2 \CQZb} \KKR~.
\label{eq:mhtree}
\end{equation}
This implies an upper bound of $m_{h, {\rm tree}} \leq \MZ$ for the
light $\cp$-even Higgs-boson mass of the MSSM. For a discussion of large
higher-order corrections to this bound, see \refse{sec:mh}. The
direct prediction of an upper bound for the mass of the light $\cp$-even
Higgs-boson mass is one of the most striking phenomenological
predictions of the MSSM. The existence of such a bound, which does not
occur in the case of the SM Higgs boson, can be related to the fact that
the quartic term in the Higgs potential of the MSSM is given in terms of
the gauge couplings, while the quartic coupling is a free parameter in
the SM.


\subsection{The scalar quark sector of the MSSM}
\label{subsec:sfermions}

The squark mass term of the MSSM Lagrangian is given by
\BE
{\cal L}_{m_{\tilde{f}}} = -\frac{1}{2} 
   \Big( \tilde{f}_L^{\dag},\tilde{f}_R^{\dag} \Big)\; {\bf Z} \; 
   \VL \tilde{f}_L \\[0.5ex] \tilde{f}_R \VR~,
\label{squarkmassmatrix}
\end{equation}
where
\BE
\renewcommand{\arraystretch}{1.5}
{\bf Z} = \ML M_{\tilde{Q}}^2 + M_Z^2\CZb(I_3^f-Q_f\sw^2) + m_f^2 
    & m_f (A_f - \mu \{\CTb;\Tb\}) \\
    m_f (A_f - \mu \{\cot\be;\tan\be\}) 
    & M_{\tilde{Q}'}^2 + M_Z^2 \CZb Q_f\sw^2 + m_f^2 \VR ,
\label{squarkmassenmatrix}
\end{equation}
and $\{\CTb;\Tb\}$ corresponds to $\{u;d\}$-type squarks.
The soft SUSY breaking term $M_{\tilde{Q}'}$ is given by:
\BEA
\label{RRmasses}
M_{\tilde{Q}'} &=& \left\{ \begin{array}{cl} 
M_{\tilde{U}} & \quad\mbox{for right handed $u$-type squarks} \\
M_{\tilde{D}} & \quad\mbox{for right handed $d$-type squarks}
\end{array} \right. .
\EEA
In order to diagonalize the mass matrix and to determine the physical mass
eigenstates the following rotation has to be performed:
\BE
\VL \sfe \\ \sfz \VR = \ML \costf & \sintf \\ -\sintf & \costf \MR 
                       \VL \sfl \\ \sfr \VR .
\label{squarkrotation}
\end{equation}
The mixing angle $\tsf$ is given for $\Tb > 1$ by:
\BEA
\costf &=& \sqrt{
    \frac{(m_{\tilde{f}_R}^2 - m_{\tilde{f}_1}^2)^2}
          {m_f^2 \, (A_f - \mu \{\CTb ; \Tb \})^2 
    + (m_{\tilde{f}_R}^2-m_{\tilde{f}_1}^2)^2} } \\
\sintf &=& \mp\; {\rm sgn} \Big[ 
    A_f - \mu \{\CTb ; \Tb \} \Big] \non \\ 
& & \times \; \sqrt{ 
    \frac{m_f^2 \, (A_f - \mu \{\CTb ; \Tb \})^2}
    {m_f^2 \, (A_f - \mu \{\CTb ; \Tb \})^2 
    + (m_{\tilde{f}_R}^2 - m_{\tilde{f}_1}^2)^2} }~. 
\label{stt}
\EEA
The negative sign in (\ref{stt}) corresponds to $u$-type squarks, the
positive sign 
to $d$-type ones. 
$m_{\tilde{f}_R}^2 \equiv M_{\tilde{Q}'}^2  + M_Z^2 \CZb Q_f\sw^2 + m_f^2 $
denotes the lower right entry in the 
squark mass matrix~(\ref{squarkmassenmatrix}).
The masses are given by the eigenvalues of the mass matrix:
\BEA
\label{Squarkmasse} 
m_{\tilde{f}_{1,2}}^2 &=& \edz \KKL M_{\tilde{Q}}^2 + M_{\tilde{Q}'}^2 \KKR
  + \frac{1}{2} M_Z^2 \CZb I^f_3 + m_f^2 \\
& & \left\{ \begin{array}{l} \displaystyle \pm\; \frac{c_f}{2}\, 
    \sqrt{ \Big[ M_{\tilde{Q}}^2 - M_{\tilde{Q}'}^2 + 
                 M_Z^2 \CZb (I^f_3-2Q_f\sw^2)\Big]^2 
    + 4 m_f^2 \Big( A_u - \mu \CTb \Big)^2} \non \\[2ex]
    \displaystyle \pm\; \frac{c_f}{2}\, 
    \sqrt{ \Big[ M_{\tilde{Q}}^2 - M_{\tilde{Q}'}^2 + 
                 M_Z^2 \CZb (I^f_3-2Q_f\sw^2)\Big]^2 
    + 4 m_f^2 \Big( A_d - \mu \Tb \Big)^2} 
    \end{array} \right. \non \\
c_f &=& {\rm sgn} \KKL M_{\tilde{Q}}^2 - M_{\tilde{Q}'}^2 + 
              M_Z^2 \CZb (I^f_3-2Q_f\sw^2) \KKR \non
\EEA
for $u$-type and $d$-type squarks, respectively.
Since the non-diagonal entry of the mass matrix
\refeq{squarkmassenmatrix} is proportional to the fermion mass, 
mixing becomes particularly important for $\tilde f = \Stop$, 
in the case of $\Tb \gg 1$ also for $\tilde f = \Sbot$.

\bigskip
For later purposes it is convenient to express the
squark mass matrix
in terms of the physical masses $\msfe, \msfz$ and the mixing angle $\tsf$:
\BE
{\bf Z} = \ML \cosQtf \msfe^2 + \sinQtf \msfz^2 & 
              \sintf \costf (\msfe^2 - \msfz^2) \\
              \sintf \costf (\msfe^2 - \msfz^2) &
              \sinQtf \msfe^2 + \cosQtf \msfz^2 
          \MR~.
\label{smm:physparam}
\end{equation}
$A_f$ can be written as follows:
\BE
A_f = \frac{\sintf \costf (\msfe^2 - \msfz^2)}{\mf} + \mu \{\CTb; \Tb\}.
\label{eq:af}
\end{equation}


\subsection{Charginos}
\label{subsec:charginos}

The charginos $\tilde{\chi}_i^+\; (i=1,2)$ are four component Dirac
fermions. The mass eigenstates are obtained from the winos
$\tilde{W}^\pm$ and the charged higgsinos $\tilde{H}^-_1$,
$\tilde{H}^+_2$:
\BE
\tilde{W}^+ = \VL -i \la^+ \\[0.5ex] i \bar{\la}^- \VR 
              \quad;\quad
\tilde{W}^- = \VL -i \la^- \\[0.5ex] i \bar{\la}^+ \VR 
              \quad;\quad
\tilde{H}^+_2 = \VL \psi^+_{H_2} \\[0.5ex] \bar{\psi}^-_{H_1} \VR
                \quad;\quad
\tilde{H}^-_1 = \VL \psi^-_{H_1} \\[0.5ex] \bar{\psi}^+_{H_2} \VR~.
\end{equation}
The chargino masses are defined as mass eigenvalues of the
diagonalized mass matrix,
\BE
{\cal L}_{\tilde{\chi}^+,{\rm mass}} = -\frac{1}{2}\,
         \Big( \psi^+,\psi^- \Big) \ML 0 & {\bf X}^T \\ {\bf X} & 0 \MR
         \VL \psi^+ \\ \psi^- \VR + \hc~,
\end{equation}
or given in terms of two-component fields
\BE
\left. \begin{array}{c} 
       \psi^+ = (-i\la^+, \psi^+_{H_2}) \\[1.5ex] 
       \psi^- = (-i\la^-, \psi^-_{H_1}) 
       \end{array} 
\right.~,
\end{equation}
where {\bf X} is given by
\BE
{\bf X} = \ML M_2 & \sqrt2\, M_W\, \sbe \\[1ex] \sqrt2\, M_W\, 
          \Cb & \mu \MR~.
\end{equation}
In the mass matrix $M_2$ is the soft SUSY-breaking parameter for the
Majorana mass term. $\mu$ is the Higgsino mass parameter from the Higgs
potential~\refeq{higgspot}. 

The physical (two-component) mass eigenstates are obtained via
unitary $(2 \times 2)$~matrices {\bf U} and {\bf V}:
\BE
\left. \begin{array}{c} 
       \chi_i^+ = V_{ij}\, \psi_j^+ \\[1.5ex] 
       \chi_i^- = U_{ij}\, \psi_j^- 
       \end{array} 
\right. \qquad i,j=1,2~.
\end{equation}
This results in a four-component Dirac spinor
\BE
\tilde{\chi}_i^+ = \VL \chi_i^+ \\[0.5ex] \bar{\chi}_i^- \VR 
                   \qquad i=1,2~,
\end{equation}
where {\bf U} and {\bf V} are given by
\BE
{\bf U} = {\bf O}_- \qquad;\qquad
{\bf V} = \left\{\begin{array}{cc} {\bf O}_+ & 
          \quad\det {\bf X}>0 \\[1ex] 
          \sigma_3\, {\bf O}_+ & \quad\det {\bf X}<0 \end{array} \right.
\end{equation}
with
\BE
{\bf O}_\pm = \ML \cos\phi_\pm & \sin\phi_\pm \\[0.5ex] 
              -\sin\phi_\pm & \cos\phi_\pm \MR~;
\end{equation}
$\cos\phi_\pm$ und $\sin\phi_\pm$ are given by
$(\epsilon = \mbox{sgn} [\det {\bf X}])$
\BEA
\tan\phi_+ = \frac{\sqrt{2}\, M_W(\sbe m_{\tilde{\chi}_1^+} 
             + \epsilon\, \Cb m_{\tilde{\chi}_2^+})}
             {(M_2\, m_{\tilde{\chi}_1^+} 
             + \epsilon\, \mu\, m_{\tilde{\chi}_2^+} )}  \non \\ 
\tan\phi_- = \frac{-\mu\, m_{\tilde{\chi}_1^+} 
             - \epsilon\, M_2\, m_{\tilde{\chi}_2^+}} 
             {\sqrt{2}\, M_W (\sbe m_{\tilde{\chi}_1^+} 
             + \epsilon\, \Cb m_{\tilde{\chi}_2^+})}~. 
\EEA
(If $\phi_+ < 0$ it has to be replaced by $\phi{_+} + \pi$.)
$m_{\tilde{\chi}_1^+}$ and $m_{\tilde{\chi}_2^+}$ are the eigenvalues
of the diagonalized matrix
\BEA
{\bf M}^2_{{\rm diag},\tilde{\chi}^+} &=& 
{\bf V\, X^{\dagger}\, X\, V}^{-1} \; = \; 
{\bf U^*\, X\, X^{\dagger}\, (U^*)}^{-1} \non \\
{\bf M}_{{\rm diag},\tilde{\chi}^+} &=& 
{\bf U^*\, X\, V}^{-1} \; =\; 
\ML m_{\tilde{\chi}_1^+} & 0 \\[0.5ex] 0 & m_{\tilde{\chi}_2^+} \MR .
\EEA
They are given by
\BEA
m^2_{\tilde{\chi}_{1,2}^+} &=& \frac{1}{2}\, \bigg\{ 
    M_2^2 + \mu^2 + 2M_W^2 \mp \Big[ (M_2^2-\mu^2)^2 \non \\ 
& & +\; 4M_W^4\CQZb + 4M_W^2(M_2^2+\mu^2+2\,\mu\, M_2\, \SZb) 
    \Big]^{\frac{1}{2}} \bigg\}~.
\label{Charmasse}
\EEA


\subsection{Neutralinos}
\label{subsec:neutralinos}

Neutralinos $\tilde{\chi}_i^0\; (i=1,2,3,4)$ are four-component
Majorana fermions. They are the mass eigenstates of the
photino,~$\tilde{\gamma}$, the zino,~$\tilde Z$, and the neutral higgsinos,
$\tilde{H}^0_1$ and $\tilde{H}^0_2$, with
\BE
\tilde{\gamma} = \VL -i \la_\gamma \\[0.5ex] 
                      i \bar{\la}_\gamma \VR  \quad;\quad
\tilde{Z} = \VL -i \la_Z \\[0.5ex] i \bar{\la}_Z \VR
            \quad;\quad
\tilde{H}^0_1 = \VL \psi^0_{H_1} \\[0.5ex] \bar{\psi}^0_{H_1} \VR
              \quad;\quad
\tilde{H}^0_2 = \VL \psi^0_{H_2} \\[0.5ex] \bar{\psi}^0_{H_2} \VR~.
\end{equation}
Analogously to the SM, the photino and zino are mixed states from the
bino,~$\tilde B$, and the wino,~$\tilde W$,
\BE
\tilde{B} = \VL -i \la^\prime \\[0.5ex] i \bar{\la}^\prime \VR 
\qquad;\qquad
\tilde{W}^3 = \VL -i \la^3 \\[0.5ex] i \bar{\la}^3 \VR~,
\end{equation}
with
\BEA
\tilde{\gamma} &=& \tilde{W}^3\, \sw + \tilde{B}\, \cw \non \\
\tilde{Z} &=& \tilde{W}^3\, \cw - \tilde{B}\, \sw~.
\EEA
The mass term in the Lagrange density is given by
\BE
{\cal L}_{\tilde{\chi}^0,{\rm mass}} = -\frac{1}{2}(\psi^0)^T\, {\bf Y}\, 
                                  \psi^0 + \hc~,
\end{equation}
with the two-component fermion fields
\BE
(\psi^0)^T = (-i\la^\prime , -i\la^3 , \psi_{H_1}^0 , 
              \psi_{H_2}^0)~.
\end{equation}
The mass matrix {\bf Y} is given by
\BE
\renewcommand{\arraystretch}{1.2}
{\bf Y} = \MLv M_1 & 0 & -M_Z\sw\Cb & M_Z\sw\sbe \\ 0 & 
          M_2 & M_Z\cw\Cb & -M_Z\cw\sbe \\
          -M_Z\sw\Cb & M_Z\cw\Cb & 0 & -\mu \\ M_Z\sw\sbe & 
          -M_Z\cw\sbe & -\mu & 0 \MR~.
\label{Y}
\end{equation}
The physical neutralino mass eigenstates are obtained with the
unitary transformation matrix~{\bf N}:
\BE
\chi_i^0 = N_{ij}\, \psi_j^0 \qquad i,j=1,\ldots,4,
\end{equation}
resulting in the four-component spinor (representing the mass
eigenstate) 
\BE
\tilde{\chi}_i^0 = \VL \chi_i^0 \\[0.5ex] \bar{\chi}_i^0 \VR 
\qquad i=1,\ldots,4~.
\end{equation} 
The diagonal mass matrix is then given by
\BE
{\bf M}_{{\rm diag},\tilde{\chi}^0} = {\bf N^*\, Y\, N}^{-1}~.
\end{equation}


\subsection{Gluinos}
\label{subsec:gluinos}

The gluino, $\gl$, is the spin~1/2 superpartner (Majorana fermion) of
the gluon. According to the 8 generators of $SU(3)_C$ (colour octet),
there are 8 gluinos, all having the same Majorana mass
\BE
\mgl = M_3~. 
\end{equation}
In SUSY GUTs $M_1$, $M_2$ and $M_3$ are not independent but connected
via 
\BE
\mgl = M_3 = \frac{g_3^2}{g_2^2}\, M_2 \; = \; 
      \frac{\al_s}{\al_{\rm em}}\, \sw^2\, M_2, \;\;
M_1 = \frac{5}{3} \frac{\sw^2}{\cw^2}\, M_2~.
\label{G-GUT}
\end{equation}


\subsection{Non-minimal flavour violation}
\label{subsec:nmfv}

The most general flavour structure of the soft SUSY-breaking
sector with flavour non-diagonal terms would
induce large flavour-changing neutral-currents, contradicting
the experimental results~\cite{pdg}. Attempts to avoid this kind of
problem include flavour-diagonal SUSY-breaking scenarios, like minimal
Supergravity (with universality assumptions) or gauge-mediated
SUSY-breaking, see the next subsection. In these 
scenarios, the sfermion-mass matrices are flavour diagonal in the same
basis as the quark matrices at the SUSY-breaking scale. However, a certain
amount of flavour mixing is generated due to the renormalization-group
evolution from the SUSY-breaking scale down to the electroweak
scale. Estimates of this radiatively induced off-diagonal squark-mass
terms indicate that the largest entries are those connected to the
SUSY partners of the left-handed quarks~\cite{NMFVestimate,savoy},
generically denoted as $\De_{LL}$. Those off-diagonal soft
SUSY-breaking terms scale with the square of diagonal soft
SUSY-breaking masses $\msusy$, whereas the
$\De_{LR}$ and $\De_{RL}$ terms scale linearly, and $\De_{RR}$ with
zero power of $\msusy$. Therefore, usually the hierarchy 
$\De_{LL} \gg \De_{LR,RL} \gg \De_{RR}$ is realized. It was also
shown in \citeres{NMFVestimate,savoy} that mixing between the third and
second generation squarks can be numerically significant due to the
involved third-generation Yukawa couplings.
On the other hand, there are strong experimental bounds on squark
mixing involving the first generation, coming from data on 
$K^0$--$\bar K^0$ and 
$D^0$--$\bar D^0$~mixing~\cite{FirstGenMix,FirstGenMix2}.

Considering the scalar quark sector with non-minimal flavour violation
(NMFV) for the second and third generation, 
the squark mass matrices in the basis of 
$(\SchaL, \StopL, \SchaR, \StopR)$ and 
$(\SstrL, \SbotL, \SstrR, \SbotR)$
are given by
\BEA
\label{eq:massup}
M_{\tiu}^2 &=& 
\MLv 
M_{\tilde L_c}^2 & \De_{LL}^t & \mc \Xc  & \De_{LR}^t \\[.3em]
\De_{LL}^t & M_{\tilde L_t}^2 & \De_{RL}^t &\mt \Xt \\[.5em]
\mc \Xc & \De_{RL}^t & M_{\tilde R_c}^2 & \De_{RR}^t \\[.3em]
\De_{LR}^t & \mt \Xt & \De_{RR}^t & M_{\tilde R_t}^2 
\MR \\[1em]
\label{eq:massdown}
M_{\tid}^2 &=& 
\MLv 
M_{\tilde L_s}^2 & \De_{LL}^b & \ms \Xs &  \De_{LR}^b \\[.3em]
\De_{LL}^b & M_{\tilde L_b}^2 & \De_{RL}^b & \mb \Xb  \\[.5em]
\ms \Xs & \De_{RL}^b & M_{\tilde R_s}^2 & \De_{RR}^b \\[.3em]
\De_{LR}^b & \mb \Xb & \De_{RR}^b & M_{\tilde R_b}^2 
\MR
\EEA
with
\BEA
\label{eq:defMXt}
M_{\tilde L_q}^2 &=& M_{\tilde Q_q}^2 + \mq^2 + 
                     \CZb \MZ^2 (T_3^q - Q_q \sw^2) \nn \\
M_{\tilde R_q}^2 &=& M_{\tilde U_q}^2 + \mq^2 + 
                     \CZb \MZ^2 Q_q \sw^2 ~(q = t,c) \nn \\
M_{\tilde R_q}^2 &=& M_{\tilde D_q}^2 + \mq^2 + 
                     \CZb \MZ^2 Q_q \sw^2 ~(q = b,s) \nn \\
X_q &=& \Aq - \mu (\tb)^{-2 T_3^q}
\EEA
where $m_q$, $Q_q$ and $T_3^q$ are the mass, electric charge and
weak isospin of the quark~$q$. 
$M_{\tilde Q_q}$, $M_{\tilde U_q}$, $M_{\tilde D_q}$  are the soft
SUSY-breaking parameters. The $SU(2)$~structure of the model requires 
$M_{\tilde Q_q}$ to be equal for $\Stop$ and $\Sbot$ as well
as for $\Scha$ and $\Sstr$. 

In order to diagonalize the two $4 \times 4$~squark mass matrices, two 
$4 \times 4$~rotation matrices, $R_{\tiu}$ and $R_{\tid}$, are needed,
\BE
\tiu_{\al} \; = \; R_{\tiu}^{\al,j} \VL \SchaL \\\StopL \\
                                      \SchaR \\ \StopR \VR_j ~,~~~~
\tid_{\al} \; = \; R_{\tid}^{\al,j} \VL \SstrL \\ \SbotL \\
                                      \SstrR \\ \SbotR \VR_j ~,
\label{newsquarks}
\end{equation}
yielding the diagonal mass-squared matrices as follows,
\BEA
{\rm diag}\{m_{\tiu_1}^2, m_{\tiu_2}^2, 
          m_{\tiu_3}^2, m_{\tiu_4}^2 \}^{\al,\be} & = &
R_{\tiu}^{\al,i} \; \KL M_{\tiu}^2 \KR_{i,j} \; 
( R_{\tiu}^{\be,j} )^\dagger ~,\\
{\rm diag}\{m_{\tid_1}^2, m_{\tid_2}^2, 
          m_{\tid_3}^2, m_{\tid_4}^2 \}^{\al,\be} & = &
R_{\tid}^{\al,i} \; \KL M_{\tid}^2 \KR_{i,j} \; 
( R_{\tid}^{\be,j} )^\dagger ~.
\EEA

For the numerical analysis we use
\begin{align}
\De_{LL}^t &= \la M_{\tilde L_t} M_{\tilde L_c} \,, &
\De_{LR}^t = \De_{RL}^t = \De_{RR}^t &= 0 \,, \nn \\
\De_{LL}^b &= \la M_{\tilde L_b} M_{\tilde L_s} \,, &
\De_{LR}^b = \De_{RL}^b = \De_{RR}^b &= 0 \, .
\label{def:lambda}
\end{align}

Feynman rules that involve two scalar quarks can be obtained from the
rules given in the $\sfl,\sfr$~basis by applying the corresponding
rotation matrix ($\tiq =\tiu,\tid$), 
\BE
V(X\tiq_{\al}\tiq_{\be}^\prime) \; = \; 
 R_{\tiq}^{\al,i} \;  R_{\tiq^\prime}^{\be,j} \; 
 V(X\tiq_i \tiq^\prime_j)~.
\end{equation}
Thereby $V(X\tiq_i \tiq^\prime_j)$ denotes a generic vertex in the 
$\sfl,\sfr$~basis, and $V(X\tiq _{\al}\tiq_{\be}^\prime)$ is the
vertex in the NMFV mass-eigenstate basis.
The Feynman rules for the vertices
needed  for our applications, i.e.\ the interaction of one
and two Higgs or gauge bosons with two squarks, can be found in 
\citere{nmfv}.


\subsection{Unconstrained MSSM versus specific models for soft SUSY
breaking}
\label{subsec:introsusybreak}

In the unconstrained MSSM no specific assumptions are made about the
underlying SUSY-breaking mechanism, and a parametrization of all
possible soft SUSY-breaking terms is used that do not alter the relation
between the dimensionless couplings (which ensures that the absence of
quadratic divergences is maintained). This parametrization has the 
advantage of being very general, but the disadvantage of introducing more 
than 100 new parameters in addition to the SM. While in principle these
parameters (masses, mixing angles, complex phases) could be chosen
independently of each other, experimental constraints from
flavour-changing neutral currents, electric dipole moments, etc.\ 
seem to favour a certain degree of universality among the soft
SUSY-breaking parameters. 

Within a specific SUSY-breaking scenario, the soft SUSY-breaking terms
can be predicted from a small set of input parameters. The most
prominent scenarios in the literature
are minimal Supergravity (mSUGRA)~\cite{Hall,mSUGRArev},
minimal Gauge Mediated SUSY Breaking (mGMSB)~\cite{GR-GMSB}
and minimal Anomaly Mediated SUSY Breaking
(mAMSB)~\cite{lr,giudice,wells}. The mSUGRA and mGMSB scenarios have
four parameters and a sign, while the mAMSB scenario can be specified in
terms of three parameters and a sign. 

Detailed experimental analyses within the multi-dimensional parameter
space of the unconstrained MSSM would clearly be very involved.
Therefore one often restricts to certain benchmark scenarios, see e.g.\
\citeres{benchmark,LHbenchmark,sps,BDEGOP}, or relies on underlying
assumptions of a specific SUSY-breaking scenario.

The EWPO can be analyzed within the unconstrained MSSM (or extensions of
it), which allows to set constraints on the SUSY parameter space in 
a rather general way. In our numerical anaysis in chapter~\ref{chapter3}
we discuss the impact of EWPO in the context of the unconstrained MSSM,
while in chapter~\ref{chapter4} we focus on the mSUGRA, mGMSB and mAMSB
scenarios as special cases.


\subsection{Experimental bounds on SUSY particles}
\label{subsec:expbounds}

The non-observation of SUSY particles at the collider experiments
carried out so far place lower bounds on the masses of SUSY particles
which are typically of \order{100 \gev}~\cite{pdg}. These bounds,
however, depend on certain assumptions on the SUSY parameter space, for
instance on the couplings and decay characteristics of the particles or 
the validity of a certain SUSY-breaking scenario. 

Relaxing some of these assumptions can result in bounds that are much 
weaker than the ones that are usually quoted. 
As an example, collider experiments do not provide any lower bound on
the mass of the lightest neutralino if the GUT
relation connecting $M_1$ and $M_2$, see \refeq{G-GUT}, is
lifted~\cite{lightneutralino}. It is interesting to investigate in how
far the results for EWPO can narrow down the parameter space where the
bounds from direct searches are very weak. Such an analysis has been
carried out, for instance, for a scenario with a light scalar bottom
quark of \order{5 \gev}. In \citere{lightsbot} it has been shown that a
light scalar bottom quark is consistent with the constraints from the
EWPO and the LEP Higgs search.


\section{Electroweak precision observables}

In general there are two possibilities for virtual effects of SUSY
particles to be large enough to be detected at present and (near future) 
experiments. On the one hand, these are rare processes, where SUSY loop 
contributions do not compete with a large SM tree-level contribution. 
Examples are rare $b$ decays like $b \ra s \gamma$, 
$B_s \to \mu^+\mu^-$, and electric dipole moments (EDMs).
For processes of this kind the SUSY prediction for the rates can be much 
larger (sometimes by orders of magnitude) than the SM one. 

On the other hand, EWPO which are known with an accuracy at the per~cent
level or better have the potential to allow a discrimination between
quantum effects of the SM and SUSY models. Examples are the $W$~boson
mass, $\MW$, and the $Z$-boson observables, like the effective leptonic weak
mixing angle, $\sweff$. 

This distinction between rare processes and EWPO is of course not a 
completely rigid one. The anomalous magnetic moment of the muon, for
instance, corresponds both to a rare process according to the above
definition and to an EWPO which has been measured with high accuracy.
In view of the prospects for precision measurements of the mass of the
lightest $\cp$-even Higgs boson, $\mh$, at the next generation of
colliders, we also treat $\mh$ as an EWPO.

In the present report we concentrate our discussion on EWPO, in
particular the observables in the $W$- and $Z$-boson sector, the
anomalous magnetic moment of the muon, and the mass of the lightest
$\cp$-even Higgs boson. We just briefly comment on
rare processes in the following section and occasionally in our
numerical discussion. For a more thorough investigation of the
constraints on the SUSY parameter space we refer to the literature.
For reviews of rare decays see \citere{Breview}, results for
EDMs in the MSSM can be found in
\citeres{EDMrev1,EDMrev2} and in references therein.


\subsection{Constraints on the SUSY parameter space from rare processes}

The branching ratio $\br(b \to s \ga)$ receives, besides the SM 
loop contribution involving the $W$~boson and the top quark, additional
contributions from chargino/stop and charged Higgs/stop
loops~\cite{bsgtheo}. The SUSY contributions are particularly large 
for light charged Higgs bosons and large $\mu$ or $\tb$. The currently
available SUSY contributions to
$\br(b \to s \ga)$ include the one-loop result and leading higher-order
corrections. The comparison
of the theory prediction with the data imposes important
constraints on the parameter space both of general two-Higgs-doublet
models and of the MSSM. In the latter case it is possible that the 
two kinds of additional contributions are individually large but
interfere destructively with each other, resulting in only a small
deviation of the decay rate from the SM prediction.

\smallskip
Another interesting channel is the decay $B_s \to \mu^+\mu^-$.
The SM contribution to this decay is tiny, resulting in a BR of about
$10^{-9}$~\cite{bsmmtheosm}. Within SUSY, however, diagrams enhanced
by $\tb^3$ can 
contribute. Thus the decay width can grow with $\tan^6\be$ and the BR
can be much larger than in the SM~\cite{bsmumu}, see
\citere{bsmumuRev} for a recent review. The
available corrections in the MSSM consist of the full one-loop
evalution and the leading two-loop QCD corrections. The current bound
from the Tevatron is 
$\br(B_s \to \mu^+\mu^-) < 2.7 \times 10^{-7}$ at the 
90\%~C.L.~\cite{bsmumuRunII}.
A substantial improvement in this bound can be expected in the
forthcoming years.

\smallskip
A different way for probing SUSY is via its contribution to EDMs of
heavy quarks~\cite{EDMDoink}, of the electron and 
the neutron (see \citeres{EDMrev2,EDMPilaftsis} and references therein), 
or of deuterium~\cite{EDMRitz}. 
While SM contributions start only at the three-loop level, due to its
complex phases the MSSM can contribute already at one-loop order. Also the
leading two-loop corrections for the electron and neutron EDMs are
available. Large phases in the first two generations of (s)fermions
can only be accomodated if these generations are assumed to be very
heavy~\cite{EDMheavy} or large cancellations occur~\cite{EDMmiracle},
see however the discussion in \citere{EDMrev1}.


\subsection{Pseudo-observables versus realistic observables}

The quantities that can be directly measured in experiments are cross
sections, line shape observables, forward--backward asymmetries etc., 
deemed ``realistic observables'' in the language of \citere{ZobsSM2lA}.
The obtained results depend on the specific set of experimental cuts that 
have been applied and are influenced by detector effects and other
details of the experimental setup. In order to determine quantities like 
masses, partial widths or couplings from the primarily measured
observables, a deconvolutiuon (unfolding) procedure is applied. 
This procedure involves manipulations like unfolding the QED
corrections, subtracting photon-exchange and interference terms,
subtracting box-diagram contributions, unfolding higher-order QCD
corrections, etc.
These secondary quantities are therefore called ``pseudo-observables''
in \citere{ZobsSM2lA}. 

The procedure of going from realistic observables to pseudo-observables
results in a slight model dependence of the pseudo-observables. 
As an example, the experimental value of the $Z$-boson mass has a slight
dependence on the value of the Higgs-boson mass in the SM, see
\citeres{LEPEWWG,Passarino:2003bs}.
The EWPO on which we focus in this report are pseudo-observables in the
sense outlined above. At the level of electroweak precision physics, it
is important to keep in mind that in order to obtain the numerical
values of the EWPO given in the literature the Standard Model has been
used in several steps for calculating the subtraction terms. 
An obvious model dependence also occurs if,
instead of performing an explicit subtraction of SM terms, parameters
like $\MZ$, $\al_s(\MZ)$, etc.\ are determined directly from a 
SM fit, containing the full set of SM corrections, to the realistic 
observables. 

Using the
same numerical values of the EWPO as input for analyses within the MSSM
(or other extensions of the SM) is obviously only justified if new
physics contributions to the subtraction terms and the implemented
higher-order corrections are negligible. As an example, the experimental
value extracted for $\al_s(\MZ)$ in the MSSM (for a given SUSY mass
spectrum) would somewhat differ from the SM value of $\al_s(\MZ)$.

A consistent treatment of the model dependence of the EWPO is necessary
in a precision analysis of the MSSM. At the current level of
experimental precision the shift induced in the EWPO from taking into
account the full MSSM particle content instead of the SM will normally
be of minor importance. In some regions of the parameter space, in
particular where some of the SUSY particles are very light, an explicit
verification of the above assumption would however be desirable.

Concerning the determination of the MSSM parameters, additional
complications arise compared to the SM case. 
In general the model dependence is relatively small for masses, since
the mass of a particle can closely be related to one particular
realistic observable. For couplings (with the exception of the
electromagnetic coupling in the Thomson limit), mixing angles, etc., on 
the other hand, the model dependence is relatively large.
In contrast to the SM, many of the MSSM parameters are not closely
related to one particular observable, e.g.\ $\tan\be$, $\mu$, the stop
and sbottom mixing angles, complex phases, etc., resulting in a
relatively large model dependence. Therefore, the approach
of extracting pseudo-observables with only a fairly small model
dependence seems not to be transferable to the case of the MSSM. 
It seems that eventually the MSSM parameters will have to be determined
in a global fit of the MSSM to a large set of observables, taking into
account higher-order corrections within the MSSM, see
\citeres{fittino,sfitter} or \citere{spa} for an attempt of a
coordinated effort.


\subsection{EWPO versus effective parameters}

In this report we focus our discussion on the EWPO, i.e.\ (pseudo-)observables 
like the $W$-boson mass, $\MW$, the effective leptonic weak mixing
angle, $\sweff$, the the leptonic width of the $Z$~boson, $\Ga_l$, the
anomalous magnetic moment of the muon, $\amu \equiv (g-2)_\mu/2$, 
the mass of the lightest $\cp$-even MSSM Higgs boson, $\mh$, etc.
In the literature virtual effects of SUSY particles are often discussed
in terms of effective parameters instead of the EWPO (see e.g.\
\citere{hagiwararev} and references therein). We do not follow this
approach, and just briefly comment about it in the following.

Since for the accuracies anticipated at future colliders, see
\refta{tab:POfuture} below, it is particularly important to have a precise
understanding of how effects of new physics can be probed in a sensible
way, the virtues and range of applicability of effective parameters 
need to be assessed.

A widely uses set of parameters are the $S$, $T$,
$U$~parameters~\cite{stu}. They are defined such that they describe the 
effects of new physics contributions that enter only via
vacuum-polarization effects (i.e.\ self-energy corrections) to the
vector boson propagators of the SM (i.e.\ the new physics contributions
are assumed to have negligible couplings to SM fermions). 
The $S$, $T$, $U$ parameters can be computed in
different models of new physics as certain combinations of one-loop
self-energies. Experimentally, their values are determined by
comparing the measured values ${\cal A}_i^{\mathrm{exp}}$ of a number of 
observables with their values predicted by the SM, ${\cal A}_i^{\SM}$, 
i.e.\
${\cal A}_i^{\mathrm{exp}} = {\cal A}_i^{\SM} +
f^{\mathrm{NP}}_i(S, T, U)$.
Here ${\cal A}_i^{\SM}$ contains all known
radiative corrections in the SM, while
$f^{\mathrm{NP}}_i(S, T, U)$ is a (linear) function of the parameters
$S$, $T$, $U$ and describes the contributions of new
physics. The SM prediction ${\cal A}_i^{\SM}$ is
evaluated for a reference value of $\mt$ and $\MH$. For most
precision observables the corrections caused by a variation of $\mt$ and
$\MH$ at one-loop order can also be absorbed into the parameters $S$, $T$,
and $U$. A non-zero result for $S$, $T$, $U$ determined in this way
indicates non-vanishing contributions of new physics (with respect to
the SM reference value).

{}From their definition, it is obvious that the $S$, $T$, $U$
parameters can only be applied for parameterizing effects of physics
{\em beyond} the SM. Taking into account the full contributions within
the SM cannot be avoided, as these contributions (containing also vertex
and box corrections) cannot consistently be absorbed into the $S$, $T$, 
$U$ parameters (for a more detailed discussion of this point, see 
\citere{ringberg}).

Examples of new physics contributions that can be described in the
framework of the $S$, $T$, $U$ parameters are contributions from a
fourth generation of heavy fermions or effects from scalar quark loops
to the $W$- and $Z$-boson observables. 
A counter example going beyond the
$S$, $T$, $U$ framework are SUSY corrections to the anomalous magnetic
moment of the muon.
According to their definition, the $S$, $T$, $U$ parameters are
restricted to leading order contributions of new physics. They should 
therefore be applied only for the description of {\em small} deviations 
from the SM pre\-dic\-tions, for which a
restriction to the leading order is permissible. It appears to be
questionable, on the other hand, to apply them to cases of very large
deviations from the SM, like extensions of the SM with a very heavy Higgs 
boson in the range of several TeV.

Other parameterizations have been suggested (see e.g.\
\citeres{eps,schild}) with no reference to the SM contribution and
which are not restricted in the possible kinds of new physics. These
parameterizations are defined as certain linear combinations of
different observables. It is however not in all cases obvious that
studying the experimental values and the theory predictions for these
parameters is of advantage compared to studying the EWPO
themselves. For a recent discussion of effective parameters, see also 
\citere{Barbieri:2004qk}.


\subsection{Current experimental status of EWPO}
\label{subsec:ewpostatus}

LEP, SLC, the Tevatron, and low-energy experiments have collected an
enormous amount of data on EWPO. 
Examples for the current experimental
status of EWPO are given in \refta{tab:POstatus},
including their relative experimental precision. The quantities in the
first three lines, $\MZ$, $\gf$, and $\mt$, are usually employed as
input parameters for the theoretical predictions. The observables $\MW$,
$\sweff$, $\Ga_Z$, on the other hand, are used
for testing the electroweak theory by comparing the
experimental results with the theory predictions. Comparing the typical
size of electroweak quantum effects, which is at the per cent level,
with the relative accuracies in \refta{tab:POstatus}, which are at the
per mille level, clearly shows the sensitivity of the electroweak
precision data to loop effects.

\begin{table}[htb]
\renewcommand{\arraystretch}{1.5}
\BC
\begin{tabular}{|c||c|c|c|}
\cline{2-4} \multicolumn{1}{c||}{}
& central value & absolute error & relative error \\
\hline\hline
$\MZ$ [GeV] & 91.1875 & $\pm 0.0021$ & $\pm 0.002\%$ \\ \hline
$\Gmu$ [GeV$^{-2}$] & $1.16637 \times 10^{-5}$ & $\pm 0.00001 \times
10^{-5}$
                   & $\pm 0.0009\%$ \\ \hline
$\mt$ [GeV] & 178.0 & $\pm 4.3$ & $\pm 2.4 \%$ \\ \hline\hline
$\MW$ [GeV] & 80.425 & $\pm 0.034$ & $\pm 0.04\%$ \\ \hline
$\sweff$ & 0.23150 & $\pm 0.00016$ & $\pm 0.07\%$ \\ \hline
$\Ga_Z$ [GeV] & 2.4952 & $\pm 0.0023$ & $\pm 0.09\%$ \\
\hline
\end{tabular}
\EC
\renewcommand{\arraystretch}{1}
\caption{Examples of EWPO with their current absolute and relative
experimental errors~\cite{pdg,LEPEWWG}.}
\label{tab:POstatus}
\end{table}

The experimental accuracy of the precision observables will further be
improved at the currently ongoing Run~II of the Tevatron, the LHC and a 
future ILC, with the
possible option of a high luminosity low-energy run,
GigaZ~\cite{teslatdr,orangebook,acfarep,gigaz}.
The most significant improvements among the EWPO can be expected for 
$\MW$ and $\sweff$. If the Higgs boson will be detected, a precise
measurement of its mass will be important for testing the electroweak
theory. Concerning the input parameters, the experimental error of the
top-quark mass is the dominant source of theoretical uncertainty in
electroweak precision tests. This will remain to be the case even with
the accuracy on $\mt$ reachable at the LHC~\cite{deltamt}. Thus, the
high-precision measurement of $\mt$ at the ILC will be crucial for
an increased sensitivity to virtual effects of new 
physics~\cite{deltamt,naturetop}.

The prospective accuracy for $\MW$, $\sweff$, $\mt$ and $\mh$ (for a
value of $\mh \approx 120 \gev$) at the Tevatron, at the LHC (combined
with the data collected at the Tevatron)
and the ILC (with and witout GigaZ option) are summarized in
\refta{tab:POfuture} (see \citere{blueband} and references therein).

\begin{table}[bht]
\renewcommand{\arraystretch}{1.5}
\begin{center}
\begin{tabular}{|c||c||c|c|c|c|}
\cline{2-6} \multicolumn{1}{c||}{}
& now & Tevatron & LHC & ~ILC~  & ILC with GigaZ \\
\hline\hline
$\de\sweff(\times 10^5)$ & 16   & --- & 14--20 & ---  & 1.3  \\
\hline
$\de\MW$ [MeV]           & 34   &  20 & 15   & 10   & 7      \\
\hline
$\de\mt$ [GeV]           &  4.3 &  2.5  &  1.5 &  0.2 & 0.1   \\
\hline
$\de\mh$ [MeV]            &  --- & --- &  200 & 50 & 50 \\
\hline
\end{tabular}
\end{center}
\renewcommand{\arraystretch}{1}
\caption{
Current and anticipated future experimental uncertainties for
$\sweff$, $\MW$, $\mt$, and $\mh$ (the latter assuming $\mh \approx 115$~GeV).
Each column represents the combined results of all detectors and
channels at a given collider, taking into account correlated
systematic uncertainties, see \citere{blueband} for details.
Updated Tevatron numbers can be found in \citere{tev4lhc}.
}
\label{tab:POfuture}
\end{table}

\bigskip
Another EWPO with a high sensitivity to virtual effects of SUSY
particles is the anomalous magnetic moment of the muon, $\amu \equiv
(g-2)_\mu/2$. The final result of the Brookhaven ``Muon $g-2$
Experiment'' (E821) for $\amu$ reads~\cite{g-2exp}
\BE
\amuexp = (11\, 659\, 208 \pm 5.8) \times 10^{-10}~.
\label{eq:amuexp}
\end{equation}
The interpretation of this measurement within SUSY strongly depends on
the corresponding SM evaluation. The SM prediction depends on the
evaluation of the 
hadronic vacuum polarization and light-by-light contributions. The
former have been evaluated by \citeres{DEHZ,g-2HMNT,Jegerlehner,Yndurain}, 
the latter by \citere{LBL}, but there is a recent
reevaluation~\cite{LBLnew}, describing a possible shift of the central
value by $5.6 \times 10^{-10}$.
Depending on which hadronic evaluation is chosen, the
difference between experiment and the SM prediction lies between the two
values (including the updated QED result from \citere{Kinoshita})
\BEA
\label{deviation1}
\amuexp-\amutheo ~(\mbox{\cite{g-2HMNT}+\cite{LBL}}) & = &
(31.7\pm9.5)\times10^{-10} ~:~3.3\,\si~,\\
\amuexp-\amutheo ~(\mbox{\cite{DEHZ}+\cite{LBLnew}}) & = &
(20.2\pm9.0)\times10^{-10} ~:~2.1\,\si~.
\label{deviation2}
\EEA
These evaluations are all obtained with a $\De\al_{\rm had}$
determination from $e^+e^-$ data. 
Recent analyses concerning $\tau$ data indicate that uncertainties due to
isospin breaking effects may have been underestimated
earlier~\cite{Jegerlehner}. Furthermore new data obtained at KLOE~\cite{kloe},
where the radiative return is used to obtain data on $\De\al_{\rm had}$, 
agrees with the older $e^+e^-$ data. 
This, together with a continuing discussion about the uncertainties
inherent in the isospin transformation from $\tau$ decay, has led to the
proposal to leave out the 
$\tau$ data in the $\De\al_{\rm had}$ determination, resulting in the
estimate~\cite{g-2ICHEP04}
\BE
\label{deviationfinalorg}
\amuexp - \amutheo = (25.2 \pm 9.2)\times10^{-10} ~:~2.7\,\si~.
\end{equation}
%


\newpage

\chapter{Theoretical evaluation of precision observables}

\section{Regularisation and renormalization of supersymmetric theories}

\subsection{Basic strategy}

  In higher-order perturbation theory the
 relations between the formal parameters and measurable quantities
 are different from the tree-level relations in general.
 Moreover, the procedure is obscured by the appearance of divergences
 in the loop integrations.
 For a mathematically
  consistent treatment one has to regularize the theory, e.g.\
 by dimensional regularization (DREG), where the regularization is
 performed by analytically continuing the space-time dimension from 4 to
$D$~\cite{dreg,ga5HV}.
But then the relations between the physical quantities
 and the parameters become cut-off-dependent. Hence, the parameters
 of the basic Lagrangian, the ``bare'' parameters, have no
 physical meaning. On the other hand, the relations between
 measurable physical quantities, where the parameters
 drop out, are finite and independent of the cut-off. It is therefore
 in principle possible to perform tests of the theory in terms of
 such relations by eliminating the bare parameters.

 Alternatively, one may replace the bare parameters by renormalized ones
 by multiplicative renormalization for each bare parameter  $a_0$,
  \beq
 \label{multiplicative}
    a_0 = Z_a\, a = a +\delta a
 \end{equation}
 with renormalization constants $Z_a$ different from 1 by a
 higher-order term.
 The renormalized parameters $a$  are finite and fixed by a set of
 renormalization conditions. The decomposition (\ref{multiplicative}) 
 is to a large
 extent arbitrary. Only the divergent parts are determined directly
 by the structure of the divergences of the loop amplitudes.
 The finite parts depend on the choice of the explicit
 renormalization conditions. These conditions determine the physical
meaning of the renormalized parameters.

 Before predictions can be made from the theory,
 a set of independent parameters has to be taken from experiment.
 In practical calculations the free SM parameters are usually fixed by 
 using $\al$, $\Gmu$, $\MZ$, $m_f$, $\als$ (and possibly entries of the 
 quark and lepton mass matrices, if the off-diagonal entries are not
 neglected)
 as physical input quantities. They have to be supplemented by the 
 empirically
 unknown input parameters for the Higgs sector and the SUSY breaking sector.
 Differences between various schemes are formally
 of higher order than the one under consideration.
 The study of the
 scheme dependence of the perturbative results, possibly after improvement by
 resummation of the leading terms, gives an indication of the possible
 size of missing
 higher-order contributions.

On the theoretical side, a thorough control of the quantization and the
renormalization of the MSSM as a supersymmetric gauge theory, with
spontaneously broken gauge symmetry and softly broken supersymmetry,
is required. This is not only a theoretical question for 
establishing a solid and consistent theoretical framework 
but also a matter of practical importance for
concrete higher-order calculations, where
the quantum contributions to the Green functions have to fulfil
the symmetry properties of the underlying theory.
An increasing number of
phenomenological applications has been carried out in the 
Wess-Zumino gauge where the  number of unphysical degrees 
of freedom is minimal, but 
where supersymmetry is no longer manifest.

Moreover, a manifestly supersymmetric and gauge-invariant 
regularization for divergent loop integrals is missing.
The prescription of DREG
preserves the Lorentz and the gauge
invariance of the theory, apart from problems related to the treatment
of $\ga_5$ in dimensions other than~4. In supersymmetric theories,
however, a $D$-dimensional treatment of vector fields leads to a
mismatch between the fermionic and bosonic degrees of freedom, which
gives rise to a breaking of the supersymmetric relations. This led to
the development of dimensional reduction (DRED)~\cite{dred}. In this scheme 
only the momenta are
treated as $D$-dimensional, while the fields and the Dirac algebra are
kept 4-dimensional. It leads to ambiguities related to the treatment of
$\ga_5$~\cite{dred2}, and therefore cannot be consistently applied at
all orders (for a review, see~\citere{jackjones}).
Hence, renormalization and the structure of counterterms
have to be adapted by exploiting the basic symmetries expressed
in terms of the supersymmetric BRS transformations~\cite{White:ai}.
An additional complication in the conventional approach
assuming an invariant regularization scheme, 
however, arises from the 
modification of the symmetry transformations themselves by
higher-order terms.    

The method of algebraic renormalization, applied 
in~\citere{Kraus:1997bi} to the electroweak SM and 
in~\citere{Hollik:2002mv} for the MSSM,
avoids the difficulties of
the conventional approach. The theory is defined 
at the classical as well as the quantum level  
by the particle content and by the basic symmetries.
The essential feature of the algebraic method is the 
combination of all symmetries into the BRS transformations 
leading to the Slavnov-Taylor (ST) identity. 
In this way, the theory is defined 
by symmetry requirements that have to be satisfied after renormalization
in all orders of perturbation theory. In the case of symmetry violation
in the course of explicitly calculating vertex functions in a given order,
additional non-invariant counterterms are uniquely determined
to restore the symmetry, besides the invariant counterterms
needed for absorbing the divergences and for the normalization of fields and
parameters. 
Examples are given in~\citere{Hollik:1999xh,Hollik:2001cz} for supersymmetric 
QED and QCD and in~\citere{Grassi:algmeth} for the SM case.
Explicit evaluations at the one-loop level in supersymmetric 
models~\cite{Hollik:1999xh,Hollik:2001cz,fischer}
have shown that DRED
yields the correct counter terms.

In the following we discuss the renormalization of several sectors of
the MSSM. We focus on the sectors that are needed for the one- and
two-loop calculations reviewed below and restrict ourselves to the order 
in perturbation theory required there. These sectors are the SM gauge
bosons, the electric charge, the quark and scalar quark sector as well
as the MSSM Higgs boson sector. 

As mentioned above, many MSSM parameters
are not closely related to one particular physical observable, so that
no obvious `best choice' exists for their renormalization. Examples
treated below are $\tb$ and the mixing angles in the scalar quark
sector. Various definitions for these parameters already exist in the
literature (the situation is similar to the case of the weak mixing
angle of the electroweak theory, where the use of several different
definitions in the literature caused some confusion in the early days of
electroweak higher-order corrections). We will briefly comment on some
of them below. In view of the large number of MSSM parameters there is
clearly a need to establish some common standards in the literature in 
order to allow for a transparent comparison of different results. 
Requirements that a renormalization scheme for the whole MSSM should
fulfil  are in particular a coherent treatment of all sectors,
applicability for both QCD and electroweak corrections, and numerical
stability. Furthermore aspects of gauge (in-)dependence need to be
addressed. When formulating renormalization prescriptions for the MSSM 
particular care has to be taken in order to respect the underlying
symmetry relations of the theory. While in the SM all masses of the
particles can be fixed by independent renormalization conditions, in a
supersymmetric theory various relations exist between different masses.
Therefore only a subset of the mass parameters of the theory can be
renormalized independently. The counterterms for the other masses are
then determined in terms of the independent counterterms.
For a discussion of these issues, see e.g.\ \citere{doink}.


\subsection{Gauge boson mass renormalization}
\label{subsec:gaugebosonren}

We discuss here the renormalization of the gauge-boson masses in the
on-shell scheme~\cite{onshell} at the one-loop level.
 Writing the $W$ and $Z$ self-energies as
 \beq
\label{gaugebosonSE}
 \Sigma^{W,Z}_{\m\nu}(q) = \left(- g_{\m\nu} +
\frac{q_{\mu}q_{\nu}}{q^2} \right) \Sigma^{W,Z}(q^2) + \cdots ,
 \end{equation}
where the scalar functions $\Sigma^{W,Z}(q^2)$ are the transverse parts
of the self-energies, and defining $\Sigma^{W,Z}_{\m\nu}$ to correspond
to $(- i)$ times the loop diagrams by convention,
 we have for the one-loop propagators ($V=W,Z$)
 \beq
  \frac{- i g^{\m\sigma}}{q^2-M_V^2} 
  \left(i \,\Sigma^V_{\rho\sigma}  \right) 
  \frac{- i g^{\rho\nu}}{q^2-M_V^2} =
  \frac{- i g^{\m\nu}}{q^2-M_V^2} 
  \left( \frac{-\Sigma^V(q^2)}{q^2-M_V^2} \right) ,
 \end{equation}
where terms proportional to $q_\mu q_\nu$ (see \refeq{gaugebosonSE})
have been omitted (they are 
suppressed if the propagator is attached to a light external fermion).

 Resumming  all self-energy terms  yields a
 geometric progression for the dressed propagators:
 \bea
 \label{Wprop}   
     &   &
 \frac{- i g_{\m\nu}}{q^2-M_V^2} \left[1 + \left(
 \frac{-\Sigma^V}{q^2-M_V^2} \right) + \left(
 \frac{-\Sigma^V}{q^2-M_V^2} \right)^2 + \cdots \right] \nn \\
  &  & = \, \frac{-i g_{\m\nu}}{q^2-M_V^2+\Sigma^V(q^2) } \, .
 \eea
 The locations of the poles in
 the propagators are shifted by the self-energies.
 Consequently, the masses in the Lagrangian can no longer be interpreted
as the physical masses of the $W$ and $Z$ bosons once loop corrections
are taken into account. The mass renormalization relates these ``bare
masses'' to the physical masses $\MW$, $\MZ$ by
 \bea
 \label{baremass}
  \mwb & = & M_W^2 + \dmmw \, , \nn \\
  \mzb & = & W_Z^2 + \dmmz  \, ,
 \eea
 with counterterms of one-loop order. The propagators
 corresponding to this prescription are given by
 \beq
 \frac{-i g_{\m\nu}}{q^2-M_V^{0\,2} + \Sigma^V(q^2) } \, =\,
 \frac{-i  g_{\m\nu}}{q^2-M_V^2-\delta M_V^2 + \Sigma^V(q^2) } 
 \end{equation}              
 instead of (\ref{Wprop}).
 The renormalization conditions which ensure that $M_{W,Z}$ are the
 physical masses fix the mass counterterms to be
  \bea
 \label{massren}
 \dmmw & = & \real\, \Sigma^W(M_W^2) \, ,  \nn \\
 \dmmz & = & \real \, \Sigma^Z(M_Z^2) \, .
 \eea
These are the on-shell renormalization conditions. In an \msbar\ (or
\drbar) renormalization, on the other hand, the counterterms $\dmmw$,
$\dmmz$ are defined such that they essentially only contain the
divergent (in the limit $D \to 4$) contribution. The renormalized mass
parameters in this case do not directly correspond to the physical
masses. They explicitly depend on the renormalization scale.

While the $Z$-boson mass is commonly used as an input parameter, $\MW$
is normally traded as an input parameter for the Fermi constant $\Gmu$,
which is precisely measured in muon decay. The prediction for $\MW$ in
terms of $\Gmu$, $\MZ$, $\al$ and the parameters of the theory that
enter via loop corrections can therefore be compared to the experimental
value of $\MW$, constituting a sensitive test of the theory (see below).

Extending the above on-shell definition to higher orders requires to
take into account that the pole of the propagator of an unstable
particle is located in the complex plane rather than on the real axis
(which is the case for stable particles). A gauge-invariant mass
parameter is obtained if the mass is defined according to the real
part of the complex pole. The expansion around the complex pole leads to
a Breit-Wigner shape with a fixed width. 
The experimental determination of the gauge-boson masses, on the other
hand, uses a Breit-Wigner parametrization with running width for
historical reasons. This needs to be corrected for by a finite shift in
$\MW$ and $\MZ$. (For a more detailed discussion, see
\citere{Freitas:2002ja} and references therein.)


\subsection{Charge renormalization}
\label{subsec:ch-renorm} 

  The electroweak charge renormalization  is very similar to that
 in pure QED. In the on-shell scheme,
 the definition of $e$ as the classical
 charge in the Thomson cross-section
 \beq
 \sigma_{\rm Th} = \frac{e^4}{6\pi\,m_e^2} 
 \end{equation}
is maintained.
  Accordingly, the Lagrangian carries the bare charge
  $ e_0 =  e+\delta e$ with the charge counterterm
 $\delta e$ of one-loop order.
 The  charge counterterm $\delta e$ has to absorb the
 electroweak loop contributions to the $ee\g$ vertex in the Thomson
 limit.
  This charge renormalization condition is simplified by the validity
 of a generalization of the QED Ward identity 
  which implies that those
 corrections related to the external particles cancel
 each other. Hence, for $\delta e$  only two universal
 contributions are left,
 \beq
 \label{chargeren}
  \frac{\delta e}{e} \,=\, \frac{1}{2}\, \Pi^{\g}(0) \, -\,
 \frac{s_W}{c_W}\,\frac{\Sigma^{\g Z}(0)}{M_Z^2}  \, ,
 \quad \Pi^{\g}(0) \equiv \frac{\partial}{\partial q^2}
\Sigma^{\g}(q^2)\Bigr|_{q^2 = 0} \, .
 \end{equation}
 The first contribution
 is given by the photon vacuum polarization, $\Pi^{\g}$, 
 for real photons, $q^2=0$.
 Besides the charged-fermion loops, it
 contains also bosonic loop diagrams from $W^+W^-$ virtual states and the
 corresponding ghosts, as well as from extra charged particles in extensions
 of the SM.
 The second term contains the mixing between photon and $Z$ boson, in general
 described as a mixing propagator, $\De^{\g Z}$, with $\sgz$
 normalized according to
  \beq
\label{gammaZmixing}
  \Delta^{\g Z} =
   \frac{-i  g_{\m\nu}}{q^2} \left(
 \frac{-\Sigma^{\g Z}(q^2)}{q^2-M_Z^2} \right) \, . 
 \end{equation}
 All loop contributions to $\Sigma^{\g Z}$ vanish at
 $q^2=0$, except 
 the non-Abelian bosonic loops yield $\Sigma^{\g Z}(0)\neq 0$.
 They are the same in the standard model and in supersymmetric
 extensions. 
 $\Sigma^{\g Z}(0)$ completely vanishes in the background-field
quantization of the electroweak theory~\cite{bfm}.

\bigskip
The fermion-loop contributions to the 
photon vacuum polarization in (\ref{chargeren}) are analogous to the
electron loop in standard QED and do not depend on the details of the
electroweak theory. They give rise to a logarithmic dependence on the
fermion masses. While for the leptonic contributions the known lepton
masses can be inserted, perturbative QCD is not applicable in this
regime, and quark masses are no reasonable input parameters. 

In order to evaluate the contribution of light fermions, i.e.\ the
leptons and the quark flavours except the top quark, it is convenient to
add and subtract the photon vacuum polarization at $p^2 = \MZ^2$ and to
consider the finite quantity (for the top quark and other heavy fermions 
$\Pi^{\g}(0)$ can be evaluated directly)
\beq
 \real \, \Pigr(\mz)  =  \real\, \Pi^{\g}(M_Z^2) - \Pi^{\g}(0)~.
\end{equation}
Splitting it into the contribution of the leptons and the five light
quarks yields the quantity
\beq
 \dal  =  \dal_{\rm lept} \, + \,\dal_{\rm had} 
       =   -\,  \real\,\Pigr_{\rm lept}(\mz)
           - \, \real\,\Pigr_{\rm had}(\mz) \, ,
\end{equation}
which represents a QED-induced shift
in the electromagnetic fine structure constant
\beq
\label{delalpha}
     \al \, \ra \, \al(1+\dal) \, . 
\end{equation}
The evaluation of the leptonic content of $\dal$ 
in terms of the known lepton masses yields
at three-loop order~\cite{steinhauser} 
\beq
  \dal_{\rm lept} \; = \; 314.97687\cdot 10^{-4} \, .
\end{equation}
The 5-flavour contribution of the light quarks to the shift in the 
fine structure constant
can be derived from experimental data with the help
of a dispersion relation
\beq
\label{dispersion}
 \dal_{\rm had} = \, -\, \frac{\al}{3\pi} \, M_Z^2 \, \real \,
 \int^{\infty}_{4m_{\pi}^2}  {\rm d}s' \, \frac{R^{\g}(s')}
 {s'(s'-M_Z^2-i\veps)}
\end{equation}
where
\[
 R^{\g}(s) = \frac{\sigma(\epm \ra\g^* \ra {\rm hadrons})}
 {\sigma(\epm\ra\g^*\ra \m^+\m^-)}
\]
is an experimental input quantity for the low energy range.
Recent compilations yield the values
$\dal=0.02761\pm0.00036$~\cite{Burkhardt:2001xp},
$\dal=0.02769\pm0.00035$~\cite{jeger}, 
$\dal=0.02755\pm0.00023$~\cite{g-2HMNT}.


\subsection{Renormalization of the quark and scalar quark sector}
\label{subsec:squarkren}

The renormalization of the quark sector can differ from the SM case,
since the quark masses also appear in the scalar quark
sector. Therefore the renormalization of the quark and of the scalar
quark sector cannot be discussed separately. Since the scalar top and
bottom quarks are most relevant for the evaluation of EWPO, we will
focus on their renormalization here. Concerning the EWPO calculation
reviewed below, only a renormalization in \order{\als} is necessary.


\subsubsection{The top and scalar top sector}

The $t/\Stop$ sector contains four independent parameters:
the top-quark mass $\mt$, the stop masses $\mste$ and $\mstz$, and
either the squark mixing angle $\tst$ or, equivalently, the
trilinear coupling $\At$. Accordingly, the renormalisation of this sector 
is performed by introducing four counterterms that are determined by
four independent renormalisation conditions.

In an on-shell scheme, the 
following renormalisation conditions are imposed
(the procedure is equivalent to that of \citere{hr}, although
there no reference is made to the mixing angle).  
\begin{itemize}
\item[(i)] 
On-shell renormalization of the top-quark mass yields the top
mass counterterm,
\begin{align}\label{dmt}
\de\mt = \edz \mt \bigl [\re \Si_{{t}_L} (\mt^2) +
                         \re \Si_{{t}_R} (\mt^2) + 
                        2\re \Si_{{t}_S} (\mt^2)
\bigr]\; ,
\end{align}
with the scalar coefficients of the
unrenormalized top-quark self-energy, $\Si_t (p)$, 
in the Lorentz decomposition
\begin{align}
\label{Fermionselbstenergiezerlegung}
 \Si_t (p) &= \pslash \OM
\Si_{{t}_L} (p^2) + \pslash \OP
\Si_{{t}_R} (p^2) + \mt \Si_{{t}_S} (p^2)\; .
\end{align}
\item[(ii)] 
On-shell renormalization of the stop masses
determines the mass counterterms
\begin{equation}
\label{dmst}
\de \mste^2 = 
\re \Si_{\Stop_{11}}(m_{{\Stop}_{1}}^2) \, , \quad
\de \mstz^2 = 
\re \Si_{\Stop_{22}}(m_{{\Stop}_{2}}^2)\; ,
\end{equation}
in terms of the diagonal squark self-energies.
\item[(iii)] 
The counterterm for the mixing angle, $\tst$, (entering
\refeq{smm:physparam}) can be fixed in the following way,
\begin{align}
\label{ZusammenhangdeltathetadeltaM}  
\de \tst = 
\frac{\re \Si_{\Stop_{12}}(m_{{\Stop}_{1}}^2)+\re 
\Si_{\Stop_{12}}(m_{{\Stop}_{2}}^2)}{2(\mste^2-\mstz^2)}\; ,
\end{align} 
involving the non-diagonal squark self-energy. (This is a convenient
choice for the treatment of \order{\als} corrections. If
electroweak contributions were included, a manifestly gauge-independent
definition would be more appropriate.)
\end{itemize}
In renormalized vertices with squark and Higgs fields,
the counterterm of the trilinear
coupling $\At$ appears. 
Having already specified $\de \tst$,
the $A_t$ counterterm cannot be defined
independently but follows from the relation
\begin{align}
\sin 2 \tst = \frac{ 2 \mt ( \At - \mu \cot \be)}{\mste^2 -\mstz^2} \, ,
\end{align} 
yielding 
\begin{align}
\de \At &=  \frac{1}{\mt}
 \Bigl[\frac{1}{2} \sin 2 \tst
 \bigl(\de \mste^2 - \de \mstz^2\bigr) 
      + \cos 2 \tst
      (\mste^2 - \mstz^2)\, \de \tst 
\non \\[1.5mm] & \quad\ \quad\ \label{deltaAt}
   - \frac{1}{2 \mt} \sin 2 \tst  (\mste^2 -\mstz^2) \, \de \mt \Bigr]~.
\end{align}
This relation is valid at \order{\als} since both $\mu$ and $\tan\beta$
do not receive one-loop contributions from the strong interaction.


\subsubsection{The bottom and scalar bottom sector}

Because of SU(2)-invariance 
the soft-breaking parameters for the left-handed {\em up}- and
{\em down}-type squarks are identical, and thus
the squark masses of a given generation are not
independent. The stop and sbottom masses are connected via the relation
\begin{align}
 \cosQtb\, \msbe^2 + \sinQtb\, \msbz^2 = 
 \cosQtt\, \mste^2 + \sinQtt\, \mstz^2 +
\mb^2 - \mt^2 - M_W^2 \cos (2 \beta)~,
\label{MSb1gen}
\end{align}
with the entries of the rotation matrix in \refeq{squarkrotation}.  
Since the stop masses have already been renormalized
on-shell, only one of the sbottom mass
counterterms can be determined independently. Following
\citere{mhiggsFDalbals}, the
$\Sbotz$~mass is chosen
as the pole mass yielding the counterterm from an on-shell
renormalization condition, i.e.\ 
\begin{align}
\de \msbz^2 &= \re \Si_{\Sbot_{22}}(\msbz^2)\; ,
\label{eq:msbz}
\end{align}
whereas the counterterm for $\msbe$ is determined 
as a combination of other counterterms, 
according to
\begin{align} 
\de \msbe^2 &= 
\frac{1}{\cos^2 \tsb} \Bigl(
  \cos^2 \tst \de \mste^2   
 + \sin^2 \tst \de \mstz^2 
 - \sin^2 \tsb \de \msbz^2
 - \sin 2 \tst (\mste^2 -\mstz^2)\de \tst \nonumber
 \\[1.5mm]
& \quad 
 + \sin 2 \tsb (\msbe^2 -\msbz^2)\de \tsb 
-  2 \mt\, \de\mt + 2 \mb\, \de \mb   \Bigr)~.
\label{ms1CT}
\end{align}
Accordingly, the numerical value of $\msbe$ does not
correspond to the pole mass. The pole mass can be obtained 
from $\msbe$ via a finite shift of $\mathcal O(\als)$
(see e.g.~\citere{dr2lA}).

There are three more parameters with counterterms to be determined: 
the $b$-quark mass $\mb$, the mixing angle $\tsb$,
and the trilinear coupling~$\Ab$. They are connected via
\begin{align}
\label{mixingangleAparametermbrelation}
\sin 2 \tsb = \frac{ 2 \mb ( \Ab - \mu\tb)}{\msbe^2 -\msbz^2} \, ,
\end{align} 
which reads in terms of counterterms
\begin{align}
\label{deltaSbot}
2 \cosZtb\; \de\tsb &= \sinZtb \frac{\de\mb}{\mb} 
                   + \frac{2\mb\,\de\Ab}{\msbe^2 - \msbz^2}
                   - \sinZtb \frac{\de\msbe^2 - \de\msbz^2}
                                  {\msbe^2 - \msbz^2}~.
\end{align}
Only two of the three counterterms, $\de\mb$, $\de\tsb$, $\de\Ab$ can
be treated as independent, which offers a variety of choices.

As discussed in \citere{mhiggsFDalbals} a convenient
choice is the ``$\mb$ \drbar'' scheme, whereas a scheme analogous to
the one in the $t/\Stop$~sector, involving a bottom pole mass, 
can lead to artificially enhanced
higher-order corrections.

Concerning the renormalization of the top and the bottom mass, 
there are important differences. 
The top-quark pole mass can be directly extracted from experiment and,
due to its large numerical value as compared to other quark masses 
and the fact that
the present experimental error is much larger than the QCD scale, 
it can be used as input for theory predictions in a well-defined way.
For the mass of the bottom quark, on the other hand, problems related to
non-perturbative effects are much more severe.
Therefore the parameter extracted from the comparison of theory and 
experiment~\cite{pdg} 
is not the bottom pole mass. Usually the value of the bottom mass 
is given in the \msbar\ renormalisation scheme,
with the renormalisation scale $\mu^{\msbarm}$ chosen as the
bottom-quark mass, i.e.\ $\mb^{\msbarm}(\mb^{\msbarm})$~\cite{pdg}.

Another important difference to the top/stop sector is the replacement
of $\cot\be \rightarrow \tan\be$. As a consequence, 
very large effects can occur in this scheme for large values of $\mu$
and $\tan\be$~\cite{deltamb1}.

Potential problems with the bottom pole mass can be avoided by adopting
a renormalisation scheme with a running bottom-quark mass. In the
context of the MSSM it seems appropriate to use the \drbar\
scheme~\cite{dred} and to include the SUSY contributions at 
$\mathcal O(\als)$ into the running. 
We denote this running bottom mass as
$\mb^{\drbarm, \text{MSSM}}(\mu^{\drbarm})$.

The ``$\mb$ \drbar'' scheme
mentioned above uses a \drbar\ renormalization for both 
$\mb$ and $A_b$. In the \drbar\ scheme, the $b$-quark mass counterterm can
be determined by the expression
 \begin{align}
 \de \mb = \edz  \mb \bigl[
 \re {\Si}_{{b}_L}^{\rm div} (\mb^2)  +
 \re  {\Si}_{{b}_R}^{\rm div} (\mb^2) + 
 2\re {\Si}_{{b}_S}^{\rm div} (\mb^2) 
 \bigr]\; ,
\label{eq:dembdrbar}
 \end{align}
where ${\Si}^{\rm div}$ means replacing the one- and two-point integrals
$A$ and $B_0$ in the quark self-energies by their divergent parts.
The counterterm for the trilinear coupling $\Ab$ in the
\drbar\ scheme reads~\cite{mhiggsFDalbals}
 \begin{align} \non
 \de \Ab &= 
 \frac{1}{\mb}\Bigl[
 -\tan \tsb
   \re \Si_{\Sbot_{22}}^{\rm div}(\msbz^2)
 + \frac{1}{2}
  (\re \Si_{\Sbot_{12}}^{\rm div}(\msbe^2)+
   \re \Si_{\Sbot_{12}}^{\rm div}(\msbz^2))\\[1.5mm]
& \non \quad\ \quad\ 
  +  \tan \tsb \Bigl(\cos^2 \tst
  \re  \Si_{\Stop_{11}}^{\rm div}(\mste^2)
  + \sin^2 \tst
  \re  \Si_{\Stop_{22}}^{\rm div}(\mstz^2) \\[1.5mm]
& \non \quad\ \quad\ 
 - \frac{1}{2}  \sin 2 \tst
  (\re \Si_{\Stop_{12}}^{\rm div}(\mste^2)+
 \re \Si_{\Stop_{12}}^{\rm div}(\mstz^2)) \Bigr) 
\\[1.5mm]
& \non \quad\ \quad\
 -  \mt^2 
 \Bigl(\re {\Si}_{{t}_L}^{\rm div}(\mt^2) + 
  \re {\Si}_{{t}_R}^{\rm div}(\mt^2) + 
 2\re {\Si}_{{t}_S}^{\rm div}(\mt^2)\Bigr)
\Bigr] \\[1.5mm]
& \non \quad\ 
  + \edz  \bigl(2 \tan \tsb \mb - \frac{1}{2 \mb}(\msbe^2-\msbz^2) \sin 2
 \tsb \bigr) \times \\[1.5mm] 
& \quad \quad
  \Bigl 
  (\re {\Si}_{{b}_L}^{\rm div} (\mb^2)
 + \re {\Si}_{{b}_R}^{\rm div} (\mb^2) 
 + 2\re {\Si}_{{b}_S}^{\rm div} (\mb^2) \Bigr)~.   
\label{AbcountertermDR}
 \end{align} 
The counterterms for the mixing angle, $\de \tsb$, and
the $\tilde{b}_1$ mass, $\de \msbe^2$, 
are dependent quantities and can be determined
as combinations of the independent counterterms,
invoking (\ref{ms1CT}) 
and~(\ref{deltaSbot}),
\begin{align}\non
 \de \tsb &= \frac{1}{\msbe^2 -\msbz^2} \Bigl[
  \mb \de \Ab 
 +\tan \tsb \de \msbz^2
 + \de\mb
  \Bigl(\frac{1}{2 \mb}(\msbe^2 -\msbz^2) \sin 2 \tsb  - 2 \tan \tsb
  \mb\Bigl)  
\\[1.5mm] &   \quad\  \label{thetabinAbMBdrbar}   
- \tan \tsb \Bigl(\cos^2 \tst \de\mste^2
 + \sin^2 \tst \de\mstz^2 
 -    \sin 2 \tst (\mste^2 -\mstz^2) \de \tst 
 - 2 \mt \de \mt \Bigr) \Bigr] ~,
\\[2mm]
\de \msbe^2 &= \tan^2 \tsb \de \msbz^2 
+ 2 \tan \tsb \mb\de\Ab 
+ 2 \Bigl( \frac{1}{\mb} \sin^2 \tsb (\msbe^2-\msbz^2)
        + (1 - \tan^2 \tsb) \mb \Bigr) \de \mb \nonumber
\\[1mm]
 &   \quad \label{msb1inAbMBdrbar}
+(1 - \tan^2 \tsb)  
 \Bigl(\cos^2 \tst \de \mste^2 
      + \sin^2 \tst \de \mstz^2 
      - \sin 2 \tst (\mste^2-\mstz^2) \de \tst - 2 \mt \de \mt  \Bigr)~. 
\end{align}
The renormalized quantities in this scheme depend on the \drbar\
renormalization scale $\mu^{\drbarm}$.

\bigskip
In order to determine the value of 
$\mb^{\drbarm, \text{MSSM}}(\mu^{\drbarm})$ from the 
value $\mb^{\msbarm}(\mu^{\msbarm})$ that is extracted from the experimental
data one has to note that by definition $\mb^{\drbarm, \text{MSSM}}$ contains 
all MSSM contributions at \order{\als}, while $\mb^{\msbarm}$ contains
only the \order{\als} SM correction, i.e.\ the gluon-exchange
contribution. Furthermore, a finite shift arises from the transition
between the \msbar\ and the \drbar\ scheme. 

The expression for $\mb^{\drbarm, \text{MSSM}}(\mu^{\drbarm})$ 
is most easily derived
by formally relating $\mb^{\drbarm, \text{MSSM}}$ to the bottom pole mass 
first and
then expressing the bottom pole mass in terms of the \msbar\ mass (the
large non-perturbative contributions affecting the bottom pole mass drop 
out in the relation of $\mb^{\drbarm, \text{MSSM}}$ to $\mb^{\msbarm}$).
Using the equality $\mb^{\rm OS} + \de \mb^{\rm OS} = \mb^{\drbarm,
\text{MSSM}} + \de \mb^{\drbarm, \text{MSSM}}$ and the expressions for 
the on-shell counterterm and
the \drbar\ counterterm 
one finds
\begin{equation}
\label{eq:mbdrbar1}
\mb^{\drbarm, \text{MSSM}}(\mu^{\drbarm}) = \mb^{\text{OS}} + \edz  \mb
  \bigl(\Si^{\rm fin}_{b_L}({\mb}^2) + \Si^{\rm fin}_{{b}_R} ({\mb}^2) \bigr) 
+ \mb\, \Si^{\rm fin}_{b_S}(\mb^2)~.
\end{equation}
Here the $\Si^{\rm fin}$ are the UV-finite parts of the bottom quark
self-energy coefficients.
They depend on the \drbar\ scale 
$\mu^{\drbarm}$ and are evaluated for
on-shell momenta, $p^2 = \mb^2$. 
Inserting $\mb^{\text{OS}} =
\mb^{\msbarm}(M_Z) b^{\text{shift}}$, where
\begin{align}\label{dregdredbmassshift}
b^{\text{shift}} \equiv \Bigl[1 +
\frac{\alpha_s}{\pi} \Bigl(\frac{4}{3} - \ln
\frac{(\mb^{\msbarm})^2}{M_Z^2} \Bigr)\Bigr]~,   
\end{align}
one finds the desired expression for $\mb^{\drbarm}$,
\begin{align}
  \label{eq:mbdrbar2}
\mb^{\drbarm, \text{MSSM}}(\mu^{\drbarm}) &= \mb^{\msbarm}(M_Z)
  b^{\text{shift}} + \edz  \mb 
  \Bigl(\Si^{\rm fin}_{b_L}({\mb}^2) + \Si^{\rm fin}_{{b}_R} ({\mb}^2)
\Bigr)
+ \mb\, \Si^{\rm fin}_{b_S}(\mb^2)~.
\end{align}


\subsection{MSSM Higgs boson sector renormalization}
\label{subsec:Higgsrenorm}

In order to perform higher-order calculations in the Higgs boson
sector, the renormalized Higgs boson
self-energies are needed (see \refse{sec:mh}). 
The parameters appearing in the Higgs
potential, see \refeq{higgspot}, are renormalized as follows:
\begin{align}
\label{rMSSM:PhysParamRenorm}
  \MZ^2 &\to \MZ^2 + \dMZsq,  & \tadh &\to \tadh +
  \dtadh, \\ 
  \MW^2 &\to \MW^2 + \dMWsq,  & \tadH &\to \tadH +
  \dtadH, \notag \\ 
  M_{\rm Higgs}^2 &\to M_{\rm Higgs}^2 + \de M_{\rm Higgs}^2, & 
  \tanb &\to \tanb (1+\dtanb). \notag \\
  \mHp^2 &\to \mHp^2 + \de\mHp^2 \notag
\end{align}
The renormalization of $\MW$ and $\MZ$ has been described in
\refse{subsec:gaugebosonren}.
$M_{\rm Higgs}^2$ denotes the tree-level Higgs boson mass matrix given
in \refeq{higgsmassmatrixtree}. $\tadh$ and $\tadH$ are the tree-level
tadpoles, i.e.\ the terms linear in $h$ and $H$ in the Higgs potential.

The field renormalization matrices of both Higgs multiplets
can be written symmetrically, 
\begin{align}
\label{rMSSM:higgsfeldren}
  \begin{pmatrix} h \\[.5em] H \end{pmatrix} \to
  \begin{pmatrix}
    1+\tfrac{1}{2} \dZ{hh} & \tfrac{1}{2} \dZ{hH} \\[.5em]
    \tfrac{1}{2} \dZ{hH} & 1+\tfrac{1}{2} \dZ{HH} 
  \end{pmatrix} 
  \begin{pmatrix} h \\[.5em] H \end{pmatrix}~,
\end{align}
and for the charged Higgs boson
\BE
\label{rMSSM:ZHp}
H^\pm \to H^\pm (1 + \dZ{H^-H^+})~.
\end{equation}

\noindent
For the mass counterterm matrices we use the definitions
\begin{align}
  \delta M_{\rm Higgs}^2 =
  \begin{pmatrix}
    \dmhsq  & \dmhHsq \\[.5em]
    \dmhHsq & \dmHsq  
  \end{pmatrix}~.
\end{align}
The renormalized self-energies, $\hSi(p^2)$, can now be expressed
through the unrenormalized self-energies, $\Si(p^2)$, the field
renormalization constants and the mass counterterms.
This reads for the $\cp$-even part,
\begin{subequations}
\label{rMSSM:renses_higgssector}
\begin{align}
\ser{hh}(p^2)  &= \se{hh}(p^2) + \dZ{hh} (p^2-\mhtree^2) - \dmhsq, \\
\ser{hH}(p^2)  &= \se{hH}(p^2) + \dZ{hH}
(p^2-\tfrac{1}{2}(\mhtree^2+\mHtree^2)) - \dmhHsq, \\ 
\ser{HH}(p^2)  &= \se{HH}(p^2) + \dZ{HH} (p^2-\mHtree^2) - \dmHsq~,
\end{align}
\end{subequations}
and for the charged Higgs boson
\BE
\label{rMSSM:SEHp}
\ser{H^-H^+}(p^2) = \se{H^-H^+}(p^2) + \dZ{H^-H^+} (p^2 - \mHp^2)
                                     - \de\mHp^2~.
\end{equation}

Inserting the renormalization transformation into the Higgs mass terms
leads to expressions for their counter terms which consequently depend
on the other counter terms introduced in~(\ref{rMSSM:PhysParamRenorm}). 

For the $\cp$-even part of the Higgs sectors, these counter terms are:
\begin{subequations}
\label{rMSSM:HiggsMassenCTs}
\begin{align}
\dmhsq &= \de\MA^2 \cos^2(\alpha-\beta) + \delta \MZ^2 \sin^2(\alpha+\beta) \\
&\quad + \tfrac{e}{2 \MZ \sw \cw} (\dtadH \cos(\alpha-\beta)
\sin^2(\alpha-\beta) + \dtadh \sin(\alpha-\beta)
(1+\cos^2(\alpha-\beta))) \notag \\ 
&\quad + \dtanb \sinb \cosb (\MA^2 \sin 2 (\alpha-\beta) + \MZ^2 \sin
2 (\alpha+\beta)), \notag \\ 
\dmhHsq &= \tfrac{1}{2} (\de\MA^2 \sin 2(\alpha-\beta) - \dMZsq \sin
2(\alpha+\beta)) \\ 
&\quad + \tfrac{e}{2 \MZ \sw \cw} (\dtadH \sin^3(\alpha-\beta) -
\dtadh \cos^3(\alpha-\beta)) \notag \\ 
&\quad - \dtanb \sinb \cosb (\MA^2 \cos 2 (\alpha-\beta) + \MZ^2 \cos
2 (\alpha+\beta)), \notag \\ 
\dmHsq &= \de\MA^2 \sin^2(\alpha-\beta) + \dMZsq \cos^2(\alpha+\beta) \\
&\quad - \tfrac{e}{2 \MZ \sw \cw} (\dtadH \cos(\alpha-\beta)
(1+\sin^2(\alpha-\beta)) + \dtadh \sin(\alpha-\beta)
\cos^2(\alpha-\beta)) \notag \\ 
&\quad - \dtanb \sinb \cosb (\MA^2 \sin 2 (\alpha-\beta) + \MZ^2 \sin
2 (\alpha+\beta))~. \notag 
\end{align}
\end{subequations}
For the charged Higgs boson it reads
\BE
\label{rMSSM:dmHp}
\de\mHp^2 = \de\MA^2 + \de\MW^2~.
\end{equation}

\bigskip
For the field renormalization it is sufficient to give each Higgs doublet one
renormalization constant,
\begin{align}
\label{rMSSM:HiggsDublettFeldren}
  \cHe \to (1 + \tfrac{1}{2} \dZ{\cHe}) \cHe, \quad
  \cHz \to (1 + \tfrac{1}{2} \dZ{\cHz}) \cHz~.
\end{align}
This leads to the following expressions for the various field
renormalization constants in \refeq{rMSSM:higgsfeldren}:
\begin{subequations}
\label{rMSSM:FeldrenI_H1H2}
\begin{align}
  \dZ{hh} &= \sinasq \dZ{\cHe} + \cosasq \dZ{\cHz}, \\[.2em]
  \dZ{hH} &= \sina \cosa (\dZ{\cHz} - \dZ{\cHe}), \\[.2em]
  \dZ{HH} &= \cosasq \dZ{\cHe} + \sinasq \dZ{\cHz}, \\[.2em]
  \dZ{H^-H^+} &= \sinbsq \dZ{\cHe} + \cosbsq \dZ{\cHz}~.
\end{align}
\end{subequations}
The counter term for $\tb$ can be expressed in terms of the vaccuum
expectation values as
\begin{equation}
\de\tb = \frac{1}{2} \KL \dZ{\cHz} - \dZ{\cHe} \KR +
\frac{\de v_2}{v_2} - \frac{\de v_1}{v_1}~,
\end{equation}
where the $\de v_i$ are the renormalization constants of the $v_i$:
\begin{equation}
v_1 \to \KL 1 + \dZ{\cHe} \KR \KL v_1 + \de v_1 \KR, \quad
v_2 \to \KL 1 + \dZ{\cHz} \KR \KL v_2 + \de v_2 \KR~.
\end{equation}

The renormalization conditions are fixed by an appropriate
renormalization scheme. For the mass counterterms besides the 
on-shell conditions for $\MW$ and $\MZ$ (see \refeq{massren}) also 
$\MA$ can be renormalized on-shell:
\begin{align}
\label{rMSSM:mass_osdefinition}
  \de\MA^2 = \re \se{AA}(\MA^2). 
\end{align}
Since the tadpole coefficients are chosen to vanish in all orders,
their counter terms follow from $T_{\{h,H\}} + \de T_{\{h,H\}} = 0$: 
\begin{align}
  \dtadh = -{\tadh}, \quad \dtadH = -{\tadH}~. 
\end{align}
For the remaining renormalization constants for $\de\tb$, $\dZ{\cHe}$
and $\dZ{\cHz}$ several choices are possible, see e.g.\
\citere{feynhiggs1.2,tanbetaren}. A convenient
choice is a \drbar\ renormalization of $\de\tb$, $\dZ{\cHe}$
and $\dZ{\cHz}$, 
\begin{subequations}
\label{rMSSM:deltaZHiggsTB}
\begin{align}
  \dtanb &= \dtanb^{\drbarm} 
       \; = \; -\ed{2\CZa} \KKL \re \Sip_{hh}(\mhtree^2) - 
                             \re \Sip_{HH}(\mHtree^2) \KKR^{\rm div}, \\[.5em]
  \dZ{\cHe} &= \dZ{\cHe}^{\drbarm}
       \; = \; - \KKL \re \Sip_{HH \; |\al = 0} \KKR^{\rm div}, \\[.5em]
  \dZ{\cHz} &= \dZ{\cHz}^{\drbarm} 
       \; = \; - \KKL \re \Sip_{hh \; |\al = 0} \KKR^{\rm div}~.
\end{align}
\end{subequations}


\section{Sources of large SUSY corrections}

\label{sec:sources}

\subsection{Possible sources}

Besides the known sources of sizable higher-order corrections in the
SM, e.g.\ contributions enhanced by powers of $\mt$ or logarithms of
light fermions, there are additional sources of possibly large
corrections within the MSSM:

\begin{itemize}

\item
Large corrections can arise not only from loops containing the top
quark, but also its scalar superpartners. In the MSSM Higgs sector,
Yukawa corrections from the top and
scalar top quark sector can be especially large.
The one-loop corrections, for instance to the upper bound on the mass of
the lightest $\cp$-even Higgs boson, can reach the level of 100\%.
The leading one-loop term from the top and scalar top sector entering
the predictions in the Higgs sector is given by~\cite{mhiggs1l}
\BE
\sim \Gmu \, \mt^4 \log\KL \frac{\mste\mstz}{\mt^2} \KR~.
\end{equation}

\item
While the Higgs sector of the MSSM is $\cp$-conserving at tree level,
large $\cp$-violating effects can be induced by the loop corrections.

\item
Effects from the $b/\Sbot$ sector of the MSSM can also be very important
for large values of $\tb$ and $\mu$.

\item
The $b$~Yukawa coupling can receive large SUSY corrections, yielding a
shift in the relation between the $b$~quark mass and the corresponding
Yukawa coupling~\cite{deltamb1},
\BE
y_b = \frac{\wz}{v\,\Cb} \frac{\mb}{1 + \De\mb}~.
\end{equation}
The quantity $\De\mb$ contains in particular a contribution involving a
gluino in the loop, which gives rise to a correction proportional to
$(\als \,\mu\,\mgl\,\tb)$, which can be large. For $\De\mb \to -1$ the
$b$~Yukawa coupling even becomes non-perturbative.
This issue is discussed in \refse{subsec:deltamb}.

\item
Besides the scalar quark sector, SUSY theories have further possible
sources of large isospin splitting, which can give large contributions
to the $\rho$~parameter~\cite{rho,dr1lA}.

\item
Soft SUSY-breaking masses can induce splittings in the supersymmetric
coupling relations~\cite{susycoup,susycoup1Lfull} (i.e.\ the equality
of a SM coupling $g_i$ with the corresponding supersymmetric coupling
$h_i$). If scalar superpartners
have masses at a high scale $M$, and all the other masses are light
with mass $m \sim M_{\rm weak}$, the resulting corrections are given by
\BE
\frac{h_i(m)}{g_i(m)} -1 \approx
\frac{g_i^2(m)}{16\,\pi^2} \, \De b_i \log\frac{M}{m}~,
\end{equation}
where $\De b_i$ is the one-loop beta function coefficient contribution
from all light particles whose superpartners are heavy. If $M \gg m$
these corrections to the SUSY coupling relation can be sizeable.

\item
Another type of possibly large corrections in supersymmetric theories
are the so-called Sudakov logs (see \citere{sudakov} and references
therein). They appear in the form of $\log(q^2/\msusy^2)$ (where $q$
is the momentum transfer)  in the production cross sections of SUSY
particles at $e^+e^-$ colliders.

\item
In general, SUSY loop contributions can become large if some of the SUSY
particles are relatively light.

\end{itemize}


\subsection{Resummation in the $b/\Sbot$ sector}
\label{subsec:deltamb}

The relation between the bottom-quark mass and the Yukawa coupling $y_b$,
which in lowest order reads $\mb = y_b v_1/\sqrt{2}$, receives radiative
corrections proportional to $y_b v_2 = y_b \tb \, v_1$. Thus, large
$\tb$-enhanced contributions can occur, which need to be properly taken
into account. As shown in \citeres{deltamb1,deltamb} the leading terms of 
\order{\alb(\als\tb)^n} can be resummed by using an appropriate
effective bottom Yukawa coupling. 

Accordingly, an effective bottom-quark mass is obtained by extracting
the UV-finite $\tb$-enhanced term $\De \mb$ from
\refeq{eq:mbdrbar2} (which enters through $\Si_{b_S}$) and writing it
as $1/(1 + \De \mb)$ into the  
denominator. In this way the leading powers of $(\als\tb)^n$ are
correctly resummed~\cite{deltamb1,deltamb}. This yields
\begin{equation}
\label{eq:mbdrbarresum}
\mb^{\drbarm, \text{MSSM}}(\mu^{\drbarm}) = 
  \frac{\mb^{\msbarm}(M_Z) b^{\text{shift}}  + \edz  \mb
  \Bigl(\Si^{\rm fin}_{b_L}({\mb}^2) + \Si^{\rm fin}_{{b}_R} ({\mb}^2)
\Bigr)
+ \mb\, \widetilde{\Si}^{\rm fin}_{b_S}(\mb^2)}{1 + \De \mb}~,
\end{equation}
where $\widetilde{\Si}_{b_S} \equiv \Si_{b_S} + \De \mb$ denotes the 
non-enhanced remainder of the scalar $b$-quark self-energy at
\order{\als}, and $b^{\text{shift}}$ is given in \eqref{dregdredbmassshift}.
The $\tb$-enhanced scalar part of the
$b$-quark self-energy, $\De \mb$, is given at \order{\als} by%
\footnote{
There are also corrections of \order{\alt} to $\De\mb$ that can be
resummed~\cite{deltamb}. These effects usually amount up to 5--10\% of
the \order{\als} corrections. Since in this report we only consider
\order{\alb\als} contributions, these
corrections have been omitted. Further corrections from subleading
resummation terms can be found in \citere{deltamb3}.
}%
\begin{align}
\label{eq:deltamb}
\De \mb =   \frac{2}{3 \pi} \als\tb\, \mu\, \mgl\,
 I (\msbe^2, \msbz^2, \mgl^2) ,
\end{align}
with
\begin{align}
I (\msbe^2, \msbz^2, \mgl^2) =
- \frac{\msbe^2 \msbz^2 \log(\msbz^2/\msbe^2) +
        \msbe^2 \mgl^2  \log(\msbe^2/\mgl^2) +
        \mgl^2 \msbz^2  \log(\mgl^2/\msbz^2)}
       {(\msbe^2 - \mgl^2) (\mgl^2 - \msbz^2) (\msbz^2 - \msbe^2)}~,
\label{eq:I}
\end{align}
and $\De \mb > 0$ for $\mu > 0$.

In the ``$\mb$ \drbar'' defined above, the
effective bottom-quark mass as given in \refeq{eq:mbdrbarresum}
should be used everywhere instead of the \drbar\ bottom quark mass.
This also applies to the bottom mass in the
sbottom-mass matrix squared, \refeq{squarkmassenmatrix}, from which
the sbottom mass eigenvalues are determined.
The effects of $\De\mb$, i.e.\ the leading effects
of \order{\als}, can be incorporated into a lowest-order result
(e.g.\ the one-loop results for the renormalized Higgs boson
self-energies, see \refse{sec:mh}) by using the effective bottom-quark
mass of \refeq{eq:mbdrbarresum}
(or the correspondingly shifted value in other renormalization
schemes).


\section{Electroweak precision observables in the MSSM}

In this section we briefly introduce the electroweak precision
observables that are discussed in this report. A description of the
current status of their theoretical evaluation
within the MSSM will be given in
the following sections and the remaining theoretical uncertainties will
be discussed. 

The current experimental status of the EWPO and prospective improvements
of their precision in the future have been summarized in
\refse{subsec:ewpostatus}. In order to fully exploit the experimental
precision of the EWPO the theoretical uncertainties should be reduced
significantly below the level of the experimental errors. 

Concerning the theoretical predictions, two kinds of uncertainties need to be
taken into account: the theoretical uncertainties from unknown
higher-order corrections (``intrinsic'' theoretical uncertainties) and 
the uncertainties induced by the experimental errors of the input
parameters (``parametric'' theoretical uncertainties). The parametric
uncertainty induced by the known input parameters (in the SM case in
particular $\mt$ and $\De\al_{\rm had}$) needs to be reduced in order to
increase the sensitivity to the unknown parameters of the model (in the
SM case $\MH$).

The EWPO discussed in the following sections are:

\begin{itemize}

\item
The $W$~boson mass can be evaluated from
\BE
\MW^2 \KL 1 - \frac{\MW^2}{\MZ^2} \KR = 
\frac{\pi\,\al}{\wz\,\Gmu} (1 + \De r)~,
\label{calcMW}
\end{equation}
where $\al$ is the fine structure constant and $\Gmu$ the Fermi constant.
This relation arises from comparing the prediction for muon decay with
the experimentally precisely known Fermi constant. The radiative
corrections are summarized in the quantity $\De r$, derived first for
the SM in \citere{sirlin}. The prediction for $\MW$ within
the SM or the MSSM is obtained from evaluating $\De r$ in these models
and solving \refeq{calcMW} in an iterative way. The theory status of the
prediction for $\MW$ is reviewed in
\refse{sec:evalMW}.

\item
Another important group of EWPO are the $Z$~boson observables, among
which we mostly concentrate on the effective leptonic weak mixing angle
at the $Z$~boson resonance,
$\sweff$. It can be defined through the form factors at the $Z$~boson
pole of the vertx coupling of the $Z$ to leptons ($l$). If this vertex
is written as $i \bar l \ga^\mu (g_{\rm V} - g_{\rm A} \ga_5) l Z_\mu$ then
\BE
\sweff = 
   \frac{1}{4} \left(
    1- \real\frac{g_{\rm V}}{g_{\rm A}} \right) \, .
\label{calcSW}
\end{equation}
At the tree level this amounts to the sine of the weak mixing angle,
$\sin^2\theta_{\rm W} = 1 - \MW^2/\MZ^2$, in the on-shell scheme.
Loop corrections enter through the form factors $g_{\rm V}$ and $g_{\rm A}$. 
The theoretical evaluation is reviewed in \refse{sec:Zobs}.

\item
The quantity $\De\rho$, 
\BE
\De\rho = \frac{\Si^Z(0)}{\MZ^2} - \frac{\Si^W(0)}{\MW^2} ,
\label{delrho}
\end{equation}
 parameterizes the leading universal corrections to the electroweak
 precision observables induced by
the mass splitting between fields in an isospin doublet~\cite{rho}.
$\Si^{Z,W}(0)$ denote the transverse parts of the 
unrenormalized $Z$ and $W$ boson
self-energies at zero momentum transfer, respectively.
The induced shifts in the two above described observables 
are given in leading order by
\BE
\de\MW \approx \frac{\MW}{2}\frac{\cw^2}{\cw^2 - \sw^2} \De\rho, \quad
\de\sweff \approx - \frac{\cw^2 \sw^2}{\cw^2 - \sw^2} \De\rho .
\label{precobs}
\end{equation}
The theoretical evaluation of $\De\rho$ is discussed in
\refse{sec:evalDelrho}.

\item
Another very powerful observable for constraining the parameter space of
the MSSM is the mass of the lightest $\cp$-even Higgs boson, $\mh$. If
the Higgs boson will be found at the next generation of colliders, its
mass will be measured with high precision. We therefore refer to $\mh$
also as an EWPO. While $\mh$ is bounded from above at tree-level by
$\mh \le \MZ$, it receives large radiative corrections. The leading
one-loop contribution, arising from the $t/\Stop$~sector,
reads~\cite{mhiggs1l} 
\BE
\De \mh^2 = \frac{3\, \Gmu}{\wz\, \pi^2\,\SQb} 
          \; \mt^4\,\log \KL \frac{\mste \mstz}{\mt^2} \KR~.
\end{equation}
The loop corrections, entering via Higgs-boson propagator corrections,
can shift $\mh$ by 50--100\%. The theoretical status is reviewed
in \refse{sec:mh}.

\item
As a further
precision observable that we investigate in detail in this report we
consider
the anomalous magnetic moment of the muon, $\amu \equiv (g-2)_\mu$.  
It is related to the 
photon--muon vertex function $\Gamma_{\mu\bar\mu A^\rho}$ as follows:
\BEA
\bar u(p')\Ga_{\mu\bar\mu A^\rho}(p,-p',q) u(p) & = &
\bar u(p')\left[\ga_\rho F_V(q^2) + (p+p')_\rho F_M(q^2) +
  \ldots\right] u(p)~, \non \\
\amu & = & -2m_\mu F_M(0)~,
\EEA
where $F_M(q^2) = 0$ at tree-level. Non-zero values are induced via
loop corrections. The theoretical evaluation is discussed
in \refse{sec:evalg-2}. 

\end{itemize}


\section{The $\rho$ parameter}
\label{sec:evalDelrho}

We start our discussion with the quantity $\De\rho$, see \refeq{delrho},
which parametrizes in particular the leading contributions from loops of
scalar quarks and leptons to the $W$-boson mass and the $Z$-boson
observables.


\subsection{One-loop results}
\label{subsec:oneloop}

In the SM the dominant contribution to $\De\rho$ at the \onel\ level
arises from the $t/b$ doublet due to its large mass splitting. With
both fermion masses non-zero it reads
\BE
\De\rho^{\SM}_0 = \frac{3\,\Gmu}{8\,\wz\,\pi^2} \; F_0(\mt^2, \mb^2) ,
\label{delrhoSM1l}
\end{equation}
with
\BE
F_0(x,y) = x + y - \frac{2\,x\,y}{x - y} \log \frac{x}{y} .
\end{equation}
$F_0$ has the properties $F_0(m_a^2, \mb^2) = F_0(\mb^2, m_a^2)$, 
$F_0(m^2, m^2) = 0$, $F_0(m^2, 0) = m^2$. Therefore for $\mt \gg \mb$
\refeq{delrhoSM1l} reduces to the well known quadratic correction 
\BE
\De\rho^{\SM}_0 = \frac{3\,\Gmu}{8\,\wz\,\pi^2} \; \mt^2 .
\end{equation}

\begin{figure}[htb]
\begin{center}
\mbox{
\psfig{figure=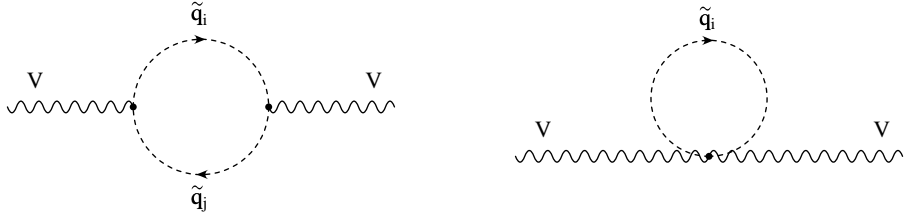,width=12cm,
       bbllx=210pt,bblly=680pt,bburx=397pt,bbury=720pt}}
\end{center}
\caption[]{Feynman diagrams for the contribution of scalar quark loops 
to the gauge boson self-energies at \onel\ order.}
\label{fig:fdgb1l}
\end{figure}

Within the MSSM the dominant SUSY correction at the \onel\
level arises from 
the scalar top and bottom contribution to \refeq{delrho}, see
\reffi{fig:fdgb1l}. 
%

\begin{figure}[htb!]
\begin{center}
\mbox{
\epsfig{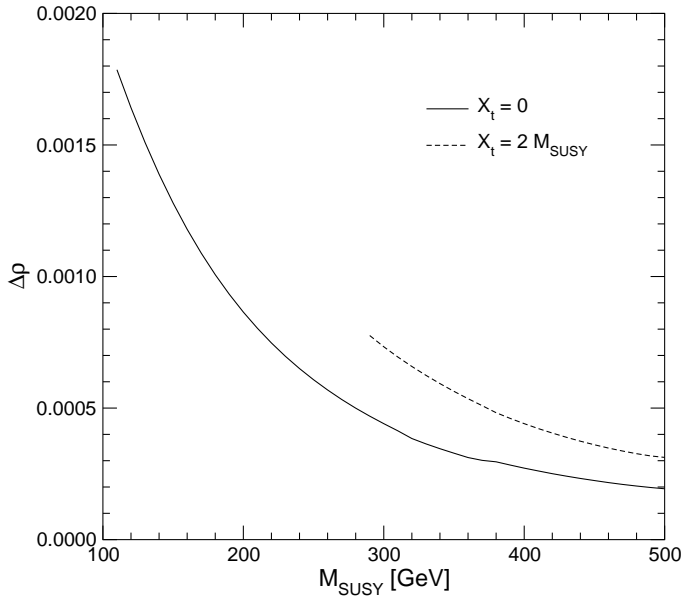}
     }
\caption{
One--loop contribution of the $(\tilde t, \tilde b)$ doublet to
$\Delta\rho$ as a function of the common squark mass $\msusy$ for
$\tb = 3$, and 
$\Xb = 0$ and $\Xt = 0 \mbox{ or } 2 \msusy$.}
\label{fig:dr1l}
\end{center}
\end{figure}

\noindent
For $\mb \neq 0$ it is given by
\BEA
\De\rho_0^\SU = \frac{3\,\Gmu}{8\, \wz\, \pi^2} 
                  \Big[ -\sinQtt\cosQtt F_0(\mste^2, \mstz^2)
                       -\sinQtb\cosQtb F_0(\msbe^2, \msbz^2) \non\\
                       +\cosQtt\cosQtb F_0(\mste^2, \msbe^2)
                       +\cosQtt\sinQtb F_0(\mste^2, \msbz^2) \non\\
                       +\sinQtt\cosQtb F_0(\mstz^2, \msbe^2)
                       +\sinQtt\sinQtb F_0(\mstz^2, \msbz^2) ~\Big] .
\label{delrhoMSSM1l}
\EEA
The size of the SUSY \onel\ contributions are shown for an exemplary
case in \reffi{fig:dr1l} as a function of $\msusy$. The parameter
$\msusy$ is defined by setting the soft SUSY-breaking parameters in the
diagonal entries of the stop and sbottom mass matrices equal to each
other for simplicity,
\begin{equation}
\msusy \equiv M_{\tilde{Q}} = M_{\tilde{U}} = M_{\tilde{D}} \, , 
\label{eq:msusy}
\end{equation}
see \refeq{squarkmassenmatrix}.
We furthermore use the shorthands
\begin{equation}
\Xt \equiv \At - \mu/\tb, \quad \Xb \equiv \Ab - \mu \tb \, .
\label{eq:XtXb}
\end{equation}
The other parameters in \reffi{fig:dr1l} are 
$\tb = 3$ and $\Xt = 0, 2 \msusy$. In this case $\De\rho_0^\SU$ can
reach values of up to $2 \times 10^{-3}$. The line for $\Xt = 2 \msusy$ 
starts only at $\msusy \approx 300 \gev$. For lower values of $\msusy$
one of the scalar top mass squares is below zero.

\subsection{Results beyond the \onel\ level}
\label{subsec:beyondoneloop}

\subsubsection{SM results}
\label{subsubsec:delrhosm}

Within the SM the \onel\ \order{\al} result from the contribution of the
$t/b$ doublet has been extended in
several ways. The dominant \twol\ corrections arise at
$\oaas$ and are given by~\cite{drSMgfals}
\BE
\De\rho^{\SM, \al\als}_1 = - \De\rho^{\SM}_0 \frac{2}{3} \frac{\als}{\pi}
                           \KL 1 + \pi^2/3 \KR .
\end{equation}
These corrections screen the \onel\ result by approximately 10\%. Also
the three-loop result at \order{\al\als^2} is known. Numerically it
reads~\cite{drSMgfals2}
\BE
\De\rho^{\SM, \al\als^2}_2 = -\frac{3\,\Gmu}{8\,\wz\,\pi^2} \, \mt^2 
                             \KL \frac{\als}{\pi} \KR^2 \cdot 14.594... ~.
\end{equation}
Furthermore the leading electroweak \twol\ contributions of 
\order{\Gmu^2\mt^4} have been calculated. 
First the result in the approximation
$\MH = 0$ had been evaluated~\cite{drSMgf2mh0},
\BEA
\De\rho^{\SM,\Gmu^2}_{1|\MH = 0} &=&
 3\,\frac{\Gmu^2}{128 \pi^4} \, \mt^4\, \times \de^{\SM}_{1|\MH = 0} \non \\
\de^{\SM}_{1|\MH = 0} &=& 19 - 2 \pi^2 .
\label{delrhoMH0}
\EEA
Later the full \order{\Gmu^2\mt^4} result for arbitrary $\MH$ 
became available~\cite{drSMgf2mt4}, where 
$\de^{\SM}_{1|\MH = 0}$ extends to
\BE
\de^{\SM}_{1|\MH \neq 0} = 19 - 2 \pi^2 + {\rm fct}(\mt, \MH) .
\label{drSMgf2}
\end{equation}
The leading two-loop contribution to $\De\rho$ in an asymptotic expansion 
for large $\MH$ of \order{\Gmu^2\MH^2\MW^2} was obtained in
\citere{ewmh2}. It turned out to be numerically small.

Leading electroweak three-loop results of \order{\Gmu^3\mt^6} and 
\order{\Gmu^2\als\mt^4} became
available more 
recently~\cite{drSMgf3mh0,drSMgf3}. Numerically they read in the case
$\MH = 0$:
\BEA
\De\rho^{\SM,\Gmu^3}_{2|\MH = 0} &=& 
\KL \frac{\Gmu}{8\,\wz\,\pi^2} \mt^2 \KR^3 \cdot 249.74 ~, \\
\De\rho^{\SM,\Gmu^2\als}_{2|\MH = 0} &=& 
\KL \frac{\Gmu}{8\,\wz\,\pi^2} \mt^2 \KR^2 
\KL \frac{\als}{\pi} \KR \cdot 2.9394 ~.
\EEA
For the case $\MH \neq 0$ the result has been obtained in several
limits, allowing a smooth interpolation, see \citere{drSMgf3} for
details. 
Most recently also the leading \order{\Gmu^3\MH^4\MW^2} contribution
was obtained~\cite{Boughezal:2004ef}. 
Besides for very large values of $\MH$ it is numerically
insignificant.


\subsubsection{The SUSY corrections at \order{\al\als}}
\label{subsubsec:delrhomssm}

\begin{figure}[ht!]
\begin{center}
\mbox{
\psfig{figure=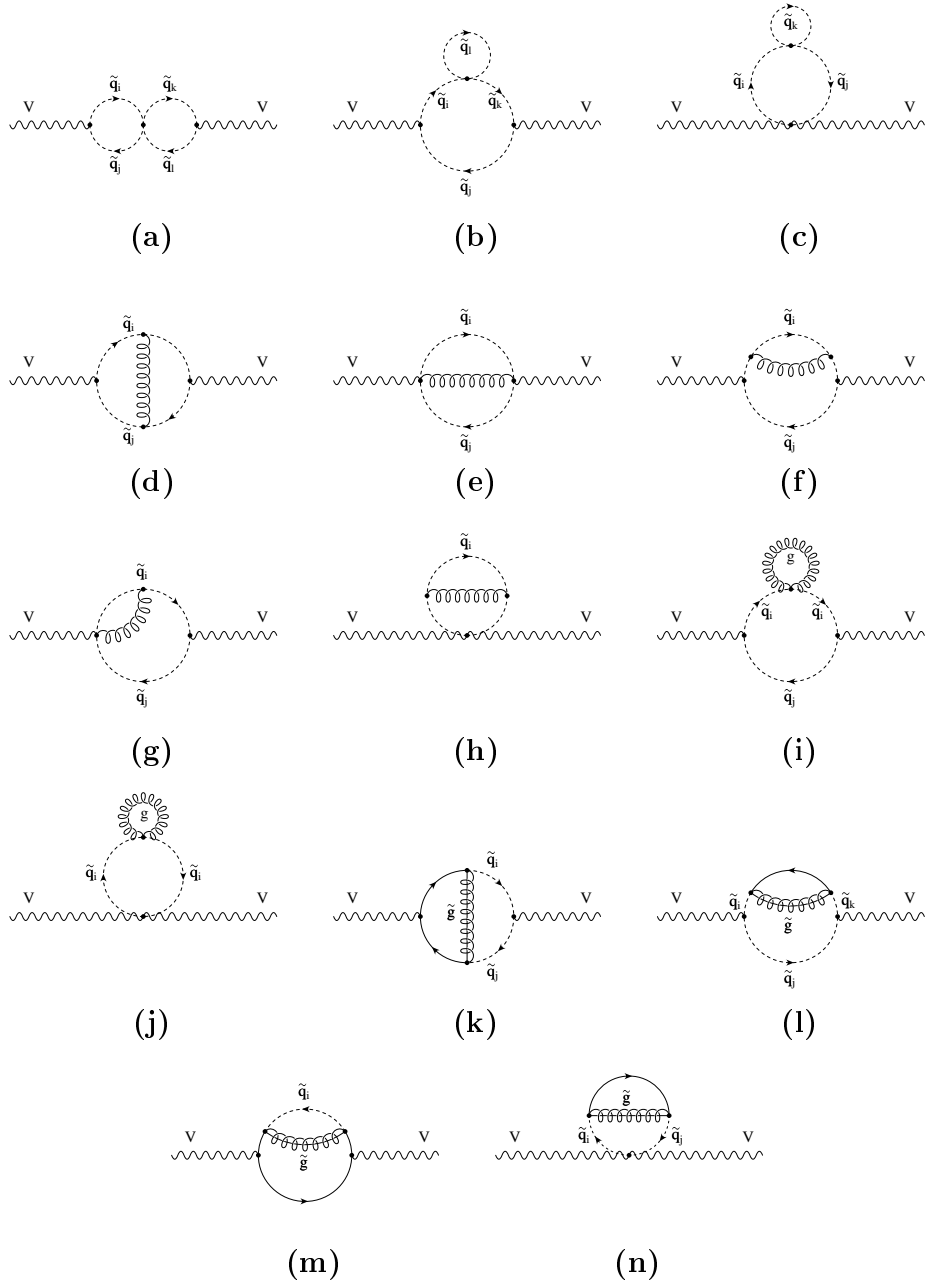,width=12cm,
       bbllx=90pt,bblly=170pt,bburx=510pt,bbury=730pt}}
\end{center}
\caption{Feynman diagrams for the contribution of scalar quark loops 
to the gauge-boson self-energies at \twol\ order.}
\label{fig:fdgb2l}
\end{figure}

The leading \twol\ corrections arising in the MSSM (beyond the SM part)
have been evaluated at \order{\al\als}~\cite{dr2lA} and
\order{\alt^2,\alt\alb,\alb^2}~\cite{drMSSMgf2A,drMSSMgf2B} (the latter
in the limit of large $\msusy$). 
The leading $\oaas$ corrections to the scalar quark
loops consist of the diagrams shown in \reffi{fig:fdgb2l}
(supplemented with the corresponding diagrams for the subloop
renormalization, see
\citere{dr2lA}). The diagrams can be divided into three groups: the
pure scalar contribution (diagrams a-c), the gluonic correction
(diagrams d-j, where the gluon-loop contribution, diagrams i-j, is
zero) and the gluino exchange correction (diagrams k-n).

The pure scalar quark diagrams give a vanishing contribution. The
gluonic correction can be cast into a compact formula~\cite{dr2lA}:
\BEA
\De\rho^\SU_{1,{\rm gluon}} 
     = \frac{\Gmu}{4\, \wz\, \pi^2} \frac{\als}{\pi}
                 \Big[ -\sinQtt\cosQtt F_1(\mste^2, \mstz^2)
                       -\sinQtb\cosQtb F_1(\msbe^2, \msbz^2) \non\\
                       +\cosQtt\cosQtb F_1(\mste^2, \msbe^2)
                       +\cosQtt\sinQtb F_1(\mste^2, \msbz^2) \non\\
                       +\sinQtt\cosQtb F_1(\mstz^2, \msbe^2)
                       +\sinQtt\sinQtb F_1(\mstz^2, \msbz^2) ~\Big] ,
\label{delrhoMSSM2l}
\EEA
with
\BEA
F_1(x, y) &=& x + y - 2\frac{x y}{x - y} 
          \log\frac{x}{y} \KKL 2 + \frac{x}{y}\ln\frac{x}{y} \KKR \non \\
 &&         + \frac{(x + y)x^2}{(x - y)^2} \log^2\frac{x}{y}
            -2 (x - y) {\rm Li}_2 \KL 1 - \frac{x}{y} \KR, 
\EEA
where $F_1$ has the properties $F_1(m_a^2, \mb^2) = F_1(\mb^2, m_a^2)$, 
$F_1(m^2, m^2) = 0$, $F_1(m^2, 0) = m^2 (1 + \pi^2/3)$.
The gluino exchange correction results in a lengthy formula, see
\citere{dr2lA}, and is not given here. It decouples for $\mgl \to \infty$. 

The analytical formula for the \order{\al\als} corrections given in 
\refeq{delrhoMSSM2l}
is expressed in terms of the physical squark masses, i.e.\ an on-shell
renormalization has been carried out for all four squark masses. 
As discussed in \refse{subsec:squarkren}, SU(2) invariance leads to a
relation between the stop and sbottom masses, so that not all four
masses can be renormalized independently. This results in a finite mass
shift of \order{\als} that is given, if expressed in terms of $\msbe$, as
the difference between the counterterm of \refeq{ms1CT} and the on-shell
counterterm. If the two-loop result is expressed in terms of the
on-shell masses, this mass shift appears in the relation between the 
physical squark masses and the (unphysical) soft SUSY-breaking mass
parameters in the squark mass matrices, see \refeq{squarkmassenmatrix}.
While this shift is formally of higher order in the evaluation of the
masses that are inserted in the two-loop result, it needs to be taken
into account in the one-loop result, 
This gives rise to an extra
contribution compared to the results discussed in
\refse{subsec:oneloop}, see \citere{dr2lA} for a more detailed
discussion.



\subsubsection{The SUSY corrections at \order{\alt^2},
\order{\alt\alb}, \order{\alb^2}}
\label{subsubsec:delrhomssmEW}

Furthermore the leading \order{\alt^2}, \order{\alt\alb}, \order{\alb^2}
corrections to $\De\rho$ have 
been evaluated in the limit $\msusy \to \infty$~\cite{drMSSMgf2A,drMSSMgf2B}.
The $\mt$ dependence of $\De\rho$ differs between the pure SM 
contribution and the additional SUSY corrections.
Within the SM, the corrections are $\sim \mt^2$ for the \onel\
and $\sim \mt^4$ 
for the \twol\ correction, leading to sizable shifts in the precision
observables. The additional SUSY corrections at the \onel\ level
(from scalar quark loops), 
on the other hand, do not contain a prefactor $\sim\mt^2$.
In the electroweak \twol\ corrections it is no longer possible to
separate out the pure SM contribution because of the extended Higgs
sector of the MSSM. The leading electroweak \twol\ corrections
in the MSSM are 
therefore of  \order{\Gmu^2\mt^4} (as in the SM case) and potentially
sizable.

The leading contributions of
\order{\alt^2}, \order{\alt \alb} and \order{\alb^2} have been derived
by extracting the
contributions proportional to $y_t^2$, $y_t y_b$ and $y_b^2$, where
\BE
y_t = \frac{\wz \, \mt}{v \, \sbe}, \quad
y_b = \frac{\wz \, \mb}{v \, \Cb} ~.
\label{ytyb}
\end{equation}
The coefficients of these terms could then be evaluated in the gauge-less 
limit, i.e.\ for $\MW, \MZ \to 0$ (keeping $\cw = \MW/\MZ$ fixed).

For the Higgs masses appearing in the two-loop diagrams the
following relations have been used, arising from the gauge-less limit
\BE
\mHp^2 = \MA^2~, \quad
\mG^2 = 0 ~, \quad
\mGp^2 = 0 ~.
\label{chargedHiggsMW0}
\end{equation}
Applying the corresponding limit also in the neutral $\cp$-even Higgs
sector would yield for the lightest $\cp$-even Higgs-boson mass
$\mh^2 = 0$ (and furthermore $\mH^2 = \MA^2$, $\Sa = -\Cb$, $\Ca = \sbe$). 
Since within the SM the limit $\MHSM \to 0$ turned out to be 
only a poor approximation of the result for arbitrary $\MHSM$, $\mh^2$
has been kept non-zero (which formally is a higher-order
effect). Keeping $\mh$ as a free parameter is also relevant in view of
the fact that the lightest MSSM Higgs boson receives large higher order
corrections,
which shift its upper bound up to $135 \gev$ (for $\msusy \le 1 \tev$
and $\mt = 175 \gev$), see \refse{sec:mh}.
These corrections can 
easily be taken into account in this way (in the Higgs contributions at 
one-loop order, however, the tree-level value of $\mh$ should be used).
Keeping $\al$ arbitrary is necessary in order to incorporate non SM-like 
couplings of the lightest $\cp$-even Higgs boson to fermions and gauge
bosons. 

On the other hand, keeping all Higgs-sector parameters
completely arbitrary is not possible, as the underlying symmetry of the
MSSM Lagrangian has to be exploited in order to ensure the
UV-finiteness of the 
two-loop corrections to $\De \rho$. Thus only those symmetry relations
have been enforced in the neutral $\cp$-even Higgs sector which are
explicitly needed in order to obtain a complete cancellation of the
UV-divergences.

It is convenient to discuss 
the \order{\alt^2 \propto \Gmu^2\mt^4} SUSY
contributions to $\De\rho$ separately, i.e.\ the case where $y_b = 0$. 
The \order{\alt^2} corrections are by far the dominant subset within
the SM, i.e.\ the \order{\alt \alb} and \order{\alb^2} corrections
can safely be neglected within the SM. The same is true within the MSSM
for not too large values of $\tb$.
It is well known~\cite{decoupling1l} that the SUSY sector of the MSSM
decouples if the general soft SUSY-breaking scale
goes to infinity (corresponding to  $\msusy \to \infty$ in the \onel\
result given above).
The leading contributions of \order{\Gmu^2\mt^4} in the case where the
scalar quarks are heavy is therefore obtained in the limit where only
the two Higgs doublet sector of the MSSM is
active~\cite{drMSSMgf2A,drMSSMgf2B}, corresponding to the limit 
$\msusy \to \infty$.

In \citere{drMSSMgf2A} the result has been obtained in the simplified
case with tree-level Higgs boson masses.
In the limit $\MW, \MZ \to 0$ the neutral $\cp$-even Higgs
boson masses at the tree-level reduce to
\BE
\mh^2 = 0 ,  \quad \mH^2 = \MA^2 .
\label{mheq0}
\end{equation}
In this limit also the relation between the angles $\al$ and
$\be$, see \refeq{alphaborn}, becomes very simple, 
$\al = \be - \pi/2$, i.e. $\Sa = -\Cb$, $\Ca = \sbe$.
The only remaining scales left are the top quark mass, $\mt$, the $\cp$-odd
Higgs boson mass, $\MA$, and $\tb$ (or $\sbe = \tb/\sqrt{1 + \TQb}$). 
In the limit of large $\tb$ (i.e.\ $(1 - \SQb) \ll 1$) the result takes
a particularly simple form. One obtains
\BE
\De\rho_{1,\hi,\mh=0}^{\SU} = 3 \, \frac{\Gmu^2}{128 \,\pi^4} \, \mt^4 \, 
\KKL \frac{19}{\SQb} - 2\,\pi^2 + \orderm{1 - \SQb} \KKR .
\label{drallmalargetb}
\end{equation}
Thus, for large $\tb$ the SM limit with $\MH^{\SM} \to 0$ (see
\refeq{delrhoMH0}) is reached.


Keeping $\tb$ arbitrary but expanding for large values of $\MA$ yields
\BEA
\De\rho_{1,\hi,\mh=0}^{\SU} &=& 3 \, \frac{\Gmu^2}{128 \,\pi^4} \, \mt^4 \, 
                      \times \non \\
&& \Bigg\{
   19 - 2 \pi^2 \non \\
&& -\frac{1 - \SQb}{\SQb} \Bigg[ \KL \log^2 A + \frac{\pi^2}{3} \KR
           \KL 8 A + 32 A^2 + 132 A^3 + 532 A^4 \KR \non \\
\label{drlargema}
&& + \log(A) \ed{30} \KL 560 A + 2825 A^2 + 11394 A^3 + 45072 A^4 \KR 
     \\
&& - \ed{1800} \KL 2800 A + 66025 A^2 + 300438 A^3 + 1265984 A^4 \KR
   + {\cal O}\KL A^5 \KR \Bigg] 
   \Bigg\} , \non
\EEA
where $A \equiv \mt^2/\MA^2$.
In the limit $A \to 0$ one obtains
\BE
\De\rho_{1,\hi,\mh=0}^{\SU} = 3 \, \frac{\Gmu^2}{128 \,\pi^4} \, \mt^4 \, 
 \KKL 19 - 2\,\pi^2 \KKR + \cO(A) ,
\end{equation}
i.e.\ exactly the SM limit for $\MH^{\SM} \to 0$ is
reached. 
This constitutes an important consistency check: in the limit $A \to 0$ 
the heavy Higgs bosons are decoupled from the theory. Thus only the
lightest $\cp$-even Higgs boson should remain, which has in the
\order{\Gmu^2\mt^4} approximation (neglecting higher-order corrections)
the mass $\mh = 0$, see
\refeq{mheq0}. As already observed in \citere{dr2lA}, 
the decoupling of the non-SM contributions
in the limit where the new scale (i.e.\ in the present case $\MA$) is
made large is explicitly seen here at the two-loop level.

Now we turn to the full \order{\alt^2} corrections. As discussed in
\citere{drMSSMgf2B}, an UV-finite result could only be obtained if the
relations in \refeq{chargedHiggsMW0} are taken into account.
The masses of the neutral Higgs bosons as well as the mixing angle
could be kept as `independent' parameters, i.e.\ they can be obtained
taking into account higher order corrections. 
The full result without the tree-level relations is rather lengthy and
can be found in \citere{drMSSMgf2B}. 

\bigskip
Now also the \order{\alt\alb}, \order{\alb^2} SUSY corrections are
considered.
The structure of the fermion doublet requires that further symmetry
relations are taken into account. Within the Higgs boson sector it is
necessary, besides using \refeq{chargedHiggsMW0}, also to use the
relations for the heavy $\cp$-even Higgs boson mass and the Higgs
mixing angle,
\BE
\mH^2 = \MA^2 ~, \quad
\Sa = -\Cb ~, \quad
\Ca = \sbe~.
\label{heavyHiggsMW0}
\end{equation}
On the other hand, $\mh$ can be kept as a
free parameter. The couplings of the lightest $\cp$-even Higgs boson
to gauge bosons and SM fermions,
however, become SM-like, once the mixing angle relations,
\refeq{heavyHiggsMW0}, are used.
Furthermore, the Yukawa couplings can no longer be treated as free
parameters, i.e.\ \refeq{ytyb} has to be employed, which ensures that
the Higgs mechanism governs the Yukawa couplings. 
Corrections enhanced by
$\tb$ thus arise only from the heavy Higgs bosons, while the
contribution from the lightest $\cp$-even Higgs boson resembles the SM
one. 


\subsection{Results in the NMFV MSSM}
\label{subsec:Delrhonmfv}

\begin{figure}[htb!]
\begin{center}
\includegraphics[width=13.5cm,clip=]{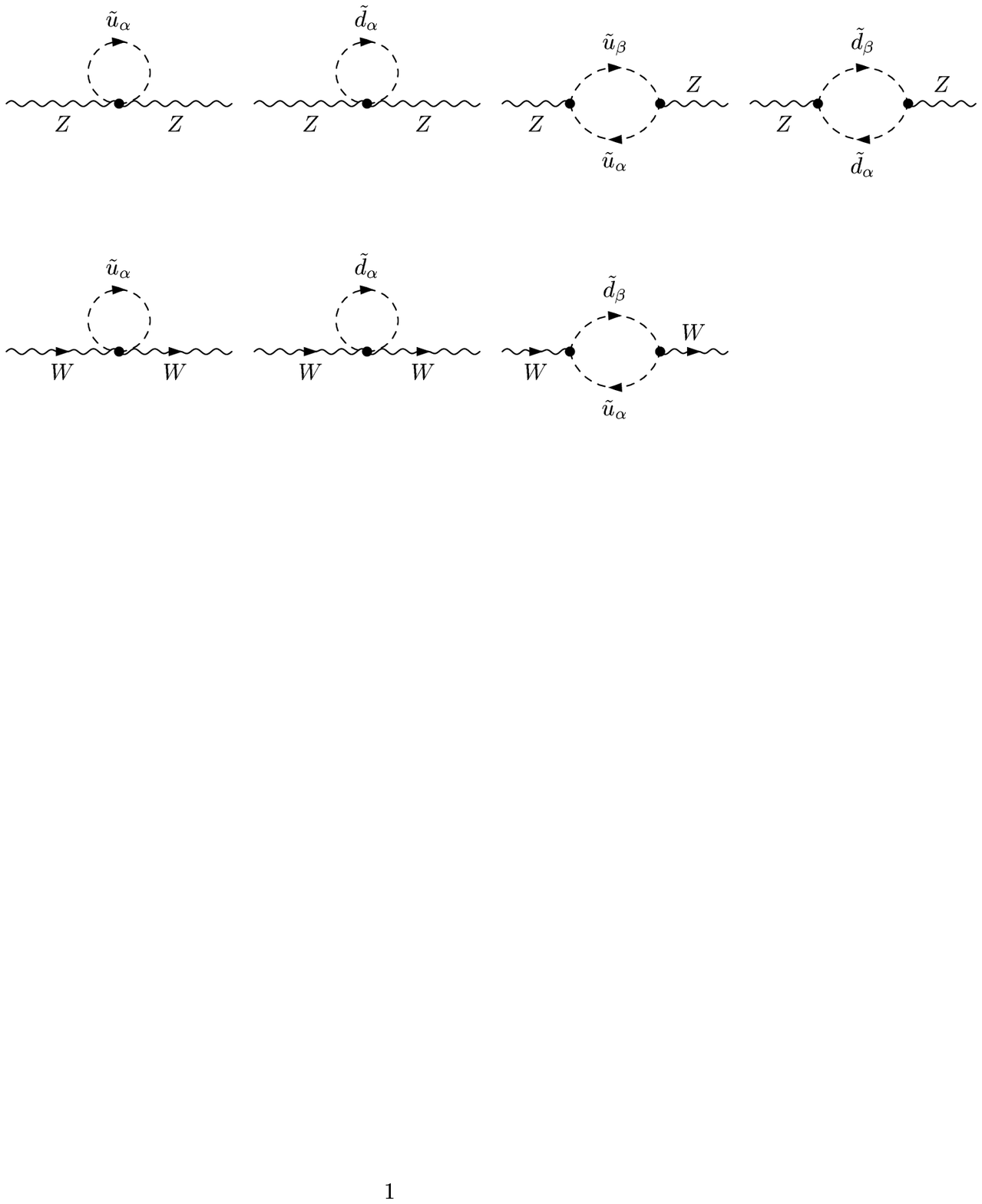}
\end{center}\vspace*{-1cm}
\caption[]{
Feynman diagrams for the squark contributions to the gauge
boson self-energies.}
\label{fig:FDVV}
\end{figure}

The existing corrections to $\De\rho$ within the NMFV MSSM~\cite{nmfv}
consist of squark contributions based on the general 
$4 \times 4$~mass matrix for both the 
$\Stop/\Scha$ and the $\Sbot/\Sstr$ sector, see \refse{subsec:nmfv}.
These corrections are visualized by  
the Feynman diagrams in~\reffi{fig:FDVV}. 
They are denoted as~$\De \rho ^{\tiq}$.

The squark contribution $\De \rho ^{\tiq}$ can be decomposed according to
\BEA
 \De \rho ^{\tiq} = \Xi_Z + \Theta_Z + \Xi_W + \Theta_W \,,
\label{rhosumme}
\EEA
where $\Xi$ and $\Theta$ correspond to different diagram
topologies, i.e.\ to diagrams with trilinear and quartic couplings, 
respectively (see \reffi{fig:FDVV}).
The explicit expressions read as follows,
\BEA
\Xi _W &=& 
 \frac{3 g^2}{8 \pi ^2 \MW^2} \sum \limits_{a,b,c,d} \sum \limits_{\al ,\be} 
 V_{\rm CKM}^{a b} V_{\rm CKM}^{c d} R_{\tiu}^{\al a} R_{\tiu}^{\al c} 
 R_{\tid}^{\be b} R_{\tid}^{\be d} 
 B_{00}(0, m_{\tiu_{\al}}^2, m_{\tid_{\be}}^2) \,,\nn \\
\Theta_W &=& 
  -\frac{3 g^2}{32 \pi ^2 \MW^2} 
  \sum \limits_a \sum \limits_{\al}
   \KKKL (R_{\tiu}^{\al a})^2  A_0(m_{\tiu_{\al}}^2) + 
         (R_{\tid}^{\al a})^2 A_0(m_{\tid_{\al}}^2)\KKKR \,,\nn \\
\Xi_Z &=& 
- \frac{3 g^2}{144 \cwq \pi^2 \MZ^2} \sum \limits_{\al,\be,\ga,\de} 
  \Big\{ \kappa_{\tid}(\ga) R_{\tid}^{\al \ga} R_{\tid}^{\be \ga} 
        \kappa_{\tid}(\de) R_{\tid}^{\al \de} R_{\tid}^{\be \de} 
  B_{00}(0, m_{\tid_{\al}}^2, m_{\tid_{\be}}^2)  \nn \\
&& +    \kappa_{\tiu}(\ga) R_{\tiu}^{\al \ga} R_{\tiu}^{\be \ga} 
        \kappa_{\tiu}(\de) R_{\tiu}^{\al \de} R_{\tiu}^{\be \de} 
  B_{00}(0, m_{\tiu_{\al}}^2, m_{\tiu_{\be}}^2) \Big\} \,,\nn \\
\Theta_Z &=& 
  \frac{3 g^2}{288 \cwq \pi^2 \MZ^2} \sum \limits_{\al,\be,\ga,\de} 
  \KKKL (\kappa_{\tid}(\ga)^2 (R_{\tid}^{\al \ga})^2 
         A_0(m_{\tid_{\al}}^2) + 
         \kappa_{\tiu}(\ga)^2 (R_{\tiu}^{\al \ga})^2 
         A_0(m_{\tiu_{\al}}^2) \KKKR~.
\EEA
Here the indices run from 1 to 2 for Latin letters, and from 1 to 4 for Greek 
letters. The expressions contain the one-point integral $A_0$ and 
the two-point integral $B_{00}$ in $B_{\mu\nu} (k) = g_{\mu\nu}\, B_{00} + 
k_{\mu} k_{\nu} B_{11}$ in the convention of~\citere{feynarts}. 
The remaining constants $\kappa_ {\tiu}$ and $\kappa_ {\tid}$
are defined as follows,
\BEA
\kappa_{\tid} = \VL
3 - 2 \,\swq \\ 3 - 2 \,\swq \\  -2\, \swq \\  -2\, \swq 
\VR~, \qquad \qquad
\kappa_{\tiu} = \VL
-3 + 4 \,\swq \\ -3 + 4 \,\swq \\ 4 \,\swq \\ 4\, \swq
\VR~.
\EEA

The CKM matrix only affects $\Xi_W$. 
Corrections from the first-generation squarks are negligible
due to their very small mass splitting.
Non-minimal flavor mixing of the first generation with the other ones has
been set to zero, 
but conventional CKM mixing is basically present. Although it is required 
for a UV finite result, it yields only negligibly small effects. Therefore,
for simplification, we drop the first generation and restore the
cancellation of UV divergences by a unitary $2 \times 2$ matrix
replacing the \{23\}-submatrix of the CKM matrix,
\BEA
V_{\rm CKM} = \ML
V_{cs} & V_{cb} \\
V_{ts} & V_{tb} \MR 
= \ML
\cos \epsilon &\sin \epsilon \\
-\sin\epsilon &\cos\epsilon
\MR~,
\label{eq:CKM}
\EEA
with $|\epsilon| \approx 0.04$ close
to the experimental entries~\cite{pdg} of the conventional CKM matrix. 

Since $\De\rho^{\tiq}$ is a finite quantity, and the CKM matrix
effects (and therefore, the $\epsilon$ dependence) only appear in $\Xi_W$, 
it has been shown~\cite{nmfv} that $\Xi_W$ (and thus $\De\rho$) is
symmetric under the simultaneous reversal of signs 
$\epsilon \to -\epsilon$, $\la \to -\la$ (see \refeq{def:lambda}),
i.e.\ only the relative sign has a physical consequence, affecting
the results for $\De\rho$ significantly.
In physical terms, non-minimal squark mixing can either strengthen or
partially compensate the CKM mixing.


\section{Evaluation of $\MW$}
\label{sec:evalMW}

One of the most important quantities for testing the SM or its extensions
is the relation between the massive gauge boson masses, $\MW$ and $\MZ$,
in terms
of the Fermi constant, $\Gmu$, and the fine structure constant,
$\al$. This relation can be derived from muon decay, where the Fermi
constant
enters the muon lifetime, $\tau_{\mu}$, via the expression
\beq
\tau_{\mu}^{-1} = \frac{\Gmu^2 \, m_\mu^5}{192 \pi^3} \;
F\left(\frac{m_{\mathrm{e}}^2}{m_\mu^2}\right)
\left(1 + \frac{3}{5} \frac{m_\mu^2}{\MW^2} \right)
\left(1 + \Delta q \right) ,
\label{eq:fermi}
\end{equation}
with $F(x) = 1 - 8 x - 12 x^2 \ln x + 8 x^3 - x^4$. By convention, this
defining equation is supplemented with the QED corrections within the
Fermi Model,
$\Delta q$. Results for $\Delta q$ have been available for a long time at
the one-loop~\cite{delqol} and, more recently, at the two-loop
level~\cite{delqtl} (the error in the two-loop term is from the hadronic
uncertainty), 
\begin{equation}
\Delta q = 1.810\, \alpi+(6.701\,\pm 0.002)
        \left(\alpi\right)^2 \, . 
\label{eq:delq}
\end{equation}
Commonly,
tree-level $W$~propagator effects giving
rise to the (numerically insignificant) term $3 m_\mu^2/(5 \MW^2)$ in
\refeq{eq:fermi} are also included in the definition of
$\Gmu$, although they are not part of the Fermi Model prediction.
With the second order term of \refeq{eq:delq} the defining equation for
$\Gmu$ in terms of the experimental muon lifetime, \refeq{eq:fermi},
yields the value of $\Gmu$ given in \refta{tab:POstatus}.

 Within a given model, $\Gmu$ can be calculated in terms of the model
 parameters.
 The Fermi
 constant is given by the  expression 
 \beq
 \label{gfermi}
 \frac{\Gmu}{\sqrt{2}} \, = \,
 \frac{e_0^2}{8\swo \mwb} \left[ 1+\frac{\Sigma^W(0)}{\MW^2}
 +\,(VB) \right] \, .
 \end{equation}
 This equation  contains the bare parameters
 with  the bare mixing angle.
 The term $(VB)$ schematically summarizes the vertex
 corrections and box diagrams in  the decay
 amplitude.
 A set of infrared-divergent ``QED correction'' graphs has
 been   removed from this class of diagrams. These left-out
 diagrams, together with the real bremsstrahlung contributions,
 reproduce the QED correction factor of the Fermi
 model result in \refeqs{eq:fermi}, (\ref{eq:delq}) 
and therefore have no influence on the
 relation between $\Gmu$ and the model parameters.

 Equation (\ref{gfermi}) contains the bare parameters $e_0, \MW^0, \sw^0$.
 Expanding the bare
 parameters  and keeping only terms of one-loop order yields the expression,
 \bea
 \label{gfermi1}
 \frac{\Gmu}{\sqrt{2}} & = & \frac{e^2}{8\sw^2 \MW^2}  \nn \\
     &   & \times \left[
  1+2\frac{\delta e}{e} -\frac{\cw^2}{\sw^2} \left(\dmz-\dmw\right)
  +\frac{\Sigma^W(0)-\dmmw}{\MW^2} +(VB) \right] \nn \\
  & \equiv & \frac{e^2}{8\sw^2 \MW^2}\, \left(
         1+ \Dr\right) \, ,
\label{eq:delr}
 \eea
which is equivalent to \refeq{calcMW}.
 The  quantity $\Dr$ is the finite combination
 of loop diagrams and counterterms in (\ref{gfermi1}).
The prediction for $\MW$ within the SM or the MSSM is obtained from
evaluating $\De r$ in these models and solving \refeq{eq:delr},
\begin{equation}
\MW^2 = \MZ^2 \left\{\frac{1}{2} + \sqrt{\frac{1}{4} -
        \frac{\pi \al}{\sqrt{2} \GF \MZ^2} \Bigl[1 + \De r
        (\MW, \MZ, \mt, \dots) \Bigr]}\, \right\} .
\label{eq:MW}
\end{equation}
In practice, this can be 
done by an iterative procedure since $\De r$ itself depends on
$\MW$.

The one-loop contributions to $\De r$ can be written as
\BE
\De r = \De\al - \frac{\cw^2}{\sw^2}\De\rho
        +(\De r)_{\rm rem},
\label{defdeltar2}
\end{equation}
where $\De\al$ is the shift in the fine structure constant due to the
light fermions of the SM, $\De\al \propto \log m_f$ (see the
discussion in \refse{subsec:ch-renorm}), and $\De\rho$ is
the leading contribution to the $\rho$ parameter from fermion and
sfermion loops. The remainder part,
$(\De r)_{\rm rem}$, contains in particular the contributions from the
Higgs sector.

In the following we will discuss the status of the theoretical
evaluation of $\MW$. After a brief review of the SM contribution, the
additional MSSM corrections are described in more detail.


\subsection{SM corrections}
\label{subsec:MWSMeval}

 In the \sm, the result for $(VB)$ in \refeq{gfermi1} is
 \beq
 \label{vertexbox}
  (VB)   =   \frac{\al}{\pi \sw^2} \left(
  \Delta -\log\frac{\MW^2}{\mu^2} \right)
  +\frac{\al}{4\pi \sw^2} \left( 6+\frac{7-4\sw^2}{2\sw^2}
   \log \cw^2 \right) .
 \end{equation}
 The singular part of this equation involving the 
 divergence $\Delta \equiv 2/(4 -D) - \gamma + \log 4 \pi$ (see 
 App.~\ref{app:loopint})
 coincides, up to a factor, with the
 non-Abelian bosonic contribution to the charge counterterm
 in \refeq{chargeren}:
 \beq
   \frac{\al}{\pi \sw^2} \left(
  \Delta -\log\frac{\MW^2}{\mu^2} \right)  \, = \,
  \frac{2}{\cw \sw}\, \frac{\Sigma^{\g Z}(0)}{\MZ^2} \, .
 \end{equation}
 Extra non-standard vertex and box diagrams do not change the singular
part; they contribute another finite term $(VB)_{\rm non-standard}$.
 Together with \refeq{chargeren} and \refeq{vertexbox}, 
 we obtain from \refeq{gfermi1} the following expression,
 \bea
 \label{deltar1loop}
 \Dr & = & \Pi^{\g}(0) -\frac{\cw^2}{\sw^2} \left(\dmz-\dmw\right)
           +\frac{\Sigma^W(0)-\dmmw}{\MW^2}   \nn    \\
     &   &\mbox{} + 2\,\frac{\cw}{\sw}\,\frac{\Sigma^{\g Z}(0)}{\MZ^2} \, +\,
 \frac{\al}{4\pi \sw^2} \left( 6+
 \frac{7-4 \sw^2}{2 \sw^2} \log \cw^2   \right) \, 
 + (VB)_{\rm non-standard}~,
 \eea
where in the on-shell renormalization the renormalization constants
are given by the on-shell self-energies, as specified in~\refeq{massren}.

Beyond the complete \onel\ result~\cite{sirlin}, resummations of the
leading \onel\
contributions $\De\al$ and $\De\rho$ are known~\cite{resum}.
They correctly take into account the terms of the form 
$(\De\al)^2$, $(\De\rho)^2$, $(\De\al\De\rho)$, and
$(\De\al\De r_{\mathrm{rem}})$ at the two-loop level and the leading
powers in $\De\al$ to all orders. 

Higher-order QCD corrections to $\De r$ are known at
\order{\al\als}~\cite{drSMgfals,deltarSMgfals,twoloopint} and
\order{\al\als^2}~\cite{drSMgfals2,drSMgfals2LF} since about 10~years. 
Recently the full electroweak two-loop result for $\De r$ has been
completed. It consists of the fermionic
contribution~\cite{Freitas:2002ja,deltarSMgf2,2lfermc}, which involves
diagrams with one or two closed fermion loops, and the purely bosonic
two-loop contribution~\cite{deltarSM2Lbos}.

Beyond two-loop order, besides higher-order contributions to $\De\rho$
(see \refse{subsec:beyondoneloop}) the results for the pure fermion-loop
corrections (i.e.\ contributions containing $n$ fermion loops at
$n$-loop order) are known up to four-loop order~\cite{floops}. They
contain in particular the leading contributions in $\De\al$ and
$\De\rho$.

Since the full result for $\MW$ is rather lenghty and 
contains numerical integrations of integrals appearing in the
electroweak two-loop contributions, a simple parametrization is given in
\citere{deMWSMtheo}.  It approximates the
full result for $\MW$ to better than 0.5~MeV for $10 \gev \leq \MH \leq
1 \tev$ if the other parameters are varied within their combined $2 \si$
region around their experimental central values.

The expected size of the unknown higher-order corrections, i.e.\ the
estimated theory uncertainties~\cite{deMWSMtheo} (for $\MH \lsim 300$~GeV) 
are summarized in \refta{higherorderunc} (see
Ref.~\cite{blueband,deMWSMtheo,epssmgw} for further details). 

\begin{table}[h!]
\renewcommand{\arraystretch}{1.5}
\begin{tabular}{|c|c|c|c|c|}
\hline
 \multicolumn{2}{|c|}{2-loop} &
 \multicolumn{3}{c|}{3-loop} \\
\hline
   ~\order{\al^2, {\rm ferm}}~ & ~\order{\al^2, {\rm bos}}~ 
 & ~\order{\al\als^2, {\rm ferm}}~ & ~\order{\Gmu^2\als \mt^2\MZ^2}~ 
 & ~\order{\al^3}~ \\ 
                                                    \hline\hline
   compl.~\cite{Freitas:2002ja,deltarSMgf2,2lfermc} & 
   compl.~\cite{deltarSM2Lbos}
 & compl.~\cite{drSMgfals2,drSMgfals2LF} & 3.0
 & 1.5 \\ \hline
\end{tabular}

\vspace{.5em}

\begin{tabular}{|c|c|}
\hline
 \multicolumn{2}{|c|}{4-loop} \\
\hline
   ~\order{\Gmu\als^3\mt^2}~ &
   ~\order{\Gmu^2\als^2\mt^4}~ \\
                                                    \hline\hline
   1.3 & 1.4 \\ \hline
\end{tabular}
\renewcommand{\arraystretch}{1}
\caption{
Estimated uncertainties from
unknown higher-order corrections to $\MW$ in MeV~\cite{deMWSMtheo}.
}
\label{higherorderunc}
\end{table}

\noindent
Currently these intrinsic uncertainties result in~\cite{deMWSMtheo} 
\BE
\de\MW^{\rm SM,intr} ~({\rm current}) = 4 \mev ~.
\end{equation}
It seems reasonable that the evaluation of further higher-order
corrections will lead to a reduction of this uncertainty by a factor of
two or more on the timescale of 5--10 years. We therefore estimate as
future intrinsic uncertainty
\BE
\de\MW^{\rm SM,intr} ~({\rm future}) = 2 \mev ~.
\label{MWfutureunc}
\end{equation}

The dominant theoretical uncertainty at present is the uncertainty
induced by the experimental errors of the input parameters.
The most important uncertainties arise from the experimental error of
the top-quark mass and the hadronic contribution to the shift in the
fine structure constant. The current
errors for $\mt$~\cite{mtexp} and $\De\al_{\rm had}$~\cite{Burkhardt:2001xp} 
induce the following parametric uncertainties
\BEA
\de\mt^{\rm current} = 4.3 \gev &\Rightarrow&
\De\MW^{{\rm para},\mt} ~({\rm current}) \approx 26 \mev,  \\[.3em]
\de(\De\al_{\rm had}^{\rm current}) = 36 \times 10^{-5} &\Rightarrow&
\De\MW^{{\rm para},\De\al_{\rm had}} ~({\rm current}) \approx 6.5 \mev .
\EEA
At the ILC, the top-quark mass will be measured with an accuracy of
about 100~MeV~\cite{teslatdr,orangebook,acfarep}. The parametric 
uncertainties induced by the
future experimental errors of $\mt$ and $\De\al_{\rm had}$~\cite{fredl}
will then be~\cite{deltamt}
\BEA
\de\mt^{\rm future} = 0.1 \gev &\Rightarrow&
\De\MW^{{\rm para},\mt} ~({\rm future}) \approx 1 \mev,  \\[.3em]
\de(\De\al_{\rm had}^{\rm future}) = 5 \times 10^{-5} &\Rightarrow&
\De\MW^{{\rm para},\De\al_{\rm had}} ~({\rm future}) \approx 1 \mev .
\EEA
Thus, the precision measurement of the top-quark mass at the ILC and 
prospective improvements in the determination of $\De\al_{\rm had}$
(see the discussion in \citere{fredl}) will reduce the parametric
uncertainties to the same level as the prospective intrinsic
uncertainties, \refeq{MWfutureunc}, allowing a very sensitive test of
the electroweak theory.


\subsection{SUSY corrections}

In this subsection we review the current status of the SUSY
corrections to $\MW$. The intrinsic uncertainties from missing
higher-order SUSY corrections will be discussed in
\refse{subsec:MWsweffuncSUSY}. 

\subsubsection{One-loop corrections}

\begin{figure}[t!]
\begin{center}
\mbox{
\psfig{figure=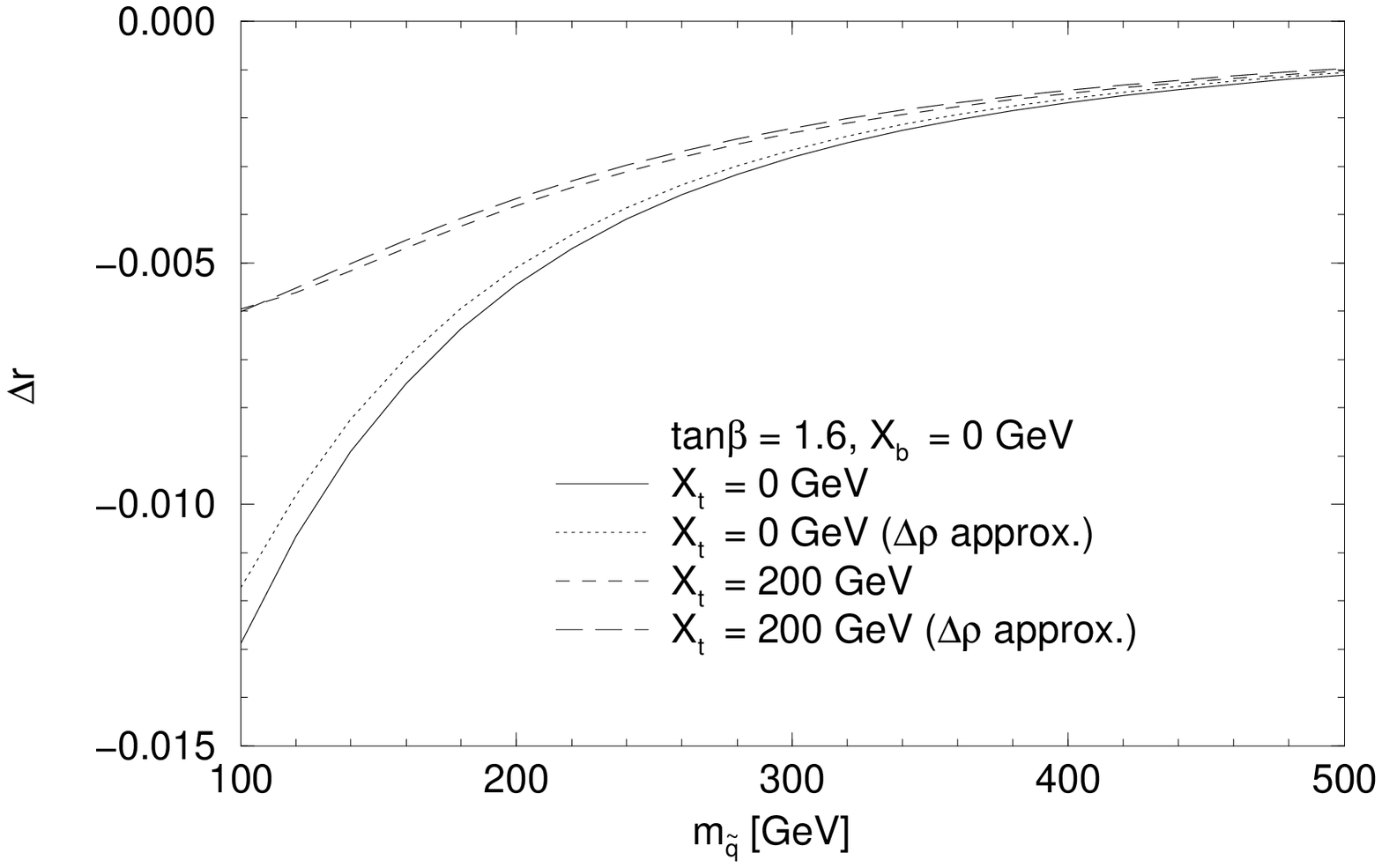,width=9cm,height=9cm,
       bbllx=120pt,bblly=70pt,bburx=450pt,bbury=420pt}
     }
\vspace{-1em}
\caption{
The $\Stop/\Sbot$ corrections to $\De r$ at the \onel\ level,
\refeq{deltarmssm},  are compared with the approximation,
\refeq{deltarapprox}. 
The results are shown as a function of $\msq (\equiv \msusy)$ for $\tb = 1.6$, 
$\Xb = 0$ and $\Xt = 0, 200 \gev$.}
\label{fig:deltar1lA}
\end{center}
\end{figure}
\begin{figure}[h!]
\begin{center}
\mbox{
\psfig{figure=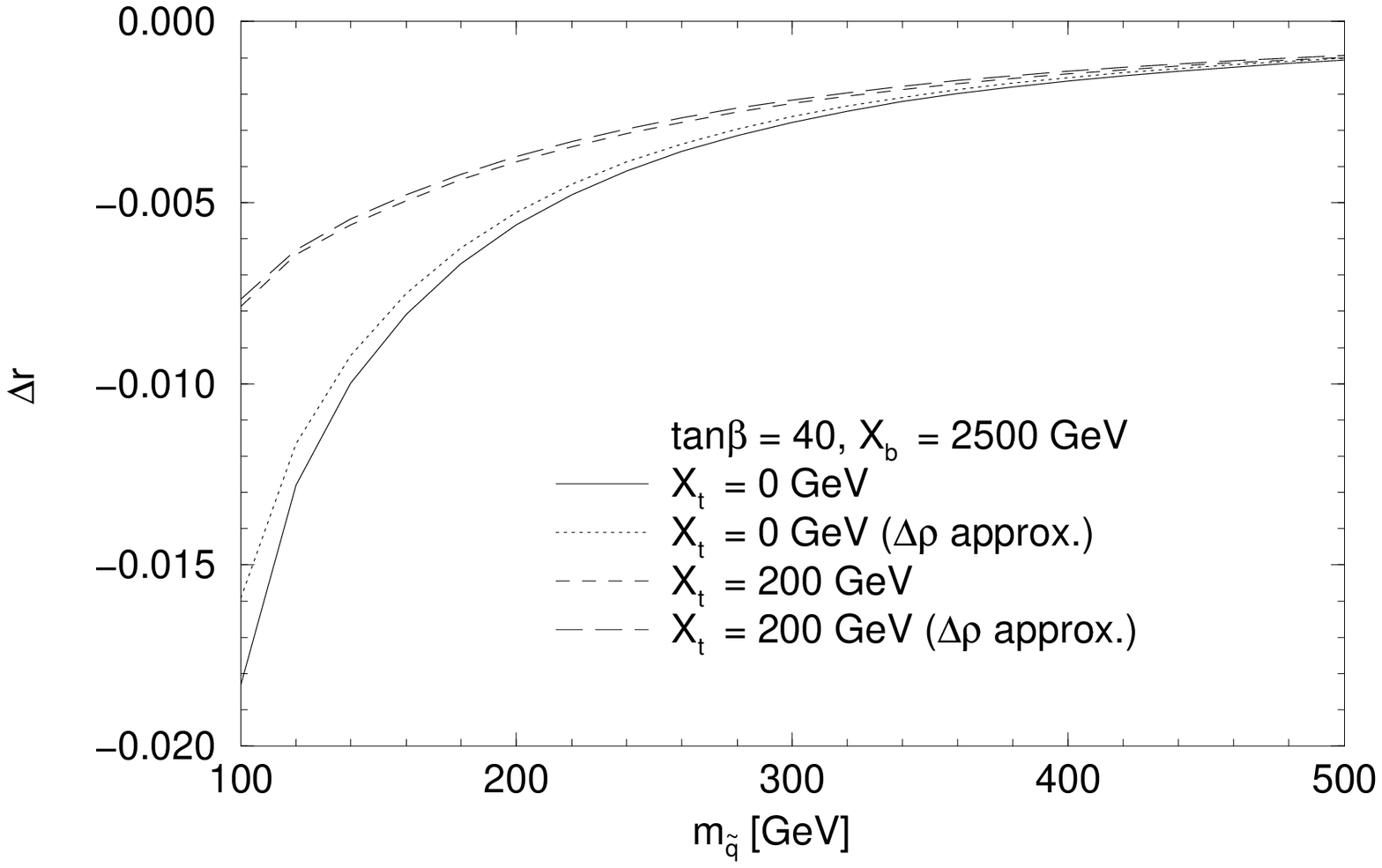,width=9cm,height=9cm,
       bbllx=120pt,bblly=70pt,bburx=450pt,bbury=420pt}
     }
\vspace{-1em}
\caption{
The $\Stop/\Sbot$ corrections to $\De r$ at the \onel\ level,
\refeq{deltarmssm},  are compared with the approximation,
\refeq{deltarapprox}. 
The results are shown as a function of $\msq (\equiv \msusy)$ for $\tb = 40$, 
$\Xb = 2500 \gev$ and $\Xt = 0, 200 \gev$.}
\label{fig:deltar1lB}
\vspace{-1em}
\end{center}
\end{figure}

The complete \onel\ corrections to $\De r_{\SU}$~%
\footnote{
From here on we drop the subscript $\mbox{}_{\SU}$ .%
}%
~were evaluated independently by two
groups~\cite{deltarMSSM1lA,deltarMSSM1lB}. The main part of the
contributions stems from the $\Stop/\Sbot$ doublet that enters at the
\onel\ level only via gauge-boson self-energies. Therefore, only Feynman 
diagrams
as depicted in \reffi{fig:fdgb1l} have to be evaluated, but contrary
to \refse{subsec:beyondoneloop} also with non-vanishing external
momentum. 
In general, all scalar-quark contributions
(yielding $\Si^{\ga Z}(0) = 0$, according to the comment
after \refeq{gammaZmixing}) are contained in
\BE
\De r^{\sq} = \Pi^{\ga}(0) - \frac{\cw^2}{\sw^2}
        \KL \frac{\de \MZ^2}{\MZ^2}
          - \frac{\de \MW^2}{\MW^2} \KR 
          + \frac{\Si^W(0) - \de \MW^2}{\MW^2}.
\label{deltarmssm}
\end{equation}
In the approximation of neglecting the external momenta in the
self-energies the second term in \refeq{deltarmssm} reduces to 
$\De\rho$, leading to a decomposition as in 
\refeq{defdeltar2}. For loops of scalar quarks the corrections mainly 
arise from the contribution to $\De\rho$ (see \refeq{delrhoMSSM1l}), 
so that $\De r^{\sq}$ can be approximated as
\BE
\De r^{\sq} \, \approx \, - \frac{\cw^2}{\sw^2} \; \De\rho 
      \, \approx \, - 3.5 \De\rho .
\label{deltarapprox}
\end{equation}
The full \onel\ result from the $\Stop/\Sbot$~sector is compared with this
approximation in \reffis{fig:deltar1lA} and \ref{fig:deltar1lB}. The
case of no-mixing in the $\Sbot$~sector is shown in
\reffi{fig:deltar1lA} for $\tb = 1.6$ and $\Xt = 0, 200 \gev$. The
full result is reproduced by the $\De\rho$~approximation within a few
per cent. The same applies for large mixing in the $\Sbot$~sector, see
\reffi{fig:deltar1lB}, with $\Xb = 2500 \gev$, $\tb = 40$ and 
$\Xt = 0, 200 \gev$. 

\bigskip
As investigated in detail in \citeres{deltarMSSM1lA,deltarMSSM1lB},
the full SUSY \onel\ contribution to $\De r$ does not exceed
\order{0.0015} (explicit formulas for the self-energy contributions
are given in \citere{MSSMSE1l}, see also \citeres{dr1lA,dr1lB}.) 
The main contribution is given by the universal corrections, see
\refeq{deltar1loop}. The corrections beyond the $\Stop/\Sbot$~sector
arise from the other scalar quarks, entering only in the universal
corrections, and the sleptons and gauginos, entering in the universal
as well as in the non-universal contributions~\cite{deltarMSSM1lA}. 

\bigskip
The full one-loop corrections from third and second generation squarks
in the NMFV MSSM, using \refeq{deltarmssm}, have been derived in
\citere{nmfv}.


\subsubsection{Corrections beyond one-loop}
\label{subsubsec:MWbeyond1LSUSY}

Since the dominant \onel\ corrections are given by the
$\Stop/\Sbot$~contributions, the existing \twol\ calculations have
focused on this sector. The only existing \twol\ calculation
for $\De r$, going beyond the $\De\rho$ approximation as presented in
\refse{subsubsec:delrhomssm}, are the gluon-exchange corrections
of \order{\al\als}~\cite{dr2lB}. This is the only result in the
$\Stop/\Sbot$~sector beyond \onel\ order that can be obtained as an
analytical formula due to the presence of the massless gluon in the
\twol\ two-point function. The gluino-exchange corrections, on the
other hand, have been shown to decouple for large $\mgl$~\cite{dr2lA},
see \refse{subsubsec:delrhomssm}. 

The \order{\al\als} gluonic corrections are evaluated from the Feynman
diagrams as shown in \reffi{fig:fdgb2l}, but taking into account the
momentum dependence.
Furthermore the derivative of the photon
self-energy is needed. It is given by ($D = 4 - 2\de$)
\BEA
\Pi^{\ga}(0) &=& -C_F \frac{\al}{\pi} \frac{\als}{\pi} 
                \frac{3}{4 \sw^2}  
                \sum_{f=1,2} \sum_{a=1,2} v_f (1 - \de) \de \non\\
  & & \KKL \frac{(1 - 2 \de) (1 - \de)^2}
                {(2 - \de) (1 - 4 \de^2)} 
           \frac{\KL A_0(\msfi) \KR^2}{\msfi^4} +
           \frac{A_0(\msfi) B_0(\msfi^2, 0, \msfi)}{\msfi^2} \KKR
\label{gpsegluonp0}
\EEA
with
\BEA
  & & v_1 = (-\frac{4}{3} \sw \cw)^2, v_2 = (\frac{2}{3} \sw \cw)^2\non\\
  & & \msfi = \KKKL \begin{array}{r@{\quad:\quad}l}
              \msti & f = 1 \\
              \msbi & f = 2 
                    \end{array} \right. \non
\EEA
The most complicated parts are the gauge boson self-energies with
non-zero external momentum. The general case is given by
\BEA
\Si^{V_1V_2}(p^2) &=& C_F \frac{\al}{\pi} \frac{\als}{\pi} 
                \frac{3}{4 \sw^2} \tilde{g}_{V_1V_2} 
                \sum_{f=1,2} \quad \sum_{a,b=1,2} 
                g_{ab}^{V_1V_2} \frac{1}{3 - 2 \de} \Bigg[ \non\\
  & & -\frac{m_a^2}{2} F_{ab} T_{11234'}(m_a^2, m_a^2, \bar \mb^2, m_a^2, 0) 
      -\frac{1}{2} F_{ab} T_{1234'}(m_a^2, \bar m_b^2, m_a^2, 0) \non\\
  & & -\frac{m_a^2 + \bar m_b^2 - p^2}{8} F_{ab}
             T_{123'45}(m_a^2, \bar m_b^2, 0, m_a^2, \bar m_b^2)
      +\frac{1}{2} (1-\de) T_{234'}(m_a^2, \bar m_b^2, 0) \non\\
  & & +\frac{m_a^2}{2} F_{ab} B'_0(p^2, m_a^2, m_a^2, \bar m_b^2) 
                              B_0(m_a^2, 0, m_a^2)
      -\frac{m_a^2 + \bar m_b^2 - p^2}{4} 
             \KL B_0(p^2, m_a^2, \bar m_b^2) \KR ^2 \non\\
  & & +\frac{m_a^2 - \bar m_b^2 - p^2}{p^2} m_a^2 
       B_0(m_a^2, 0, m_a^2) B_0(p^2; m_a^2, \bar m_b^2) \non\\
  & & -\frac{1}{2} \KL \frac{m_a^2}{p^2} (A_0(m_a) - A_0(\bar m_b))
                        - 2 (1 - \de)^2 A_0(m_a)
                        + (1 - \de) \frac{m_a^2 - \bar m_b^2}{p^2}
                          A_0(m_a) \KR \non\\
  & & \qquad B_0(m_a^2, 0, m_a^2) \non\\
  & & -\frac{1}{2} \KL 2 (1 - \de) - \frac{m_a^2 - \bar m_b^2}{p^2} \KR
        \frac{(1 - \de)^2}{(1 - 2 \de) m_a^2} A_0^2(m_a) 
      \Bigg],
\label{wzsegluonp2}
\EEA
with
$$
F_{ab} = \frac{((m_a - \bar m_b)^2 - p^2) 
               ((m_a + \bar m_b)^2 - p^2)}{p^2}
$$

\noindent
and
\begin{itemize}
\item $V_1, V_2 \in \KKKL \gamma, Z^0 \KKKR $ \\
\BEA
g_{ab}^{V_1V_2} &=& \KKL (2 t_a t_b - \de_{ab}) a_{V_1}^f 
                          - \de_{ab} v_{V_1}^f \KKR
                    \KKL (2 t_a t_b - \de_{ab}) a_{V_2}^f
                          - \de_{ab} v_{V_2}^f \KKR \non\\
 && \mbox{with} \non \\
 && t_i = \KKKL \begin{array}{r@{\quad:\quad}l}
              \costf & i = 1 \\
              \sintf & i = 2 
              \end{array} \right. , \non\\
 && a_Z^f = I_3^f, \quad v_Z^f = I_3^f - 2 \sw^2 Q_f, \non\\
 && a_\ga^f = 0, \quad v_\ga^f = 2 \sw\cw Q_f, \non \\
\tilde{g}_{V_1V_2} &=& 1, \non\\
m_i &=& \bar m_i = \KKKL \begin{array}{r@{\quad:\quad}l}
              \msti & f = 1 \\
              \msbi & f = 2 
              \end{array} \right. \non
\EEA

\item $V_1 = V_2 = W $
\BEA
g_{ab}^{WW} &=& s_a^2 t_b^2, \non\\
 && \mbox{with} \non \\
 && s_a = \KKKL \begin{array}{r@{\quad:\quad}l}
              \costt & a = 1 \\
              \sintt & a = 2 
              \end{array} \right. ,\quad
t_b  =  \KKKL \begin{array}{r@{\quad:\quad}l}
              \costb & b = 1 \\
              \sintb & b = 2 
              \end{array} \right. ,\non\\
\tilde{g}_{WW} &=& \frac{1}{\cw^2}, \non\\
m_i &=& \KKKL \begin{array}{r@{\quad:\quad}l}
              \msti & f = 1 \\
              \msbi & f = 2 
              \end{array} \right. ,\quad
\bar m_j  =  \KKKL \begin{array}{r@{\quad:\quad}l}
              \msbj & f = 1 \\
              \mstj & f = 2 
              \end{array} \right. .\non
\EEA
\end{itemize}
The functions $A_0$, $B_0$~\cite{a0b0c0d0}, $T_{234'}$, $T_{123'4}$,
$T_{1123'4}$ and $T_{123'45}$~\cite{twoloopint,t134} can be found in
the Appendix.


\section{Evaluation of $Z$~boson observables}
\label{sec:Zobs}

The measurement of the $Z$-boson mass from the $Z$ lineshape at LEP
has provided us with an additional precise input parameter besides
 $\al$ and $\Gmu$. Other observable quantities from the $Z$ peak
allow us to perform precision tests
of the electroweak theory by comparison with the theoretical predictions
given by specific models.
At the $Z$~boson resonance in $\epm$ annihilation, 
two classes of precision observables are
available: 
\begin{itemize}
\item[a)] inclusive quantities:
  \begin{itemize}
   \item[$\bullet$] the partial leptonic and hadronic decay width 
                    $\Ga_f$,
   \item[$\bullet$] the total decay width $\Ga_Z$,
   \item[$\bullet$] the hadronic peak cross section $\si_h$,
   \item[$\bullet$] the ratio of the hadronic to the electronic decay
                  width of the $Z$~boson, $R_h$,
   \item[$\bullet$] the ratio of the partial decay width for 
                    $Z\to c\bar{c} \, ( b \bar{b} )$ 
                   to the hadronic width, $R_{c(b)}$.
  \end{itemize}
\item[b)] asymmetries and weak mixing angles:
  \begin{itemize}
   \item[$\bullet$] the \it forward-backward \rm asymmetries $\AFB^f$,
   \item[$\bullet$] the \it left-right \rm  asymmetries $\ALR^f$,
   \item[$\bullet$] the $\tau$ polarization $P_\tau$,
   \item[$\bullet$] the effective weak mixing angle $\sweff$.
  \end{itemize}
\end{itemize}
All these quantities can be written in a transparent way with the 
help of effective
vector and axial vector couplings, which comprise
the genuine electroweak loop contributions, besides 
those from the QED virtual-photon corrections, which are
the same in the SM and in supersymmetric extensions.


\subsection{The effective $Z f \bar{f}$ couplings}

 The structure of the resonating $Z$ amplitude  allows us to define
 neutral-current 
 (NC) vertices at the $Z$ peak
 with effective coupling constants
 $g_{\rm V,A}^f$, equivalently to the use of $\rho_f, \kappa_f$:
 \bea
 \label{nccoup}
  \Ga_{\m}^{\rm NC} & = & \left( \sqrt{2}\Gmu\mz \rho_f \right)^{1/2}
 \left[ (I_3^f-2Q_f\sw^2\kappa_f)\gamu-I_3^f\gamu\gafi \right] \nn\\
   & = & \left( \sqrt{2}\Gmu\mz \right)^{1/2} \,
   \left( g_{\rm V}^f \,\gamu -  g_{\rm A}^f \,\gamu\gafi \right)  \, .
 \eea
  The complete expressions for the
  effective couplings read as follows:
 \bea
 \label{effcoup}
  g_{\rm V}^f & = &  \left( v_f +2\sw \cw \, Q_f \frac{\Pgzrz}{1+\Pigr(\mz)}
              + \fvzf \right) \left( \frac
               {1-\Dr}{1+\Pizr(\mz)} \right)^{1/2}, \nn \\
  g_{\rm A}^f & = &  \left( a_f
              + \fazf \right) \left( \frac
               {1-\Dr}{1+\Pizr(\mz)} \right)^{1/2} \, .
 \eea
 Besides $\Dr$,
 the building blocks
 are the following finite combinations of two-point
 functions evaluated at $s=\mz$:
 \bea
 \Pir^Z(s) & = & \frac{\real\,\Sigma^Z(s)-\dmmz}{s-\mz}- \Pi^{\g}(0)
                                             \nn            \\
     & &     +\,\frac{\cw^2-\sw^2}{\sw^2} \left(\dmz-\dmw
         - 2\,\frac{\sw}{\cw} \,\frac{\Sigma^{\g Z}(0)}{\mz}
        \right) \, ,  \nn \\[0.2cm]
 \hat{\Pi}^{\g Z}(s)
     & = & \frac{\sgz(s)-\sgz(0)}{s}\,-\,\frac{\cw}{\sw}
    \left(\dmz-\dmw\right) +\,2\,\frac{\sgz(0)}{\mz}        \nn \\
 \eea
 and the finite form factors $F_{\rm V,A}$
 at $s=\mz$ from the vertex corrections 
$\Lambda_{\mu}^{\mbox{\scriptsize (1-loop)}}$ (including the
 external-fermion wave function  renormalizations), 
 \bea
  \Lambda_{\mu}^{\mbox{\scriptsize (1-loop)}} 
  & = & \frac{e}{2\sw\cw}  
    \left( \gamu\fvzf(s)
                - \gamu\gafi\fazf(s) +\,I_3^f \gamu(1-\gafi) \,
                \frac{\cw}{\sw} \,\frac{\sgz(0)}{\mz} \right) \, . 
              \\  & &  \nn
 \eea
  For the explicit expressions for the self-energies and the vertex 
 corrections including the MSSM contributions see
  \citeres{MSSMSE1l,hollik1990,pierce,Zobsfull1l}. 
 Owing to the imaginary parts of the self-energies and vertices,
 the form factors and the effective couplings, respectively, are
 complex quantities.

 \bigskip \noi
 {\em Effective Mixing Angles.}
 We can define 
 effective mixing angles for a given fermion species $f$ 
 according to
 \beq
 \label{sineffective}
  \sin^2\theta_f = 
   \frac{1}{4\mid Q_f\mid} \left(
    1-\re\frac{g_{\rm V}^f}{g_{\rm A}^f} \right) \, ,
 \end{equation}
 from the effective coupling constants in (\ref{effcoup}).
 They are of particular interest since they determine the on-resonance
 asymmetries.
A special case is the effective mixing angle for the light leptons
($f =\ell$), which is commonly denoted as the effective mixing angle
(assuming lepton universality),
\beq
 \sin^2\theta_{\rm eff} =  \sin^2\theta_\ell \, ,
\end{equation}
as, e.g., in the analysis of experimental data from LEP 
and SLC~\cite{lepewwg}. 

\subsection{$Z$~boson observables}
\label{subsec:Zobs}

 From lineshape measurements one obtains the parameters
 $M_Z,\, \Ga_Z,\, \sigma_0$, or the partial widths.
 Here $M_Z$ will be used as a precise input parameter, together
 with $\al$ and $\Gmu$; the width and partial widths are
specific model predictions.

 \medskip 
 The total
 $Z$ width $\Gamma_Z$ can be calculated
 as the sum over the partial decay widths
 \beq
\label{totalwidth}
  \Gamma_Z = \sum_f  \Ga_f\, , \quad \Ga_f =
  \Gamma  (Z\ra f\bar{f}) 
 \end{equation}
 (other decay channels are not significant).
  The fermionic partial
 widths,
  when
 expressed in terms of the effective coupling constants defined
 in (\ref{effcoup}), read
 \bea
 \label{partialwidth} 
 \Gamma_f  & = &  \Ga_0 \,\sqrt{1-\frac{4m_f^2}{\mz}}
  \, \left[
  \mid g_{\rm V}^f \mid ^2 \left(1+\frac{2m_f^2}{\mz}\right) +
  \mid g_{\rm A}^f\mid ^2 \left(1-\frac{4m_f^2}{\mz}\right)
                            \right] 
                          \nn \\[0.2cm]
   &   &   \times \,  (1+\delta_{\rm QED})
      \, +\, \Delta\Ga^f_{\rm QCD}     \\[0.2cm]
   & \simeq & \Ga_0
  \, \left[
  \mid g_{\rm V}^f\mid ^2  +
  \mid g_{\rm A}^f\mid ^2 \left(1-\frac{6m_f^2}{\mz}\right)
                            \right]
  \,  (1+\delta_{\rm QED})
      \, +\, \Delta\Ga^f_{\rm QCD} \nn
 \eea
 with
 \beq
 \Ga_0 \, =\,
   N_{\rm C}^f\,\frac{\sqrt{2}\Gmu M_Z^3}{12\pi} \, .
 \end{equation}
 The photonic QED correction, given at one-loop order by
 \beq
 \delta_{\rm QED} = Q_f^2\, \frac{3\al}{4\pi} \, ,
 \end{equation}
  is 
 small, at most 0.17\% for charged leptons.

 The factorizable SM (i.e.\ gluonic)
QCD corrections for hadronic final states can be written as
 follows:
 \beq
 \label{QCD}
  \Delta\Ga^f_{\rm QCD}\, =\, \Ga_0
   \left( \mid g_{\rm V}^f\mid ^2+\mid g_{\rm A}^f\mid ^2 \right)
   \, K_{\rm QCD} \, ,
 \end{equation}
 where \cite{qcdq}
 \bea
 K_{\rm QCD}  & = & \frac{\als}{\pi} +1.41 \left(
   \frac{\als}{\pi}\right)^2 -12.8 \left(
   \frac{\als}{\pi}\right)^3 
  - \frac{Q_f^2}{4} \frac{\al\als}{\pi^2}
 \eea
  for the light quarks with $m_q\simeq 0$, with
 $\als = \als(\mz)$.
 \index{QCD corrections!to the Z width}

 \smallskip
 For $b$ quarks
 the QCD corrections are different owing to  finite $b$ mass terms
 and to top-quark-dependent two-loop diagrams
  for the axial part:
 \bea
  \Delta\Ga_{\rm QCD}^b & = & \Ga_0 \left( \,
              |g_{\rm V}^b|^2+ |g_{\rm A}^b| ^2 \right) 
               \,  K_{\rm QCD}   \nn \\
     &  &     +\, \Ga_0 \left[
            |g_{\rm V}^b|^2  \, R_{\rm V} \,+\,
            |g_{\rm A}^b|^2 \, R_{\rm A}  \right]  \, .
 \eea
 The coefficients in the perturbative expansions
 \bea
  R_{\rm V} &=& c_1^{\rm V} \aspi + c_2^{\rm V} \left(\aspi\right)^2 
           + c_3^{\rm V} \left(\aspi\right)^3 + \cdots,
  \nn \\
  R_{\rm A} &=&  c_1^{\rm A} \aspi 
              + c_2^{\rm A} \left(\aspi\right)^2 + \cdots \, ,
  \nn
 \eea
 depending on $\mb$ and $\mt$,
 have been calculated up to third order in $\als$, except for the 
 $\mb$-dependent singlet terms,
 which are known to $O(\als^2)$ \cite{qcdb,qcdb1}. 
 For a review of the QCD 
 corrections to the $Z^0$ width,
 with the explicit expressions for $R_{\rm V,A}$,
 see \citere{qcdq1}.

 The partial decay rate into $b$ quarks, in particular the
 ratio $R_b = \Ga_b/\Ga_{\rm had}$, is an observable with
 special sensitivity to the top quark mass. Therefore, 
 beyond the pure QCD corrections, the two-loop contributions
 of the mixed QCD--electroweak type are also important.
 The QCD corrections were first derived for 
 the leading term of $O(\als\Gmu \mt^2)$
 \cite{jeg}
 and were subsequently completed by the 
  $O(\als)$ correction to the $\log \mt/\MW$ term 
 \cite{log} and the residual terms of $O(\al\als)$ \cite{harlander}.

 At the same time, the complete two-loop 
 $O(\al\als)$ corrections  
 to the partial widths for decay into the light quarks 
 have also been obtained, beyond those that are already contained in
 the factorized expression (\ref{QCD}) with the electroweak
 one-loop couplings \cite{czarnecki}. These 
 ``non-factorizable'' corrections  
 yield an extra negative contribution of 
 $-0.55(3)$ MeV to the total hadronic $Z^0$ width.
 
Besides the standard gluonic QCD corrections, there are
supersymmetric QCD corrections involving virtual gluinos  
and squarks, which turned out to be very small~\cite{SUSYQCD,junger},
for masses of the SUSY partners in accordance with the bounds
from direct experimental searches.

\medskip \noi
From the partial widths and the total width (\ref{totalwidth})
the following set of combinations can be formed,

\smallskip 
the hadronic peak cross section, with the hadronic width
$\Ga_{\rm had} = \sum_{q} \Ga_{q}$,
\BE
 \si_h = \frac{12 \pi}{\MZ^2} \frac{\Ga_{e} \Ga_{\rm had}}{\Ga_Z^2} \
{},
\end{equation}

the ratio of the hadronic to the electronic decay width,
\BE
 R_e = \frac{\Ga_{\rm had}}{\Ga_{e}} \ ,
\end{equation}

the ratio of the partial decay width for $Z \to b \bar{b} \, (c \bar{c})$ 
to the total hadronic decay width,
\BE
 R_{b (c)} = \frac{\Ga_{b(c)}}{\Ga_{\rm had}} \ .
\end{equation}

\medskip \noi
The various asymmetries
depend on the ratios of the
vector to the axial vector coupling and thus on the effective
mixing angles defined in \refeq{sineffective}, in terms of the 
combinations
\BE
\cA^f = \frac{2 \, (1- 4 |Q_f| \sin^2\theta_f)}
             {1+(1- 4|Q_f| \sin^2\theta_f)^2}\, ,
\end{equation}
yielding

\medskip
the {\em left-right} asymmetry and the $\tau$ {\em polarization},
\BE
 \ALR = \cA^e \, , \quad P_\tau = \cA^\tau   \, ,
\end{equation}

the {\em forward-backward} asymmetries,
\BE
 \AFB^f = \frac{3}{4} \, \cA^e \, \cA^f .
\end{equation}
 Final-state QCD corrections, in the case of quark pair production,
are important for the forward-backward asymmetries, at the  
one-loop level given by
 \beq
 \label{finalAFBQCD}
 \afb^q \ra  \afb^q \, \left( 1- \frac{\al_{\rm s}(\mz)}{\pi}\right)\, , 
 \end{equation}
 in the absence of cuts. 
 Finite-mass effects have to be considered for $b$ quarks only;
 they are discussed  in \citere{laerman}.
 Two-loop QCD corrections in the massless approximation
 are also available~\cite{seymour}.  
 The SUSY-QCD corrections again turn out to be small for
realistic values for squark and gluino masses~\cite{junger}.


\subsection{The effective leptonic mixing angle}

Since $\sweff$ is a precision observable with high
sensitivity for testing the electroweak theory, we discuss in this
section the status of the theoretical predictions for $\sweff$ in the SM
and the MSSM.

\subsubsection{SM corrections}

Recently the complete result for the fermionic two-loop corrections has
been obtained~\cite{sweff2lferm}, improving the prediction compared to
the previously known \order{\Gmu^2 \mt^2\MZ^2} term~\cite{sw2effSMmt2}.
Contrary to the case of the $W$-boson mass, see
\refse{subsec:MWSMeval}, the purely bosonic \twol\ corrections are not
yet completely known. 

Beyond two-loop order, the same kind of corrections are known as for
$\MW$, i.e.\ QCD corrections of 
\order{\al\als}~\cite{drSMgfals,deltarSMgfals,twoloopint} and
\order{\al\als^2}~\cite{drSMgfals2,drSMgfals2LF}, pure fermion-loop
corrections up to four-loop order~\cite{floops}, and three-loop
corrections entering via $\De\rho$
(see \refse{subsec:beyondoneloop}).

A simple parametrization of the SM result for $\sweff$ containing all
relevant higher-order corrections can be found in \citere{sweff2lferm}.
It reproduces the exact calculation with a maximal deviation of 
$4.5 \times 10^{-6}$ for $10 \gev \leq \MH \leq 1 \tev$ if the other
parameters are varied within their combined $2 \si$
region around their experimental central values.

The estimated theory uncertainties for different parts of the unknown
higher-order corrections 
are summarized in \refta{higherorderunc2} (see
\citeres{sweff2lferm,blueband} for further details). 

\begin{table}[h!]
\renewcommand{\arraystretch}{1.5}
\begin{tabular}{|c|c|c|c|c|}
\hline
 \multicolumn{2}{|c|}{2-loop} &
 \multicolumn{3}{c|}{3-loop} \\
\hline
   ~\order{\al^2, {\rm ferm}}~ & ~\order{\al^2, {\rm bos}}~ 
 & ~\order{\al\als^2, {\rm ferm}}~ & ~\order{\Gmu^2\als \mt^2\MZ^2}~ 
 & ~\order{\al^3}~ \\
                                                    \hline\hline
   compl.~\cite{sweff2lferm} & 
   1.2
 & compl.~\cite{drSMgfals2,drSMgfals2LF} & 2.3 & 2.5 \\ \hline
\end{tabular}

\vspace{.5em}

\begin{tabular}{|c|c|}
\hline
 \multicolumn{2}{|c|}{4-loop} \\
\hline
   ~\order{\Gmu\als^3\mt^2}~ &
   ~\order{\Gmu^2\als^2\mt^4}~ \\
                                                    \hline\hline
   1.1 & 2.4 \\ \hline
\end{tabular}
\renewcommand{\arraystretch}{1}
\caption{
Estimated uncertainties from
unknown higher-order corrections to $\sweff$ 
in~$[10^{-5}]$~\cite{sweff2lferm,blueband}
}
\label{higherorderunc2}
\end{table}

\noindent
Currently these intrinsic uncertainties result in~\cite{sweff2lferm}
\BE
\de\sweff^{\rm SM,intr} ~({\rm current}) = 5 \times 10^{-5} ~.
\end{equation}
In the future, an improvement down to about
\BE
\de\sweff^{\rm SM,intr} ~({\rm future}) = 2 \times 10^{-5} 
\label{sw2efffutureunc}
\end{equation}
seems achievable.

Concerning the parametric uncertainties, the current
errors for $\mt$~\cite{mtexp} and $\De\al_{\rm had}$~\cite{Burkhardt:2001xp} 
give rise to
\BEA
\de\mt^{\rm current} = 4.3 \gev &\Rightarrow&
\De\sweff^{{\rm para},\mt} ~({\rm current}) \approx 14 \times 10^{-5}, \\[.3em]
\de(\De\al_{\rm had}^{\rm current}) = 36 \times 10^{-5} &\Rightarrow&
\De\sweff^{{\rm para},\De\al_{\rm had}} ~({\rm current}) \approx 13
\times 10^{-5} .
\EEA
The parametric uncertainties induced by the
future experimental errors of $\mt$ and $\De\al_{\rm had}$
are~\cite{deltamt}
\BEA
\de\mt^{\rm future} = 0.1 \gev &\Rightarrow&
\De\sweff^{{\rm para},\mt} ~({\rm future}) \approx 0.4 \times 10^{-5}, \\[.3em]
\de(\De\al_{\rm had}^{\rm future}) = 5 \times 10^{-5} &\Rightarrow&
\De\sweff^{{\rm para},\De\al_{\rm had}} ~({\rm future}) \approx 1.8
\times 10^{-5} .
\EEA
Compared to the GigaZ accuracy (see \refta{tab:POfuture}) 
on $\sweff$ also the parametric
uncertainty induced by the experimental error of $\MZ$ is 
non-negligible~\cite{deltamt}
\begin{equation}
\de\MZ = 2.1 \mev \;\; \Rightarrow \;\; \De\sweff^{{\rm
para},\MZ} \approx 1.4 \times 10^{-5} .
\end{equation}
As in the case of $\MW$, 
the precision measurement of the top-quark mass at the ILC and 
prospective improvements in the determination of $\De\al_{\rm had}$
will reduce the parametric
uncertainties to the same level as the prospective intrinsic
uncertainties, \refeq{sw2efffutureunc}.


\subsubsection{MSSM corrections}

As for $\MW$, the largest correction to $\sweff$ in the MSSM can be
expected from scalar quark contributions. The shift in $\sweff$ is
then given by
\BE
\De\sweff^{\sq} = \frac{\cw^2 \sw^2}{\cw^2 - \sw^2} \, \De r^{\sq}
                    - \sw\cw \hat\Pi^{\ga Z}(\MZ^2) ,
\label{sweffmssm}
\end{equation}
with
\BE
\hat\Pi^{\ga Z}(\MZ^2) = \frac{\Si^{\ga Z}(\MZ^2)}{\MZ^2}
                           - \frac{\cw}{\sw}
                             \KL \frac{\de\MZ^2}{\MZ^2} 
                                -\frac{\de\MW^2}{\MW^2} \KR ,
\end{equation}
and $\De r^{\sq}$ from \refeq{deltarmssm}.

In the MSSM the complete one-loop corrections to $\sweff$ have been
evaluated as described in \refse{subsec:Zobs}. Beyond \onel\ order the
leading term can be included via the $\rho$~parameter approximation,
\refeq{precobs}, where $\De\rho$ at the \twol\ level is given in
\refse{subsec:beyondoneloop}. 
The intrinsic uncertainties from missing
higher-order SUSY corrections will be discussed in
\refse{subsec:MWsweffuncSUSY}.

\bigskip
The full one-loop corrections from third and second generation squarks
in the NMFV MSSM, using \refeq{sweffmssm}, have been derived in
\citere{nmfv}.


\section{The lightest Higgs boson mass as a precision observable}
\label{sec:mh}

A striking prediction of the MSSM is the existence of at least one
light Higgs boson. The search for this particle 
is one of the main goals at the present and the next generation
of colliders. Direct searches at LEP have already ruled out a
considerable fraction of the MSSM parameter
space~\cite{LEPHiggsearly,LEPHiggsSM}. With the forthcoming data
from the Tevatron, the LHC and the ILC
either a light Higgs boson will be discovered or the MSSM will be ruled
out as a viable theory
for physics at the weak scale. Furthermore, if one or more Higgs
bosons are discovered, their masses and couplings will be determined
with high accuracy at the ILC. Thus, a
precise knowledge of the dependence of the masses and mixing angles of
the MSSM Higgs sector on the relevant supersymmetric parameters is of
utmost importance to reliably compare the predictions of the MSSM with
the (present and future) experimental results.

The Higgs sector of the MSSM has been described in
\refse{subsec:higgssector} at tree-level, leading to the prediction for
the lightest MSSM Higgs boson, $\mhtree \le \MZ$, see \refeq{eq:mhtree}. 
However, this mass bound, which arises from the gauge sector of the
theory, is subject to large radiative
corrections in particular from the Yukawa sector of the 
theory~\cite{mhiggs1l}. Because of the importance of the higher-order
corrections, a lot of effort has been devoted to obtain higher-order
results in the MSSM Higgs sector.
Results for the complete \onel\ contributions are
available~\cite{mhiggsFD1l,pierce}. Corrections beyond one-loop order
have been obtained with different methods.
Leading and subleading \twol\ corrections have been obtained in
the Effective Potential (EP) approach~\cite{mhiggsEP,effpotfull},
the Renormalization Group (RG) improved EP approach~\cite{mhiggsRG},
and with the
Feynman-diagrammatic method~\cite{mhiggsletter,mhiggslong,mhiggslle}.
Detailed comparisons of the different methods have been
performed~\cite{bse,mhiggsComp}. 
The higher-order corrections shift
the upper bound on $\mh$ to 
$\mh \lsim 136 \gev$~\cite{mhiggslong,mhiggsAEC}
(for $\mt = 178 \gev$ and $\msusy \le 1 \tev$, neglecting uncertainties
from unknown higher-order corrections).

In the case that the MSSM parameters possess non-vanishing complex phases
(cMSSM), the upper bound on $\mh$ remains the same as for the MSSM with
real parameters, but the Higgs-boson couplings can vary significantly 
compared to the case with real parameters. Complex phases are possible
for the trilinear couplings, $A_f, f = t, b, \tau, \ldots$, for the
Higgsino mass parameter, $\mu$, and for the gaugino mass terms, $M_1$,
$M_2$, $M_3 = \mgl$ (where one of the latter ones can be rotated away
by a redefinition of the corresponding fields). Recently
the different methods for the evaluation of higher-order corrections
in the MSSM Higgs sector have even been extended to the cMSSM, 
reaching nearly the precision as in the real 
MSSM~\cite{mhiggsCPXgen,mhiggsCPXEP,mhiggsCPXRG1,mhiggsCPXFD,habilSH}. 
In the following, however, we will focus on the real case.


\subsection{Higher-order corrections to $\mh$}
\label{subsec:mhcorr}

The tree-level bound on $\mh$, being obtained from the gauge
couplings, receives 
large corrections from SUSY-breaking effects in the Yukawa sector of the
theory. The leading one-loop correction is proportional to $\mt^4$.
The leading logarithmic one-loop term (for vanishing mixing
between the scalar top quarks) reads~\cite{mhiggs1l} 
\BE
\De \mh^2 = \frac{3 \Gmu \mt^4}{\wz\, \pi^2\,\SQb}
          \ln \KL \frac{\mste \mstz}{\mt^2} \KR~.
\label{deltamhmt4}
\end{equation}
Corrections of this kind have drastic effects on the predicted value of
$\mh$ and many other observables in the MSSM Higgs sector. The one-loop
corrections can shift $\mh$ by 50--100\%. 

In the Feynman diagrammatic (FD) approach the higher-order
corrected Higgs boson masses are derived by finding the
poles of the $h,H$-propagator 
matrix. Its inverse is given by
\BE
\left(\Delta_{\rm Higgs}\right)^{-1}
= - i \ML p^2 -  \mHtree^2 + \hSi_{HH}(p^2) &  \hSi_{hH}(p^2) \\
     \hSi_{hH}(p^2) & p^2 -  \mhtree^2 + \hSi_{hh}(p^2) \MR,
\label{higgsmassmatrixnondiag}
\end{equation}
where the $\hSi(p^2)$ denote the renormalized Higgs-boson
self-energies (see \refeq{rMSSM:renses_higgssector}),
and $p$ is the external momentum.
Determining the poles of the matrix $\Delta_{\rm Higgs}$ in
\refeq{higgsmassmatrixnondiag} is equivalent to solving
the equation
\begin{equation}
\left[p^2 - \mhtree^2 + \hSi_{hh}(p^2) \right]
\left[p^2 - \mHtree^2 + \hSi_{HH}(p^2) \right] -
\left[\hSi_{hH}(p^2)\right]^2 = 0\,.
\label{eq:proppole}
\end{equation}

The status of the available results for the self-energy contributions to
\refeq{higgsmassmatrixnondiag} can be summarized as follows. For the
one-loop part, the complete result within the MSSM is 
known~\cite{mhiggs1l,mhiggsFD1l}. The by far dominant
one-loop contribution is the \order{\alt} term due to top and stop 
loops ($\alt \equiv y_t^2 / (4 \pi)$, where $y_t$ has been defined in
\refeq{ytyb}).

The evaluation of two-loop corrections is quite advanced and it has now 
reached a stage where all the presumably dominant
contributions are known. They include the strong corrections, usually
indicated as \order{\alt\als}, and Yukawa corrections, \order{\alt^2},
to the dominant one-loop \order{\alt} term, as well as the strong
corrections to the bottom/sbottom one-loop \order{\alb} term ($\alb
\equiv y_b^2 / (4\pi)$), i.e.\ the \order{\alb\als} contribution. The
latter can be relevant for large values of $\tb$. Presently, the
\order{\alt\als}~\cite{mhiggsEP,mhiggsRG,mhiggsletter,mhiggslong},
\order{\alt^2}~\cite{mhiggsEP,mhiggsRG,mhiggsEP1,mhiggsEP2},
\order{\alb\als}~\cite{mhiggsEP4,mhiggsFDalbals}, 
\order{\alt\alb}, \order{\alb^2}~\cite{mhiggsEP4b} contributions to
the self-energies 
are known for vanishing external momenta.  In the (s)bottom
corrections the all-order resummation of the $\tb$-enhanced terms,
\order{\alb(\als\tb)^n}, is also performed \cite{deltamb1,deltamb}
(see \refse{subsec:deltamb}).
The above results have been implemented into the program
\fh~\cite{feynhiggs,feynhiggs1.2,feynhiggs2.2}, which
evaluates observables in the MSSM Higgs sector (including also results
with complex phases).

Recently, also the full electroweak \twol\ corrections in the
approximation of vanishing external momentum~\cite{effpotfull} and the
leading two-loop momentum dependent effects~\cite{mhiggsQ2} have been 
published. For these corrections no public code is available yet. In
order to apply this result for expressing $\mh$ in terms of physical
masses, a transition of the parameters $\MZ$ and $\MA$ in 
\citeres{effpotfull,mhiggsQ2} to their on-shell values will be required
at the two-loop level.

\bigskip
Besides the masses of the Higgs bosons, also their couplings are
affected by large higher-order corrections. For the MSSM with real
parameters, leading corrections can conveniently be absorbed into the
couplings by introducing an effective mixing angle $\aeff$. It is
obtained from the higher-order corrected Higgs-boson mass matrix in the
approximation where the momentum dependence of the Higgs-boson
self-energies is neglected.

The Higgs-boson mass matrix in the $\Pe$-$\Pz$ basis reads in this case
\BE
M^{2}_{\rm Higgs}
= \VL \mpe^2 - \hSi_{\Pe}(0)\;\;\;\;\;\; \mpez^2 - \hSi_{\PePz}(0) \\
     \mpez^2 - \hSi_{\PePz}(0)\;\;\;\;\;\; \mpz^2 - \hSi_{\Pz}(0) \VR~,
\label{higgsmassmatrixunrot}
\end{equation}
 where the $\hSi_s(0)~(s = \Pe, \PePz, \Pz)$ denote the
 renormalized Higgs-boson self-energies (in the $\Pe,\Pz$ basis),
 including one- and \twol\ (and possibly higher-order) corrections.
These self-energies (at zero external momentum)
have to be inserted into \refeq{higgsmassmatrixunrot}.
Diagonalizing this higher-order corrected Higgs-boson mass matrix 
\BE
M_{\rm Higgs}^2 \stackrel{\aeff}{\longrightarrow}
\ML \mH^2 & 0 \\ 0 & \mh^2 \MR
\end{equation}
yields the effective mixing angle $\aeff$:
\BE
 \aeff = {\rm arctan}\KKL 
  \frac{-(\MA^2 + \MZ^2) \sbe \Cb - \hSi_{\PePz}}
       {\MZ^2 \CQb + \MA^2 \SQb - \hSi_{\Pe} - \mh^2} \KKR~,~~
  -\frac{\pi}{2} < \aeff < \frac{\pi}{2}~.
\end{equation}

Replacing in the Higgs-boson couplings 
the tree-level mixing angle $\al$ by the  higher-order
corrected effective mixing angle $\aeff$ leads to the inclusion of the 
leading higher-order corrections that enter via Higgs-boson propagator
corrections~\cite{hff,eehZhA}.


\subsection{Remaining intrinsic and parametric uncertainties}
\label{subsec:mhunc}

If the MSSM is realised in nature, the light $\cp$-even Higgs-boson
mass will be measured with high precision at the next generation of
colliders. 
The prospective accuracies for a light SM-like Higgs boson 
that can be obtained in the experimental
determination of $\mh$ at the LHC~\cite{mhdetLHC} and at the
ILC~\cite{teslatdr,orangebook,acfarep} are:
\BEA
\de\mh^{\rm exp} &\approx& 200 \mev {\rm ~~(LHC)} ~, \\
\de\mh^{\rm exp} &\approx& 50 \mev {\rm ~~(ILC)} ~.
\EEA
Since $\mh$ depends sensitively on the other sectors of the
MSSM, in particular on the $\Stop$~sector (see \refeq{deltamhmt4}), the
light $\cp$-even Higgs-boson mass will be very important for precision
tests of the MSSM.

The remaining theoretical uncertainties in the prediction for $\mh$ have
been discussed in \citeres{mhiggsAEC,deltamt,susyewpo,lcwsParisHiggs}. 
For recent reviews on the
current status of the theoretical prediction see also
\citeres{habilSH,mhiggsWN}.

\bigskip
We begin with the discussion of the parametric uncertainties. 
Since the leading one-loop corrections to $\mh$ are proportional to
the fourth power of the top quark mass, the predictions for $\mh$ and
many other observables in the MSSM Higgs sector sensitively depend 
on the numerical value of $\mt$. As a rule of thumb~\cite{tbexcl}, a shift of 
$\de \mt = 1\gev$ induces a parametric theoretical uncertainty of $\mh$
of also about $1 \gev$, i.e.\ 
\begin{equation}
\De\mh^{{\rm para},\mt} \approx \de\mt~.
\label{eq:thumb}
\end{equation}

The uncertainties induced by the experimental error of $\mt$ at the 
LHC~\cite{mtdetLHC} and the ILC~\cite{teslatdr,orangebook,acfarep},
\BEA
\de\mt^{\rm exp} &\approx& \mbox{1--2} \gev {\rm ~~(LHC)} ~, \\
\de\mt^{\rm exp} &\approx& 0.1 \gev {\rm ~~(ILC)} ~,
\EEA
can be compared with the parametric uncertainties induced by the other
SM input parameters.
Besides $\mt$, the other SM input parameters whose experimental errors
can be relevant for the prediction of $\mh$ are $\MW$, $\als$, and
$\mb$. 
The $W$~boson mass enters only in higher orders through the
quantum corrections to muon decay (since $\GF$ is used for the 
parametrization, see \refeq{eq:delr}). 

The present experimental error of $\de\MW^{\rm exp} = 34 \mev$ leads to
a parametric theoretical uncertainty of $\mh$ below $0.1 \gev$. In view
of the prospective improvements in the experimental accuracy of $\MW$, 
the parametric uncertainty induced by $\MW$ will be smaller than the one
induced by $\mt$, even for $\de\mt = 0.1 \gev$. 

The current experimental error of the strong coupling constant, 
$\de\als(\MZ) = 0.002$~\cite{pdg}, induces a parametric
theoretical uncertainty of $\mh$ of about $0.3 \gev$. Since a future
improvement of the error of $\als(\MZ)$ by about a factor of 2 can be
envisaged~\cite{pdg,gigaz,alsdet}, the parametric uncertainty
induced by $\mt$ 
will dominate over the one induced by $\als(\MZ)$ down to the level of 
$\de\mt = 0.1$--$0.2 \gev$.

The mass of the bottom quark currently has an experimental error of
about $\de\mb = 0.1 \gev$~\cite{pdg,mbhoang}. A future improvement of this
error by about a factor of 2 seems to be feasible~\cite{mbhoang}.
The influence of the bottom and sbottom  loops on $\mh$
depends on the parameter region, in particular on the values of $\tb$
and $\mu$ (the Higgsino mass parameter). For small $\tb$ and/or $\mu$ the
contribution from bottom and 
sbottom loops to $\mh$ is typically below $1 \gev$, in which case
the uncertainty induced by the current experimental error on $\mb$ is
completely negligible. For large values of $\tb$ and $\mu$, the effect of 
bottom/sbottom loops can exceed $10 \gev$ in
$\mh$~\cite{mhiggsEP4,mhiggsAEC,mhiggsFDalbals}. Even in these cases 
we find that the uncertainty in $\mh$ induced by $\de\mb = 0.1 \gev$
rarely exceeds the level of $0.1 \gev$, since 
higher-order QCD corrections effectively reduce the bottom quark
contributions. Thus, the parametric uncertainty induced by $\mt$ will in
general dominate over the one induced by $\mb$, even for 
$\de\mt \approx 0.1 \gev$.

\smallskip
The comparison of the parametric uncertainties of $\mh$ induced by the 
experimental errors of $\MW$, $\als(\MZ)$ and $\mb$ with the one induced
by the experimental error of the top quark mass shows that an
uncertainty of $\de\mt \approx 1 \gev$, corresponding to the
accuracy achievable at the LHC, will be the dominant parametric uncertainty
of $\mh$. The accuracy of $\de\mt \approx 0.1 \gev$ achievable at
the ILC, on the other hand, will allow a reduction of the parametric
theoretical uncertainty induced by $\de\mt$ to about the same level as
the uncertainty induced by the other SM input parameters.

\bigskip
We now turn to the intrinsic theoretical uncertainties in the prediction
for $\mh$ from unknown higher-order corrections.
Even if all the available higher-order corrections described above are
taken into account, the intrinsic uncertainty in $\mh$ from unknown
higher-order corrections is still estimated to be quite 
substantial~\cite{feynhiggs1.2,mhiggsAEC}%
\footnote{%
For codes that do not include all the existing higher-order
corrections the intrinsic theoretical uncertainties can be much larger.
}%
.
The numerical relevance of the unknown higher-order corrections
depends on the region of MSSM parameter space that one considers. An
overall estimate of the intrinsic uncertainty can therefore be only a
rough guidance for ``typical'' MSSM parameter regions. In regions where
higher-order corrections are particularly enhanced (for instance very
large mixing in the stop sector or regions where the bottom Yukawa
coupling is close to being non-perturbative) the theoretical
uncertainties can be significantly larger.

\begin{figure}[htb!]
\newcommand{\psfragtextscale}{0.75}
\psfrag{5x-1}[][][\psfragtextscale]{0.5}
\psfrag{1x0}[][][\psfragtextscale]{1}
\psfrag{5x0}[][][\psfragtextscale]{5}
\psfrag{1x1}[][][\psfragtextscale]{10}
\psfrag{5x1}[][][\psfragtextscale]{50}
\psfrag{0}[][][\psfragtextscale]{0}
\psfrag{60}[][][\psfragtextscale]{60}
\psfrag{70}[][][\psfragtextscale]{70}
\psfrag{80}[][][\psfragtextscale]{80}
\psfrag{90}[][][\psfragtextscale]{90}
\psfrag{100}[][][\psfragtextscale]{100}
\psfrag{110}[][][\psfragtextscale]{110}
\psfrag{120}[][][\psfragtextscale]{120}
\psfrag{130}[][][\psfragtextscale]{130}
\psfrag{140}[][][\psfragtextscale]{140}
\psfrag{150}[][][\psfragtextscale]{150}
\psfrag{500}[][][\psfragtextscale]{500}
\psfrag{1000}[][][\psfragtextscale]{1000}
\psfrag{1500}[][][\psfragtextscale]{1500}
\psfrag{TB}[][][\psfragtextscale]{$\tb$}
\psfrag{MA0}[][][\psfragtextscale]{$M_A [\mathrm{GeV}]$}
\psfrag{Mh0}[][][\psfragtextscale]{$m_h [\mathrm{GeV}]$}
\begin{center}
\epsfig{figure=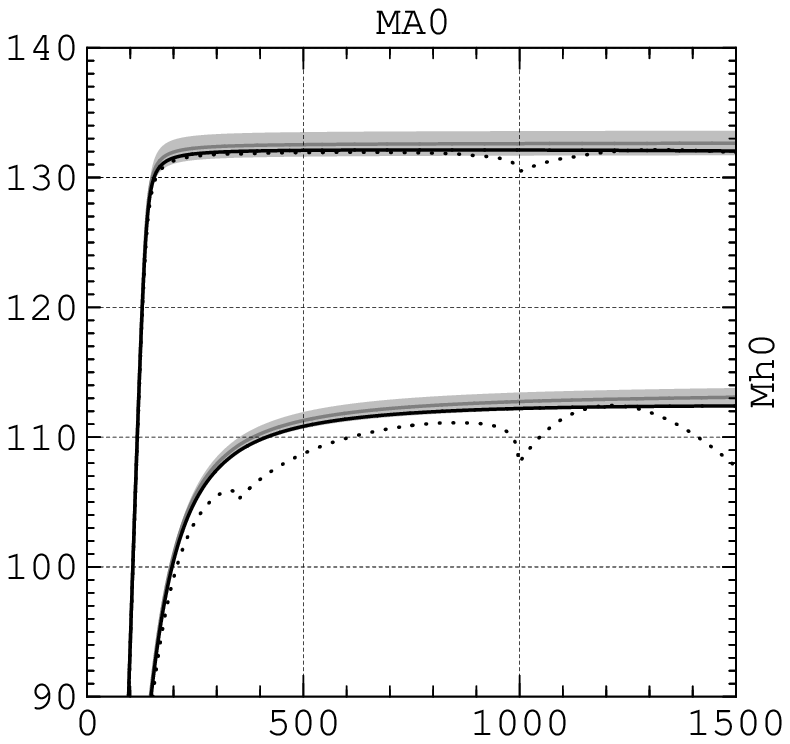,width=7cm}
\epsfig{figure=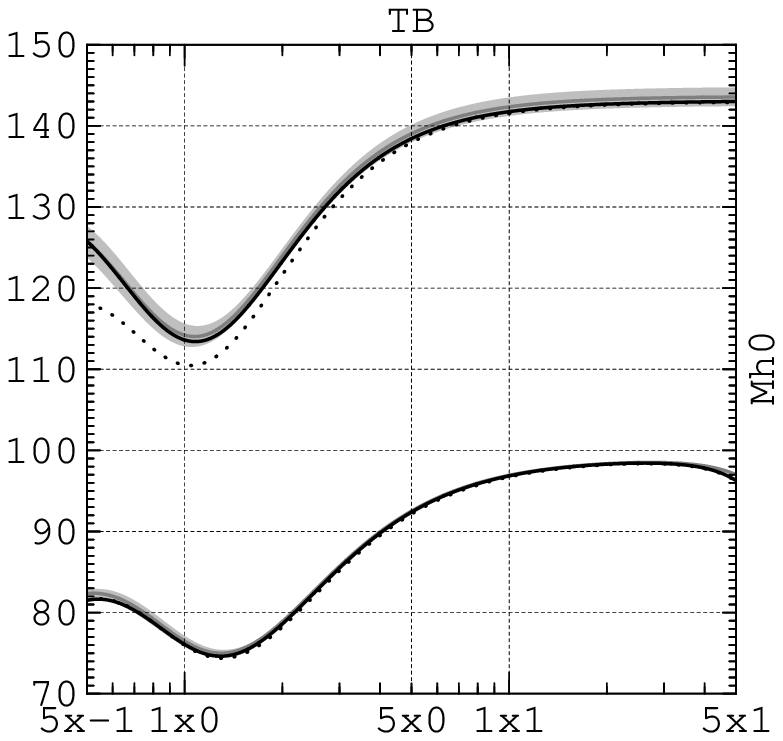,width=7cm}
\end{center}
\caption{ 
The renormalization scale dependence of $\mh$  introduced via
a \drbar\ definition of $\tb$ and the Higgs field
renormalization constants is shown as a function of $\MA$ (left plot)
and $\tb$ (right). The lower curves correspond to $\tb = 2$ (left) and
$\MA = 100 \gev$ (right). For the upper curves we have set $\tb = 20$
(left) and $\MA = 500 \gev$ (right). 
$\mu_{\drbarm}$ has been varied from $\mt/2$ to
$2\mt$. The other parameters are $\msusy = 500 (1000) \gev$, 
$\Xt = 2\, \msusy$, $M_2 = \mu = 500 \gev$.
The dotted line corresponds to a full on-shell scheme, for more
details see \citere{feynhiggs1.2}.
}
\label{fig:RenComp}
\end{figure}

At the two-loop level, various genuine electroweak two-loop corrections
from different sectors of the MSSM
are not yet included in the publicly available codes. A rough estimate
of their numerical impact can be obtained from the relative importance
of the corresponding contributions at the \onel\ level. 
This has been performed in \citere{mhiggsAEC} and yielded an estimate of
the remaining uncertainty of unknown two-loop corrections of about $2 \gev$. 
Another way of estimating the effect of unknown two-loop corrections is
to apply different renormalization schemes at the \onel\ level and to vary the 
renormalization scale of quantities that are renormalized according to
the \drbar\ scheme~\cite{feynhiggs1.2}. As an example for the latter
approach, \reffi{fig:RenComp} shows the effect of varying the renormalization
scale that enters via the renormalization of $\tb$ and the Higgs field
renormalization constants at the one-loop level for ``typical'' MSSM
parameters (see caption). The corresponding shift
in the one-loop result for $\mh$, which is of the order of genuine
two-loop corrections that are not included in the current prediction for
$\mh$, is indicated by the grey areas. The uncertainty in $\mh$ from
varying $\mu_{\drbarm}$ from $\mt/2$ to
$2\mt$ is in accordance with the above estimate of the uncertainty from
missing two-loop corrections of about $\pm 2 \gev$.

Beyond two-loop order, corrections that effectively shift the value of
the top-quark mass entering the calculation are particularly important
because of the sensitive dependence of $\mh$ on $\mt$. Corrections of
this kind of \order{\alt\als^2} can be estimated by varying the
renormalization scheme of the top-quark mass at the \twol\ level.
Another possibility for estimating the size of three-loop corrections is
to analyze the numerical impact of the leading logarithmic three-loop
term that can easily be obtained with renormalization group
methods~\cite{mhiggsAEC,ahoang}. Both possibilities have been
investigated in detail in
\citere{mhiggsAEC}, yielding an estimate of the intrinsic 
theoretical uncertainty beyond the \twol\ level
of about~$1.5 \gev$.
Similar strategies in the case of the \order{\alb\als^2}
correction~\cite{mhiggsFDalbals} lead to an intrinsic uncertainty of
up to $2 \gev$ in the case of $\mu < 0$ (in regions where the effects of the
bottom/sbottom sector are strongly enhanced), and of about $\sim 100 \mev$
for $\mu > 0$.

As an overall estimate for the
current intrinsic uncertainties in the prediction of $\mh$ we obtain
\BE
\De\mh^{\rm intr} ~({\rm current}) = 3 \gev .
\label{eq:mhintrcurr}
\end{equation}
On the timescale of 5--10 years it seems reasonable to expect that the 
complete two-loop calculation (which is already technically feasible
with the currently existing tools) can be incorporated into efficient 
codes and
that the higher-order uncertainties can be reduced by at least a factor
of two, leading to the estimate
\BE
\De\mh^{\rm intr} ~({\rm future}) = 0.5 \gev .
\end{equation}


\subsection{Higgs sector corrections in the NMFV MSSM}
\label{subsec:mhnmfv}

Within the MSSM with MFV, the dominant \onel\ contributions to the 
self-energies in~(\ref{higgsmassmatrixunrot})
result from the Yukawa part of the theory (i.e.\
neglecting the gauge couplings); 
they are described  by loop diagrams involving 
third-generation quarks and squarks.
Within the MSSM with NMFV, the squark loops have to be modified
by introducing the generation-mixed squarks, as given in
\refse{subsec:nmfv}. 
The leading terms are obtained by evaluating the contributions to 
the renormalized Higgs-boson self-energies at zero external momentum, 
$\hSi_s(0), s = hh, hH, HH$. 
The evaluation has been restricted to the dominant Yukawa
contributions resulting from the top and $t/\Stop$ (and $c/\Scha$) sector.
Corrections from $b$ and $b/\Sbot$ (and
$s/\Sstr$) could only be important for very large values of
$\tb$, $\tb \gsim \mt/\mb$ and have not been considered so far.
The analytical result of the renormalized Higgs boson self-energies,
based on the general $4 \times 4$~structure of the $\Stop/\Scha$~mass
matrix, has been derived in \citere{nmfv}. 
However, as has also been shown in \citere{nmfv}, the corrections for
$\Mh$ are not significant for moderate generation mixing.


\section{The anomalous magnetic moment of the muon}
\label{sec:evalg-2}

Another observable which is important in the context of precision tests
of the electroweak theory is the anomalous magnetic moment of the muon, 
$\amu \equiv (g-2)_\mu/2$. For the interpretation of the $\amu$ results
in the context of Supersymmetry (or other models of new physics)
the current status of the comparison of the SM prediction with the
experimental result is crucial, see \citeres{g-2review,g-2review2}
for reviews and the discussion in \refse{subsec:ewpostatus}.
It currently results in a deviation of~\cite{g-2ICHEP04} 
\BE
\label{deviationfinal}
\amuexp - \amutheo = (25.2 \pm 9.2)\times10^{-10} ~:~2.7\,\si~.
\end{equation}


\subsection{MSSM one-loop calculation}

The anomalous magnetic moment $\amu$ of the muon is related to the 
photon--muon vertex function $\Gamma_{\mu\bar\mu A^\rho}$ as follows:
\BEA
\label{covdecomp}
\bar u(p')\Ga_{\mu\bar\mu A^\rho}(p,-p',q) u(p) & = &
\bar u(p')\left[\ga_\rho F_V(q^2) + (p+p')_\rho F_M(q^2) +
  \ldots\right] u(p),\\
\amu & = & -2m_\mu F_M(0).
\EEA
It can be extracted from the regularized vertex function using the
projector~\cite{g-2SM2lA,g-2SM2lB}
\BEA
\label{projector}
\amu & = & \frac{1}{2(D-1)(D-2)m_\mu^2} {\rm Tr}\Bigg\{
            \frac{D-2}{2}\left[m_\mu^2\ga_\rho - D p_\rho\pslash -
            (D-1)m_\mu p_\rho\right]V^\rho\nonumber\\
&&\qquad\quad +\frac{m_\mu}{4}
  \left(\pslash+m_\mu\right)
\left(\ga_\nu\ga_\rho-\ga_\rho\ga_\nu\right)
  \left(\pslash+m_\mu\right)T^{\rho\nu}\Bigg\},\\
V_\rho & = & \Ga_{\mu\bar\mu A^\rho}(p,-p,0),\\
T_{\rho\nu} & = & \frac{\partial}{\partial q^\rho}
\Ga_{\mu\bar\mu A^\nu}(p-(q/2),-p-(q/2),q)\bigg|_{q=0}.
\EEA
Here the muon momentum is on-shell, $p^2=m_\mu^2$, and $D$ is the
dimension of space-time. For more details see
\citeres{g-2SM2lA,g-2SM2lB,g-2CNH}.

\begin{figure}[htb!]
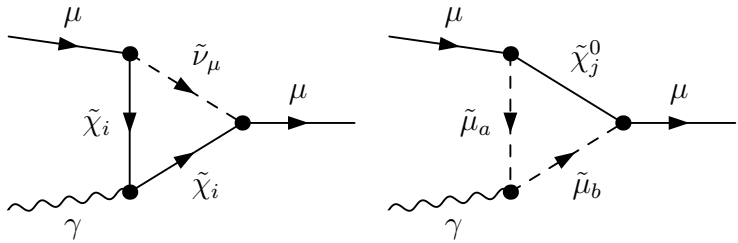

\begin{center}
\begin{feynartspicture}(288,168)(2,1)

\FADiagram{}
\FAProp(0.,15.)(7.,14.)(0.,){/Straight}{1}
\FALabel(3.7192,15.5544)[b]{$\mu$}
\FAProp(0.,5.)(7.,6.)(0.,){/Sine}{0}
\FALabel(3.7192,4.44558)[t]{$\gamma$}
\FAProp(20.,10.)(13.5,10.)(0.,){/Straight}{-1}
\FALabel(16.75,11.07)[b]{$\mu$}
\FAProp(7.,14.)(7.,6.)(0.,){/Straight}{1}
\FALabel(5.93,10.)[r]{$\tilde \chi_i$}
\FAProp(7.,14.)(13.5,10.)(0.,){/ScalarDash}{1}
\FALabel(10.5824,12.8401)[bl]{$\tilde \nu_\mu$}
\FAProp(7.,6.)(13.5,10.)(0.,){/Straight}{1}
\FALabel(10.5824,7.15993)[tl]{$\tilde \chi_i$}
\FAVert(7.,14.){0}
\FAVert(7.,6.){0}
\FAVert(13.5,10.){0}

\FADiagram{}
\FAProp(0.,15.)(7.,14.)(0.,){/Straight}{1}
\FALabel(3.7192,15.5544)[b]{$\mu$}
\FAProp(0.,5.)(7.,6.)(0.,){/Sine}{0}
\FALabel(3.7192,4.44558)[t]{$\gamma$}
\FAProp(20.,10.)(13.5,10.)(0.,){/Straight}{-1}
\FALabel(16.75,11.07)[b]{$\mu$}
\FAProp(7.,14.)(7.,6.)(0.,){/ScalarDash}{1}
\FALabel(5.93,10.)[r]{$\tilde \mu_a$}
\FAProp(7.,14.)(13.5,10.)(0.,){/Straight}{0}
\FALabel(10.4513,12.6272)[bl]{$\tilde \chi_j^0$}
\FAProp(7.,6.)(13.5,10.)(0.,){/ScalarDash}{1}
\FALabel(10.5824,7.15993)[tl]{$\tilde \mu_b$}
\FAVert(7.,14.){0}
\FAVert(7.,6.){0}
\FAVert(13.5,10.){0}

\end{feynartspicture}

\end{center}
\vspace{-2em}
\caption[]{
The generic one-loop diagrams for the MSSM contribution to $\amu$:
diagram with a sneutrino-chargino loop (left) and the diagram with a
smuon-neutralino loop (right).
}
\label{fig:g-21L}
\end{figure}

\bigskip
The complete one-loop contribution to $\amu$ 
can be devided into
contributions from diagrams with a smuon-neutralino loop and with a
sneutrino-chargino loop, see \reffi{fig:g-21L}, leading to
\BE
\De\amuSUoL = \De\amu^{\tilde \chi^\pm \tilde \nu_\mu} +
               \De\amu^{\tilde \chi^0 \tilde \mu}
\end{equation}

The full one-loop expression can be found in~\cite{g-2MSSMf1l}, see
\citere{g-2early} for earlier evaluations. If all SUSY
mass scales are set to a common value, 
\BE
\msusy = m_{\cha{}} = m_{\neu{}} = m_{\Smu} = m_{\Sneum}
\end{equation}
the result is given by
\BE
\amu^{\SU,{\rm 1L}} = 13 \times 10^{-10}
             \KL \frac{100 \gev}{\msusy} \KR^2 \tb\;
 {\rm  sign}(\mu) ~.
\label{susy1loop}
\end{equation}
Obviously, supersymmetric effects can easily account for a
$(20\ldots30)\times10^{-10}$ deviation, if $\mu$ is positive and
$\msusy$ lies roughly between 100 GeV (for small $\tb$) and
600 GeV (for large $\tb$).
Eq.\ (\ref{susy1loop}) also shows that for certain parameter choices
the supersymmetric contributions could have values of either
$\amu^{\SU} \gsim 55\times10^{-10}$ or $\amu^{\SU} \lsim -5\times 10^{-10}$,
both outside the $3\si$ band of the allowed range according to
\refeq{deviationfinal}. This means
that the $(g-2)_\mu$ measurement places strong bounds on the supersymmetric
parameter space.


\subsection{MSSM two-loop calculation}

In order to fully exploit the precision of the $(g-2)_\mu$ experiment 
within SUSY, see e.g.\
\citeres{g-2appl1,g-2appl2,g-2scan1L,recentfit} for discussions of the
resulting constraints on the parameter space,
the theoretical 
uncertainty of the SUSY loop contributions from unknown higher-order
corrections needs to be under control.
It should be significantly lower than the experimental error
given in \refeq{eq:amuexp} and the hadronic
uncertainties in the SM
prediction, leading to the combined uncertainty given in 
\refeq{deviationfinal}.

For the electroweak part of the SM prediction the desired level of
accuracy has been
reached with the computation of the complete two-loop 
result~\cite{g-2SM2lA,g-2SM2lB}, which reduced the intrinsic uncertainty
from QED and electroweak effects below the level of about
$1\times10^{-10}$~\cite{g-2ICHEP04}. 
For the SUSY contributions, a similar level of accuracy has not been
reached yet, since the corresponding
two-loop corrections are 
partially unknown. Four parts of the two-loop contribution have been
evaluated up to now that will be reviewed in the next subsections.


\subsubsection{Two-loop QED corrections}

The
first part are the leading $\log \KL m_\mu/\msusy\KR$-terms of
supersymmetric one-loop diagrams with a photon in the second loop. 
They are given by~\cite{g-2MSSMlog2l}
\BE
\De\amu^{\SU,{\rm 2L,QED}} = \De\amuSUoL \times
\KL \frac{4\,\al}{\pi} \log\KL \frac{\msusy}{m_\mu} \KR \KR~.
\end{equation}
They amount to about $-8\%$ of the supersymmetric one-loop contribution
(for a SUSY mass scale of $\msusy = 500 \gev$).


\subsubsection{Two-loop Two-Higgs-doublet contributions}

\begin{figure}[htb!]
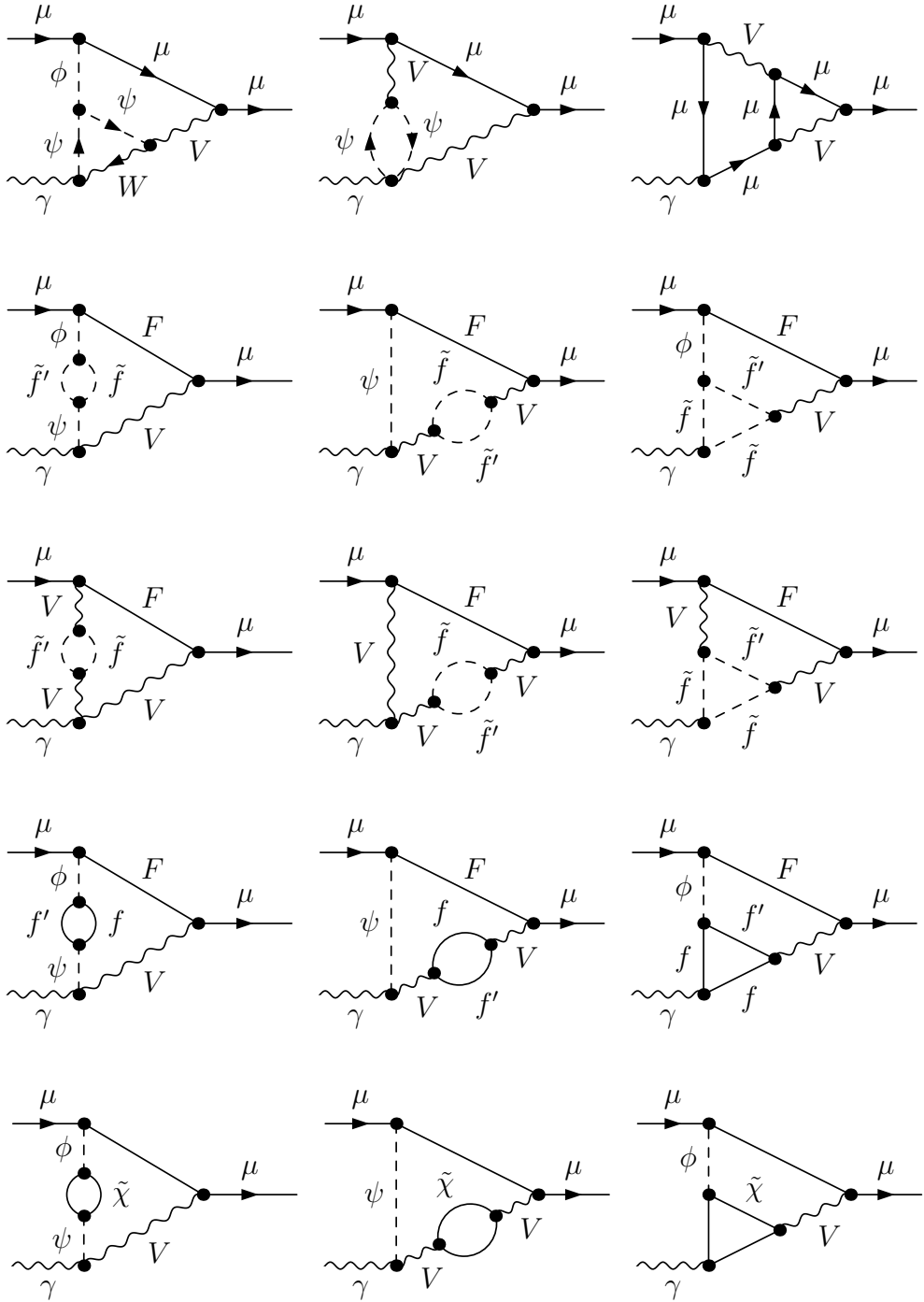

\begin{center}
\unitlength=0.9bp%

\begin{feynartspicture}(432,140)(3,1)

\FADiagram{}
\FAProp(0.,15.)(5.,15.)(0.,){/Straight}{1}
\FALabel(2.5,16.07)[b]{$\mu$}
\FAProp(0.,5.)(5.,5.)(0.,){/Sine}{0}
\FALabel(2.5,3.93)[t]{$\gamma$}
\FAProp(20.,10.)(15.,10.)(0.,){/Straight}{-1}
\FALabel(17.5,11.07)[b]{$\mu$}
\FAProp(5.,15.)(15.,10.)(0.,){/Straight}{1}
\FALabel(10.2132,13.4064)[bl]{$\mu$}
\FAProp(5.,10.)(5.,15.)(0.,){/ScalarDash}{0}
\FALabel(4.18,12.5)[r]{$\phi$}
\FAProp(5.,10.)(5.,5.)(0.,){/ScalarDash}{-1}
\FALabel(3.93,7.5)[r]{$\psi$}
\FAProp(10.,7.5)(5.,10.)(0.,){/ScalarDash}{-1}
\FALabel(7.71318,9.65636)[bl]{$\psi$}
\FAProp(10.,7.5)(5.,5.)(0.,){/Sine}{1}
\FALabel(7.71318,5.34364)[tl]{$W$}
\FAProp(10.,7.5)(15.,10.)(0.,){/Sine}{0}
\FALabel(12.7132,7.84364)[tl]{$V$}
\FAVert(10.,7.5){0}
\FAVert(5.,10.){0}
\FAVert(5.,15.){0}
\FAVert(5.,5.){0}
\FAVert(15.,10.){0}

\FADiagram{}
\FAProp(0.,15.)(5.,15.)(0.,){/Straight}{1}
\FALabel(2.5,16.07)[b]{$\mu$}
\FAProp(0.,5.)(5.,5.)(0.,){/Sine}{0}
\FALabel(2.5,3.93)[t]{$\gamma$}
\FAProp(20.,10.)(15.,10.)(0.,){/Straight}{-1}
\FALabel(17.5,11.07)[b]{$\mu$}
\FAProp(5.,15.)(15.,10.)(0.,){/Straight}{1}
\FALabel(10.2132,13.4064)[bl]{$\mu$}
\FAProp(5.,10.5)(5.,15.)(0.,){/Sine}{0}
\FALabel(6.07,12.75)[l]{$V$}
\FAProp(5.,10.5)(5.,5.)(0.545455,){/ScalarDash}{-1}
\FALabel(2.43,7.75)[r]{$\psi$}
\FAProp(5.,10.5)(5.,5.)(-0.545455,){/ScalarDash}{1}
\FALabel(7.27,9.)[l]{$\psi$}
\FAProp(15.,10.)(5.,5.)(0.,){/Sine}{0}
\FALabel(10.2132,6.59364)[tl]{$V$}
\FAVert(5.,10.5){0}
\FAVert(5.,15.){0}
\FAVert(15.,10.){0}
\FAVert(5.,5.){0}

\FADiagram{}
\FAProp(0.,15.)(5.,15.)(0.,){/Straight}{1}
\FALabel(2.5,16.07)[b]{$\mu$}
\FAProp(0.,5.)(5.,5.)(0.,){/Sine}{0}
\FALabel(2.5,3.93)[t]{$\gamma$}
\FAProp(20.,10.)(15.,10.)(0.,){/Straight}{-1}
\FALabel(17.5,11.07)[b]{$\mu$}
\FAProp(5.,15.)(5.,5.)(0.,){/Straight}{1}
\FALabel(3.93,10.)[r]{$\mu$}
\FAProp(10.,12.5)(5.,15.)(0.,){/Sine}{0}
\FALabel(7.71318,14.6564)[bl]{$V$}
\FAProp(10.,12.5)(15.,10.)(0.,){/Straight}{1}
\FALabel(12.7132,12.1564)[bl]{$\mu$}
\FAProp(10.,7.5)(10.,12.5)(0.,){/Straight}{1}
\FALabel(8.93,10.)[r]{$\mu$}
\FAProp(10.,7.5)(5.,5.)(0.,){/Straight}{-1}
\FALabel(7.71318,5.34364)[tl]{$\mu$}
\FAProp(10.,7.5)(15.,10.)(0.,){/Sine}{0}
\FALabel(12.7132,7.84364)[tl]{$V$}
\FAVert(10.,7.5){0}
\FAVert(10.,12.5){0}
\FAVert(5.,15.){0}
\FAVert(5.,5.){0}
\FAVert(15.,10.){0}

\end{feynartspicture}
\vspace{-1.5em}


\begin{feynartspicture}(432,140)(3,1)

\FADiagram{}
\FAProp(0.,15.)(5.,15.)(0.,){/Straight}{1}
\FALabel(2.5,16.07)[b]{$\mu$}
\FAProp(0.,5.)(5.,5.)(0.,){/Sine}{0}
\FALabel(2.5,3.93)[t]{$\gamma$}
\FAProp(20.,10.)(13.4,10.)(0.,){/Straight}{-1}
\FALabel(16.7,11.07)[b]{$\mu$}
\FAProp(5.,15.)(13.4,10.)(0.,){/Straight}{0}
\FALabel(9.38493,13.1371)[bl]{$F$}
\FAProp(5.,11.5)(5.,15.)(0.,){/ScalarDash}{0}
\FALabel(4.18,13.25)[r]{$\phi$}
\FAProp(5.,8.5)(5.,11.5)(0.8,){/ScalarDash}{0}
\FALabel(7.02,10.)[l]{$\Sferm$}
\FAProp(5.,8.5)(5.,11.5)(-0.8,){/ScalarDash}{0}
\FALabel(2.98,10.)[r]{$\Sfermp$}
\FAProp(5.,8.5)(5.,5.)(0.,){/ScalarDash}{0}
\FALabel(4.18,6.75)[r]{$\psi$}
\FAProp(5.,5.)(13.4,10.)(0.,){/Sine}{0}
\FALabel(9.5128,6.6481)[tl]{$V$}
\FAVert(5.,8.5){0}
\FAVert(5.,11.5){0}
\FAVert(5.,15.){0}
\FAVert(5.,5.){0}
\FAVert(13.4,10.){0}

\FADiagram{}
\FAProp(0.,15.)(5.,15.)(0.,){/Straight}{1}
\FALabel(2.5,16.07)[b]{$\mu$}
\FAProp(0.,5.)(5.,5.)(0.,){/Sine}{0}
\FALabel(2.5,3.93)[t]{$\gamma$}
\FAProp(20.,10.)(15.,10.)(0.,){/Straight}{-1}
\FALabel(17.5,11.07)[b]{$\mu$}
\FAProp(5.,15.)(5.,5.)(0.,){/ScalarDash}{0}
\FALabel(4.18,10.)[r]{$\psi$}
\FAProp(5.,15.)(15.,10.)(0.,){/Straight}{0}
\FALabel(10.1014,13.1828)[bl]{$F$}
\FAProp(8.,6.5)(5.,5.)(0.,){/Sine}{0}
\FALabel(6.71318,4.84364)[tl]{$V$}
\FAProp(12.,8.5)(8.,6.5)(0.8,){/ScalarDash}{0}
\FALabel(9.09862,9.78276)[br]{$\Sferm$}
\FAProp(12.,8.5)(8.,6.5)(-0.8,){/ScalarDash}{0}
\FALabel(10.9014,5.21724)[tl]{$\Sfermp$}
\FAProp(12.,8.5)(15.,10.)(0.,){/Sine}{0}
\FALabel(13.7132,8.34364)[tl]{$V$}
\FAVert(12.,8.5){0}
\FAVert(8.,6.5){0}
\FAVert(5.,15.){0}
\FAVert(5.,5.){0}
\FAVert(15.,10.){0}

\FADiagram{}
\FAProp(0.,15.)(5.,15.)(0.,){/Straight}{1}
\FALabel(2.5,16.07)[b]{$\mu$}
\FAProp(0.,5.)(5.,5.)(0.,){/Sine}{0}
\FALabel(2.5,3.93)[t]{$\gamma$}
\FAProp(20.,10.)(15.,10.)(0.,){/Straight}{-1}
\FALabel(17.5,11.07)[b]{$\mu$}
\FAProp(5.,15.)(15.,10.)(0.,){/Straight}{0}
\FALabel(10.1014,13.1828)[bl]{$F$}
\FAProp(5.,10.)(5.,15.)(0.,){/ScalarDash}{0}
\FALabel(4.18,12.5)[r]{$\phi$}
\FAProp(5.,10.)(5.,5.)(0.,){/ScalarDash}{0}
\FALabel(4.18,7.5)[r]{$\Sferm$}
\FAProp(10.,7.5)(5.,10.)(0.,){/ScalarDash}{0}
\FALabel(7.60138,9.43276)[bl]{$\Sfermp$}
\FAProp(10.,7.5)(5.,5.)(0.,){/ScalarDash}{0}
\FALabel(7.60138,5.56724)[tl]{$\Sferm$}
\FAProp(10.,7.5)(15.,10.)(0.,){/Sine}{0}
\FALabel(12.7132,7.84364)[tl]{$V$}
\FAVert(10.,7.5){0}
\FAVert(5.,10.){0}
\FAVert(5.,15.){0}
\FAVert(5.,5.){0}
\FAVert(15.,10.){0}

\end{feynartspicture}
\vspace{-1.5em}


\begin{feynartspicture}(432,140)(3,1)

\FADiagram{}
\FAProp(0.,15.)(5.,15.)(0.,){/Straight}{1}
\FALabel(2.5,16.07)[b]{$\mu$}
\FAProp(0.,5.)(5.,5.)(0.,){/Sine}{0}
\FALabel(2.5,3.93)[t]{$\gamma$}
\FAProp(20.,10.)(13.4,10.)(0.,){/Straight}{-1}
\FALabel(16.7,11.07)[b]{$\mu$}
\FAProp(5.,15.)(13.4,10.)(0.,){/Straight}{0}
\FALabel(9.38493,13.1371)[bl]{$F$}
\FAProp(5.,11.5)(5.,15.)(0.,){/Sine}{0}
\FALabel(3.93,13.25)[r]{$V$}
\FAProp(5.,8.5)(5.,11.5)(0.8,){/ScalarDash}{0}
\FALabel(7.02,10.)[l]{$\Sferm$}
\FAProp(5.,8.5)(5.,11.5)(-0.8,){/ScalarDash}{0}
\FALabel(2.98,10.)[r]{$\Sfermp$}
\FAProp(5.,8.5)(5.,5.)(0.,){/Sine}{0}
\FALabel(3.93,6.75)[r]{$V$}
\FAProp(5.,5.)(13.4,10.)(0.,){/Sine}{0}
\FALabel(9.5128,6.6481)[tl]{$V$}
\FAVert(5.,8.5){0}
\FAVert(5.,11.5){0}
\FAVert(5.,15.){0}
\FAVert(5.,5.){0}
\FAVert(13.4,10.){0}

\FADiagram{}
\FAProp(0.,15.)(5.,15.)(0.,){/Straight}{1}
\FALabel(2.5,16.07)[b]{$\mu$}
\FAProp(0.,5.)(5.,5.)(0.,){/Sine}{0}
\FALabel(2.5,3.93)[t]{$\gamma$}
\FAProp(20.,10.)(15.,10.)(0.,){/Straight}{-1}
\FALabel(17.5,11.07)[b]{$\mu$}
\FAProp(5.,15.)(5.,5.)(0.,){/Sine}{0}
\FALabel(3.93,10.)[r]{$V$}
\FAProp(5.,15.)(15.,10.)(0.,){/Straight}{0}
\FALabel(10.1014,13.1828)[bl]{$F$}
\FAProp(8.,6.5)(5.,5.)(0.,){/Sine}{0}
\FALabel(6.71318,4.84364)[tl]{$V$}
\FAProp(12.,8.5)(8.,6.5)(0.8,){/ScalarDash}{0}
\FALabel(9.09862,9.78276)[br]{$\Sferm$}
\FAProp(12.,8.5)(8.,6.5)(-0.8,){/ScalarDash}{0}
\FALabel(10.9014,5.21724)[tl]{$\Sfermp$}
\FAProp(12.,8.5)(15.,10.)(0.,){/Sine}{0}
\FALabel(13.7132,8.34364)[tl]{$V$}
\FAVert(12.,8.5){0}
\FAVert(8.,6.5){0}
\FAVert(5.,15.){0}
\FAVert(5.,5.){0}
\FAVert(15.,10.){0}

\FADiagram{}
\FAProp(0.,15.)(5.,15.)(0.,){/Straight}{1}
\FALabel(2.5,16.07)[b]{$\mu$}
\FAProp(0.,5.)(5.,5.)(0.,){/Sine}{0}
\FALabel(2.5,3.93)[t]{$\gamma$}
\FAProp(20.,10.)(15.,10.)(0.,){/Straight}{-1}
\FALabel(17.5,11.07)[b]{$\mu$}
\FAProp(5.,15.)(15.,10.)(0.,){/Straight}{0}
\FALabel(10.1014,13.1828)[bl]{$F$}
\FAProp(5.,10.)(5.,15.)(0.,){/Sine}{0}
\FALabel(3.93,12.5)[r]{$V$}
\FAProp(5.,10.)(5.,5.)(0.,){/ScalarDash}{0}
\FALabel(4.18,7.5)[r]{$\Sferm$}
\FAProp(10.,7.5)(5.,10.)(0.,){/ScalarDash}{0}
\FALabel(7.60138,9.43276)[bl]{$\Sfermp$}
\FAProp(10.,7.5)(5.,5.)(0.,){/ScalarDash}{0}
\FALabel(7.60138,5.56724)[tl]{$\Sferm$}
\FAProp(10.,7.5)(15.,10.)(0.,){/Sine}{0}
\FALabel(12.7132,7.84364)[tl]{$V$}
\FAVert(10.,7.5){0}
\FAVert(5.,10.){0}
\FAVert(5.,15.){0}
\FAVert(5.,5.){0}
\FAVert(15.,10.){0}

\end{feynartspicture}
\vspace{-1.5em}


\begin{feynartspicture}(432,140)(3,1)

\FADiagram{}
\FAProp(0.,15.)(5.,15.)(0.,){/Straight}{1}
\FALabel(2.5,16.07)[b]{$\mu$}
\FAProp(0.,5.)(5.,5.)(0.,){/Sine}{0}
\FALabel(2.5,3.93)[t]{$\gamma$}
\FAProp(20.,10.)(13.4,10.)(0.,){/Straight}{-1}
\FALabel(16.7,11.07)[b]{$\mu$}
\FAProp(5.,15.)(13.4,10.)(0.,){/Straight}{0}
\FALabel(9.38493,13.1371)[bl]{$F$}
\FAProp(5.,11.5)(5.,15.)(0.,){/ScalarDash}{0}
\FALabel(4.18,13.25)[r]{$\phi$}
\FAProp(5.,8.5)(5.,11.5)(0.8,){/Straight}{0}
\FALabel(7.02,10.)[l]{$f$}
\FAProp(5.,8.5)(5.,11.5)(-0.8,){/Straight}{0}
\FALabel(2.98,10.)[r]{$f'$}
\FAProp(5.,8.5)(5.,5.)(0.,){/ScalarDash}{0}
\FALabel(4.18,6.75)[r]{$\psi$}
\FAProp(5.,5.)(13.4,10.)(0.,){/Sine}{0}
\FALabel(9.5128,6.6481)[tl]{$V$}
\FAVert(5.,8.5){0}
\FAVert(5.,11.5){0}
\FAVert(5.,15.){0}
\FAVert(5.,5.){0}
\FAVert(13.4,10.){0}

\FADiagram{}
\FAProp(0.,15.)(5.,15.)(0.,){/Straight}{1}
\FALabel(2.5,16.07)[b]{$\mu$}
\FAProp(0.,5.)(5.,5.)(0.,){/Sine}{0}
\FALabel(2.5,3.93)[t]{$\gamma$}
\FAProp(20.,10.)(15.,10.)(0.,){/Straight}{-1}
\FALabel(17.5,11.07)[b]{$\mu$}
\FAProp(5.,15.)(5.,5.)(0.,){/ScalarDash}{0}
\FALabel(4.18,10.)[r]{$\psi$}
\FAProp(5.,15.)(15.,10.)(0.,){/Straight}{0}
\FALabel(10.1014,13.1828)[bl]{$F$}
\FAProp(8.,6.5)(5.,5.)(0.,){/Sine}{0}
\FALabel(6.71318,4.84364)[tl]{$V$}
\FAProp(12.,8.5)(8.,6.5)(0.8,){/Straight}{0}
\FALabel(9.09862,9.78276)[br]{$f$}
\FAProp(12.,8.5)(8.,6.5)(-0.8,){/Straight}{0}
\FALabel(10.9014,5.21724)[tl]{$f'$}
\FAProp(12.,8.5)(15.,10.)(0.,){/Sine}{0}
\FALabel(13.7132,8.34364)[tl]{$V$}
\FAVert(12.,8.5){0}
\FAVert(8.,6.5){0}
\FAVert(5.,15.){0}
\FAVert(5.,5.){0}
\FAVert(15.,10.){0}

\FADiagram{}
\FAProp(0.,15.)(5.,15.)(0.,){/Straight}{1}
\FALabel(2.5,16.07)[b]{$\mu$}
\FAProp(0.,5.)(5.,5.)(0.,){/Sine}{0}
\FALabel(2.5,3.93)[t]{$\gamma$}
\FAProp(20.,10.)(15.,10.)(0.,){/Straight}{-1}
\FALabel(17.5,11.07)[b]{$\mu$}
\FAProp(5.,15.)(15.,10.)(0.,){/Straight}{0}
\FALabel(10.1014,13.1828)[bl]{$F$}
\FAProp(5.,10.)(5.,15.)(0.,){/ScalarDash}{0}
\FALabel(4.18,12.5)[r]{$\phi$}
\FAProp(5.,10.)(5.,5.)(0.,){/Straight}{0}
\FALabel(4.18,7.5)[r]{$f$}
\FAProp(10.,7.5)(5.,10.)(0.,){/Straight}{0}
\FALabel(7.60138,9.43276)[bl]{$f'$}
\FAProp(10.,7.5)(5.,5.)(0.,){/Straight}{0}
\FALabel(7.60138,5.56724)[tl]{$f$}
\FAProp(10.,7.5)(15.,10.)(0.,){/Sine}{0}
\FALabel(12.7132,7.84364)[tl]{$V$}
\FAVert(10.,7.5){0}
\FAVert(5.,10.){0}
\FAVert(5.,15.){0}
\FAVert(5.,5.){0}
\FAVert(15.,10.){0}

\end{feynartspicture}
\vspace{-1.5em}


\begin{feynartspicture}(432,140)(3,1)

\FADiagram{}
\FAProp(0.,15.)(5.,15.)(0.,){/Straight}{1}
\FALabel(2.5,16.07)[b]{$\mu$}
\FAProp(0.,5.)(5.,5.)(0.,){/Sine}{0}
\FALabel(2.5,3.93)[t]{$\gamma$}
\FAProp(20.,10.)(13.4,10.)(0.,){/Straight}{-1}
\FALabel(16.7,11.07)[b]{$\mu$}
\FAProp(5.,15.)(13.4,10.)(0.,){/Straight}{0}
\FAProp(5.,11.5)(5.,15.)(0.,){/ScalarDash}{0}
\FALabel(4.18,13.25)[r]{$\phi$}
\FAProp(5.,8.5)(5.,11.5)(0.8,){/Straight}{0}
\FALabel(7.02,10.)[l]{$\tilde{\chi}$}
\FAProp(5.,8.5)(5.,11.5)(-0.8,){/Straight}{0}
\FAProp(5.,8.5)(5.,5.)(0.,){/ScalarDash}{0}
\FALabel(4.18,6.75)[r]{$\psi$}
\FAProp(5.,5.)(13.4,10.)(0.,){/Sine}{0}
\FALabel(9.5128,6.6481)[tl]{$V$}
\FAVert(5.,8.5){0}
\FAVert(5.,11.5){0}
\FAVert(5.,15.){0}
\FAVert(5.,5.){0}
\FAVert(13.4,10.){0}

\FADiagram{}
\FAProp(0.,15.)(5.,15.)(0.,){/Straight}{1}
\FALabel(2.5,16.07)[b]{$\mu$}
\FAProp(0.,5.)(5.,5.)(0.,){/Sine}{0}
\FALabel(2.5,3.93)[t]{$\gamma$}
\FAProp(20.,10.)(15.,10.)(0.,){/Straight}{-1}
\FALabel(17.5,11.07)[b]{$\mu$}
\FAProp(5.,15.)(5.,5.)(0.,){/ScalarDash}{0}
\FALabel(4.18,10.)[r]{$\psi$}
\FAProp(5.,15.)(15.,10.)(0.,){/Straight}{0}
\FAProp(8.,6.5)(5.,5.)(0.,){/Sine}{0}
\FALabel(6.71318,4.84364)[tl]{$V$}
\FAProp(12.,8.5)(8.,6.5)(0.8,){/Straight}{0}
\FALabel(9.09862,9.78276)[br]{$\tilde{\chi}$}
\FAProp(12.,8.5)(8.,6.5)(-0.8,){/Straight}{0}
\FAProp(12.,8.5)(15.,10.)(0.,){/Sine}{0}
\FALabel(13.7132,8.34364)[tl]{$V$}
\FAVert(12.,8.5){0}
\FAVert(8.,6.5){0}
\FAVert(5.,15.){0}
\FAVert(5.,5.){0}
\FAVert(15.,10.){0}

\FADiagram{}
\FAProp(0.,15.)(5.,15.)(0.,){/Straight}{1}
\FALabel(2.5,16.07)[b]{$\mu$}
\FAProp(0.,5.)(5.,5.)(0.,){/Sine}{0}
\FALabel(2.5,3.93)[t]{$\gamma$}
\FAProp(20.,10.)(15.,10.)(0.,){/Straight}{-1}
\FALabel(17.5,11.07)[b]{$\mu$}
\FAProp(5.,15.)(15.,10.)(0.,){/Straight}{0}
\FAProp(5.,10.)(5.,15.)(0.,){/ScalarDash}{0}
\FALabel(4.18,12.5)[r]{$\phi$}
\FAProp(5.,10.)(5.,5.)(0.,){/Straight}{0}
\FAProp(10.,7.5)(5.,10.)(0.,){/Straight}{0}
\FALabel(7.60138,9.43276)[bl]{$\tilde{\chi}$}
\FAProp(10.,7.5)(5.,5.)(0.,){/Straight}{0}
\FAProp(10.,7.5)(15.,10.)(0.,){/Sine}{0}
\FALabel(12.7132,7.84364)[tl]{$V$}
\FAVert(10.,7.5){0}
\FAVert(5.,10.){0}
\FAVert(5.,15.){0}
\FAVert(5.,5.){0}
\FAVert(15.,10.){0}

\end{feynartspicture}


\caption{
Some MSSM two-loop diagrams for $\amu$ with (depending on
the diagram) 
$F = \mu, \bar\nu_\mu$; 
$f, f' = t, b, \tau, \nu_\tau$; 
$\phi = h, H, A, H^\pm, G, G^\pm$; 
$\psi = G^\pm, H^\pm$; 
$\Sferm, \Sfermp = \Stop, \Sbot, \Stau, \Snet$;
$V = \ga, Z, W$;
$\tilde\chi^\pm_{1,2}$;
$\tilde\chi^0_{1,2,3,4}$.
}
\vspace{-2em}
\label{fig:g-22L}
\end{center}
\end{figure}

In the MSSM, the bosonic electroweak two-loop contributions 
differ from the SM because of the extended MSSM Higgs
sector. 
This class is defined by selecting all MSSM two-loop
diagrams without a closed loop of fermions or sfermions and without
pure QED-diagrams,
see the first line in \reffi{fig:g-22L}. 
The results presented in this section have been obtained in
\citere{g-2CNH}. 

The result $\amuSUbos$ reads
\BEA
\amuSUbos & = & 
\frac{5}{3}\frac{\Gmu m_\mu^2}{8 \pi^2\sqrt{2}}
 \;\frac{\al}{\pi}
\KL c^\SUbos_L \log\frac{m_\mu^2}{\MW^2} 
+ c^\SUbos_0 \KR~,
\label{resTHDMbosonic}
\EEA
where the coefficient of the logarithm is given by
\BEA
\label{cLTHDM}
c^\SUbos_L & = & 
\frac{1}{30}\left[98+9c_L^{h}+23(1-4\sw^2)^2\right],\\
\label{cLh}
c_L^{h} & = & \frac{c_{2\be} \MZ^2}{c_\be} 
\KKL \frac{c_\al c_{\al+\be}}{\mH^2} + \frac{s_\al s_{\al+\be}}{\mh^2} \KKR~.
\EEA
Here $c_\al\equiv \cos\al$, etc. 
Using the tree-level relations in the Higgs sector, it can be shown
that $c_L^h = 1$, and thus the logarithms in the
SM and the MSSM are identical.
The coefficient $c^\SUbos_0$ is more complicated and not given here,
see \citere{g-2CNH}.


\subsubsection{Two-loop corrections with a closed SM fermion/sfermion loop}

The third known part are the diagrams
with a closed loop of SM fermions or scalar fermions calculated in
\citere{g-2FSf}, extending previous results of
\citeres{g-2BarrZee1,g-2BarrZee2}. 

The two-loop diagrams discussed in this subsection can be subdivided
into three classes (all diagrams are understood to include the corresponding 
subloop renormalization):
\\{($\sfn V \phi$)} 
diagrams with a sfermion ($\Stop$, $\Sbot$, $\Stau$,
$\Snet$) loop, where at least one gauge and one Higgs boson is
exchanged, see the second line of \reffi{fig:g-22L};
\\{($\sfn V V$)} diagrams with a
sfermion loop, where only gauge bosons appear in the second loop,
see the third line of \reffi{fig:g-22L};
\\{($f V \phi$)} diagrams with a fermion ($t$, $b$, $\tau$, $\nu_\tau$)
loop, where at least one gauge and one Higgs boson are
present in the other loop, see the fourth line of 
\reffi{fig:g-22L}. The corresponding
diagrams with only gauge bosons are identical to the SM diagrams and
give no genuine SUSY contribution.
The difference between the SM and the MSSM 
originates from the extended Higgs
sector of the MSSM. Diagrams where two Higgs bosons couple to
the external muon are suppressed by an extra factor of $m_\mu^2/\MW^2$ and
hence negligible.

The counterterm diagrams contain the renormalization
constants $\de M^2_{W,Z}$, $\de Z_e$, $\de t_{h,H}$ corresponding
to mass, charge and tadpole renormalization and can be easily
evaluated. For the evaluation the on-shell renormalization scheme has
been chosen. This leads to 
$\de M^2_{W,Z} = {\rm Re}\Si^{W,Z}(M^2_{W,Z})$, where 
$\Si^{W,Z}$ denote the transverse parts of the gauge-boson
self-energies, see \refse{subsec:gaugebosonren}.
The charge renormalization is given by $\de Z_e = - 1/2\; \Pi^\ga(0)$,
see \refeq{chargeren}. 
The tadpoles are renormalized such that the sum of the tadpole
contribution $T$ and the counterterm vanishes, i.e.\
$\de t_{h,H} = -T_{h,H}$, see \refse{subsec:Higgsrenorm}.

Numerically the most important contribution comes from the diagrams
with a Higgs boson and photon exchange.
This type of
contributions can be particularly enhanced by the ratio of the mass
scale of the dimensionful Higgs--Sfermion coupling
divided by the mass scale of the particles running in the loop, i.e.\
by ratios of the form
$\{\mu,A,\frac{\mt^2}{\MW}\}/\{m_{\sfn},m_{h,H}\}$, which can be much
larger than one. For
large $\tb$ and large sfermion mixing, the leading 
terms are typically given by the parts of the couplings with the
highest power of $\tb$ 
and by the 
loop with the lightest sfermion. 
These contributions involve only $H$-exchange, since the
$h$-couplings approach the SM-Higgs coupling for not too small $\MA$.
They can be well  approximated by the formulas~\cite{g-2FSf}
\BEA
\label{stopcontrib}
\De\amu^{\Stop,{\rm 2L}} &=&
-0.013\times10^{-10}\;\frac{\mt\, \mu \tb}{\mst \mH}{\rm\ sign}(\At), \\
\De\amu^{\Sbot,{\rm 2L}} &=&
-0.0032\times10^{-10}\;\frac{\mb\, \Ab \tan^2\beta}{\msb \mH}{\rm\ sign}(\mu),
\label{sbotcontrib}
\EEA
where $\mst$ and $\msb$ are the masses of the lighter $\Stop$ and
$\Sbot$, respectively, and $\mH$ is the mass of the heavy $\cp$-even
Higgs boson. The formulas  holds up to few percent if the respective
sfermion mass fulfils $m_{\Stop,\Sbot}\lsim\mH$. 
Since the heavier sfermions also contribute and
tend to cancel the contributions of the lighter sfermions, these
formulas do not approximate the full result very 
precisely, but they do provide the right sign and order of magnitude.


\subsubsection{Two-loop contributions with a closed
               chargino/neutralino loop}

The 2-loop contributions to $\amu$ containing a closed
chargino/neutralino loop constitute a separately UV-finite and
gauge-independent class and have been evaluated in \citere{g-2CNH}. 
Corresponding diagrams are shown in the last line of 
\reffi{fig:g-22L}.

The chargino/neutralino two-loop contributions, $\amuX$,
depend on the mass parameters for the charginos and neutralinos $\mu$,
$M_{1,2}$, the $\cp$-odd Higgs mass $\MA$, and $\tb$. 
It is interesting to
note that, contrary to \citere{drMSSMgf2B},  no tree-level relations in
the Higgs sector were needed in order to find a UV-finite result.
This is due to the fact that each two-loop diagram contributing to
$(g-2)_\mu$ together with its corresponding subloop renormalization is
finite. 

The parameter dependence of
$\amuX$ is quite straightforward~\cite{g-2CNH}. 
If all supersymmetric mass scales
are set equal, $\mu=M_2=\MA\equiv \msusy$ (with the only exception that
$M_1=5/3\;\sw^2/\cw^2 \; M_2$), the approximate leading behaviour of 
$\amuX$ is simply given by $\tb/\msusy^2$, and the following relation
holds, 
\BE
\amuX \approx 11\times10^{-10}
\left(\frac{\tb}{50}\right)
\left(\frac{100 \gev}{M_{\rm SUSY}}\right)^2\;{\mbox{sign}}(\mu)~.
\label{chaneuapproxorg}
\end{equation}
As shown in \citere{g-2CNH}, the approximation is very
good except for very small $\msusy$ and small $\tb$, where
the leading term is suppressed by the small $\mu$, 
and subleading terms begin to dominate.


\subsubsection{Remaining intrinsic uncertainties}

So far, at the two-loop level, the MSSM corrections to the
Two-Higgs-Doublet model (THDM) one-loop diagrams have been
evaluated. The only exception here are the diagrams that contain as a
second loop an additional closed smuon-neutrino or muon-sneutrino
loop. However, these corrections are expected to be small. 

The remaining two-loop corrections that are not yet available are\\
$-$ the contributions with a mixed SM fermion/sfermion loop
    attached to a SUSY one-loop diagram.\\
$-$ the full THDM corrections to the SUSY one-loop
diagrams. This will include as a subset also the QED corrections
evaluated in \citere{g-2MSSMlog2l}, where, however, all SUSY masses
had been set equal to $\msusy$. 

The first missing class of mixed SM fermion/sfermion contributions
might in principle be as large as the SM fermion or scalar fermion
corrections obtained in \citere{g-2FSf}, see above. This leaves an
intrinsic uncertainty of about $\sim 3 \times 10^{-10}$. The second
class gives corrections smaller than 10\% to the MSSM one-loop
result. Assuming that the corresponding intrinsic uncertainties are
less than half of the evaluated corrections, the combined effect of the
unknown two-loop corrections can be
estimated to be about
\BE
\De\amu^{\rm intr} ~({\rm current}) = 6 \times 10^{-10}~.
\end{equation}
After a full two-loop calculation will be available, the intrinsic
theoretical uncertainty from unknown
QED and electroweak higher-order corrections should be at the level of
\BE
\De\amu^{\rm intr} ~({\rm future}) = 1 \times 10^{-10}~.
\end{equation}


\section{Tools and codes for the evaluation of electroweak precision
observables} 
\label{sec:toolsandcodes}

The large number of different fields in the MSSM gives rise to a
plethora of possible interaction vertices. Calculations at the one-loop
level and beyond therefore usually involve a lot of Feynman diagrams. The
diagrams in general contain several mass scales, making their evaluation
(in particular beyond one-loop order) increasingly difficult.
Since the necessary steps can be structured in a strictly algorithmic
way, they can be facilitated with the help of computer algebra tools
and numerical programs.

Computer algebra tools have heavily been used in deriving the results
discussed above. Because of the multitude of scales involved in SUSY
higher-order corrections, in most cases the result cannot be expressed
in a compact form. Instead, the results presented above have been transformed
into public computer codes (also being used for the numerical
evaluation in \refses{chapter3} and \ref{chapter4}).  


\subsection{Tools for the calculation of EWPO}
\label{subsec:tools}

The calculation of higher-order SUSY Feynman-diagrams consists of
several steps. First the topologically different diagrams
for the given loop order and the number of external legs need to be
generated. Inserting the fields of the model
under consideration into the topologies in all possible ways leads to
the Feynman diagrams. The Feynman rules translate these graphical
representations into mathematical expressions. Since the loop integrals
in general lead to divergences, the expressions
need to be regularized and renormalized. The evaluation of the Feynman
amplitudes involves a treatment of the
Lorentz structure of the amplitude, calculation of Dirac traces etc.
At the one-loop level it is possible to reduce all tensor integrals
to a set of standard scalar integrals,
which can be expressed in terms of known analytic functions.
In contrast to the one-loop case, no general algorithm exists so far for
the evaluation of two-loop corrections in the electroweak theory.
The main obstacle in two-loop calculations in massive gauge theories
is the complicated structure of the two-loop integrals, which makes both
the tensor integral reduction and the evaluation of scalar integrals
very difficult. In general the occurring integrals are not expressible
in terms of polylogarithmic functions~\cite{scharf}. For the
evaluation of some types 
of integrals that do not permit an analytic solution numerical
methods and expansions in their kinematical variables have been
developed. Computer-algebraic methods can facilitate most of the
above-mentioned steps. There are computer algebra packages available based
on {\em FORM}~\cite{form}, {\em Mathematica}~\cite{mathematica} or both.

A package for the generation of SUSY amplitudes and drawing the
corresponding diagrams is 
{\em FeynArts}~\cite{feynarts,fa-mssm}. As a feature of particular
importance for higher-order
calculations in the electroweak theory, {\em FeynArts} generates not only the
unrenormalized diagrams at a given loop order but also the counterterm
contributions at this order and the counterterm diagrams needed for the
subloop renormalization. For one-loop calculations with
up to four external legs (the inclusion of five external legs is
currently under way) the package {\em FormCalc}~\cite{formcalc} can be
used, where for numerics the {\em LoopTools}~\cite{looptools} package
can easily be linked. For the evaluation of two-loop diagrams with up to
two external legs the program {\em TwoCalc}~\cite{twocalc} can be used. 
It is based on an algorithm for the tensor reduction of
general two-loop 2-point functions and can be used
for an automatic reduction of Feynman
amplitudes for two-loop self-energies with arbitrary masses, external
momenta, and gauge parameters to a set of standard scalar integrals.
The above computer algebra codes evaluate
the multi-loop diagrams analytically without performing expansions for
small parameters etc.

The program {\em QGRAPH}~\cite{qgraph} is an efficient generator
for Feynman diagrams (so far restricted to the SM, see however
\citere{qgraphSusy}). As output the diagrams  
are encoded in a symbolic
notation. Being optimized for high speed, {\em QGRAPH} is particularly
useful for applications involving a very large number (i.e.~\order{10^4})
of diagrams. Its output, depending on the number of
scales and external legs can then be passed to
{\em MATAD}~\cite{matad}, {\em MINCER}~\cite{mincer} or 
{\em EXP}~\cite{exp}, where expansions for small parameters are
performed. 

An alternative package for SM and SUSY one-loop calculations is the
{\em GRACE} system~\cite{grace}.

Overviews about codes for higher-loop and -leg calculations can be
found in \citeres{RHreview,Weiglein:2001ci,SDAmsterdam}.


\subsection{Public codes for the numerical evaluation of EWPO}
\label{subsec:codes}

The results presented in \refses{sec:evalDelrho}, \ref{sec:evalMW} and
\ref{sec:Zobs} have 
been implemented in the code {\em POMSSM}%
\footnote{
A new version of {\em POMSSM} is currently prepared
and will be available from the authors.
}%
, which
has been used for the numerical evaluation in \refses{chapter3} and
\ref{chapter4}. The Higgs boson sector evaluations have been done with
the code {\em FeynHiggs}~\cite{feynhiggs,feynhiggs1.2,habilSH,feynhiggs2.2},
including the corrections described in \refse{sec:mh}. This code also
performs an evaluation of all Higgs boson decay widths as well as
production cross sections for photon colliders. 
Also the results for $\De\rho$ as
described in \refse{sec:evalDelrho} are included as a subroutine. 
Other codes for evaluations of Higgs sector observables are 
{\em Hdecay}~\cite{hdecay} and {\em CPsuperH}~\cite{cpsh}. 
The results for the anomalous magnetic
moment of the muon, described in \refse{sec:evalg-2}, are available as
a subroutine for the code {\em FeynHiggs}. 

\newpage

\chapter{MSSM predictions versus experimental data}
\label{chapter3}

Now we study the impact of the higher-order corrections 
to the
electroweak precision observables discussed above. The MSSM predictions
are compared with the current experimental results and constraints on
the parameter space of the unconstrained MSSM are discussed. We
furthermore investigate how the improved electroweak precision
measurements at the next generation of colliders enhance the 
sensitivity of testing the electroweak theory.


\section{MSSM predictions for $\MW$ and $\sweff$}
\label{sec:MWSWpred}

\subsection{Numerical analysis in the MSSM}
\label{subsec:MWSWnumanal}

\subsubsection{Results for $\De\rho$}

\begin{figure}[h!]
\begin{center}
\mbox{
\epsfig{figure=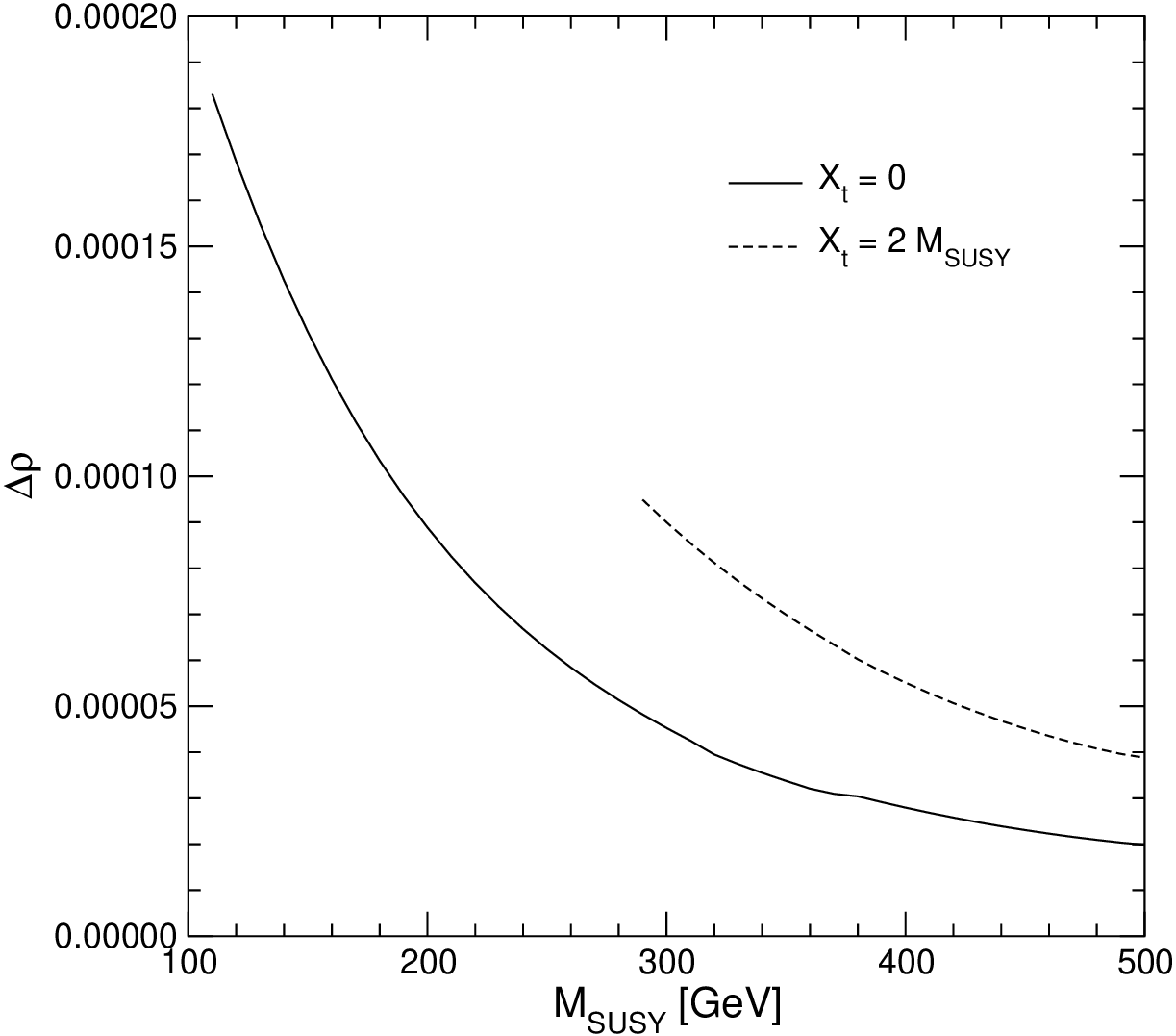,width=9cm,height=7cm}
     }
\caption{
$\De\rho_{1,{\rm gluon}}^{\SU}$ 
as a function of the common squark mass $\msusy$ for
$\tb = 3$, $\Xb = 0$ and $\Xt = 0, 2 \msusy$.
}
\label{fig:dr2lgluon}
\end{center}
\vspace{-1em}
\end{figure}

\begin{figure}[htb!]
\vspace{2em}
\begin{center}
\mbox{
\epsfig{figure=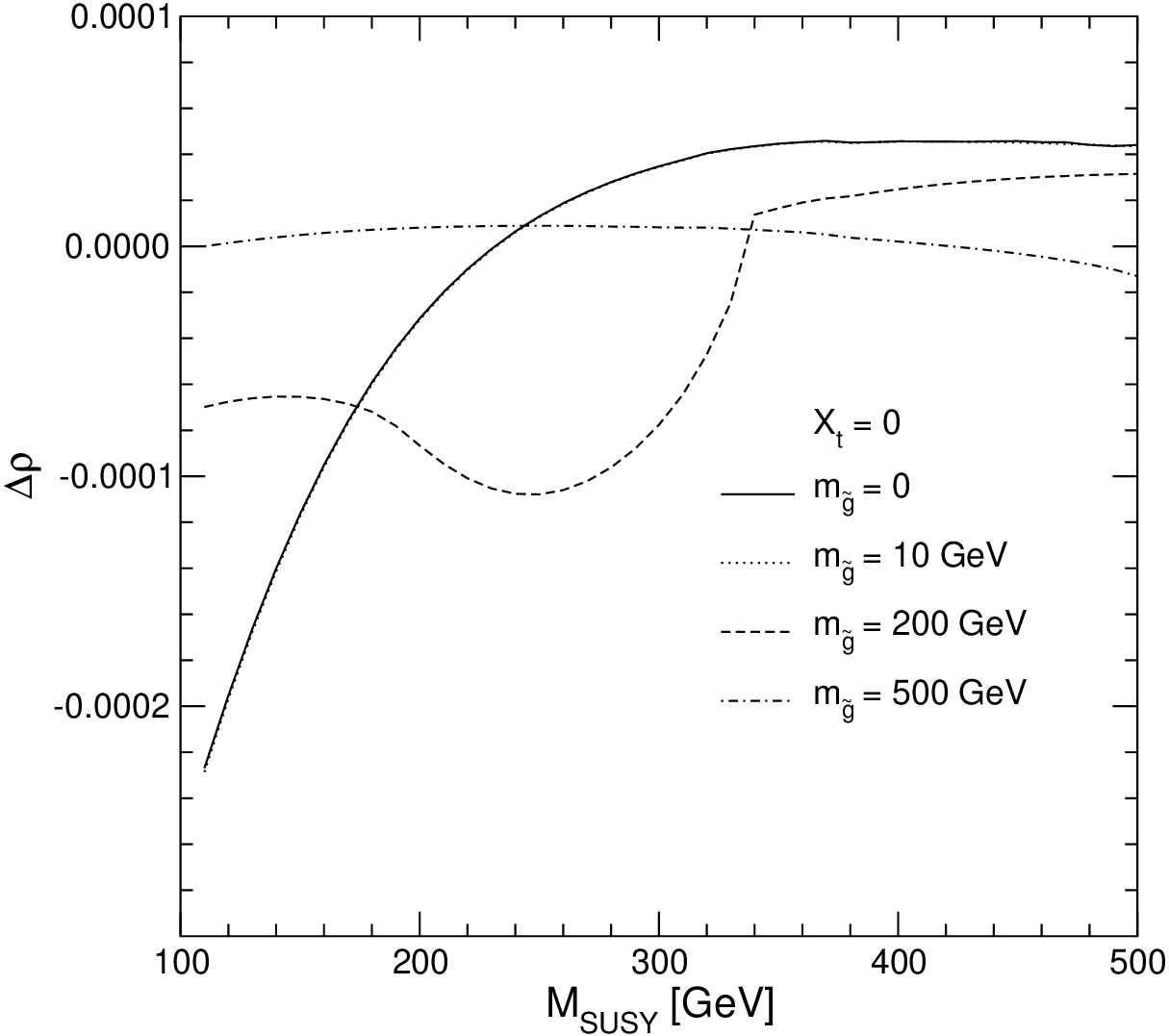,width=9cm,height=7cm}
     }
\vspace{1em}
\caption{
$\De\rho_{1,{\rm gluino}}^{\SU}$ 
as a function of the common squark mass $\msusy$ for
$\tb = 3$, $\Xb = 0$, $\Xt = 0$ and $\mgl = 0, 10$ 
(the curves are indistinguishable), $200, 500 \gev$
\cite{dr2lA}.
}
\label{fig:dr2lgluino1}
\end{center}
\end{figure}

\begin{figure}[htb!]
\begin{center}
\mbox{
\psfig{figure=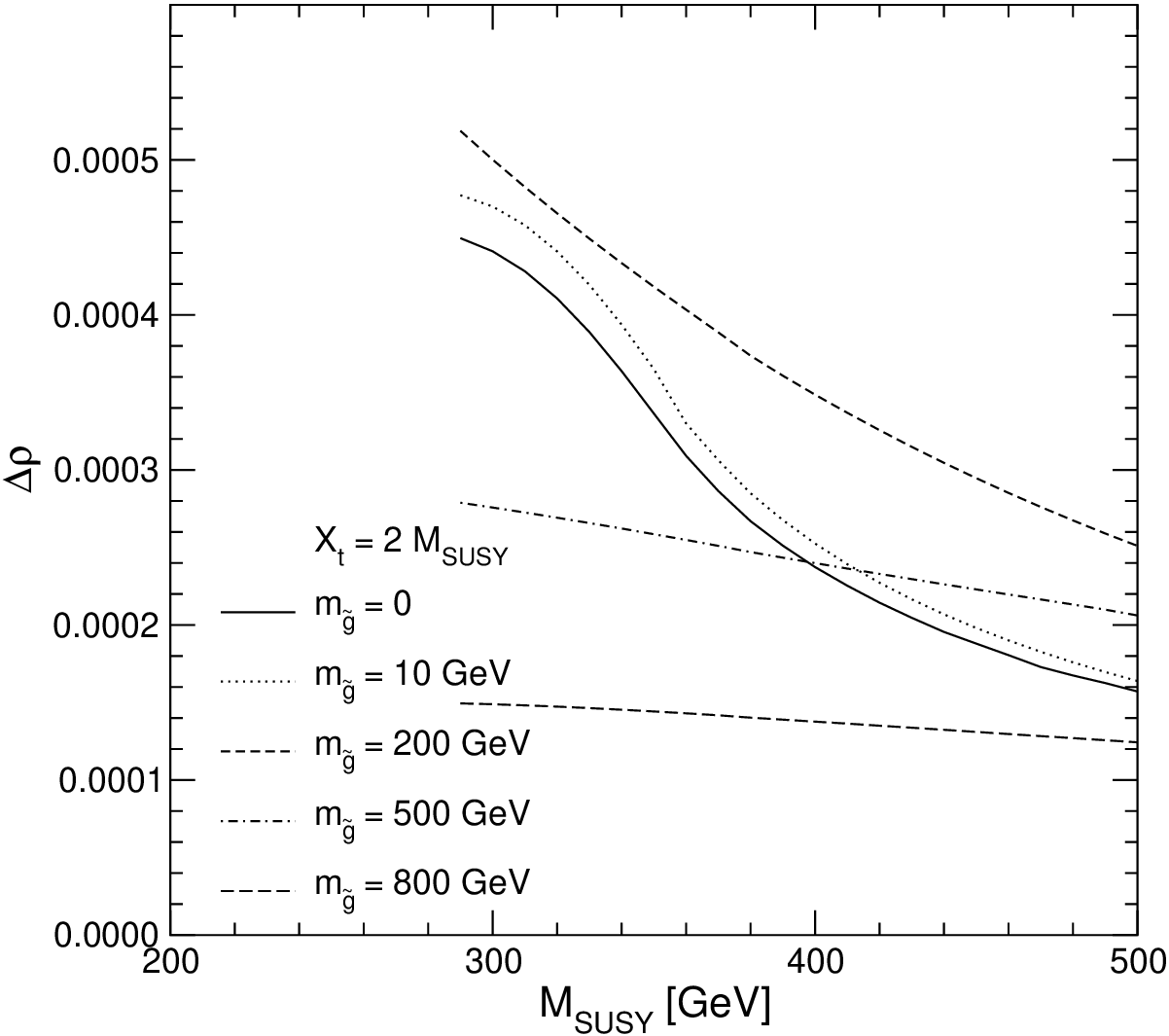,width=9cm,height=7cm}
     }
\vspace{1em}
\caption{
$\De\rho_{1,{\rm gluino}}^{\SU}$ 
as a function of the common squark mass $\msusy$ for
$\tb = 3$, 
$\Xb = 0$, $\Xt = 2 \msusy$ and $\mgl = 0, 10, 200, 500, 800 \gev$.}
\label{fig:dr2lgluino2}
\end{center}
\end{figure}

We start our discussion of the numerical results with the quantity
$\De\rho$, which parametrizes leading SUSY contributions to the
$W$-boson mass and the $Z$-boson
observables, see \refse{sec:evalDelrho}.
The effect of the gluonic SUSY \twol\ contributions as given in
\refeq{delrhoMSSM2l} (the four squark masses are renormalized on-shell;
the mass shift arising from the SU(2) relation is understood to be
absorbed into the one-loop result, see \refse{subsubsec:delrhomssm})
is shown for an exemplary
case in \reffi{fig:dr2lgluon} as a function of $\msusy$. The other parameters
are $\tb = 3$ and $\Xt = 0, 2 \msusy$. 
The line for $\Xt = 2 \msusy$ 
starts only at $\msusy \approx 300 \gev$. For lower values of $\msusy$
one of the scalar top mass squares is below zero.
$\De\rho_{1,{\rm gluon}}^\SU$ can
reach values of up to $0.2 \times 10^{-3}$. The results for the
gluino-exchange contribution are shown in \reffi{fig:dr2lgluino1} 
($\Xt = 0$) and \reffi{fig:dr2lgluino2} ($\Xt = 2 \msusy$) for
$\mgl = 0, 10, 200, 500 \gev$ (and $\mgl = 800 \gev$ in the latter) as
a function of $\msusy$. The results for 
$\mgl = 0$ and $10 \gev$ are indistinguishable for $\Xt = 0$. The
decoupling for large $\mgl$ is visible already for $\mgl = 500 \gev$.
In the case of $\Xt = 2 \msusy$, see \reffi{fig:dr2lgluino2},
$\De\rho_{1,{\rm gluino}}^\SU$ is in general positive and can reach
values up to $0.5 \times 10^{-3}$ for $\mgl = 200 \gev$. As can be
seen in the figure, for larger values of $\mgl$ the contribution to
$\De\rho$ decouples as expected.
Contrary to the SM case where the strong \twol\ corrections screen the
\onel\ result, the $\oaas$ corrections in the MSSM increase the \onel\
contributions by up to 35\%, thus enhancing the sensitivity to scalar
quark effects. 

\begin{figure}[htb!]
\vspace{1em}
\begin{center}
\mbox{
\epsfig{figure=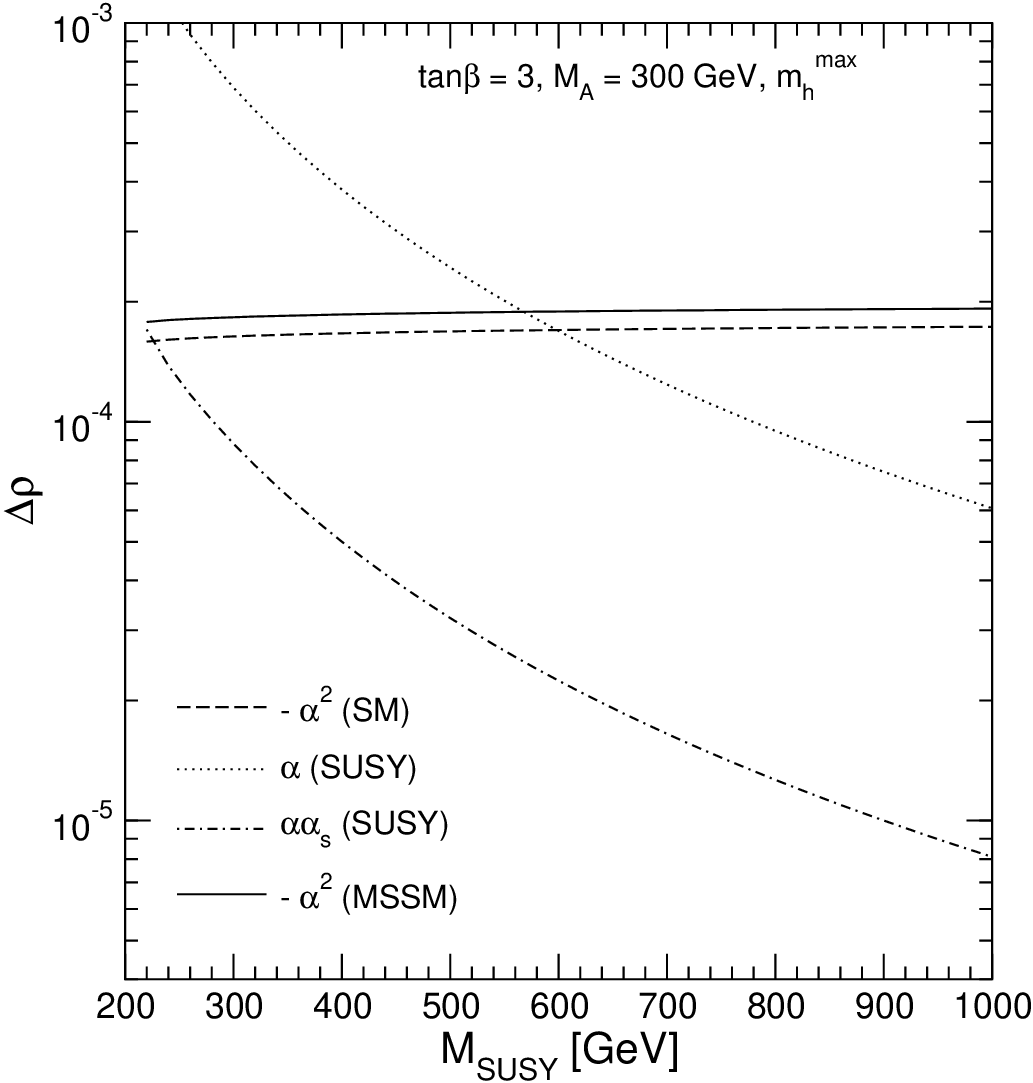,width=7cm,height=6.3cm}  
\hspace{1em}
\epsfig{figure=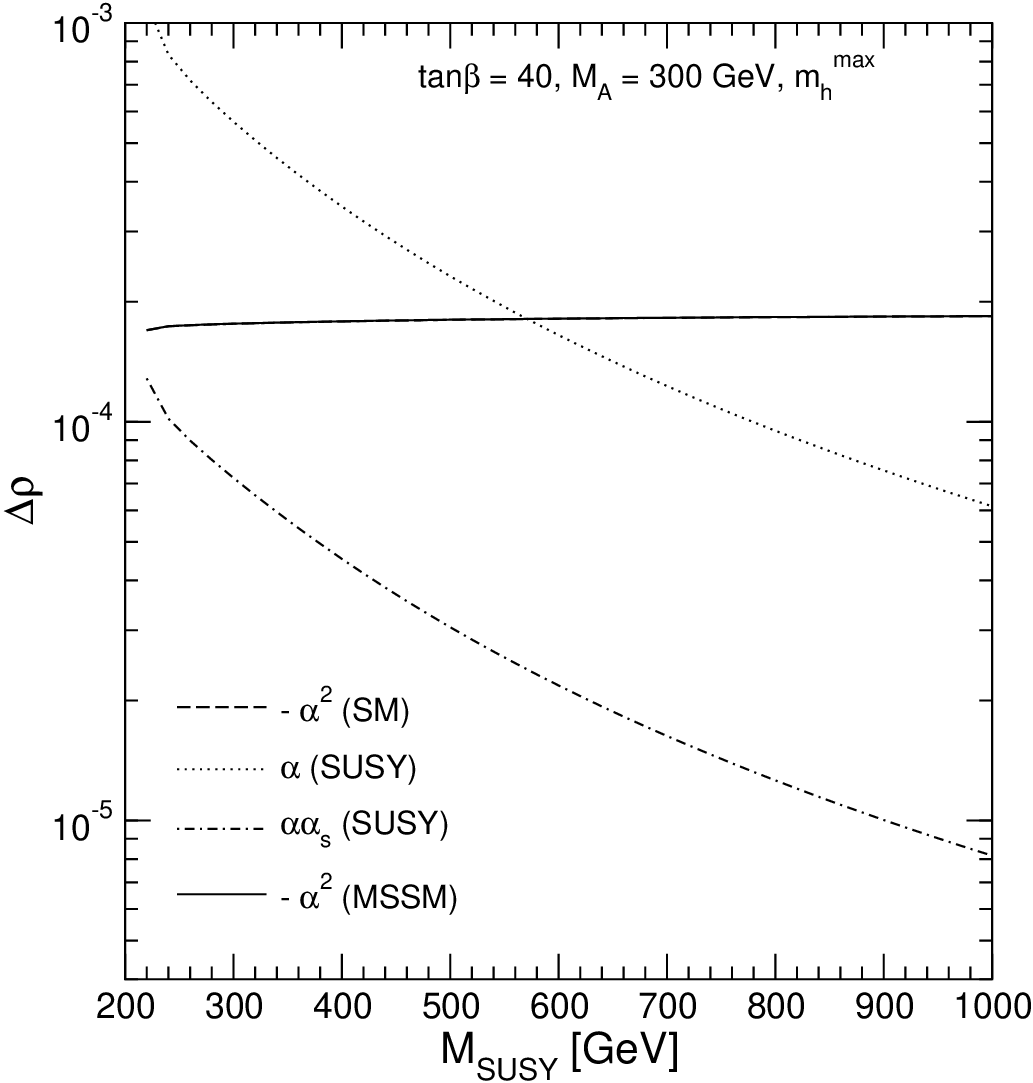,width=7cm,height=6.3cm} 
}
\end{center}
\caption[]{
The contribution of the leading \order{\alt^2} MSSM corrections in the
limit of large $\msusy$,
$\De\rho_{1,\hi}^{\SU}$, is shown as a function of $\msusy$ for
$\MA = 300 \gev$ and 
$\tb = 3$ (left plot) or $\tb = 40$ (right plot) in the case of
the $\mhmax$~scenario, see \refapp{chap:benchmark}.
$\De\rho_{1,\hi}^{\SU}$ is compared with the leading
\order{\alt^2} SM contribution and with the
leading MSSM corrections originating from the $\Stop/\Sbot$ sector of
\order{\al} and \order{\al\als}.
Both \order{\alt^2} contributions are
negative and are for comparison shown with reversed sign.
In the right plot the \order{\alt^2} corrections differ by about
$1.5 \times 10^{-7}$, which is indistinguishable in the plot.
}
\label{fig:dr2lgf2}
\vspace{1em}
\end{figure}

In \reffi{fig:dr2lgf2} the numerical result of the leading
\order{\alt^2} MSSM corrections in the limit 
of large $\msusy$ (see \refse{subsubsec:delrhomssm})
is shown. It is compared with the other
contributions to $\De\rho$: the \order{\alt^2} SM correction (with 
$\MH^{\SM} = \mh$) and the SUSY contributions from the scalar
quark sector at \order{\al} and \order{\al\als}. The results are shown
as a function of $\msusy$, which enters the
\order{\alt^2} corrections indirectly via its effect on $\mh$.
For small $\tb$ and $\MA = 300 \gev$, see the left plot of 
\reffi{fig:dr2lgf2}, the effective change arising from the new
genuine MSSM corrections compared to the \order{\alt^2} SM contribution
with $\MH^{\SM} = \mh$ is sizable. While the full \order{\alt^2}
result is larger than the $\oa$ corrections for 
$\msusy \gsim 600 \gev$, it is larger than the $\oaas$ corrections for
all $\msusy$. However, the genuine MSSM corrections 
are always smaller than the MSSM $\oa$ contributions but they are of
equal size as 
the $\oaas$ corrections for $\msusy \approx 600 \gev$.
(Note that for smaller $\msusy$
the approximation of neglecting the scalar-quark contributions in
the \order{\alt^2} result may no longer be valid.)
Since they enter with a different sign into $\De\rho$, they can
compensate each other.
Similar results are found in 
the no-mixing case, which is not shown here.

The case of large $\tb$ and $\MA = 300 \gev$ is shown in the right
plot of \reffi{fig:dr2lgf2}. 
The curve for the \order{\alt^2} MSSM corrections in the limit 
$\msusy \to \infty$ is indistinguishable in the plot
from the \order{\alt^2} SM contribution with $\MH^{\SM} = \mh$.
The difference between these two corrections is
approximately $1.5 \times 10^{-7}$, while the $\oaas$ corrections are
about $10^{-5}$ even for $\msusy = 1000 \gev$. The purely electroweak
corrections decouple much faster for large $\tb$ than the $\oaas$
corrections, see also \refeq{drallmalargetb}.

\begin{figure}[htb!]
\begin{center}
\mbox{
\epsfig{figure=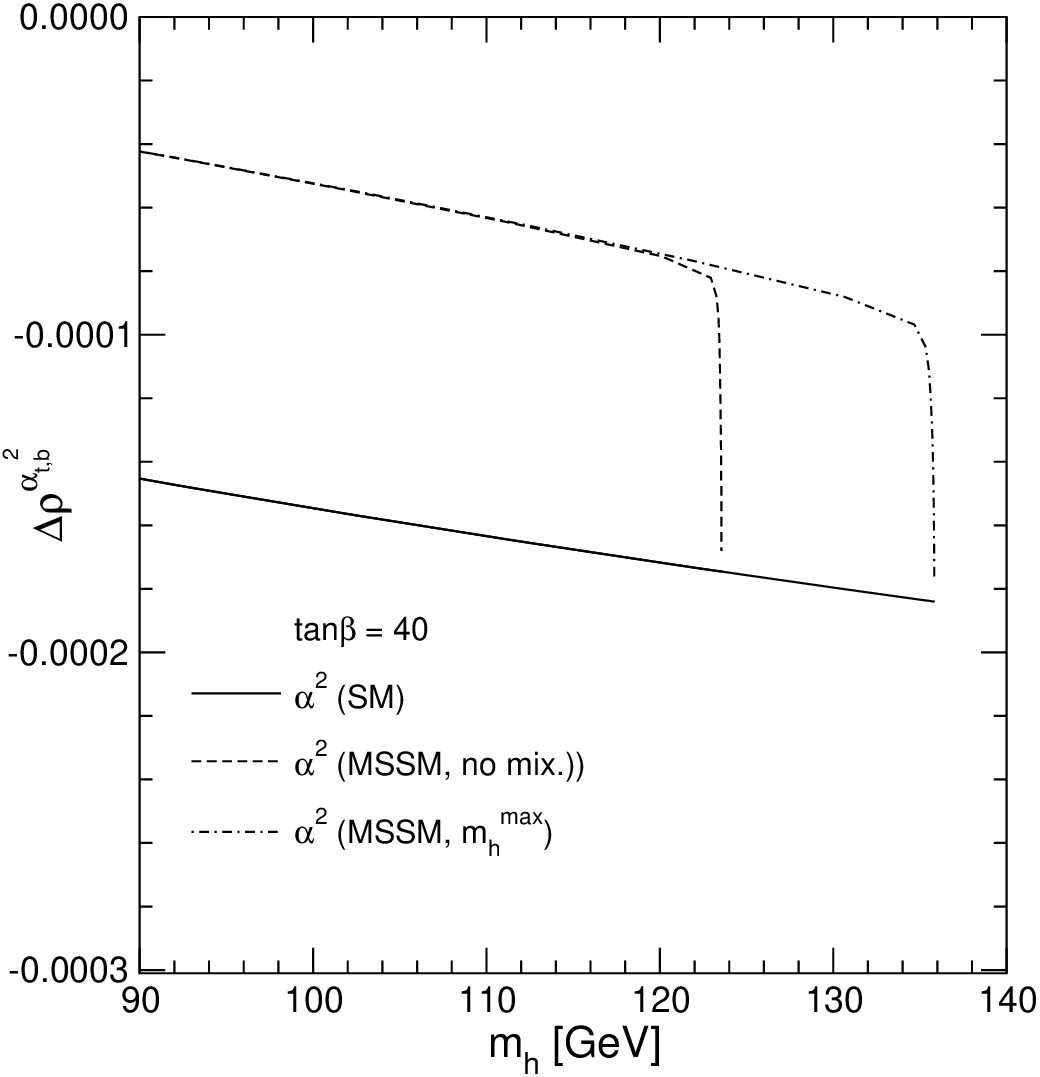,width=7cm,height=6.8cm} 
\hspace{1em}
\epsfig{figure=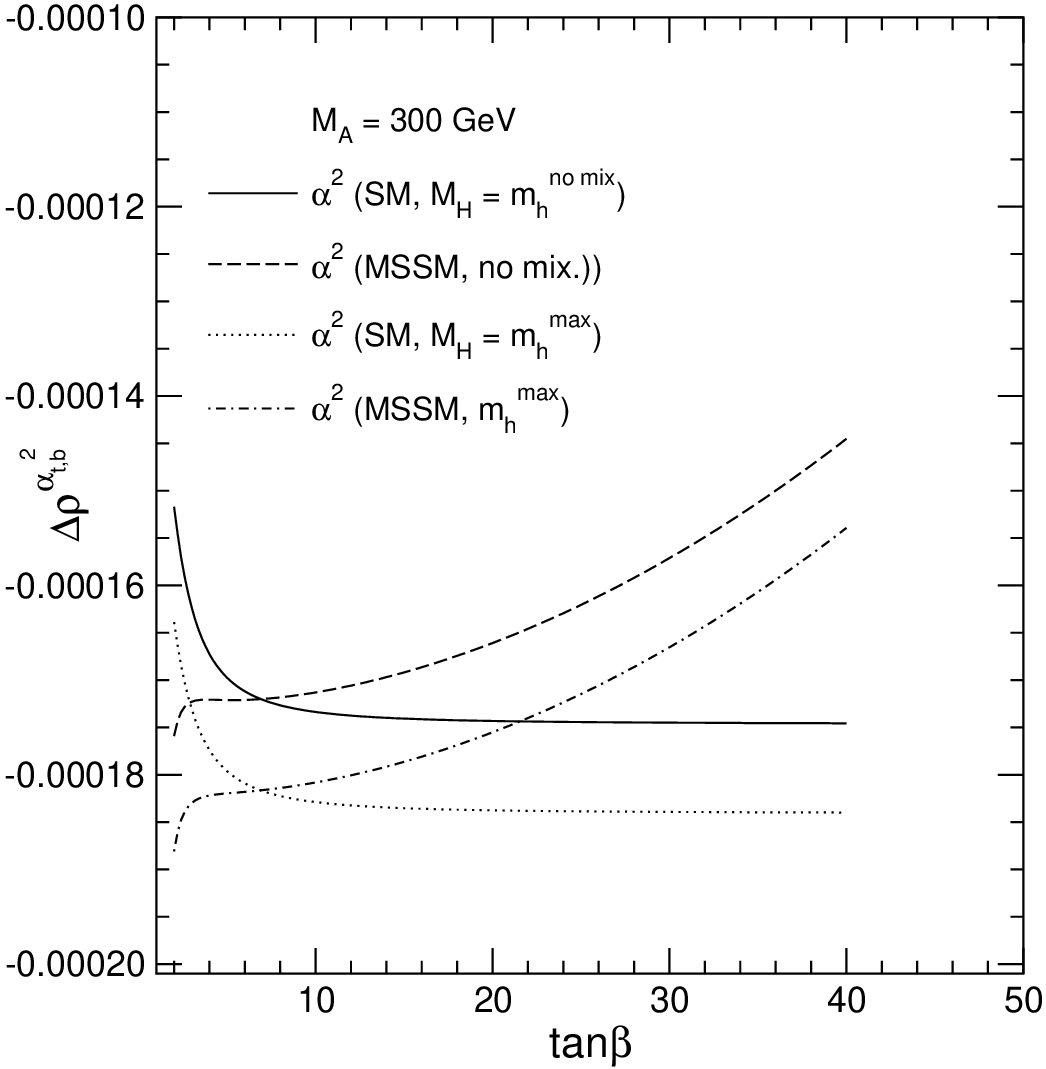,width=7cm,height=6.8cm} 
}
\end{center}
\caption[]{
The \order{\alt^2}, \order{\alt \alb}, and \order{\alb^2} MSSM
contributions to $\De\rho$ in the $\mhmax$ and the no-mixing scenarios
(see \refapp{chap:benchmark}) are compared with the corresponding SM
result with $\MHSM = \mh$.  
In the left plot $\tb$ is fixed to $\tb = 40$, while $\MA$
is varied from $50 \gev$ to $1000 \gev$. In the right plot $\MA$ is
set to $300 \gev$, while $\tb$ is varied. The bottom quark mass is set
to $\mb = 4.25 \gev$.
}
\label{fig:dr2lgf2tb}
\vspace{-1em}
\end{figure}

In \reffi{fig:dr2lgf2tb} we show the result for the \order{\al_t^2},
\order{\al_t \al_b}, and \order{\al_b^2} MSSM contributions to 
$\De\rho$ in the $\mhmax$ and the no-mixing
scenarios~\cite{benchmark,LHbenchmark} (see also \refapp{chap:benchmark}) 
compared with the corresponding SM result with $\MHSM = \mh$. 
In the left plot $\tb$ is fixed to $\tb = 40$ and $\MA$ is varied from
$50 \gev$ to $1000 \gev$. In the right plot $\MA$
is fixed to $\MA = 300 \gev$, while $\tb$ is varied. 

For large $\tb$ the \order{\al_t \al_b} and \order{\al_b^2}
contributions yield a significant effect caused by the heavy Higgs bosons
in the loops, entering with the other sign than the \order{\al_t^2}
corrections, while the contribution of the lightest Higgs boson is
SM-like. As one can see in \reffi{fig:dr2lgf2tb}, for large $\tb$ the
MSSM contribution to $\De\rho$ is smaller than the SM value. For large
values of $\MA$, the SM result is recovered.

\begin{figure}[t!]
\begin{center}
\mbox{
\psfig{figure=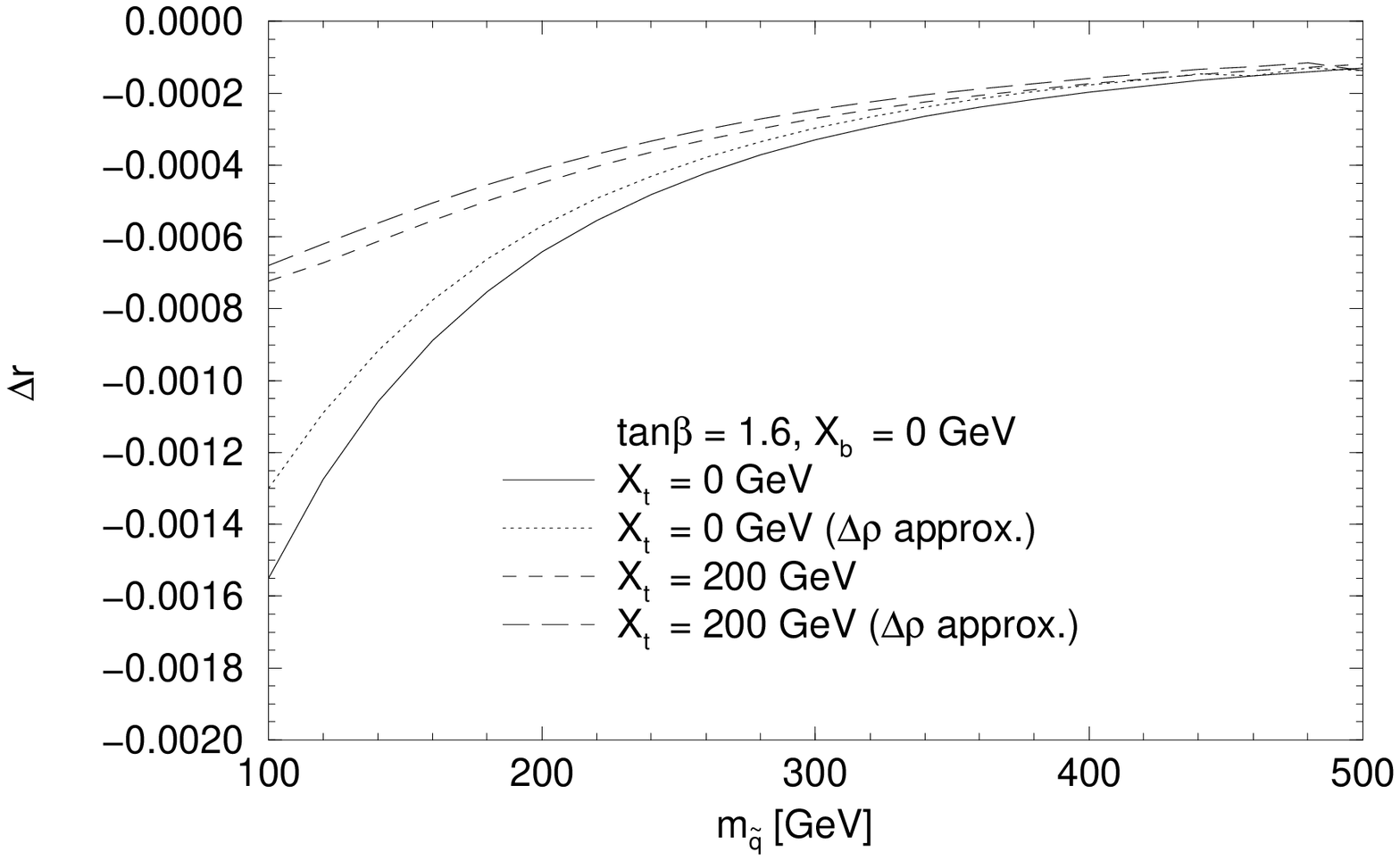,width=9cm,height=9cm,
       bbllx=120pt,bblly=70pt,bburx=450pt,bbury=420pt}
     }
\vspace{-1em}
\caption{
The $\Stop/\Sbot$ corrections to $\De r$ at the \twol\ level,
\refeq{deltarmssm},  are compared with the $\De\rho$ approximation,
\refeq{deltarapprox}. 
The results are shown as a function of $\msusy$ for $\tb = 1.6$, 
$\Xb = 0$ and $\Xt = 0, 200 \gev$.}
\label{fig:deltar2lA}
\end{center}
\end{figure}
\begin{figure}[h!]
\begin{center}
\mbox{
\psfig{figure=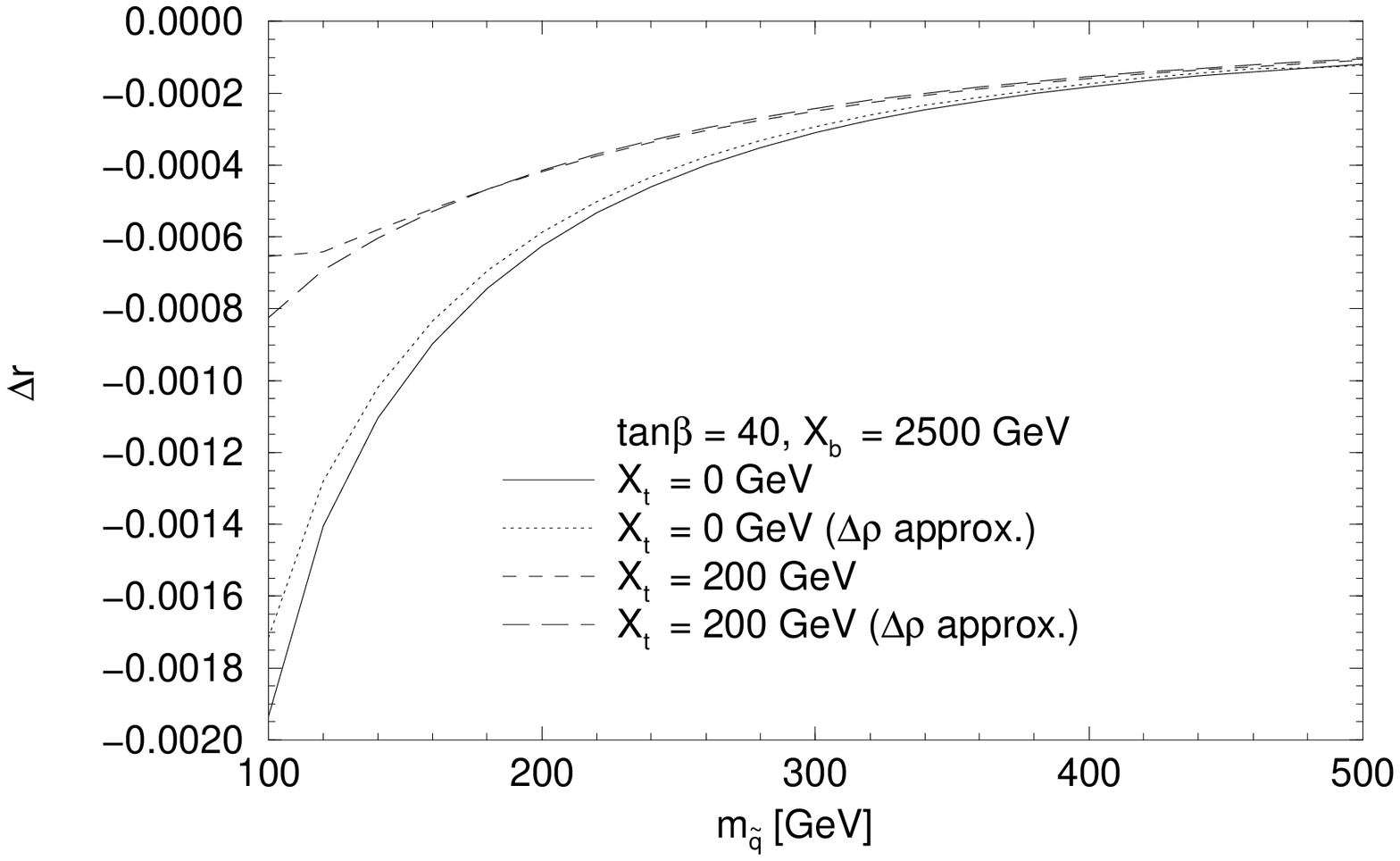,width=9cm,height=9cm,
       bbllx=120pt,bblly=70pt,bburx=450pt,bbury=420pt}
     }
\vspace{-1em}
\caption{
The $\Stop/\Sbot$ corrections to $\De r$ at the \twol\ level,
\refeq{deltarmssm},  are compared with the $\De\rho$ approximation,
\refeq{deltarapprox}. 
The results are shown as a function of $\msusy$ for $\tb = 40$, 
$\Xb = 2500 \gev$ and $\Xt = 0, 200 \gev$.}
\label{fig:deltar2lB}
\vspace{-1em}
\end{center}
\end{figure}


\subsubsection{Quality of the $\De\rho$ approximation}

We now turn to the numerical effects on $\MW$ and $\sweff$
induced by $\De\rho$. As a first step the quality of the $\De\rho$
approximation, using \refeq{precobs}, is analyzed~\cite{dr2lB}. 
We show the comparison
of the $\De\rho$ approximation with the full evaluation at the
two-loop level, where such a calculation is available. As described in
\refse{subsubsec:MWbeyond1LSUSY}, only the two-loop gluonic corrections
to $\De r$ have been calculated so far.
In \reffis{fig:deltar2lA}, \ref{fig:deltar2lB} we show the full
gluonic two-loop contribution to $\De r$ together with the
corresponding $\De\rho$ approximation. 
The no-mixing case in the $\Sbot$~sector is presented in
\reffi{fig:deltar2lA} with $\tb = 1.6$ and $\Xt = 0, 200 \gev$. The
case with $\Xb = 2500 \gev$ is shown in \reffi{fig:deltar2lB} with
$\tb = 40$ and $\Xt = 0, 200 \gev$. As for the \onel\ case, see
\reffis{fig:deltar1lA}, \ref{fig:deltar1lB}, also in the \twol\ case
the $\De\rho$ approximation reproduces the full result to better
than~10\%.


\subsubsection{Corrections to $\MW$ and $\sweff$ induced by $\De\rho$}
\label{subsubsec:MWsweffshift}

We illustrate the effects of the corrections to $\De\rho$ discussed above 
on the observables $\MW$ and $\sweff$ for the example of the
\order{\alt^2} MSSM contributions in the limit of large $\msusy$.

\begin{figure}[htb!]
\mbox{}\vspace{1em}
\begin{center}
\mbox{
\epsfig{figure=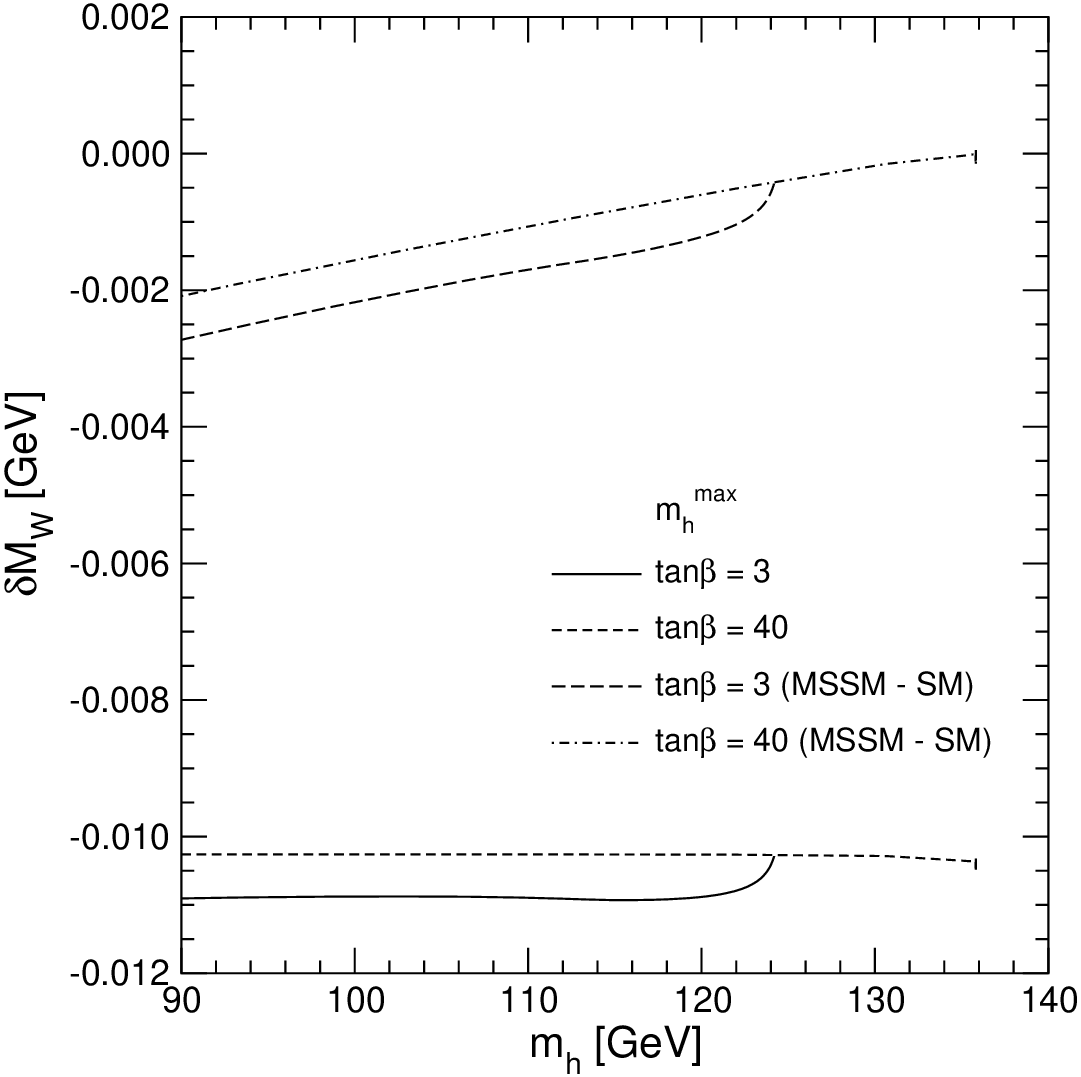,width=7cm,height=6.3cm}  
\hspace{1em}
\epsfig{figure=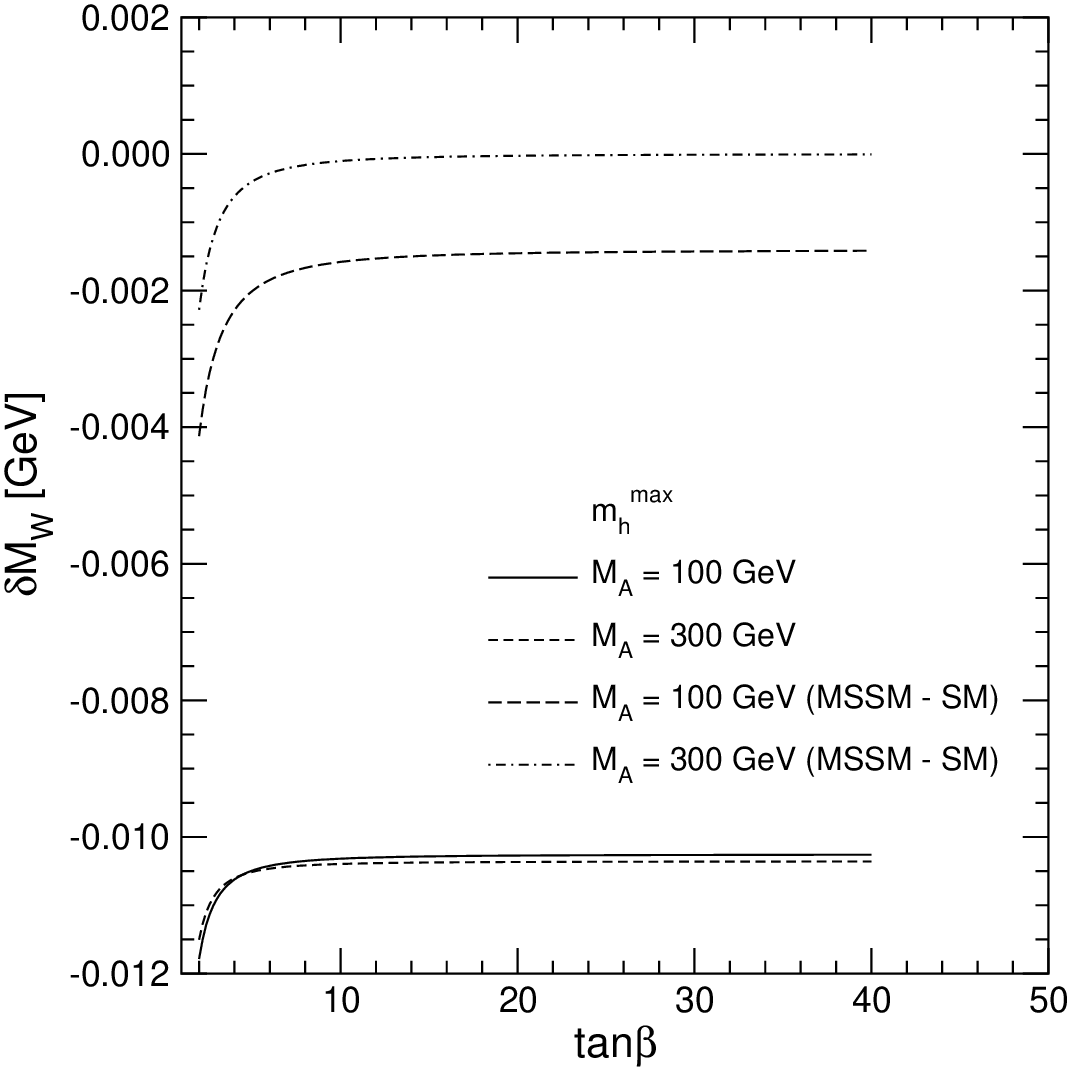,width=7cm,height=6.3cm} 
}
\end{center}
\caption[]{
The shift $\de\MW$ induced by the 
\order{\alt^2} MSSM contribution and the effective change 
compared with the SM result 
are shown for $\msusy = 1000 \gev$ in
the $\mhmax$~scenario. 
The other parameters are $\mu = 200 \gev, \Ab = \At$. $\mh$ is
obtained in the left (right) plot from varying $\MA$ from 
$50 \gev$ to $1000 \gev$, while keeping $\tb$ fixed at $\tb = 3, 40$
(from varying $\tb$ from 2 to 40, while keeping $\MA$ fixed at
$\MA = 100, 300 \gev$).
}
\label{fig:MW2lgf2}
\end{figure}

\reffi{fig:MW2lgf2} shows the shift $\de\MW$ induced by the 
\order{\alt^2} MSSM contribution for $\msusy = 1000 \gev$ in the
$\mhmax$~scenario, see \refapp{chap:benchmark}.
The other parameters are $\mu = 200 \gev, \Ab = \At$. $\mh$ is
obtained in the left (right) plot from varying $\MA$ from 
$50 \gev$ to $1000 \gev$, while keeping $\tb$ fixed at $\tb = 3, 40$
(from varying $\tb$ from 2 to 40, while keeping $\MA$ fixed at
$\MA = 100, 300 \gev$).
Besides the absolute \order{\alt^2} MSSM contribution (solid and
short-dashed lines) also the ``effective change'' compared to the SM is
shown, i.e.\ the difference between the \order{\alt^2} MSSM contribution
and the \order{\alt^2} SM contribution with $\MHSM = \mh$ (long-dashed
and dot-dashed lines). 
While the full result shows contributions to $\MW$ of up to
$11 \mev$, the effective change is much smaller, mostly below the
level of $2 \mev$. 

For large $\tb$ the \order{\alt \alb} and \order{\alb^2}
contributions yield a significant effect from the heavy Higgs bosons
in the loops, entering with the other sign than the \order{\al_t^2}
corrections, while the contribution of the lightest Higgs boson is
SM-like, see \refse{subsec:beyondoneloop}.
The effective change in the predictions for the precision observables
from the \order{\alt \alb} and \order{\alb^2} corrections can
exceed the one from the \order{\alt^2} corrections. It can amount up
to $\de\MW \approx +5 \mev$ for $\tb = 40$.

\begin{figure}[htb!]
\mbox{}\vspace{3em}
\begin{center}
\mbox{
\epsfig{figure=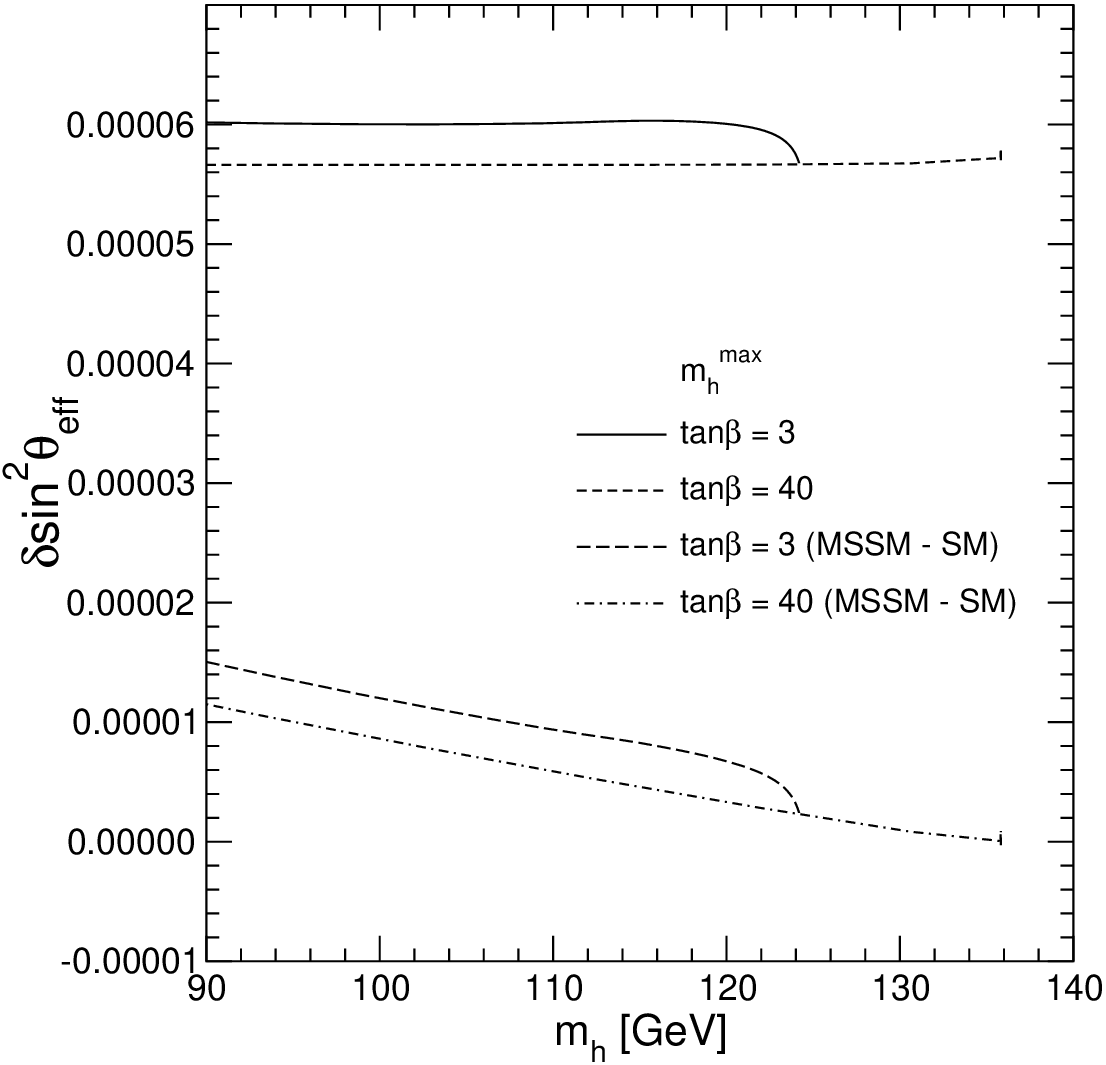,width=7cm,height=6.3cm}  
\hspace{1em}
\epsfig{figure=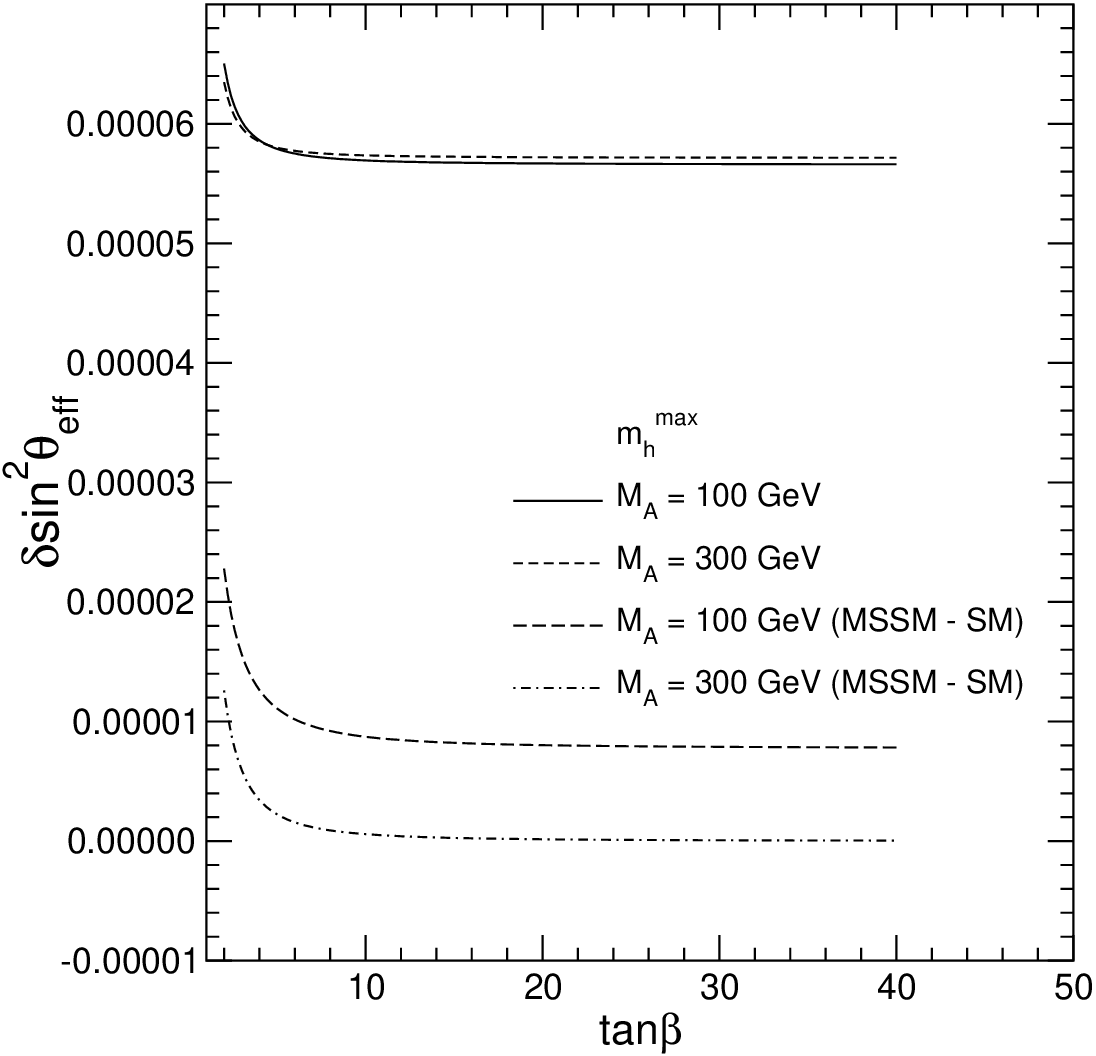,width=7cm,height=6.3cm} 
}
\end{center}
\caption[]{
The shift $\de\sweff$ induced by the
\order{\alt^2} MSSM contribution and the effective change
compared with the SM result
are shown for $\msusy = 1000 \gev$ in
the $\mhmax$~scenario.
The other parameters are $\mu = 200 \gev, \Ab = \At$. $\mh$ is
obtained in the left (right) plot from varying $\MA$ from 
$50 \gev$ to $1000 \gev$, while keeping $\tb$ fixed at $\tb = 3, 40$
(from varying $\tb$ from 2 to 40, while keeping $\MA$ fixed at
$\MA = 100, 300 \gev$).
}
\label{fig:sw2eff2lgf2}
\end{figure}

\reffi{fig:sw2eff2lgf2} shows the shift $\de\sweff$ induced by the
absolute \order{\alt^2} MSSM contribution (solid and short-dashed lines)
and the effective change (long-dashed and dot-dashed lines)
for $\msusy = 1000 \gev$ in the $\mhmax$~scenario.
The other parameters are $\mu = 200 \gev, \Ab = \At$. $\mh$ is
obtained in the left (right) plot from varying $\MA$ from 
$50 \gev$ to $1000 \gev$, while keeping $\tb$ fixed at $\tb = 3, 40$
(from varying $\tb$ from 2 to 40, while keeping $\MA$ fixed at
$\MA = 100, 300 \gev$).
While the full result shows contributions to
$\sweff$ of up to $6 \times 10^{-5}$, the effective change is much
smaller, mostly below the level of $1 \times 10^{-5}$. 

For large $\tb$, the effective change in the predictions for the precision 
observables from the \order{\alt \alb} and \order{\alb^2} corrections can
exceed the one from the \order{\alt^2} corrections. It can amount up
$\de\sweff \approx -3\times 10^{-5}$ for $\tb = 40$.


\subsubsection{MSSM predictions for $\MW$ and $\sweff$ in comparison
with present and future experimental precisions}
\label{subsubsec:numanalMWsweff}

Now we focus on the comparison of the $\MW$ and $\sweff$ prediction
with the present data and the prospective experimental precision at the
next generation of colliders.

\begin{figure}[htb!]
\begin{center}
\epsfig{figure=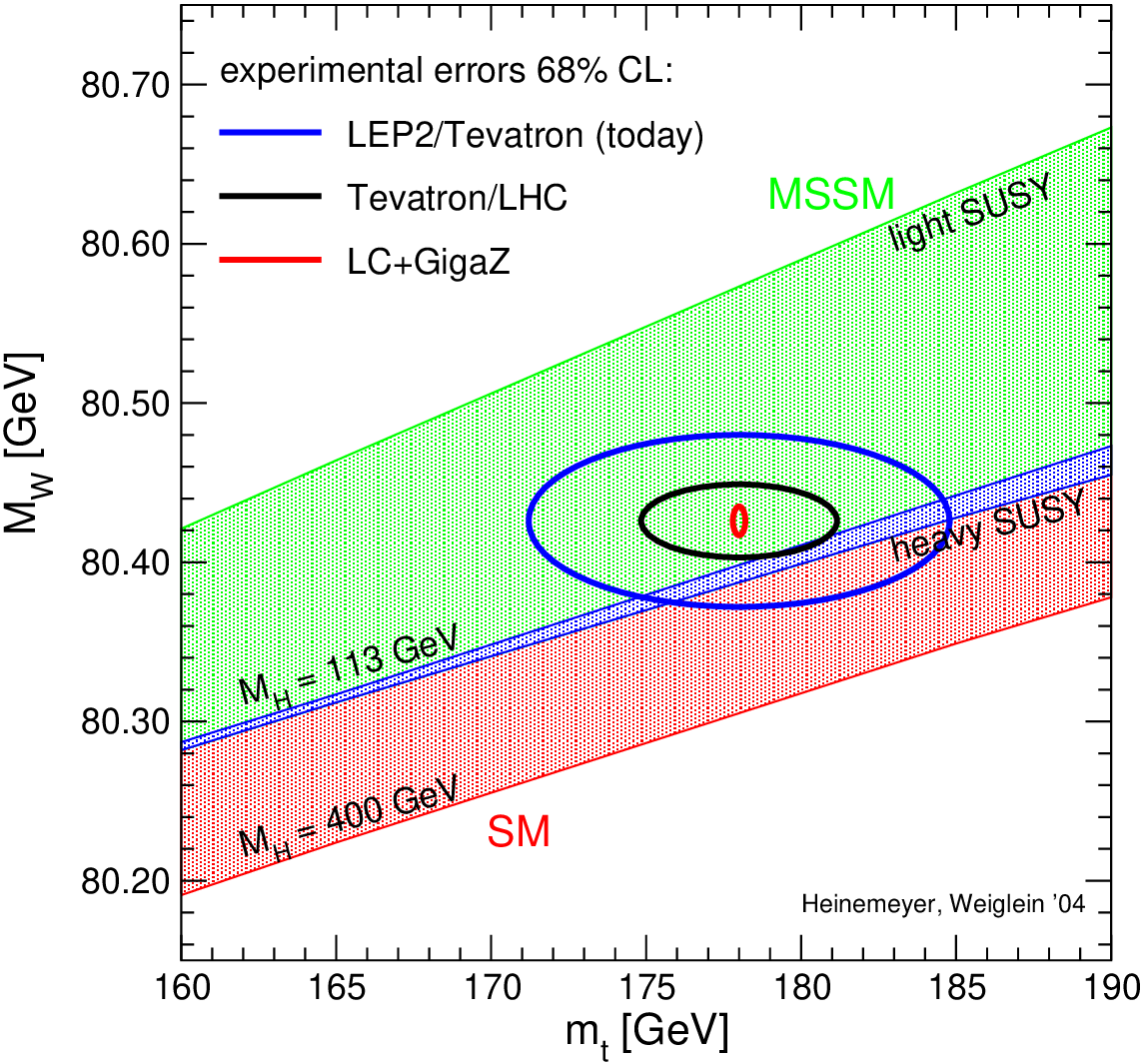,width=12cm}
\begin{picture}(0,0)
\CBox(-110,40)(-20,51){White}{White}
\end{picture}
\vspace{1em}
\caption[]{The current experimental results for $\MW$ and $\mt$ and the
prospective accuracies at the next generation of colliders are shown in
  comparison with the SM prediction (medium-shaded and dark-shaded 
(red and blue) bands) and the MSSM
prediction (light-shaded and dark-shaded (green and blue) bands).
}
\label{fig:MWMT}
\end{center}
\end{figure}

In \reffi{fig:MWMT} we compare the SM and the MSSM predictions for $\MW$
as a function of $\mt$. The predictions within the two models 
give rise to two bands in the $\mt$--$\MW$ plane with only a relatively small
overlap region (indicated by a dark-shaded (blue) area in \reffi{fig:MWMT}). 
The allowed parameter region in the SM (the medium-shaded (red)
and dark-shaded (blue) bands) arises from varying the only free parameter 
of the
model, the mass of the SM Higgs boson, from $\MH = 113 \gev$ (upper edge
of the dark-shaded (blue) area) to $400 \gev$ (lower edge of the
medium-shaded (red) area).
The light-shaded (green) and the dark-shaded (blue) areas indicate
allowed regions for the unconstrained MSSM. SUSY masses close
to their experimental lower limit are assumed for the upper edge of the
light-shaded (green)
area, while the decoupling limit with SUSY masses of \order{2 \tev}
yields the lower edge of the dark-shaded (blue) area. Thus, the overlap 
region between
the predictions of the two models corresponds in the SM to the region
where the Higgs boson is light, i.e.\ in the MSSM allowed region ($\mh
\lsim 140 \gev$). In the MSSM it corresponds to the case where all
superpartners are heavy, i.e.\ the decoupling region of the MSSM. 
The current 68\%~C.L.\ experimental results%
\footnote{
The plot shown here is an update of 
\citeres{deltarMSSM1lA,mondplot,susyewpo}.  
}%
~for $\mt$ and $\MW$ slightly favor the MSSM over the SM. 
The prospective accuracies for the LHC and the ILC with GigaZ option,
see \refta{tab:POfuture}, are also shown in the plot (using the current
central values), indicating the
potential for a significant improvement of the sensitivity of the
electroweak precision tests~\cite{gigaz}.

\begin{figure}[htb!]
\begin{center}
\epsfig{figure=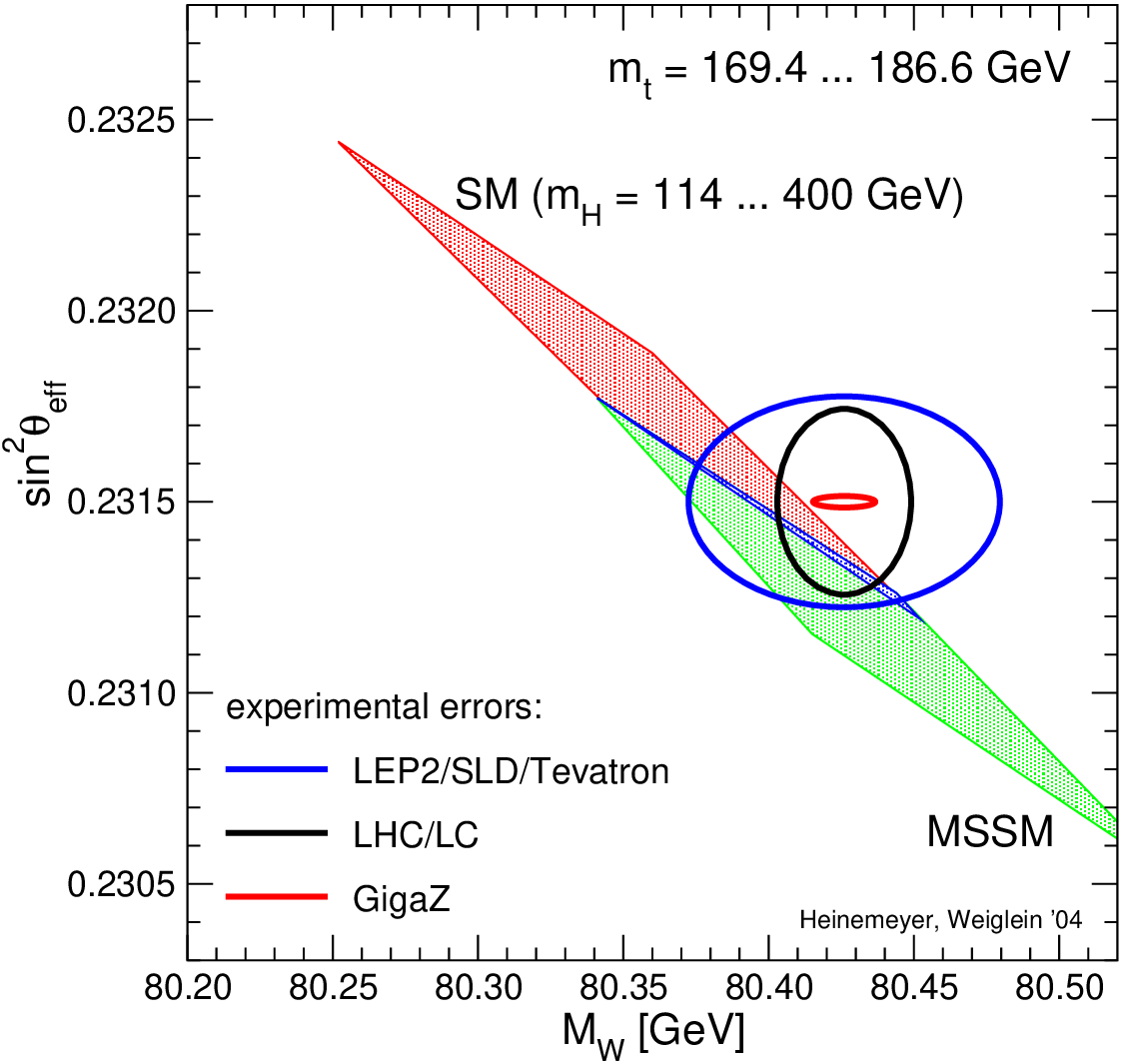,width=12cm}
\begin{picture}(0,0)
\CBox(-110,39)(-10,50){White}{White}
\end{picture}
\vspace{1em}
\caption[]{The current experimental results for $\MW$ and $\sweff$ and the
prospective accuracies at the next generation of colliders are shown in
  comparison with the SM prediction (medium-shaded and dark-shaded 
(red and blue) bands) and the MSSM
prediction (light-shaded and dark-shaded (green and blue) bands).
}
\label{fig:MWsw2eff}
\end{center}
\end{figure}

In \reffi{fig:MWsw2eff} the comparison between the SM and the MSSM is
shown in the $\MW$--$\sweff$ plane (see also
\citeres{mondplot,susyewpo}). As above, the predictions in the SM
(medium-shaded and dark-shaded (red and blue) bands) and possible
MSSM regions (light-shaded and dark-shaded (green and blue) bands) are shown 
together with the current 68\%~C.L.\ experimental results and the 
prospective accuracies for the LHC and the ILC with GigaZ option. Again
the MSSM is slightly favored over the SM. It should be noted that the
prospective improvements in the experimental accuracies, in particular
at the ILC with GigaZ option, will provide a high sensitivity to deviations
both from the SM and the MSSM.

\smallskip
The central value for the experimental value of $\sweff$ in
\reffi{fig:MWsw2eff} is based on
both leptonic and hadronic data. The two most precise
measurements, $A_{\rm LR}$ from SLD and $A^{\rm b}_{\rm FB}$ from 
LEP, differ from each other by about $3\si$ (see \citere{LEPEWWG}).
This, together with the NuTeV anomaly (see below), gave rise to a
relatively low fit probability of the SM global fit in the past years,
and had caused
considerable attention in the literature. In particular, several
analyses have been performed where the hadronic data on $A_{\rm FB}$
have been excluded from the global fit 
(see e.g.\ \citeres{chanowitz,lightsf}). It has been noted that in this
case the SM global fit, possessing a significantly
higher fit probability, yields
an upper bound on $\MH$ which is rather low in view of the experimental
lower bound on $\MH$ of $\MH > 114.4$~GeV~\cite{LEPHiggsSM}. The value of 
$\sweff$ corresponding to the measurement of $A_{\rm LR}({\rm SLD})$
alone is $\sweff = 0.23098 \pm 0.00026$~\cite{LEPEWWG}. \reffi{fig:MWsw2eff}
shows that adopting the latter value of $\sweff$ makes the agreement between
the data and the SM prediction much worse, while the MSSM provides a
very good description of the data. In accordance with this result, in
\citere{lightsf} it has been found that the contribution of light
gauginos and scalar leptons in the MSSM (in a scenario with vanishing
SUSY contribution to $\De\rho$) gives rise to a shift in $\MW$
and $\sweff$ as compared to the SM case which brings the MSSM prediction
in better agreement with the experimental values of $\MW$ and 
$A_{\rm LR}({\rm SLD})$. 

On the other hand, it has also been investigated whether the discrepancy
between $A_{\rm LR}$ and $A^{\rm b}_{\rm FB}$ could be explained in
terms of contributions of some kind of new physics. The (loop-induced)
contributions from SUSY particles in the MSSM are however too small to
account for the $3\si$ difference between the two observables (see e.g.\
\citere{lightsf}). Thus, the quality of the fit to $A_{\rm LR}$ and
$A^{\rm b}_{\rm FB}$ in the MSSM is similar to the one in the SM.

With the latest experimental values of the precision observables and the
most up-to-date theory predictions the probability of the global fit in
the SM is about 26\%~\cite{LEPEWWG} (if the NuTeV result is not
included). In particular, the most recent 
experimental value
of the top-quark mass, $\mt = 178.0 \pm 4.3 \gev$~\cite{mtexp}, and a
slight shift in the experimental value of $\MW$ have
led to an improvement of the overall fit quality. Although the
discrepancy between $A_{\rm LR}$ from SLD and 
$A^{\rm b}_{\rm FB}$ from LEP remains, it seems not well motivated to
discard any of the two measurements.

As mentioned above,
another observable for which the SM prediction shows a large deviation
by about $3 \si$ from the experimental value is the neutrino--nucleon 
cross section measured at NuTeV~\cite{nutev}. Also in this case loop
effects of SUSY particles in the MSSM are too small to account for a
sizable fraction of the discrepancy (see e.g.\
\citeres{Davidson:2001ji,nutevSUSY2}).


\subsection{Intrinsic uncertainty in $\MW$ and $\sweff$
from SUSY corrections}
\label{subsec:MWsweffuncSUSY}

The remaining theoretical uncertainties in the prediction for $\MW$ and 
$\sweff$ from unknown higher-order corrections in the MSSM (i.e.\ loop
corrections from SM particles and superpartners) are considerably larger
than in the SM, since the results for higher-order corrections in the
MSSM are not quite as advanced yet as in the SM. The current intrinsic 
uncertainties in the MSSM can roughly be estimated by
comparing the size of the known corrections in the MSSM (see above) to
the corresponding corrections in the SM and by assuming that the unknown 
higher-order corrections in the MSSM enter with the same relative weight
as the corresponding corrections in the SM, whose numerical effects are
known. This kind of estimate does not take into account specific
enhancement factors in the MSSM, like for instance corrections that grow
with powers of $\tb$. In general, the additional contributions from
superpartners in the loops will be bigger the smaller the SUSY mass
scale is. As in the case of $\mh$, the estimate for the
intrinsic uncertainty of $\MW$ and $\sweff$ should be understood to
refer to ``typical'' regions of the MSSM parameter space. In parts of
the parameter space where certain corrections are particularly 
enhanced (see the discussion in \refse{sec:sources}) the intrinsic 
uncertainties can be larger.

Taking the above considerations into account, a crude estimate of the
current intrinsic uncertainties yields~\cite{susyewpo}
\BE
\mbox{MSSM: } \quad
\de\MW^{\rm intr} ~({\rm current}) = 10 \mev ~, \quad
\de\sweff^{\rm intr} ~({\rm current}) = 12 \times 10^{-5} ~,
\label{eq:MWSWintrSUSYnow}
\end{equation}
i.e.\ uncertainties that are roughly twice as large as the current
uncertainties in the SM.

With sufficient effort on higher-order calculations in the MSSM, it
should be possible in the future to reduce the intrinsic uncertainties 
to the same level as we had estimated for the SM (see
\refeqs{MWfutureunc}, (\ref{sw2efffutureunc})):
\BE
\mbox{MSSM: } \quad
\de\MW^{\rm intr} ~({\rm future}) = 2 \mev ~, \quad 
\de\sweff^{\rm intr} ~({\rm future}) = 2 \times 10^{-5} ~.
\label{eq:MWSWintrSUSYfuture}
\end{equation}


\subsection{Results in the NMFV MSSM}

The analytical results obtained for the EWPO in the NMFV MSSM have
been derived for the general case of mixing between
the third and second generation of squarks,
i.e.\ all NMFV contributions, $\De_{LL,LR,RL,RR}$, can be chosen
independently in the $\Stop/\Scha$ and in the $\Sbot/\Sstr$ sector,
see \refse{subsec:nmfv}. 
Corrections from the first-generation squarks are not considered, for
reasons discussed in \refse{subsec:nmfv}. 
The numerical analysis of NMFV effects for the EWPO, however, 
have been performed for the simpler, but  
well motivated, scenario 
where only mixing between $\StopL$ and $\SchaL$ as well as between
$\SbotL$ and $\SstrL$ is considered. 
The only flavor off-diagonal entries 
in the squark-mass matrices are normalized according to  
$\De_{LL}^{t,b} = \la^{t,b} M_{\tilde Q_3} M_{\tilde Q_2}$,
following~\cite{NMFVestimate,FirstGenMix,FirstGenMix2} %
\footnote{
The parameters $\la^t$ and
$\la^b$ introduced here are denoted by 
$(\delta_{{\scriptsize{LL}}}^u)_{23}$ and 
$(\delta_{LL}^d)_{23}$ 
in~\cite{NMFVestimate,FirstGenMix,FirstGenMix2}. 
}%
, where $M_{\tilde Q_3, \tilde Q_2}$ are the soft SUSY-breaking
masses for the $SU(2)$ squark doublet in the third and second
generation. NMFV is thus parametrized in terms of the dimensionless
quantities $\la^t$ and $\la^b$ (see
\cite{FirstGenMix,FirstGenMix2,expconst2,expconst3} for experimentally
allowed ranges).
The case of $\la^t = \la^b = 0$ corresponds to the MSSM with
minimal flavor violation (MFV). In detail, it has been set
\begin{align}
\De_{LL}^t &= \la^t M_{\tilde L_t} M_{\tilde L_c} \,, &
\De_{LR}^t = \De_{RL}^t = \De_{RR}^t &= 0 \,, \nn \\
\De_{LL}^b &= \la^b M_{\tilde L_b} M_{\tilde L_s} \,, &
\De_{LR}^b = \De_{RL}^b = \De_{RR}^b &= 0 \, ,
\end{align}
for the entries in the matrices (\ref{eq:massup}) and (\ref{eq:massdown}).

For the sake of simplicity, the same flavour mixing parameter
has been assumed in the numerical analysis for
the $\Stop - \Scha$ and $\Sbot - \Sstr$ sectors, $\la = \la^t = \la^b$. 
It should be noted in this context that the LL
blocks of the up-squark and down-squark mass matrices are not
independent because of the $SU(2)$ gauge invariance; they are related
trough the CKM mass matrix~\cite{FirstGenMix2}, which also implies that
a large difference between these two parameters is not allowed.


\subsubsection{Results for $\De\rho$}
\label{subsubsec:Delrhonmfv}

For the numerical evaluation~\cite{nmfv}, the $\mhmax$ 
and the no-mixing scenario have been used~\cite{LHbenchmark}, 
but with a free scale $\msusy$, see \refapp{chap:benchmark}.
The results are independent of $\MA$.
The numerical values of the SUSY parameters are
\BE
\msusy = 1\tev \; {\rm and}\; 2 \tev, \quad \tb = 30, 
\quad \mu=200 \gev, \, \quad \epsilon = 0.04,
\label{eq:inputs}
\end{equation}
if not explicitly stated otherwise. 
The variation with $\mu$ and $\tb$ is very weak,
since they do not enter the squark couplings to the vector bosons. 

\begin{figure}[[htb!]
\begin{center}
\vspace*{0.3cm}
\epsfig{figure=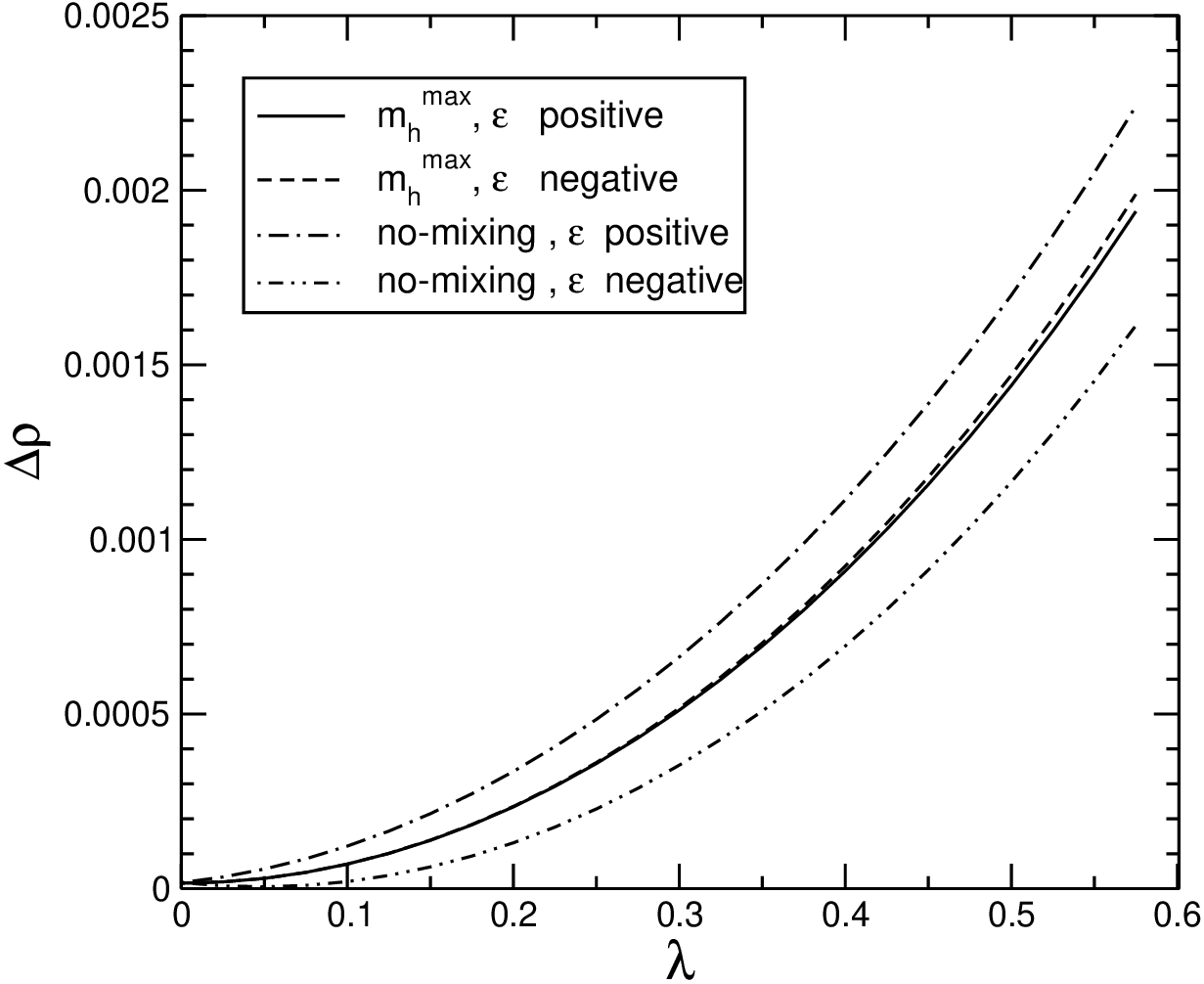,width=8.3cm}
\end{center}\vspace*{-0.7cm}
\caption[]{
The variation of $\De\rho^{\tiq}$ with $\la (= \la^t = \la^b)$ in the
$\mhmax$ and no-mixing~scenarios for different relative signs of
$\epsilon$ and $\la$ \cite{nmfv}). $\msusy = 2 \tev$, 
the other SUSY parameters are given in \refeq{eq:inputs}.}
\label{fig:eps}
\end{figure}

The behaviour with the sign of $\epsilon$ is shown in
\reffi{fig:eps} for the corrections to $\De\rho^{\tiq}$ as a function
of $\la (= \la^t = \la^b)$. The results are shown
for different relative signs of $\epsilon$ and
$\la$, choosing $\la > 0$, and fixing $|\epsilon| =0.04$. 
$\msusy$ has been set to $\msusy = 2 \tev$. For the $\mhmax$~scenario 
the effect is small, but in the no-mixing~scenario the results
are affected significantly by the sign of $\epsilon$. 
The squark contribution to $\De\rho^{\tiq}$ can become 
of \order{10^{-3}} for $\la \geq 0.5$.

\psfrag{e}{{\small{$\epsilon$}}}

\begin{figure}[htb!]
\begin{center}
\epsfig{figure=figs/deltarhol.eps,width=8cm}
\end{center}\vspace*{-0.6cm}
\caption[]{
The variation of $\De\rho^{\tiq}$ with $\la = \la^t = \la^b$ in the
$\mhmax$ and no-mixing~scenarios. $\msusy$ has been fixed to
$1 \tev$ and $2 \tev$ \cite{nmfv}. 
}
\label{fig:rholam}
\end{figure}

\begin{figure}[htb!]
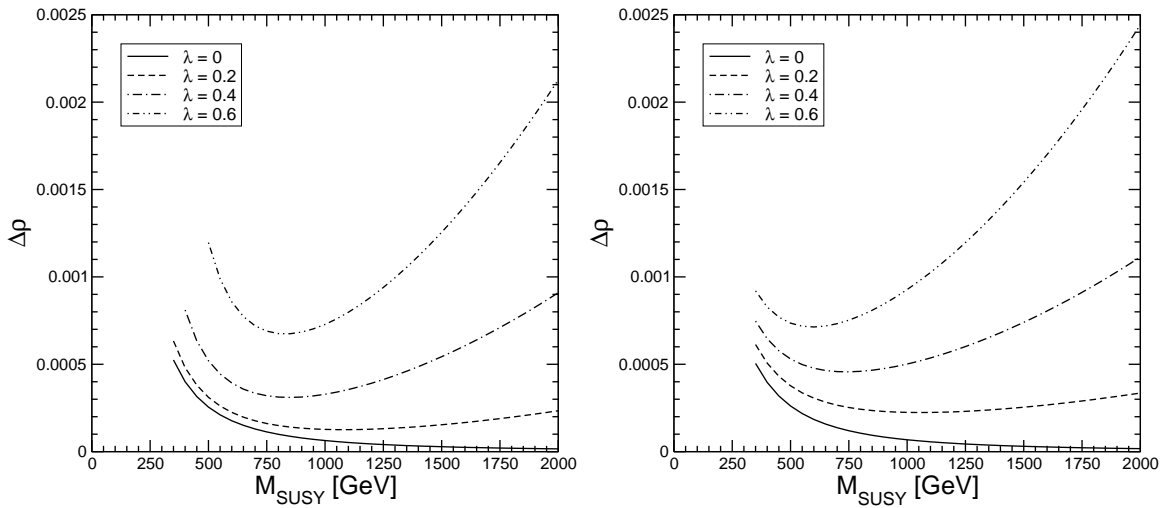

\begin{center}
\vspace*{0.3cm}
\epsfig{figure=figs/drhoMsusy_mhmax.eps,width=7.6cm}
\epsfig{figure=figs/drhoMsusy_nomix.eps,width=7.6cm}
\end{center}\vspace*{-0.6cm}
\caption[]{
The variation of $\De\rho^{\tiq}$ with $\msusy$ 
in the $\mhmax$~scenario (left panel) and
no-mixing~scenario (right panel), for different values of $\la$
\cite{nmfv}. }
\label{fig:rhoMSUSY}
\end{figure}

In \reffi{fig:rholam} we show the dependence of $\De\rho^{\tiq}$ on
$\la (= \la^t = \la^b)$ for both the~$\mhmax$ and
no-mixing~scenario and for two values of the SUSY mass scale,
$\msusy=1 \tev$ and $\msusy=2 \tev$. It is clear that
$\De\rho^{\tiq}$ grows with the $\la$ parameter, being close to zero
for $\la = 0$ and $\msusy=2 \tev$. 
One can also see that the effects on $\De\rho^{\tiq}$ are in general
larger for the no-mixing~scenario (see also
the results shown in \citere{dr2lA}).
For large values of $\msusy$ the correction increases with
increasing~$\la$ since the splitting in the squark sector increases.

The behavior of the corrections with the SUSY mass scale is shown 
in \reffi{fig:rhoMSUSY} for different values of $\la$ 
in the $\mhmax$~scenario (left panel) and in the no-mixing~scenario
(right panel). The region below $\msusy \lsim 400 \gev$ (depending on the
scenario) implies too low and hence forbidden values for the
squark masses. The curves are only for the allowed regions.
For $\la = 0$, $\De\rho^{\tiq}$ decreases, being zero
for large $\msusy$ values, in agreement with the results shown in
\citere{dr2lA}. We have also found that, for $\la \neq 0$ and small 
values of $\msusy$, $\De\rho^{\tiq}$ decreases until it reaches 
a minimum and then increases for largest values of the SUSY scale. 
This increasing behavior is more pronounced for larger $\la$ values,
reaching the level of a few per mill.
The reason lies once again in the increasing mass splitting.


\subsubsection{Numerical evaluation for $\MW$ and $\sweff$}
\label{subsubsec:numevalMWsw2eff}

\begin{figure}[htb!]
\begin{center}
\hspace*{-1cm}
\epsfig{figure=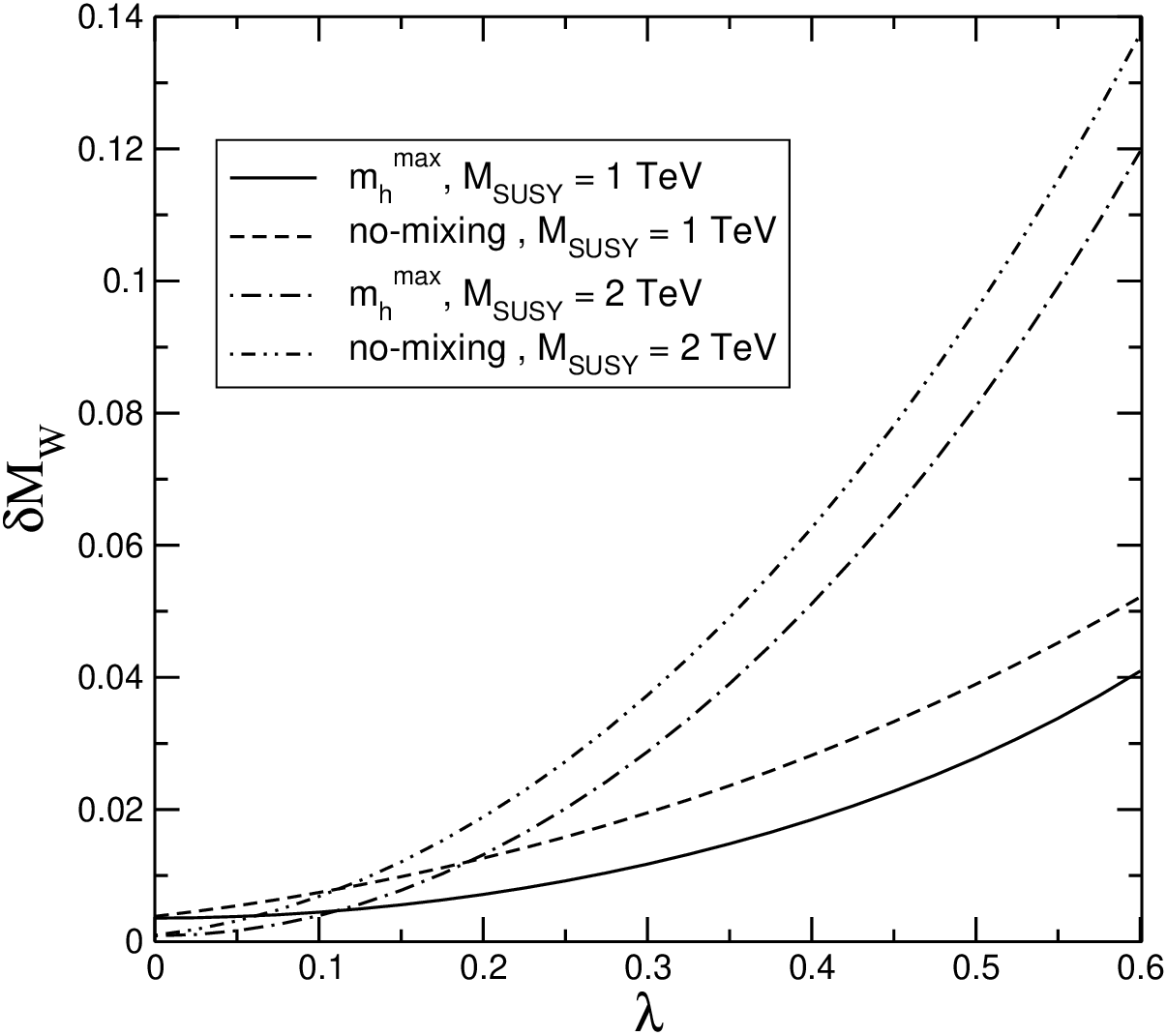,width=7.5cm}\hspace*{0.4cm}
\epsfig{figure=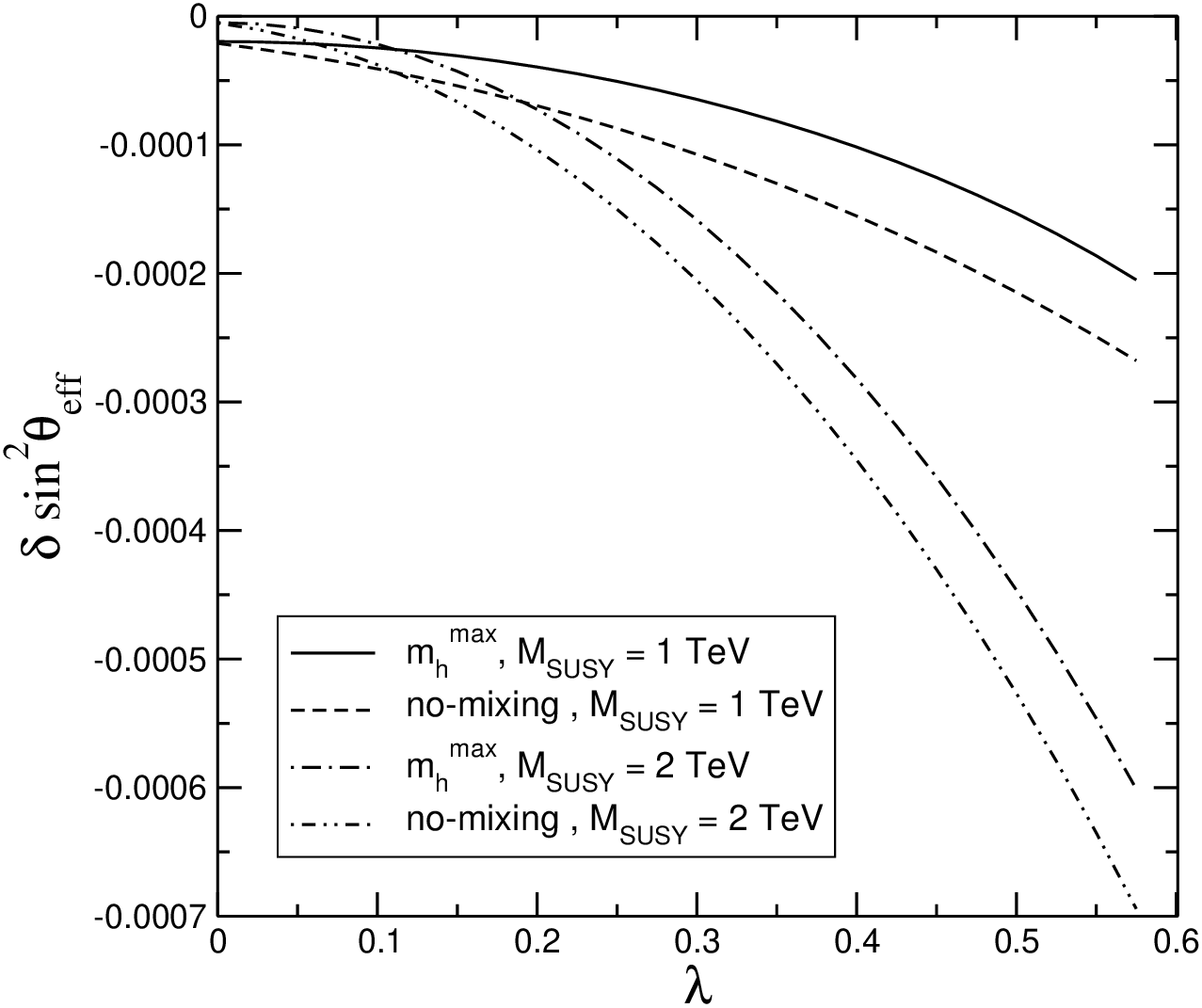,width=7.8cm}
\end{center}\vspace*{-0.6cm}
\caption[]{
The variation of $\de \MW$ and $\de \sweff$ as a function of 
$\la = \la^t = \la^b$, for the $\mhmax$ and
no-mixing~scenarios and different choices of $\msusy$ obtained with
\refeq{precobs}  \cite{nmfv}. Using the complete
expressions, \refeq{deltarmssm} and \refeq{sweffmssm}, 
yields practically indistinguishable results.} 
\label{fig:Ewprecision}
\end{figure}

Here the numerical effects of the NMFV contributions on the electroweak
precision observables, $\de \MW$ and $\de \sweff$, are briefly
analyzed~\cite{nmfv}. 
The shifts in $\MW$ and $\sweff$ have been evaluated both from the
complete expressions for the 
scalar quark contributions, \refeq{deltarmssm} and \refeq{sweffmssm}, 
and using the  $\De\rho^{\tiq}$ approximation (\ref{precobs}).
The corrections to these two observables based on \refeq{precobs} as a
function of  
$\la (= \la^t = \la^b)$ are presented in \reffi{fig:Ewprecision} with
the other parameters chosen according to~(\ref{eq:inputs}).
The $\mhmax$~scenario and no-mixing~scenario are selected for both
plots, with two values of $\msusy$, as before. 
The induced shifts in $\MW$ can become as
large as $0.14 \gev$ for the extreme case, i.e.\ when 
$\msusy=2 \tev$, $\la=0.6$ and the case of no-mixing is considered.
In the $\mhmax$~scenario $\de\MW$ is smaller, $\de \MW \lsim 0.05 \gev$,
but still sizeable.
Using the complete expressions, \refeq{deltarmssm} and
\refeq{sweffmssm}, yields 
results practically indistinguishable from those shown in
\reffi{fig:Ewprecision}~\cite{nmfv}. Thus \refeq{precobs} is a sufficiently
accurate, simple approximation for  squark-mixing effects in
the electroweak precision observables.

The shifts $\de \sweff$, shown in the right plot of
\reffi{fig:Ewprecision}, can reach values up to
$7 \times 10^{-4}$ for $\msusy=2$ TeV and $\la=0.6$ in the
no-mixing~scenario, being smaller (but still sizeable) 
for the other scenarios considered here. 

Extreme parts of the NMFV parameter space (especially for $\la^t \neq \la^b$) 
can be excluded already with today's precision. But even small values
of $\la = \la^t = \la^b$ could be probed with the future precision on
$\sweff$, provided that theoretical uncertainties will be sufficiently
under control~\cite{susyewpo}.


\section{The lightest MSSM Higgs boson mass}
\label{sec:mhpred}

\begin{figure}[htb!]
\begin{center}
\epsfig{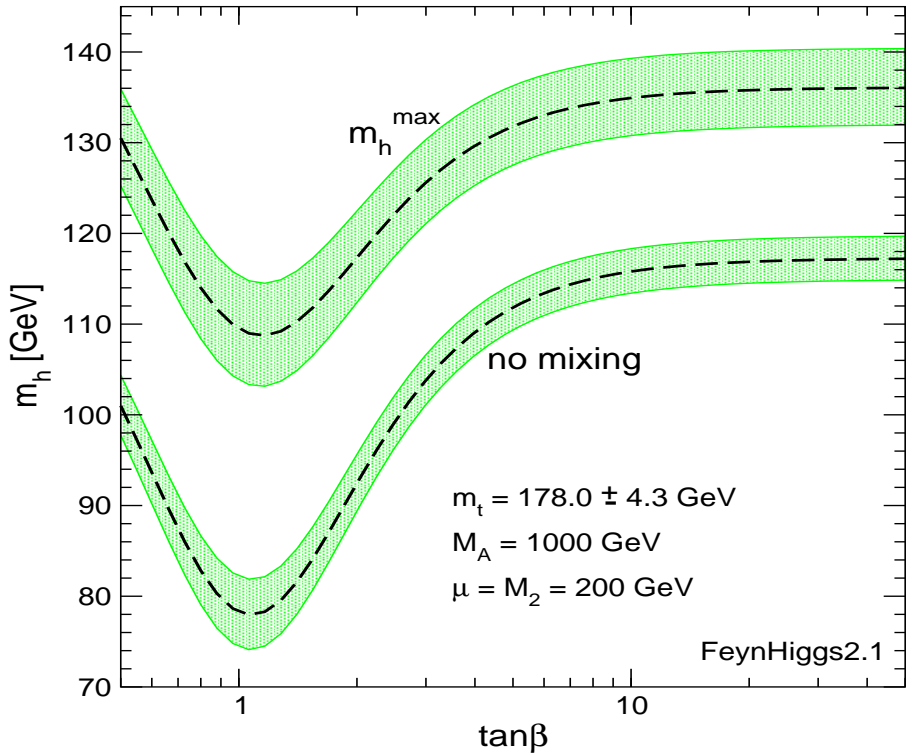}
\end{center}
\caption[]{
$\mh$ is shown as a function of $\tb$ in the $\mhmax$ and the
no-mixing scenario. $\mt$ has been varied in the interval 
$\mt = 178.0 \pm 4.3 \gev$. 
}
\label{fig:tbmh}
\end{figure}

The light $\cp$-even
MSSM Higgs boson mass, $\mh$, depends at tree-level
on $\MA$ and $\tb$. Via loop corrections, see \refse{subsec:mhcorr},
it depends most strongly on the top quark mass and on the parameters
of the scalar top sector. As an example, in \reffi{fig:tbmh} we show
$\mh$ as a function of $\tb$ in two benchmark scenarios, the $\mhmax$
and the no-mixing scenario~\cite{LHbenchmark}, see
\refapp{chap:benchmark}. $\mh$ is shown for a central value of 
$\mt = 178.0 \gev$ (dashed curves), and the variation with $\mt$ by
$\pm 4.3 \gev$ is shown as the shaded (green) band. Higher $\mh$
values are obtained for larger $\mt$. 
(All results in this section have been obtained with 
{\em FeynHiggs2.1}~\cite{feynhiggs1.2,habilSH,feynhiggs,feynhiggs2.2}.)

{}From the result for the $\mhmax$ scenario in \reffi{fig:tbmh} the
upper bound of $\mh \lsim 136 \gev$ for $\mt = 178 \gev$ and 
$\msusy = 1 \tev$ (neglecting the intrinsic theoretical uncertainties)
can be read off that was mentioned in \refse{sec:mh}. Allowing a 1~$\si$
variation of $\mt$ shifts the upper bound on $\mh$ to about $140 \gev$. 
The variation of the $\mh$ prediction with $\mt$ is even larger in the
region of small $\tb$. \reffi{fig:tbmh} shows that a 1~$\si$ upward
fluctuation of $\mt$ shifts the minimum of $\mh$ in the $\mhmax$
scenario to a value above $114 \gev$. Thus, in this case the exclusion
bound from LEP does not rule out any value of $\tb$. The comparison of
the MSSM prediction with the LEP exclusion bound is shown in
more detail in \reffi{fig:mhtb} below.

\begin{figure}[htb!]
\begin{center}
\epsfig{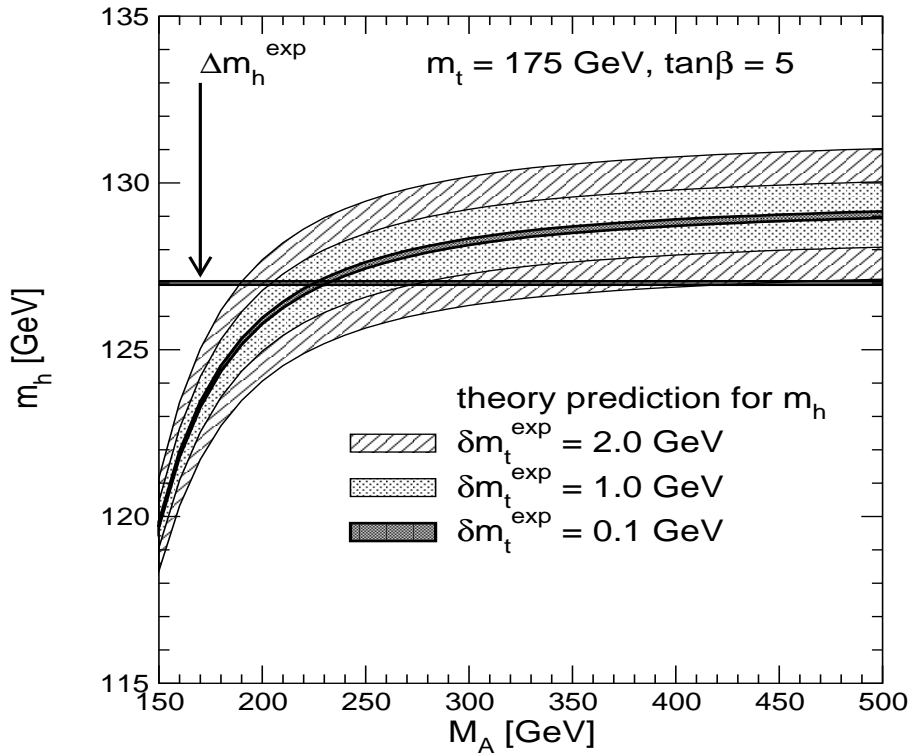}
\end{center}
\caption[]{
$\mh$ is shown as a function of $\MA$ in the $\mhmax$ scenario for
$\tb = 5$ \cite{deltamt}.
Three different precisions for $\mt$ are indicated (with a
central value of $\mt = 175 \gev$). 
The anticipated experimental error on $\mh$ at the ILC of $0.05 \gev$ is
indicated by a horizontal band.
}
\label{fig:mhMA}
\end{figure}

The relevance of the parametric uncertainty in $\mh$ induced by
different experimental errors on $\mt$ is emphasized in
\reffi{fig:mhMA}~\cite{deltamt}, 
where the prediction for $\mh$ is shown as a function of $\MA$ in the 
$\mhmax$ benchmark scenario.  
The evaluation of $\mh$ has been done for 
a central value of the top quark mass of $\mt = 175 \gev$ and for  
$\tb = 5$.
The figure shows that a reduction of the experimental error from
$\de\mtexp = 1$--$2 \gev$ (LHC) to $\de\mtexp = 0.1 \gev$ (ILC) has a
drastic effect on the prediction for $\mh$.

The prospective experimental error on $\mh$ is also shown in \reffi{fig:mhMA},
while no intrinsic theoretical uncertainty from unknown
higher-order corrections is included. If this intrinsic uncertainty can
be reduced to a level of $\de\mh^{\rm intr,future} \approx 0.1 \gev$, 
its effect in the plot would be roughly as big as the one induced by 
$\de\mtexp = 0.1 \gev$. An intrinsic uncertainty of 
$\de\mh^{\rm intr,future} \approx 1 \gev$, on the other hand, would lead
to a significant widening of the band of predicted $\mh$ values (similar
to the effect of $\de\mtexp = 1 \gev$). In this case the intrinsic
uncertainty would dominate, implying that a reduction of 
$\de\mtexp = 1 \gev$ to $\de\mtexp = 0.1 \gev$ would lead to an only 
moderate improvement of the overall theoretical uncertainty of $\mh$.

Confronting the theoretical prediction for $\mh$ with a precise
measurement of the Higgs-boson mass constitutes a very sensitive test of
the MSSM, which allows one to obtain constraints on the model parameters. 
The sensitivity of the $\mh$ prediction to $\MA$ shown in \reffi{fig:mhMA}
cannot directly be translated into a prospective indirect determination
of $\MA$, however, since \reffi{fig:mhMA} shows the situation in a
particular benchmark scenario~\cite{LHbenchmark} where, by definition,
certain fixed values of all other SUSY parameters are assumed. In a
realistic situation the anticipated experimental errors of the other
SUSY parameters and possible effects of intrinsic theoretical uncertainties
have to be taken into account. In \refse{sec:futureexp}
the prospects for an indirect determination of SUSY parameters from 
precision physics in the MSSM Higgs sector will be discussed.

As another example we demonstrate the impact of the current theory
uncertainty of $\de\mh^{\rm intr} \approx 3 \gev$~\cite{mhiggsAEC} on
the exclusion bound of $\tb$, see \citere{tbexcl} for a detailed
discussion. 
The $\mhmax$ benchmark
scenario~\cite{LHbenchmark} has been designed such that for fixed values
of $\mt$ and $\msusy$ the predicted value of the lightest $\cp$-even
Higgs boson mass is maximized for each value of $\MA$ and $\tb$. 
In \reffi{fig:mhtb} we show again $\mh$ as a function of $\tb$,
together with 
the LEP exclusion bound for the mass of a SM-like Higgs~\cite{LEPHiggsSM},
$\MHSM \ge 114.4 \gev$, as a vertical long--dashed line. 
The solid thick line shows the result in the $\mhmax$ scenario for
$\mt = 178.0 \gev$. 

\begin{figure}[htb!]
\begin{center}
\epsfig{figure=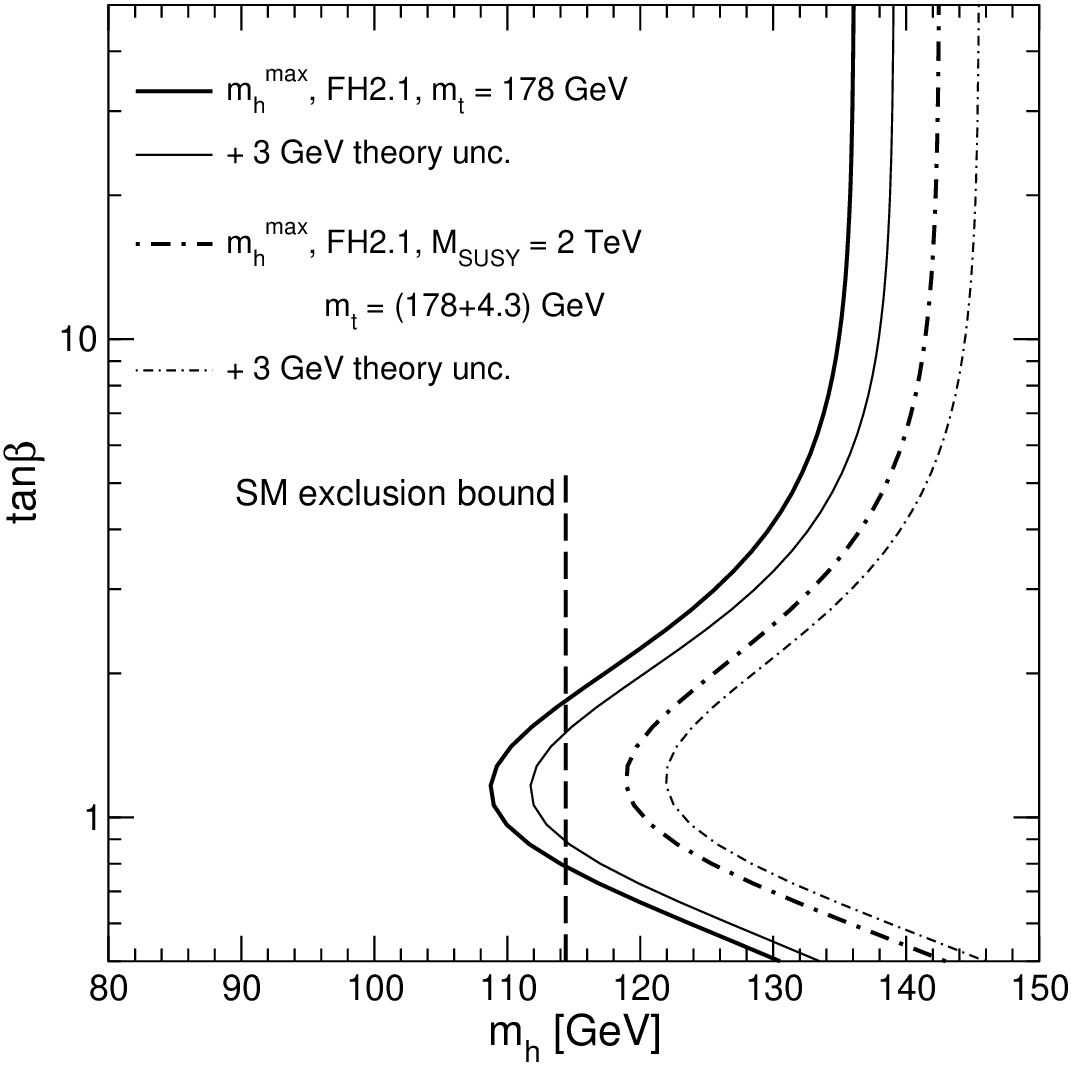,width=12.5cm}
\end{center}
\caption[]{
$\mh$ is shown as a function of $\tb$ in the $\mhmax$ scenario (solid)
and for $\mt = 178.0 + 4.3 \gev$, $\msusy = 2 \tev$ (dot-dashed). A
theory uncertainty from unknown higher-order uncertainties of 
$\de\mh^{\rm intr} = 3 \gev$ has been neglected (thick lines) or
included (thin). The SM exclusion bound of $\MH = 114.4 \gev$, which for
small values of $\tb$ also roughly applies for the MSSM, is indicated by
a dashed line.
}
\label{fig:mhtb}
\end{figure}

While in general a detailed investigation of a variety of different
possible production and decay modes is necessary in order to determine
whether a particular point of the MSSM parameter space can be excluded
via the Higgs searches or not, the situation simplifies considerably in
the region of small $\tb$ values. In this parameter region the lightest
$\cp$-even Higgs boson of the MSSM couples to the $Z$~boson with
SM-like strength, and its decay into a $b\bar b$ pair is not significantly
suppressed. Thus, within good approximation, constraints on $\tb$ can 
be obtained in this parameter region by confronting the exclusion bound
on the SM Higgs boson with the upper limit on $\mh$ within the MSSM.
From the intersection of the theoretical upper
bound in the $\mhmax$ scenario (solid thick line) with the
experimentally excluded region for $\mh$ the 
experimentally excluded region for $\tb$ can be read off.
For comparison we also show the same upper bound including the
theory uncertainty from unknown higher order corrections, 
$\de\mh^{\rm intr} \approx 3 \gev$~\cite{mhiggsAEC} (solid thin
line). Taking the theory uncertainty into account, the bound on $\tb$
is considerably weakened (see also \citere{mhiggsWN}).
Furthermore we show the $\mhmax$ scenario with the top-quark mass
shifted upwards by one standard deviation, $\mt = 182.3 \gev$ and with
$\msusy = 2 \tev$
(dot-dashed thick line), also including the $3 \gev$ intrinsic
theoretical uncertainty (dot-dashed thin line). Even without taking into 
account the intrinsic theoretical uncertainty, 
in this case no region of $\tb$ can
be excluded from the Higgs search at LEP.
This example shows that both a reduction of the experimental error on
$\mt$ and of the intrinsic theoretical uncertainty will be crucial in
order to obtain reliable bounds on the SUSY parameters from measurements
in the Higgs sector (see also \refse{sec:futureexp}).


\section{MSSM predictions for $(g-2)_\mu$}
\label{sec:g-2pred}

In our numerical discussion of SUSY contributions to the anomalous
magnetic moment of the muon we first analyze the one-loop results from a
scan over the MSSM parameter space~\cite{g-2scan1L} and then focus on
two recently obtained two-loop 
corrections: the corrections involving a closed SM fermion/sfermion
loop~\cite{g-2FSf}, and the ones involving closed chargino/neutralino
loops~\cite{g-2CNH}.


\subsection{One-loop results from a MSSM parameter scan}

\begin{figure}[htb!]
\BC
\epsfig{figure=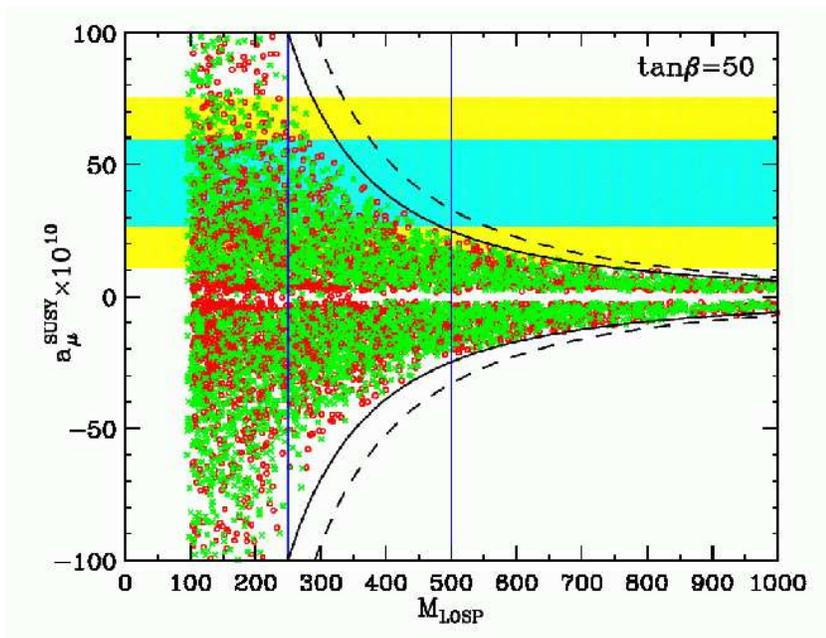, width=11cm}
\caption{%
MSSM one-loop contributions to $\amu$ are shown as a function of the
mass of the lightest observable SUSY particle (LOSP), obtained from a
scan over the MSSM parameter space \cite{g-2scan1L}.
The lighter (green) dots correspond to a $\tilde \mu$ LOSP, darker
(red) dots represent charginos/neutralinos as LOSP. The dashed lines
show the allowed contours if $|A_\mu|$ is allowed to vary up to 
$100 \tev$. The shaded bands correspond to the one/two~$\sigma$
allowed ranges in the year 2001.
}
\label{fig:g-2scan1L}
\EC
\end{figure}

The possible size of the MSSM one-loop contributions to $\amu$ can be
assessed by a parameter scan. In \reffi{fig:g-2scan1L} (from
\citere{g-2scan1L}) the possible MSSM contributions to $\amu$ are shown
as a function of the lightest observable SUSY particle (LOSP). 
The lighter (green) dots correspond to a $\tilde \mu$ LOSP, darker
(red) dots represent charginos/neutralinos as LOSP. The dashed lines
show the allowed contours if $|A_\mu|$ is allowed to vary up to 
$100 \tev$. The shaded bands correspond to the one/two~$\sigma$
allowed ranges in the year 2001.
One can see that the MSSM can easily explain the discrepancy in
\refeq{deviationfinal}. On the other hand, $\amu$ can place stringent
constraints on the allowed MSSM parameter space. In order to set
reliable bounds in the MSSM the theoretical uncertainties have to be
under control. This requires the evaluation of higher-order
contributions. The existing two-loop corrections are reviewd in the
following subsections. 

\subsection{Contributions from closed SM fermion/sfermion loops}
\label{subsec:g-2FSf}

The two-loop corrections to $(g-2)_\mu$ involving a closed SM
fermion/sfermion loop, corresponding to the diagrams in 
lines 2--4 of
\reffi{fig:g-22L}, have been evaluated in \citere{g-2FSf}, extending
earlier analyses of \citeres{g-2BarrZee1,g-2BarrZee2}. 
These two-loop corrections have a complicated parameter
dependence. Therefore a parameter scan has been performed.
$\tb$ was set to $\tb = 50$, and universal soft SUSY-breaking
parameters in the scalar fermion mass matrices were assumed.
It turned out to be crucial to take experimental constraints from $\mh$,
$\De\rho$,  $\br(b \to s \ga)$ and $\br(B_s \to \mu^+\mu^-)$ into
account (for details see \citere{g-2FSf}).
It was shown that the diagrams involving a photon
and a Higgs boson (diagram no.\ 12 in \reffi{fig:g-22L}) give the by
far largest contribution. 

\begin{figure}[htb!]
\BC
\epsfig{figure=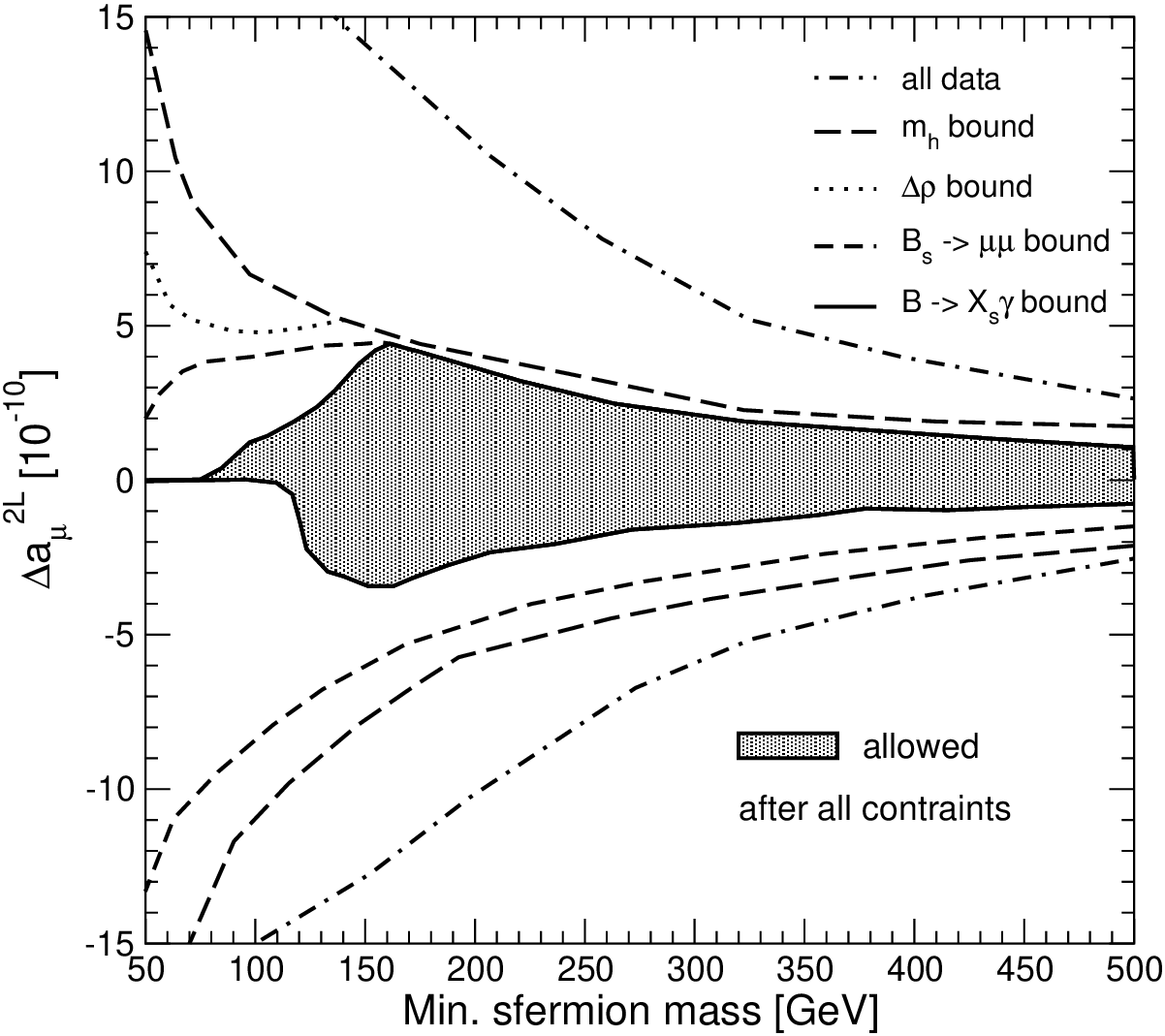, width=11cm}
\caption{%
Maximum contributions of the diagrams with a closed SM
fermion/sfermion loop to $\delamu$ as a function of the lightest squark mass, 
min\{$\mste$, $\mstz$, $\msbe$, $\msbz$\}. The constraints from $\mh$,
$\De\rho$,  $\br(b \to s \ga)$ and $\br(B_s \to \mu^+\mu^-)$ have been
taken into account (for details see \citere{g-2FSf}).
}
\label{fig:g-2FSf}
\EC
\end{figure}

The whole contribution of this set of diagrams is shown in
\reffi{fig:g-2FSf}. The results shown in the figure are the
following (see \citere{g-2FSf} for further details):
\begin{itemize}

\item The outer lines show the largest possible results if all
  experimental constraints are ignored. They show a steep rise of $\delamu$
  for decreasing $m_{\sfn_1}$; for $m_{\sfn_1} < 150 \gev$ contributions
  larger than $15 \times 10^{-10}$, corresponding to two standard
  deviations of the experimental error on $\amu$, are possible.

\item The next two lines show the possible results if the bound
  $\mh > 106.4\gev$ (it results from the experimental bound of $114.4
  \gev$ by taking into account a $5 \gev$ parametric uncertainty from
  the experimental error of $\mt$ and a $3 \gev$ intrinsic uncertainty,
  see \refse{subsec:mhunc})
  and then in addition the bound on $\De\rho$ are
  satisfied. The maximum contributions are very much reduced already
  by the $\mh$ bound, and the $\De\rho$ bound reduces further the
  positive region for small sfermion masses. If both bounds are taken
  into account,   $\delamu > 5 \times 10^{-10}$ and  $\delamu < -10 \times
  10^{-10}$  is excluded for $m_{\sfn_1} \gsim 100 \gev$.

\item The two innermost lines correspond to taking into account in
  addition the bound on $\br(B_s\to\mu^+\mu^-)$ and finally also on 
  $\br(b \to s \ga)$, resulting in the shaded area. In particular
  taking into account the $\br(b \to s \ga)$ bound 
  eliminates most data points with $m_{\sfn_1} \lsim 150 \gev$ and thus
  leads to a strong reduction of the possible size of the
  contributions (see however the discussion in \citere{bsgneubert}). 
  The largest 
  contributions of $\pm 4 \times10^{-10}$ to $\delamu$, corresponding to 
  $\sim 0.7\si$ of the experimental error, are possible for
  $m_{\sfn_1} \approx 150 \ldots 200 \gev$.

\end{itemize}
%

It should be kept in mind that the size of the corrections shown in
\reffi{fig:g-2FSf} depend on the assumption of the univesality of the
soft SUSY-breaking parameters. It has been shown in \citere{g-2FSf}
that lifting this universality assumption
can lead to substantially larger
contributions. As an example, for $M_{\tilde D}/M_{\tilde U} = 10$ 
(see \refeq{RRmasses}),
$\delamu > 15 \times 10^{-10}$ could be achieved without violating any
experimental constraint.


\subsection{Contributions from closed chargino/neutralino loops}
\label{subsec:g-2CNH}

The 2-loop contributions to $\amu$ containing a closed
chargino/neutralino loop~\cite{g-2CNH} constitute a separately
UV-finite and gauge-independent class. The corresponding diagrams were
shown in the last line of \reffi{fig:g-22L}. 
The chargino/neutralino two-loop contributions, $\amuX$,
depend on the mass parameters for the charginos and neutralinos $\mu$,
$M_{1,2}$, the $\cp$-odd Higgs mass $\MA$, and $\tb$.

\begin{figure}[htb!]
\begin{center}
\unitlength=1.25bp
\epsfig{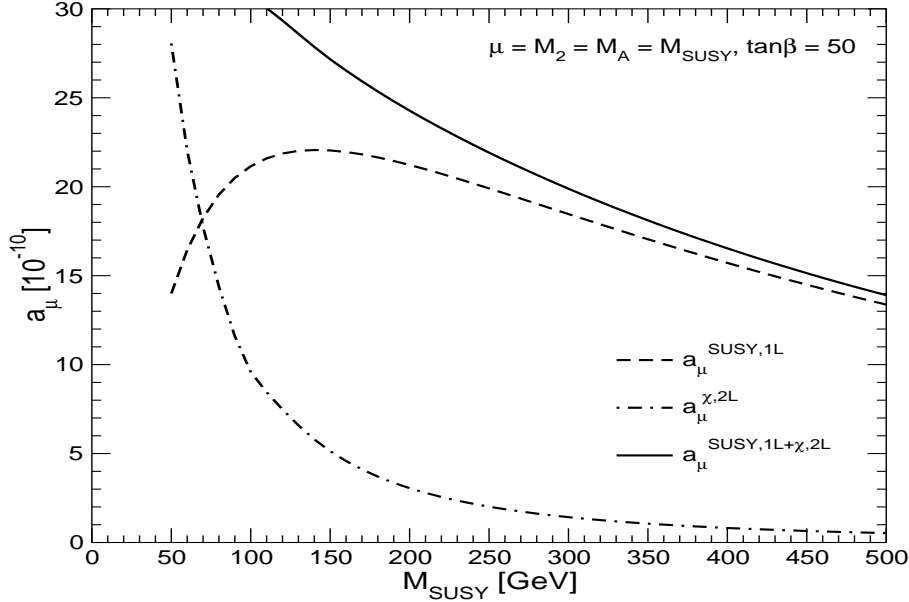}
\caption{Comparison of the supersymmetric one-loop result $\amu^{\rm
    SUSY,1L}$ (dashed) with the two-loop chargino/neutralino contributions
    $\amuX$ (dash-dotted) and the sum (full line) \cite{g-2CNH}.
     The parameters are
    $\mu=M_2=\MA\equiv \msusy$, 
    $\tb=50$, and the sfermion mass parameters are set to 1TeV.
 }
\label{fig:chaneueqmassplot}
\end{center}
\end{figure}

The chargino/neutralino sector does not only contribute to
$\amuX$ but already to $\amuSUoL$, so it is
interesting to compare the one- and two-loop contributions. 
For the case that all masses, including the smuon and
sneutrino masses, are set equal to $\msusy$, the one-loop and
two-loop contributions can be trivially compared using eqs.\
(\ref{susy1loop}), (\ref{chaneuapproxorg}), showing that the two-loop
contribution shifts the one-loop result by about~2\%. 

However, the chargino/neutralino sector might very well be significantly
lighter than the slepton sector of the second generation, in
particular in the light of FCNC and $\cp$-violating constraints, which
are more easily satisfied for heavy 1st and 2nd generation sfermions.
In \reffi{fig:chaneueqmassplot} the chargino/neutralino two-loop
contributions are therefore compared with the supersymmetric one-loop
contribution $\amuSUoL$ at fixed high smuon and sneutrino
masses $\Msl = 1 \tev$.
The other masses are again set equal, 
$\mu=M_2=\MA\equiv \msusy$. Furthermore, 
we use a large $\tb$ value, $\tb = 50$, which enhances the SUSY
contributions to $\amu$. 

It has been found that for $\msusy\lsim400 \gev$ the
two-loop contributions become more and more important. For 
$\msusy \approx100 \gev$ they amount to 50\% of the one-loop
contributions, which are suppressed by the large smuon and sneutrino
masses.

\begin{figure}[htb!]
\mbox{}\vspace{-3em}
\begin{center}
\begin{picture}(440,440)
\psfrag{MSl1000TB50}{\footnotesize \hspace{-1em}$\Msl = 1000 \gev, \tb = 50$}
\psfrag{MSl1500TB50}{\footnotesize \hspace{-1.7em}$\Msl = 1500 \gev, \tb = 50$}
\psfrag{MSl750TB25}{\footnotesize \hspace{-1em}$\Msl = 750 \gev, \tb = 25$}
\psfrag{MSl1000TB25}{\footnotesize \hspace{-1em}$\Msl = 1000 \gev, \tb = 25$}
\epsfxsize=6.5cm
\put(000,000){\epsfbox{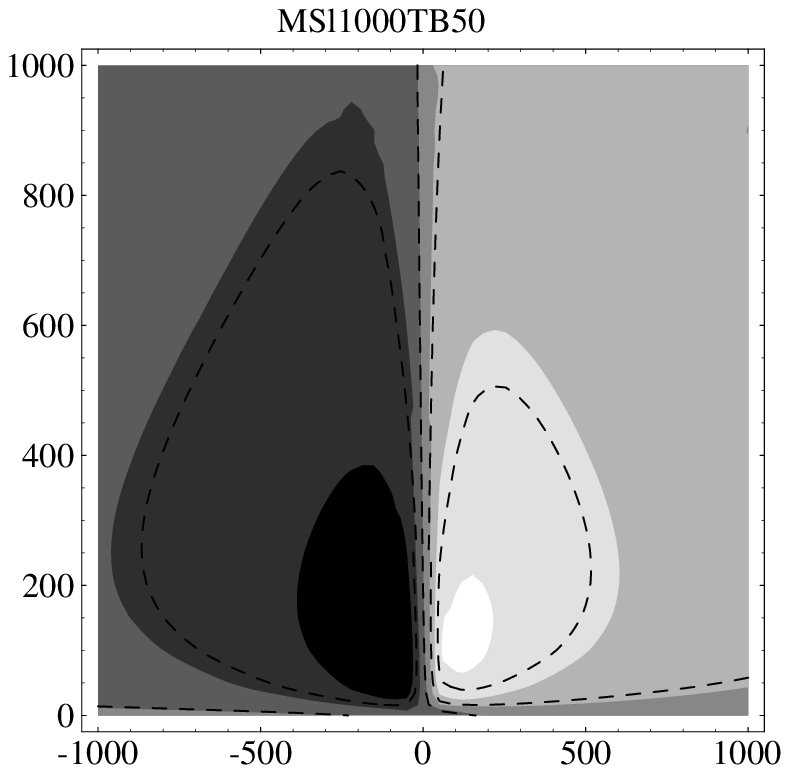}}
\CBox(50,173)(150,180){White}{White}
\put(55,175){\footnotesize $\Msl = 1000 \gev, \tb = 50$}
\epsfxsize=6.5cm
\put(220,000){\epsfbox{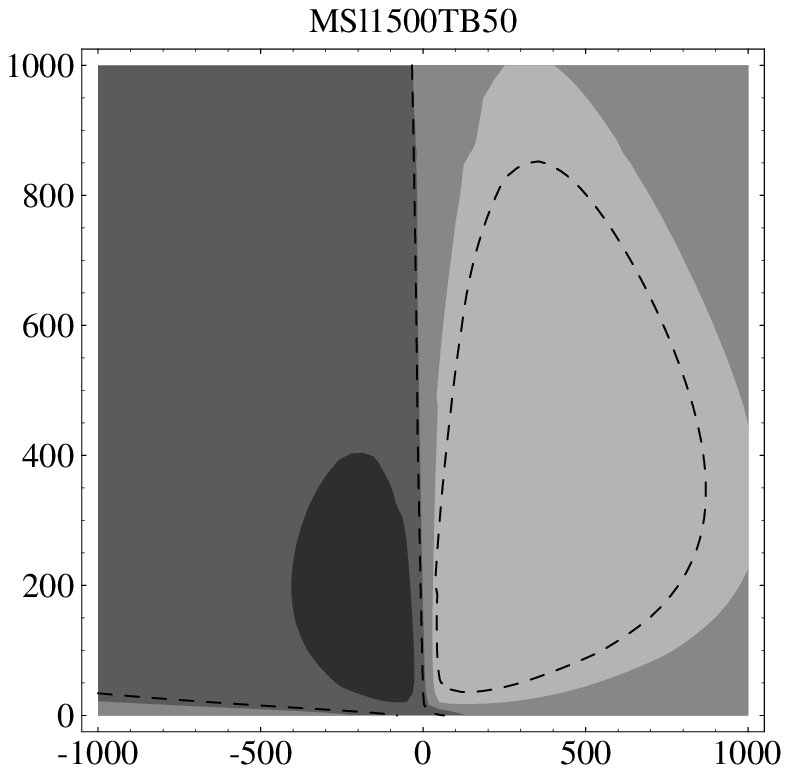}}
\CBox(270,173)(370,180){White}{White}
\put(275,175){\footnotesize $\Msl = 1500 \gev, \tb = 50$}
\epsfxsize=6.5cm
\put(000,220){\epsfbox{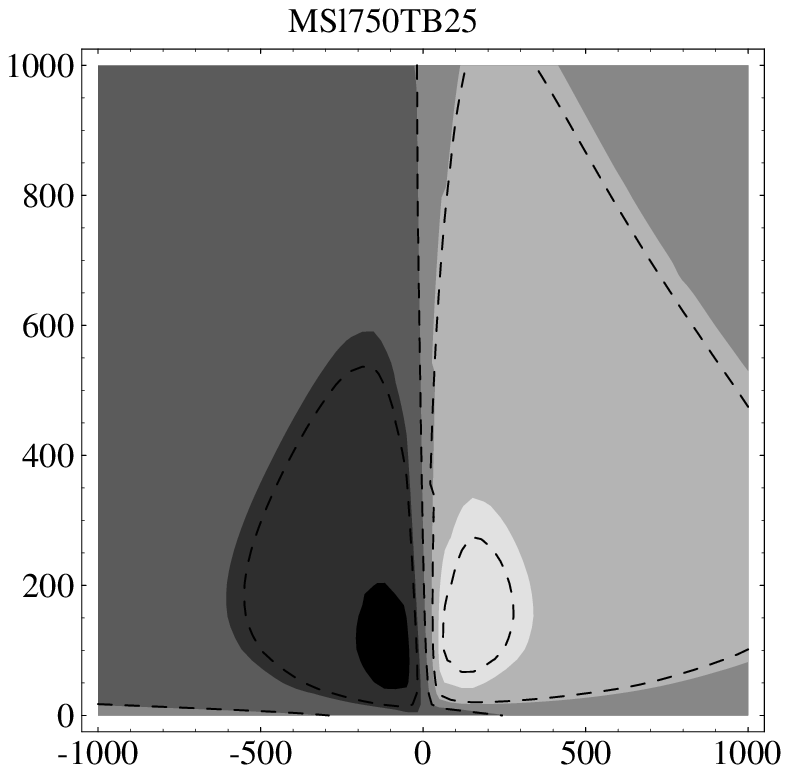}}
\CBox(50,393)(150,400){White}{White}
\put(55,395){\footnotesize $\Msl = 750 \gev, \tb = 25$}
\epsfxsize=6.5cm
\put(220,220){\epsfbox{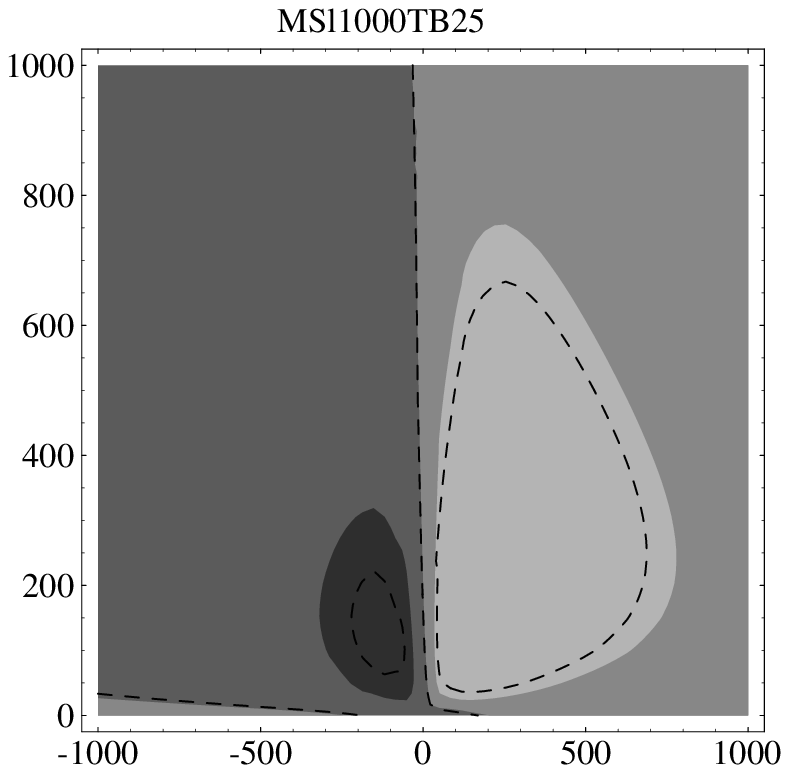}}
\CBox(270,393)(370,400){White}{White}
\put(275,395){\footnotesize $\Msl = 1000 \gev, \tb = 25$}
\put(005,180){\footnotesize $M_2$ [GeV]}
\put(225,180){\footnotesize $M_2$ [GeV]}
\put(005,400){\footnotesize $M_2$ [GeV]}
\put(225,400){\footnotesize $M_2$ [GeV]}
\put(150, -5){\footnotesize $\mu$ [GeV]}
\put(370, -5){\footnotesize $\mu$ [GeV]}
\put(150,215){\footnotesize $\mu$ [GeV]}
\put(370,215){\footnotesize $\mu$ [GeV]}
\put(110,060){\footnotesize $<1\si$}
\put(130,140){\footnotesize $1-2\si$}
\put(040,150){\footnotesize $3-4\si$}
\put(335,070){\footnotesize $1-2\si$}
\put(260,150){\footnotesize $3-4\si$}
\put(105,260){\scriptsize $<1\si$}
\put(120,300){\footnotesize $1-2\si$}
\put(040,370){\footnotesize $3-4\si$}
\put(330,300){\footnotesize $1-2\si$}
\put(350,370){\footnotesize $2-3\si$}
\put(260,370){\footnotesize $3-4\si$}
\end{picture}
\vspace{0.2em}
\caption{Constraints on the MSSM parameter space in the
  $\mu$--$M_2$-plane for $\MA = 200 \gev$ from comparing the MSSM
  prediction with the data. The different regions resulting from
  the MSSM prediction based on $\amuSUoL+\amuX$ (contours with solid
  border) and from the prediction based on $\amuSUoL$ alone (dashed 
  contours) are shown.
  The slepton mass scale (which enters only the
  one-loop prediction) and $\tb$ are indicated for each plot.
  The contours are at $(24.5,15.5,6.5,-2.5,-11.5,-20.5)\times10^{-10}$ 
  corresponding to the central value of 
  $\amuexp-\amu^{\rm theo,SM} = (24.5 \pm 9.0)\times10^{-10}$ 
  and intervals of 1--5$\sigma$ \cite{g-2CNH}).
}
\label{fig:mum2contour}
\end{center}
\end{figure}

The two-loop corrections have an important impact on 
constraints on the MSSM parameter space obtained from confronting the MSSM 
prediction with the experimental value. This is shown in
\reffi{fig:mum2contour}, where the regions in the $\mu$--$M_2$-plane
resulting from the MSSM
prediction including the two-loop correction, $\amuSUoL+\amuX$, are 
compared with the corresponding regions obtained by neglecting the
two-loop correction, i.e.\ with $\amuSUoL$ alone. 
The different panels correspond to different values of $\tb$ and 
the common smuon and sneutrino mass $\Msl$ (the latter has an impact 
only on the one-loop contribution),
while $\MA$ has been fixed to $\MA=200 \gev$.
  These parameter choices are allowed essentially in the entire
  $\mu$--$M_2$-plane by the current experimental
  constraints mentioned above, provided the $\Stop$~and $\Sbot$~mass
  parameters are of \order{1 \tev}. 
The contours drawn in 
\reffi{fig:mum2contour} correspond to the $1\si$, $2\si$, \ldots regions
around the value 
$\amuexp - \amu^{\rm theo,SM}=(24.5\pm9.0)\times10^{-10}$, 
based on \citeres{g-2HMNT,LBLnew}. 
We find that for the investigated parameter space the SUSY prediction 
for $\amu$ lies mostly in the $0-2\,\si$ region if $\mu$ is positive.
However, the new two-loop corrections shift the $1\,\si$ and $2\,\si$
contours considerably. This effect is more pronounced for smaller
$\tb$ and larger~$\Msl$.


\section{MSSM fits and constraints from existing data}
\label{sec:constraints}

There have been many studies of the sensitivity of low-energy observables 
to the scale of supersymmetry, including the precision electroweak
observables~\cite{gigaz,recentfit,lightsf,oldfits,oldfits2,oldfits3,deBoer:2003xm,recent2}. 
Such analyses face the problem of the large dimensionality of the
MSSM parameter space. In this section we discuss global fits in the
unconstrained MSSM (for real parameters and using certain universality
assumptions). Analyses in specific soft SUSY-breaking scenarios, such as 
mSUGRA, will be discussed in \refse{chapter4}.
An overview of non-supersymmetric analyses of precision observables
and resulting constraints can be found in \citere{peskinwells}.

\begin{figure}[htb!]
\begin{center}
\epsfig{figure=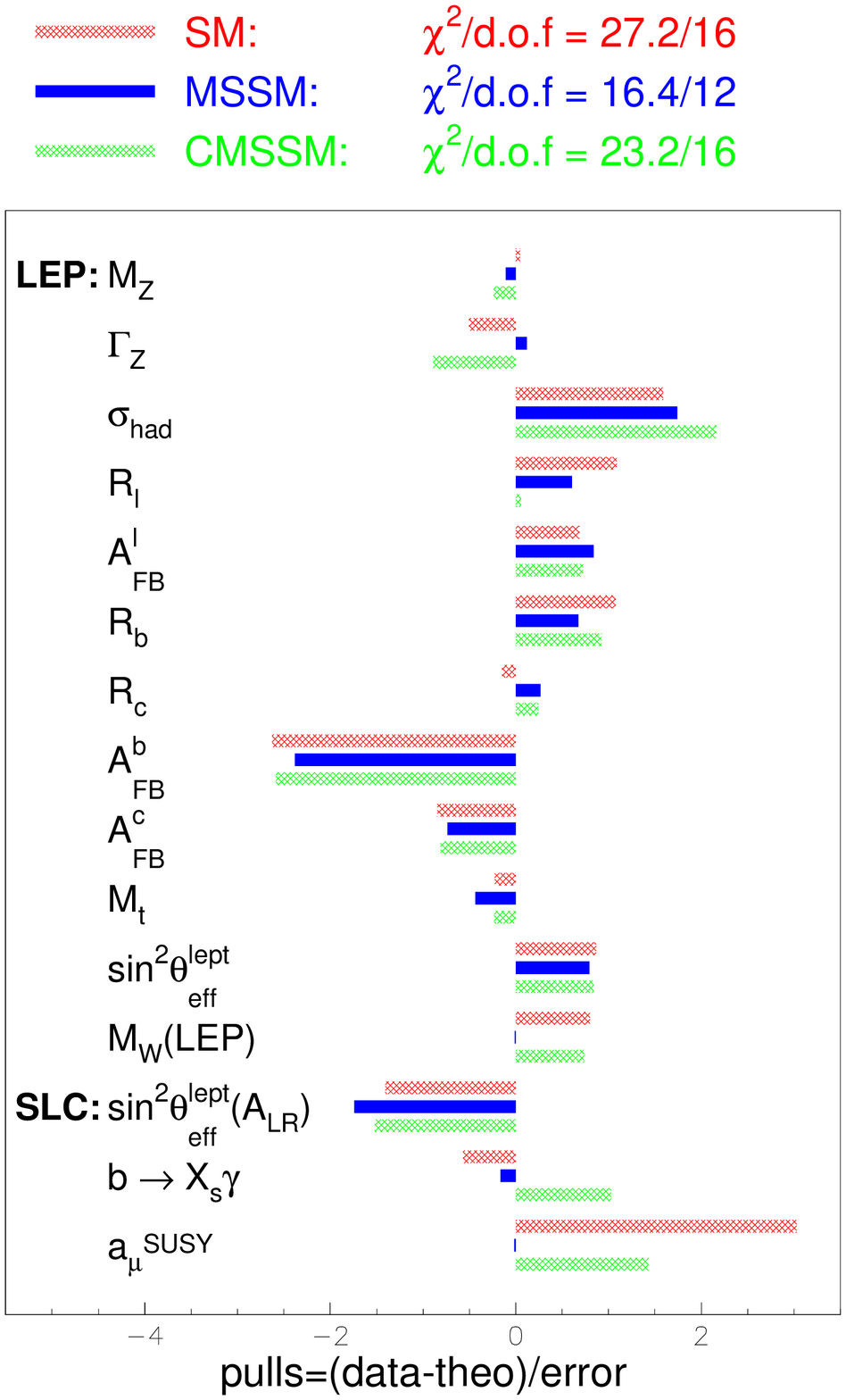,width=12cm,height=15cm}
\end{center}
\caption[]{The predictions in the SM, the MSSM and the mSUGRA scenario
(CMSSM) are compared with the data \cite{deBoer:2003xm}.
Deviations between theory and
experiment are indicated in units of one standard deviation of the
experimental results.
}
\label{fig:ewpoMSSM}
\end{figure}
 
The most recent global fit of the MSSM to the electroweak precison data
has been performed in \citere{deBoer:2003xm} (for previous analyses, see 
\citeres{hagiwararev,oldfits,oldfits2,oldfits3}).
The results are shown in \reffi{fig:ewpoMSSM}, where 
the predictions in the SM, the MSSM and the constrained MSSM (i.e.\ the
mSUGRA scenario) are compared with the experimental data (the SUSY
predictions are for $\tan\be = 35$).
\reffi{fig:ewpoMSSM} shows the features discussed above: the MSSM
predictions for $\MW$ and (for large $\tan\be$) $(g - 2)_\mu$ are in
better agreement with the
data than in the SM (slight improvements also occur for the
total width of the $Z$~boson, $\Gamma_Z$, and for $B \to X_s \ga$).
On the other hand, for the observables with the largest deviations
between theory and experiment, namely $A^{\rm b}_{\rm FB}$ and the
neutrino--nucleon cross section measured at NuTeV (the latter is not
shown in \reffi{fig:ewpoMSSM}), the MSSM does not yield a significant
improvement compared to the SM. The global fit in the MSSM has a lower
$\chi^2$ value than in the SM. Since the MSSM fit has less degrees of
freedom than the SM one, the overall fit probability in the MSSM is only
slightly better than in the SM.


\section{Future expectations}
\label{sec:futureexp}

In this section we give a few examples of the possible physics gain 
obtainable with the anticipated improvements of the accuracies of the
experimental results and the theoretical predictions for the precision
observables (see \refta{tab:POfuture} and the discussion in chapter~2).
We focus here on the effects from $\MW$, $\sweff$ and $\mh$. For a
discussion of $(g-2)_\mu$ in the framework of the mSUGRA scenario, see 
chapter~4 below.

Two examples of future prospects were already presented in
\refse{subsubsec:numanalMWsweff}. In \reffi{fig:MWMT} the SM and MSSM
predictions in the $\mt$--$\MW$ plane are shown and compared with the
current and future experimental precisions. Likewise, in
\reffi{fig:MWsw2eff} the results for the $\MW$--$\sweff$ plane are
given. It becomes aparent that the
prospective improvements in the experimental accuracies, in particular
at the ILC with GigaZ option, will provide a high sensitivity to deviations
both from the SM and the MSSM.

\begin{figure}[htb!]
\begin{center}
\epsfig{figure=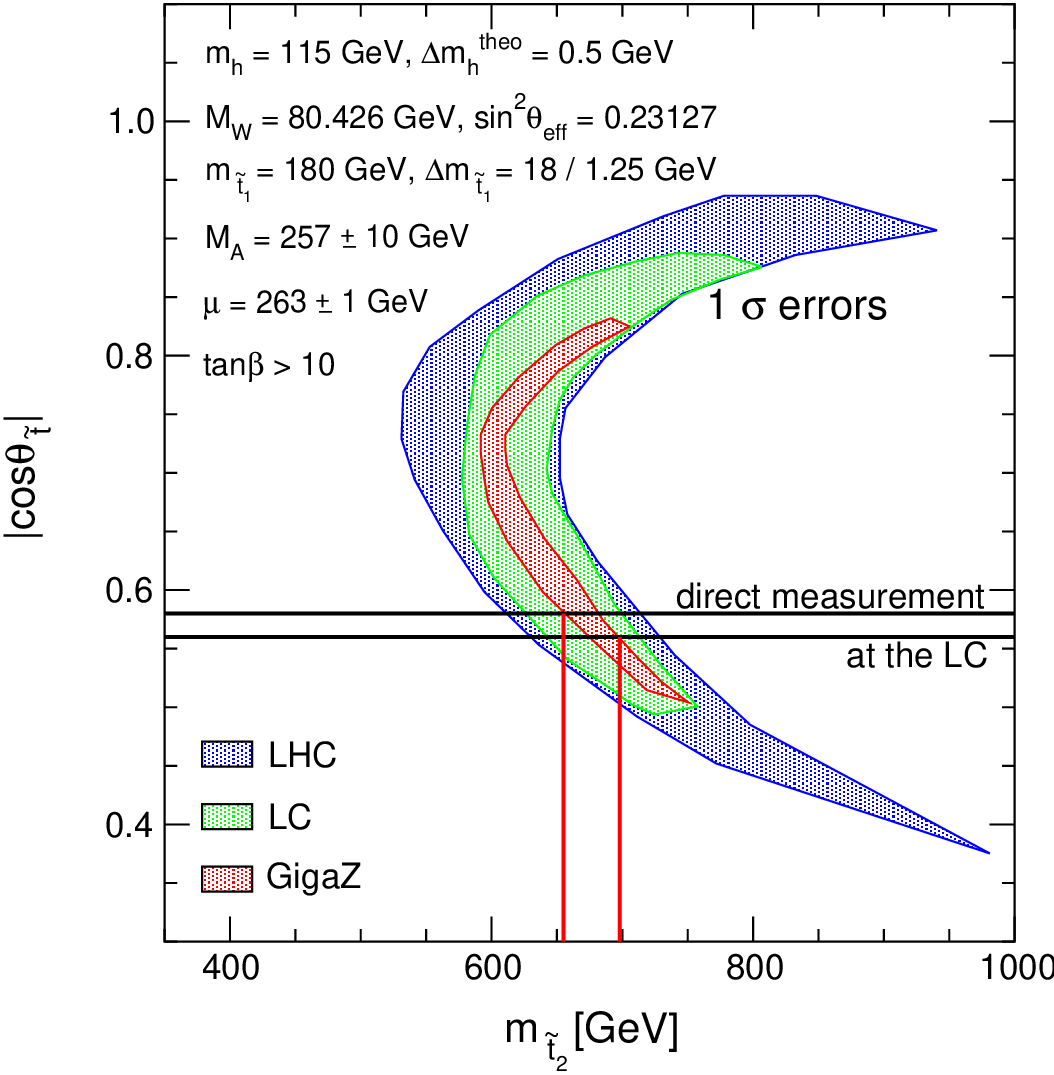,width=12cm}
\end{center}
\caption[]{
Indirect constraints on the MSSM parameter space in the 
$\mstz$--$|\costt|$ plane from measurements of 
$\mh$, $\MW$, $\sweff$, $\mt$ and $\mste$ in view of the 
prospective accuracies for
these observables at the ILC with and
without GigaZ option and at the LHC. The direct information on the
mixing angle from a measurement at the ILC is indicated together with the
corresponding indirect determination of $\mstz$.
}
\label{fig:zerwas}
\end{figure}

\bigskip
The indirect constraints on supersymmetric models from electroweak
precision tests, in particular with GigaZ accuracy, will yield
complementary information to that obtained from the direct observation
of supersymmetric particles at the Tevatron, the LHC or the ILC
(for a comprehensive overview on the prospects of the LHC and the ILC
and the potential for combined analyses using LHC and ILC data, see
\citere{lhclc}).
As an example we present an analysis in the scalar top
sector~\cite{mondplot}. 
Direct information on the stop sector parameters $\mste$ and
$\tst$ can be obtained at the ILC
from the process $e^+e^- \to \tilde t_1 \tilde t_1$, yielding a
precision of \order{1\%}~\cite{lcstop}. These direct measurements can be
combined with the indirect information from requiring consistency of the
MSSM with a precise measurement of $\mh$, $\MW$ and $\sweff$.
This is shown in Fig.~\ref{fig:zerwas}, where the allowed 
parameter space according to measurements of $\mh$, $\MW$ and $\sweff$
is displayed in the plane of the heavier stop mass, $\mstz$,
and $|\costt|$ for the accuracies at the ILC with
and without the GigaZ option and at the LHC (see \refta{tab:POfuture}). 
For $\mste$ (with an assumed central value of $180 \gev$) a precision at
the ILC of $1.25 \gev$ is taken~\cite{lcstop}, while 
for the LHC an (optimistic) uncertainty of 10\% in $\mste$ is assumed.
For the other parameters the following central values and prospective
experimental errors have been used:
$\MA = 257 \pm 10$~GeV, $\mu = 263 \pm 1$~GeV,
$M_2 = 150 \pm 1$~GeV, $\mgl = 496 \pm 10$~GeV. For the
top-quark mass an error of 0.2~GeV has been used for GigaZ/ILC and 
of 2~GeV for the LHC. For $\tb$ a lower bound of $\tan\beta > 10$
has been taken. 
For the future theory uncertainty of $\mh$ from unknown higher-order
corrections an  
error of $0.5$~GeV has been assumed. 
The central values for $\MW$ and $\sweff$ have been chosen in accordance
with a non-zero contribution to the precision observables from SUSY
loops. 
For the experimental errors at the different colliders the 
values given in sect.~\refta{tab:POfuture} have been used. For the future
intrinsic theoretical uncertainties the estimates of 
\refeq{eq:MWSWintrSUSYfuture} have been taken.

As one can see in Fig.~\ref{fig:zerwas}, the allowed parameter space in the
$\mstz$--$|\costt|$ plane is significantly reduced
from the LHC to the ILC, in particular in the GigaZ scenario (i.e.\
precision measurements of $\MW$ and $\sweff$). Using 
the information on $|\costt|$ from the direct measurement~\cite{lcstop}
allows an indirect determination of $\mstz$ with 
a precision of better than 5\% in the GigaZ case. By comparing this indirect 
prediction for $\mstz$ with direct experimental information on
the mass of this particle, the MSSM could be tested at its quantum level
in a sensitive and highly non-trivial way.

\begin{figure}[htb!]
\begin{center}
\epsfig{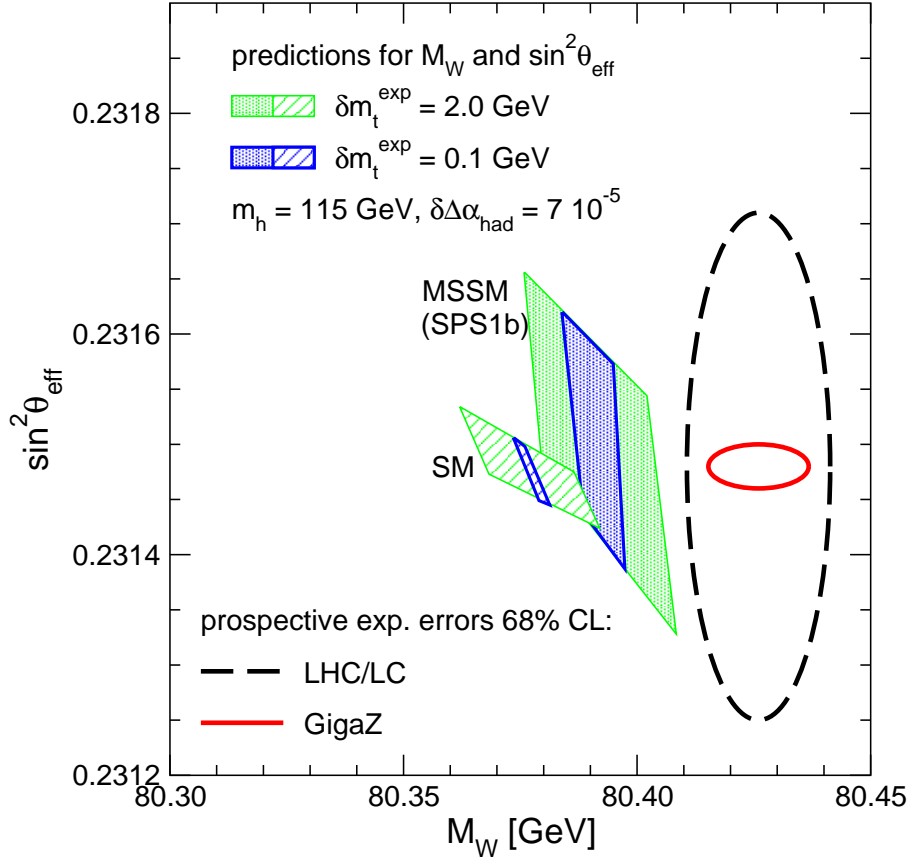}
\end{center}
\caption[]{
The predictions for $\MW$ and $\sweff$ in the SM and the MSSM
(SPS1b) \cite{deltamt}.   The inner 
(blue) areas correspond to $\de\mtexp = 0.1 \gev$ (ILC), while
the outer (green) areas arise from $\de\mtexp = 2 \gev$ (LHC). 
The anticipated experimental errors on $\MW$ and $\sweff$ at the
LHC/ILC and at the ILC with GigaZ option are indicated.
}
\label{fig:MWSW_SMvsSPS}
\end{figure}

\bigskip
As a further example~\cite{deltamt} 
for the potential of a precise measurement of the EWPO
to explore the effects of new physics, we show in \reffi{fig:MWSW_SMvsSPS} 
the predictions for $\MW$ and $\sweff$ in the SM and the MSSM in
comparison with the prospective experimental accuracy obtainable at
the LHC and the ILC without GigaZ option (labelled as LHC/LC) and with the 
accuracy obtainable at the ILC with GigaZ option (labelled as GigaZ).
For the assumed experimental central values of $\MW$ and $\sweff$ 
the current central values~\cite{LEPEWWG} are used.
For the Higgs-boson mass a future measured value of $\mh = 115 \gev$
has been assumed. 
The MSSM parameters have been chosen in this example according to the
reference point SPS1b~\cite{sps}.
In \reffi{fig:MWSW_SMvsSPS} the inner 
(blue) areas correspond to $\de\mtexp = 0.1 \gev$ (ILC), while
the outer (green) areas arise from $\de\mtexp = 2 \gev$ (LHC). 
For the error of $\De\al_{\rm had}$ we have assumed a future
determination of $7 \times 10^{-5}$. In the SM, this is the only
relevant uncertainty apart from $\de\mt$ (the remaining effects of
future intrinsic uncertainties have been neglected in this figure). The 
future experimental uncertainty of $\mh$ is insignificant for
this kind of electroweak precision tests.
For the experimental errors on the SUSY parameters
we have assumed a 5\% uncertainty for $\mste, \mstz, \msbe, \msbz$
around their values given by SPS1b. The mixing angles in the
$\Stop$~and $\Sbot$~sectors have been left unconstrained. 
The mass of the $\cp$-odd Higgs boson $\MA$ is
assumed to be determined to about 10\%, and it is assumed that
$\tb \approx 30 \pm 4.5$.

The figure shows that the improvement in $\de\mt$ from 
$\de\mt = 2 \gev$ to $\de\mt = 0.1 \gev$ strongly reduces the
parametric uncertainty in the prediction for the EWPO.  
In the SM case it leads to a reduction by about a factor of 10 
in the allowed parameter space of the $\MW-\sweff$ plane.  
In the MSSM case, where many additional parametric uncertainties enter, 
a reduction by a factor of more than 2 is obtained in this example. 
The comparison of the theoretical prediction in both models with the
GigaZ accuracy on $\sweff$ and $\MW$ illustrates how sensitively the
electroweak theory will be tested via EWPO (for a comparison with the
current experimental errors, which are not shown in
\reffi{fig:MWSW_SMvsSPS}, see \reffi{fig:MWsw2eff}).
The simultaneous improvement of the precision on $\mt$, $\sweff$ (by an
order of magnitude compared to the situation at the LHC) and $\MW$ (by a
factor of two compared to the LHC case) will greatly enhance the
potential for establishing effects of new physics via EWPO.

\begin{figure}[htb!]
\begin{center}
\epsfig{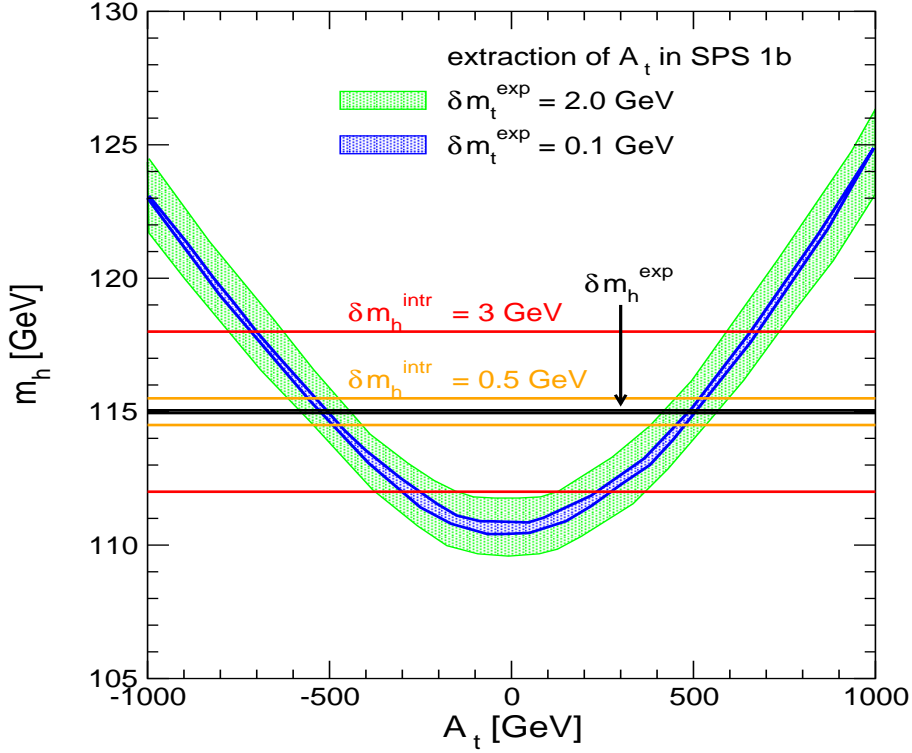}
\end{center}
\caption[]{
The prediction for $\mh$ within the SPS~1b scenario, assuming
experimental information from the LHC and the ILC on the
SUSY spectrum with experimental errors according to \citere{lhclc}, 
is shown as a function of $\At$. The light shaded (green) band 
indicates the uncertainty induced by the experimental errors of all 
MSSM input parameters (except $\At$) and an assumed error on the top-quark 
mass of $\de\mt^{\rm exp} = 2 \gev$. The dark shaded (blue) band shows
the parametric uncertainty induced by the experimental errors of all
input parameters for the case of $\de\mt^{\rm exp} = 0.1 \gev$. The
experimental error of a prospective measurement of $\mh$ is shown 
as a horizontal band. Two
further bands are shown, demonstrating the effect of an intrinsic
theoretical uncertainty on $\mh$ of $3 \gev$ (today) and 
$0.5 \gev$ (future).
}
\label{fig:mhAt}
\end{figure}

\bigskip
As mentioned above, the precision observable $\mh$ will allow to set
very stringent constraints on the MSSM parameters, in particular in the
scalar top sector (for large values of $\tb$ also in the scalar bottom
sector). This can be crucial for determining the mixing angle in the 
scalar top sector, and (as a related quantity) the trilinear Higgs-stop
coupling, $\At$. If the scalar top quarks are too heavy to be directly
produced at the ILC, only rather limited information on the mixing in
the stop sector will be available from the LHC~\cite{lhclc}.
The prospects for an indirect determination of $\At$ within the MSSM
from a precision measurement of $\mh$ are illustrated in
\reffi{fig:mhAt}.
A precise knowledge of the parameter $\At$ turned out to be
crucial for global fits of the MSSM to the data~\cite{fittino,sfitter}, 
which will be necessary in order to determine the low-energy SUSY
Lagrangian parameters, and for an extrapolation of the results
obtainable at the next generation of colliders to physics at high
scales~\cite{deltamt}.

\reffi{fig:mhAt} shows the prediction for $\mh$ as a function of $\At$,
where the parametric uncertainties induced by all other MSSM input parameters
are taken into account according to the prospective experimental
information on the SUSY spectrum from the LHC and the ILC in the SPS~1b
scenario~\cite{sps} (see \citere{lhclc}). The impact of the LHC and the ILC
precision on the top-quark mass is indicated. The sensitivity for an
indirect determination of $\At$ follows from intersecting the MSSM
prediction for $\mh$ with the experimental value. This comparison is
affected, however, by the intrinsic theoretical uncertainties of the
$\mh$ prediction. The effect of the intrinsic theoretical uncertainties 
is shown by two horizontal bands illustrating the present intrinsic
uncertainty of $3 \gev$ and a prospective uncertainty of $0.5 \gev$.
While the present intrinsic uncertainty on $\mh$ would not allow to
obtain a reliable indirect determination of $\At$, a future theoretical
uncertainty of $0.5 \gev$ together with a precision measurement of $\mt$
at the ILC would allow an indirect determination of $\At$ to better than 
about $10\%$, up to a sign ambiguity. The sign ambiguity can be resolved
using precision measurements of Higgs branching ratios at the ILC, see
\citere{deschi}.

Likewise, it has been shown in \citere{deschi} that an indirect
determination of $\MA$ can be performed (investigated in the case of
the SPS~1a scenario~\cite{sps}) from Higgs boson branching ratio
measurements at the ILC combined with a precision measurement of $\mt$
and information on the SUSY spectrum from the LHC and ILC.



\newpage

\chapter{Implications in soft SUSY-breaking scenarios}
\label{chapter4}

The fact that no SUSY partners of the SM particles have so far been
observed means that low-energy SUSY cannot be realized as an unbroken
symmetry in nature, and SUSY models thus have to incorporate
additional Supersymmetry breaking contributions. 
This is achieved by adding to the Lagrangian (defined by the 
${\rm SU(3)}_C\times {\rm SU(2)}_L \times  {\rm U(1)}_Y$ gauge
symmetry and the superpotential $W$)
further terms that respect the gauge symmetry but break 
SUSY (softly, i.e.\ no quadratic divergences appear), so
called ``soft SUSY-breaking'' (SSB) terms.
The assumption made in the MSSM that the $R$-parity symmetry is conserved
reduces the amount of 
new soft terms allowed in the Lagrangian.

In the previous sections the EWPO have been discussed within the
unconstrained MSSM. In the MSSM, no further assumptions are made on the
structure of the soft SUSY-breaking parameters, and  a
parametrization of all possible SUSY-breaking terms is
used~\cite{allSSB,moreSSB}. This 
gives rise to the huge number of 
more than 100 new parameters in addition to the SM, 
which in principle can be chosen
independently of each other. A phenomenological analysis of the EWPO
in this model in full generality would clearly be very involved, and
one usually restricts to certain benchmark scenarios, see e.g.\
\citeres{benchmark,LHbenchmark,sps,BDEGOP}.
On the other hand, models in which all the low-energy parameters are
determined in terms of a few parameters at the Grand Unification
scale (or another high-energy scale), 
employing a specific soft SUSY-breaking scenario, provide an attractive
framework for investigating SUSY phenomenology.
The most prominent scenarios in the literature
are minimal Supergravity (mSUGRA)~\cite{Hall,mSUGRArev}, 
minimal Gauge Mediated SUSY Breaking (mGMSB)~\cite{GR-GMSB} 
and minimal Anomaly Mediated SUSY Breaking 
(mAMSB)~\cite{lr,giudice,wells}. 

The Higgs boson sector has been analyzed in all three soft
SUSY-breaking scenarios, see
\citeres{ehow,asbs2,mhiggsWN,asbs,ehow2} and  
references therein. For a comprehensive analysis of EWPO 
within the mSUGRA scenario see \citere{ehow3}. 


\section{The soft SUSY-breaking scenarios}
\label{sec:susybreak}

The three most commonly studied soft SUSY-breaking scenarios are

\begin{itemize}

\item 
{\bf mSUGRA} (minimal Super Gravity scenario)~\cite{Hall,mSUGRArev}:\\
Apart from the SM parameters (for the experimental values of the SM 
input parameters we use~\citere{pdg}), 
4~parameters and a sign are required to define the mSUGRA scenario:
\BE
\{\; m_0\;,\;m_{1/2}\;,\;A_0\;,\;\tb \;,\; {\rm sign}(\mu)\; \} \;.
\label{msugraparams}
\end{equation}
The parameter $m_0$ is a common scalar mass, $m_{1/2}$ a common fermion
mass and $A_0$ a common trilinear couplings, all defined at the GUT
scale ($\sim 10^{16} \gev$). On the other hand, 
$\tb$ (the ratio of the two vacuum expectation
values) and sign($\mu$) 
are defined at the low-energy scale.%
\footnote{More precisely, the scenario where universality of the soft
SUSY-breaking parameters $m_0$, $m_{1/2}$ and $A_0$ at the GUT scale is
assumed should be called the constrained MSSM (CMSSM). An economical way
to ensure this universality is by gravity-mediated SUSY breaking in a
minimal supergravity (mSUGRA) scenario, but there are other ways to
validate the CMSSM assumptions. The mSUGRA scenario predicts in
particular a relation between the gravitino mass and $m_0$, which is not
necessarily filfilled in the CMSSM. For simplicity, we do not make the
distinction between the CMSSM and the mSUGRA scenario but use the phrase
``mSUGRA'' for both.}

\item 
{\bf mGMSB} (minimal Gauge Mediated SUSY-Breaking)~\cite{GR-GMSB}:\\
An interesting alternative to mSUGRA is based on the hypothesis
that the soft SUSY-breaking occurs at relatively low energy scales 
and is mediated mainly by gauge interactions through the so-called
``messenger sector''~\cite{GR-GMSB,oldGMSB,newGMSB}. 
Also in this scenario, the low-energy phenomenology is characterized
in terms of 4~parameters and a sign,
\BE
\{\; M_{\rm mess}, \; N_{\rm mess}, \; \Lambda, \; 
     \tb, \; {\rm sign}(\mu) \; \} \;,
\label{gmsbparams}
\end{equation}
where $M_{\rm mess}$ is the overall messenger mass scale; 
$N_{\rm mess}$ is a number called the 
messenger index, parametrizing the structure of the messenger
sector; $\Lambda$ is the universal soft SUSY-breaking mass scale felt by the
low-energy sector.
The phenomenology of mGMSB is characterized by the presence of a very
light gravitino  $\tilde{G}$ with mass given by  
$m_{3/2} = m_{\tilde{G}} = \frac{F}{\sqrt{3}M'_P} \simeq 
\left(\frac{\sqrt{F}}{100 \tev}\right)^2 2.37 \; {\rm eV}$~\cite{Fayet},  
where $\sqrt{F}$ is the fundamental scale of SUSY breaking and 
$M'_P = 2.44 \times 10^{18} \gev$ is the reduced Planck mass.
Since $\sqrt{F}$ is typically of order 100 TeV, the $\tilde{G}$ is always the 
LSP in the GMSB scenario.

\item
{\bf mAMSB} (minimal Anomaly Mediated SUSY-Breaking)~\cite{lr,giudice,wells}:\\
In this model, SUSY breaking happens on a separate brane and is  
communicated to the visible world via the super-Weyl anomaly. 
The particle spectrum is determined by 3~parameters and a sign:  
\BE
\{m_{\rm aux},\ m_{0},\ \tb,\ {\rm sign}(\mu) \} .
\label{amsbparams}
\end{equation}
The overall scale of SUSY particle masses is set by $m_{\rm aux}$, 
which is the VEV of the auxiliary field in the supergravity multiplet.
$m_0$ is introduced as a phenomenological parameter 
to avoid negative slepton mass squares, for other
approaches to this problem see \citeres{lr,negative,clm,kss,jjw}.

\end{itemize}


\section{$\De\rho$ in mSUGRA, mGMSB, mAMSB}

\label{sec:delrhoasbs}

In order to compare the prediction for $\De\rho$ in three soft
SUSY-breaking scenarios, a scan has been performed over the parameters
defined in \refeqs{msugraparams}--(\ref{amsbparams}). 
For our numerical analysis, the scan has been done in the following ranges:
\begin{itemize}
\item mSUGRA:
\BEA
50~{\rm GeV} \le &m_0& \le 1~{\rm TeV} \;, \nonumber \\
50~{\rm GeV} \le  &m_{1/2}& \le 1~{\rm TeV} \;, \nonumber \\
-3~{\rm TeV} \le &A_0& \le 3~{\rm TeV} \;, \non \\
1.5 \le &\tan\beta & \le 60 \;, \non \\
 &{\rm sign}\, \mu& = +1 .
\label{msugraparam}
\EEA

\item GMSB:
\BEA
10^4 \gev \le &\La& \le 2\,\times\,10^5 \gev \;, \non \\
1.01\,\La \le &M_{\rm mess}& \le 10^5\,\La \;, \nonumber \\
1 \le  &N_{\rm mess}& \le 8 \;, \nonumber \\
1.5 \le &\tan\beta & \le 60 \;, \non \\
 &{\rm sign}\, \mu& = +1 .
\label{gmsbparam}
\EEA

\item AMSB:
\BEA
20 \tev \le &m_{\rm aux}& \le 100 \tev , \non \\
0 \le &m_0& \le 2 \tev , \non \\
1.5 \le &\tb& \le 60 , \non \\
 &{\rm sign}\, \mu& = +1 . 
\label{amsbparam}
\EEA

\end{itemize}

For each scan point the full low-energy spectrum of the MSSM has been
evaluated. It has been checked that the low-energy result
respects the existing experimental constraints (for a more
detailed discussion, see \citere{asbs}):

\begin{itemize}

\item
{\bf LEP Higgs bounds}:

The results from the Higgs search at LEP have excluded a considerable 
part of the MSSM parameter space~\cite{LEPHiggsearly,LEPHiggsSM}. 
The results of the search for the MSSM Higgs bosons are usually
interpreted in three different benchmark scenarios~\cite{benchmark}.
The 95\% C.L.\ exclusion limit for the SM Higgs boson of 
$\MHSM > 114.4 \gev$~\cite{LEPHiggsSM} 
applies also for the lightest $\cp$-even Higgs boson of the MSSM
except for the parameter region with small $\MA$ and large $\tb$. In the
unconstrained MSSM this bound is reduced to 
$\mh > 91.0 \gev$~\cite{LEPHiggsearly} for 
$\MA \lsim 150 \gev$ and $\tb \gsim 8$ as a consequence of a reduced
coupling of the Higgs to the $Z$~boson. For the $\cp$-odd Higgs boson a
lower bound of $\MA > 91.9 \gev$ has been obtained~\cite{LEPHiggsearly}.
In order to correctly interpolate between the parameter regions where
the SM lower bound%
\footnote{
Instead of the actual experimental lower bound, 
$\MHSM \gsim 114.4 \gev$~\cite{LEPHiggsSM},  
we use the value of $113 \gev$ in order to take into account 
some effect of the uncertainty
in the theoretical evaluation of $\mh$ from unknown higher-order
corrections, which is currently estimated
to be $\sim 3 \gev$ in the unconstrained MSSM (see \refeq{eq:mhintrcurr}).
}%
~of $\MHSM \gsim 113 \gev$ and the bound $\mh \gsim 91 \gev$
apply, we use the result for the Higgs-mass exclusion given with respect
to the reduced $ZZh$ coupling squared 
(i.e.\ $\sin^2(\be-\aeff)$)~\cite{Barate:2000zr}. We have compared the
excluded region with the theoretical prediction 
obtained at the \twol\ level for $\mh$ and $\sqbaeff$ for each parameter
set (using $\mt = 175 \gev$).

\item
{\bf Experimental bounds on SUSY particle masses}

In order to restrict the allowed parameter space in the three
soft SUSY-breaking scenarios the current experimental 
constraints on their low-energy mass 
spectrum~\cite{pdg} have been employed. 
The precise values of the bounds that we have applied can be found in
\citere{asbs}. 

\item
{\bf Other restrictions}

As mentioned above, the top-quark mass is fixed to $\mt = 175 \gev$
in our analysis.
While $\De\rho^{\SM}$ depends quadratically on $\mt$ at the one-loop
level, the impact on $\De\rho^{\SU}$ is relatively mild. 

We briefly list here the further restrictions that we have taken
into account for the analysis in this section. 
For a detailed discussion see \citere{asbs}.

\begin{itemize}

\item 
The GUT or high-energy scale parameters are taken to be real, no SUSY
$\cp$-violating phases are assumed.

\item
In all models under consideration the $R$-parity symmetry
is taken to be conserved.

\item
Parameter sets that do not fulfil the condition
of radiative electroweak symmetry breaking (REWSB) are discarded (already
at the level of generating the model parameters).

\item
Parameter sets that do not fulfil the constraints that there should
be no charge or color breaking minima are
discarded (already 
at the level of generating the model parameters).

\item
We demand that the lightest SUSY particle (LSP) is uncolored and uncharged.
In the mGMSB scenario the LSP is always the gravitino, so this 
condition is automatically fulfilled.
Within the mSUGRA and mAMSB scenario, the LSP
is required to be the lightest neutralino. 
Parameter sets that result in a different LSP are excluded.

\item
We do not apply any further cosmological constraints, i.e.\ we do not
demand a relic density in the region favored by dark matter
constraints~\cite{cdm}. 

\item
The scan has been stopped at high squark masses, since the
contributions of heavy particles to $\De\rho^{\SU}$ 
decouple~\cite{dr2lA,drMSSMgf2B}. No   
parameter points with $\msq \gsim 1.5 \tev$ have been considered.

\end{itemize}

\end{itemize}

\begin{figure}[htb!]
\begin{center}
\epsfig{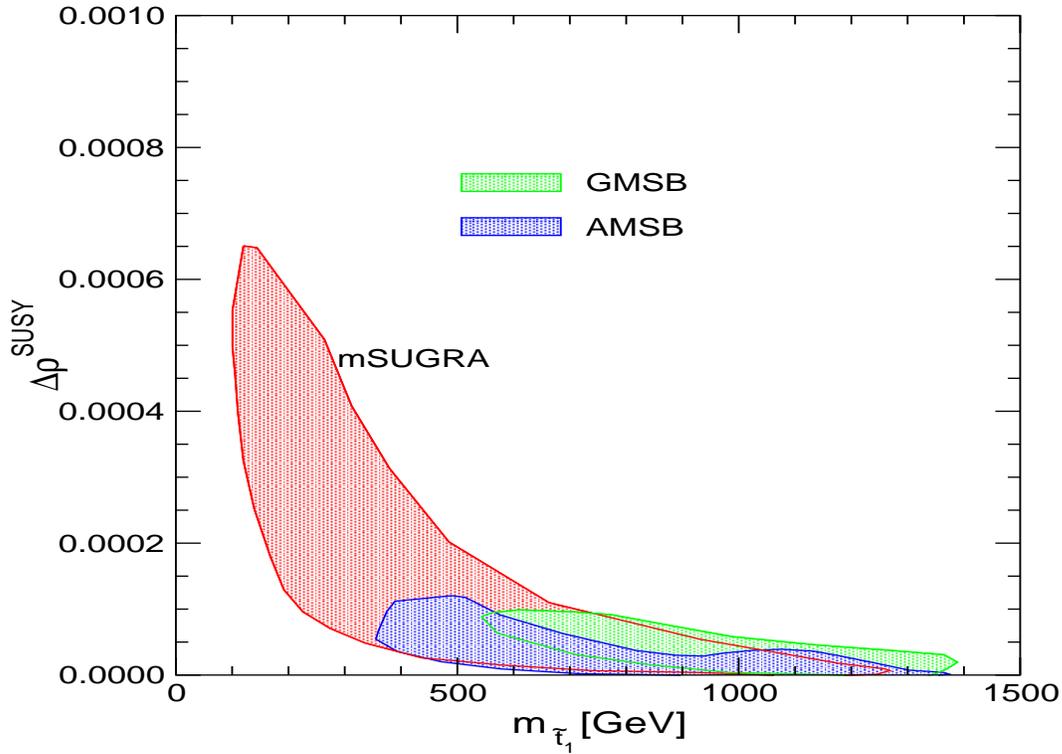}  
\end{center}
\caption[]{
$\De\rho^{\SU}$ is shown in the three soft SUSY-breaking scenarios as a
function of the lightest scalar top quark mass.
}
\label{fig:ASBSdelrho}
\end{figure}

If a point has passed all constraints, the results for the masses and
mixing angles have 
been used to determine $\De\rho$, based on the one-loop
result given in \refeq{delrhoMSSM1l} and the SUSY two-loop
contributions described in \refse{subsubsec:delrhomssm}. 
The result is shown in \reffi{fig:ASBSdelrho}, where $\De\rho^{\SU}$ is
plotted as a function of the lightest scalar top quark mass, $\mste$. 
In general, mSUGRA allows smaller scalar quark masses than mGMSB
and mAMSB, and correspondingly larger values of $\De\rho^{\SU}$  can
be realized. For $\mste \lsim 300 \gev$ values of 
$\De\rho^{\rm mSUGRA} \lsim 7 \times 10^{-4}$ can be reached. 
For larger $\mste$ values all three soft SUSY-breaking scenarios
result in $\De\rho^{\SU} \lsim 1 \times 10^{-4}$ (a shift in
$\De\rho^{\SU}$ of $1 \times 10^{-4}$ corresponds to
shifts in $\MW$ and $\sweff$ of about $\De\MW = 6 \mev$ and 
$\De\sweff = - 3 \times 10^{-5}$, respectively).
No part of the mSUGRA, mGMSB, or mAMSB
parameter space that fulfils all other phenomenological constraints
(see above) can be excluded with the current precision on the EWPO. 
On the other hand, for $\mste \gsim 500 \gev$ all three scenarios
result in roughly the same prediction, i.e.\ it would be very
challenging in this case
to obtain information on the soft SUSY-breaking scenario
with the help of $\De\rho$.

Using \refeq{precobs} the SUSY contribution to $\De\rho$ can be
translated into a shift in the prediction of $\MW$ and $\sweff$. For
$\mste \lsim 300 \gev$ the shift induced within the mSUGRA scenario
can amount up to 
\BE
\de\MW^{\rm mSUGRA} \lsim 35 \mev, \quad
|\de\sweff^{\rm mSUGRA}| \lsim 2 \times 10^{-4}~,
\end{equation}
which corresponds roughly to one standard deviation
of the current experimental
uncertainties. For larger $\mste$, $\mste \gsim 500 \gev$,
the shifts induced in $\MW$ and
$\sweff$ for all three soft SUSY-breaking scenarios fulfil
\BE
\de\MW^{\SU} \lsim 6 \mev, \quad
|\de\sweff^{\SU}| \lsim 3.5 \times 10^{-5}~.
\end{equation}
While for $\MW$ the possible shift in this case
is about one standard deviation of the GigaZ
precision, for $\sweff$ deviations of 2-3$\,\si$ of the GigaZ
precision could be realized.


\section{Prediction for $\mh$ in mSUGRA, mGMSB, mAMSB}

\begin{figure}[htb!]
\begin{center}
\epsfig{figure=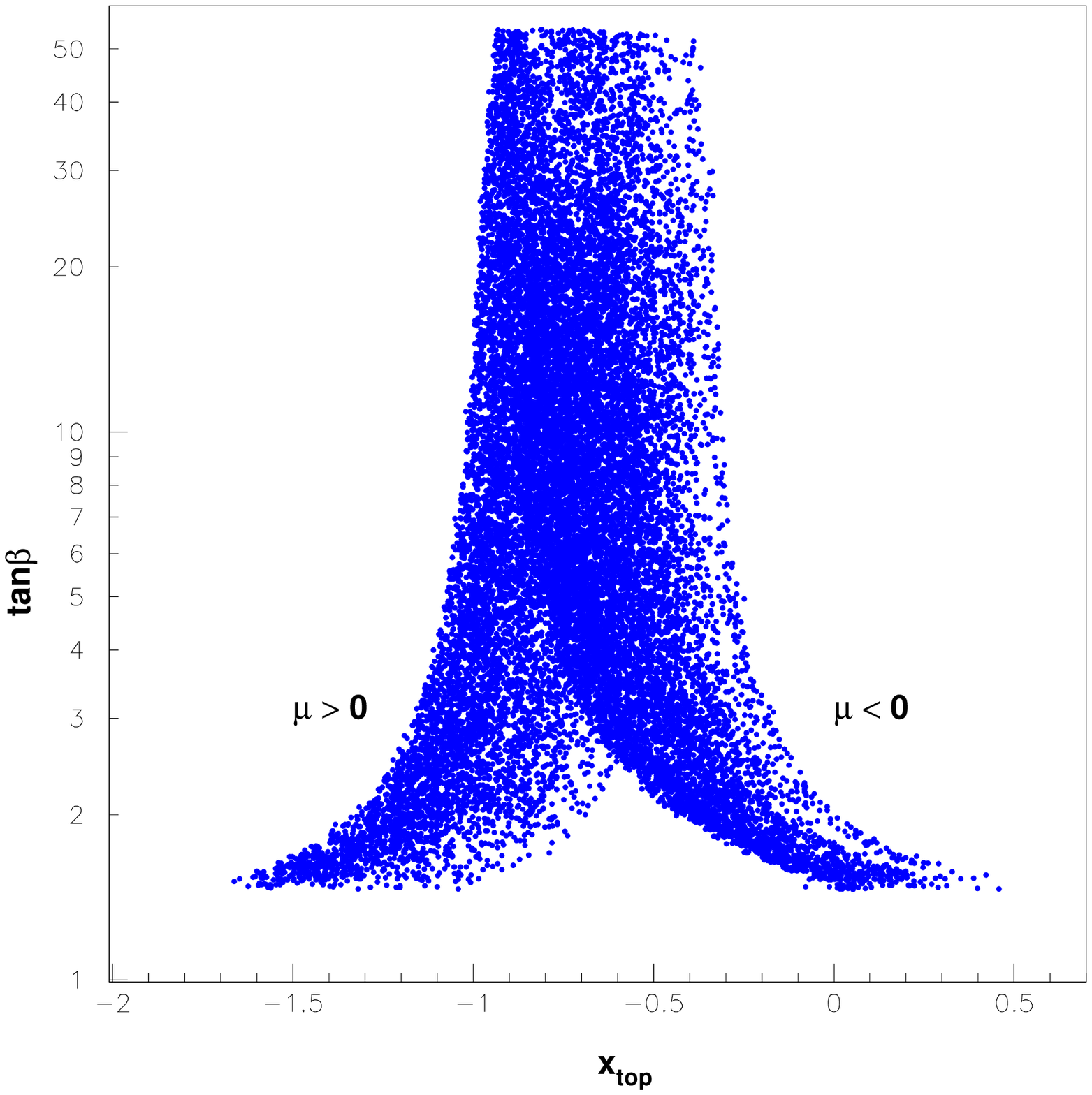,width=10cm,height=8.0cm}  
\end{center}
\caption[]{
Allowed $\tb$ values as a function of $x_{\rm top} = \Xt/\msusy$ in
the mGMSB scenario \cite{gmsbMh}.
The high-energy scan parameters are chosen as in
\refeq{gmsbparam} (but with both signs of~$\mu$).
}
\label{fig:tbvsxt}
\end{figure}

We now turn to the prediction of the lightest Higgs-boson mass for the
case where the low-energy parameters are obtained from high-scale
parameters within specific soft SUSY-breaking scenarios. Since the
low-energy parameters are connected to each other via the
renormalization group equations, they cannot be chosen independently. 
This results in a reduction of the upper bound on $\mh$ compared to the
unconstrained MSSM. As an example, we show in \reffi{fig:tbvsxt}
the allowed values of $\tb$ as function of $x_{\rm top} \equiv \Xt/\msusy$
in the mGMSB scenario~\cite{gmsbMh}. The high-energy scan parameters
are chosen as in \refeq{gmsbparam} (but with both signs of~$\mu$). It
can be seen that large values 
of $\tb$, which are necessary for large $\mh$ values, can only be
realized for $\Xt/\msusy$ between $-0.3$ and $-1$. On the other hand,
the largest values for $\mh$ are obtained for 
$\Xt/\msusy \approx +2$~\cite{mhiggslong,mhiggslle}, which cannot be
realized in the mGMSB. Similarly,
also the variation of the upper bound on $\mh$ with $\mt$ turns out to
be somewhat different in the soft SUSY-breaking scenarios compared to
the unconstrained MSSM (see below).

In the following we refer to the results of \citere{mhiggsWN}, which
are in agreement with the previous
results in \citeres{asbs2,asbs},
but use the most recent experimental value of the top-quark mass. 
In \refta{tab:ASBSmh} the
maximum values of $\mh$ for $\mt = 178.0 \gev$ in mSUGRA, mGMSB and
mAMSB are compared. In order to have comparable numbers an upper limit on the
scalar top masses in all scenarios has been chosen, 
$\sqrt{\mste\,\mstz} \le 2 \tev$. No theoretical uncertainties are
included. One can see that all three scenarios
result in significantly
lower maximum $\mh$ values than the uncontrained MSSM,
where masses up to $\sim 138 \gev$ can be realized for $\msusy \lsim 2 \tev$ 
and $\mt = 178.0 \gev$ (see \reffis{fig:tbmh}, \ref{fig:mhtb}).
The variation of this maximum $\mh$ value with $\mt$ is also shown. In
the unconstrained MSSM one has $\de\mh/\de\mt \approx 1$~\cite{tbexcl}.
In the mSUGRA, mGMSB and mAMSB scenarios this is reduced down to 
$\sim 0.58$--$0.7$. 

\begin{table}[h!]
\renewcommand{\arraystretch}{1.5}
\begin{center}
\begin{tabular}{|c||c|c|}
\cline{2-3} \multicolumn{1}{c||}{}
& maximum $\mh$ [GeV] & $\de\mh/\de\mt$ \\
\hline\hline
mSUGRA & 129.0 & 0.65 \\ \hline
mGMSB  & 123.7 & 0.70 \\ \hline
mAMSB  & 124.6 & 0.58 \\ \hline
\hline
\end{tabular}
\end{center}
\renewcommand{\arraystretch}{1}
\caption{
The maximum $\mh$ values (for $\mt = 178.0 \gev$ and 
$\sqrt{\mste\,\mstz} \le 2 \tev$) and the variation of
this maximum value with $\mt$ are shown in the three soft
SUSY-breaking scenarios. No theoretical uncertainties are included.
See \citeres{asbs2,mhiggsWN,asbs}.
}
\label{tab:ASBSmh}
\end{table}

These results have an interesting consequence for the Higgs search at
the Tevatron. The Tevatron has the potential to exclude a SM-like
Higgs boson with a mass of $\MH^{\rm SM} \lsim 130 \gev$ with an
integrated luminosity of 4--8~fb$^{-1}$~\cite{higgstev} per experiment
(and it will
furthermore reduce the experimental error on $\mt$). Since the coupling 
of the lightest
$\cp$-even Higgs boson to gauge bosons is close to 
the SM value for essentially all the parameter space of the three
soft SUSY-breaking scenarios~\cite{asbs}, the Tevatron should either
observe an excess of Higgs-like events over the background expectation
or rule out the mSUGRA, the mGMSB and the mAMSB scenarios.


\section{EWPO in mSUGRA}

In this section we review the prediction 
for $\MW$, $\sweff$, the lightest Higgs boson mass, 
the anomalous magnetic moment of the muon,
$a_\mu \equiv (g-2)_\mu/2$, and $\br(b \to s \ga)$
within the mSUGRA scenario, taking into account constraints on the cold
dark matter (CDM) 
relic density from WMAP and other cosmological data~\cite{cdm}. More details
can be found in~\citere{ehow3}.
The results have been obtained by scanning 
the universal soft supersymmetry-breaking gaugino mass $m_{1/2}$ and 
scalar mass $m_0$ for different representative values of $\tb$ and the
trilinear soft supersymmetry-breaking parameter $A_0$. The sign of
the supersymmetric Higgs parameter $\mu$ has been chosen to be
positive. 

We require the cosmological relic density $\ohsq$ due to the neutralino 
LSP to fall into the range
\BE
0.094 < \ohsq < 0.124~.
\end{equation}
Lower values of $\ohsq$ would be allowed if not all the
cosmological dark matter is composed of neutralinos. However, larger
values of $\ohsq$ are excluded by cosmology.
The CDM constraints have the effect within the mSUGRA scenario,
assuming that the dark matter consists largely of neutralinos, of
restricting $m_0$ to very narrow allowed strips for any specific choice
of $A_0$, $\tb$ and the sign of $\mu$~\cite{WMAPstrips,wmapothers}. 
Thus, the dimensionality of the
mSUGRA scenario is effectively reduced, and one may explore SUSY
phenomenology along these ``WMAP strips''.
We furthermore take into account the constraints on the parameter space 
from the direct search for supersymmetric particles~\cite{pdg} and Higgs 
bosons~\cite{LEPHiggsearly,LEPHiggsSM}.

For $\tb$ two values have been chosen, $\tb = 10, 50$, representing
values in the
lower and the upper part of the (experimentally and theoretically)
allowed parameter space. For the GUT-scale
parameter $A_0$ five different values have been investigated (below also
a scan over $A_0$ is performed), 
$A_0 = (-2,\, -1,\, 0,\, 1,\, 2) \times m_{1/2}$, in order to cover the
allowed parameter space. The top-quark mass has been fixed to 
$\mt = 178 \gev$. Since the results are analyzed along the WMAP strips,
they are given as a function of $m_{1/2}$. The corresponding $m_0$ values
(for fixed $A_0$ and $\tb$) follow from the CDM constraint.
The non-excluded values for $m_{1/2}$ start at around 
$m_{1/2} \approx 200 \gev$ for both values of $\tb$. While for $\tb = 10$
$m_{1/2}$ is restricted by the CDM constraint
to be $m_{1/2} \lsim 900 \gev$, for $\tb = 50$
the allowed values exceed $m_{1/2} \gsim 1500 \gev$.

\begin{figure}[htb!]
\begin{center}
\includegraphics[width=.48\textwidth]{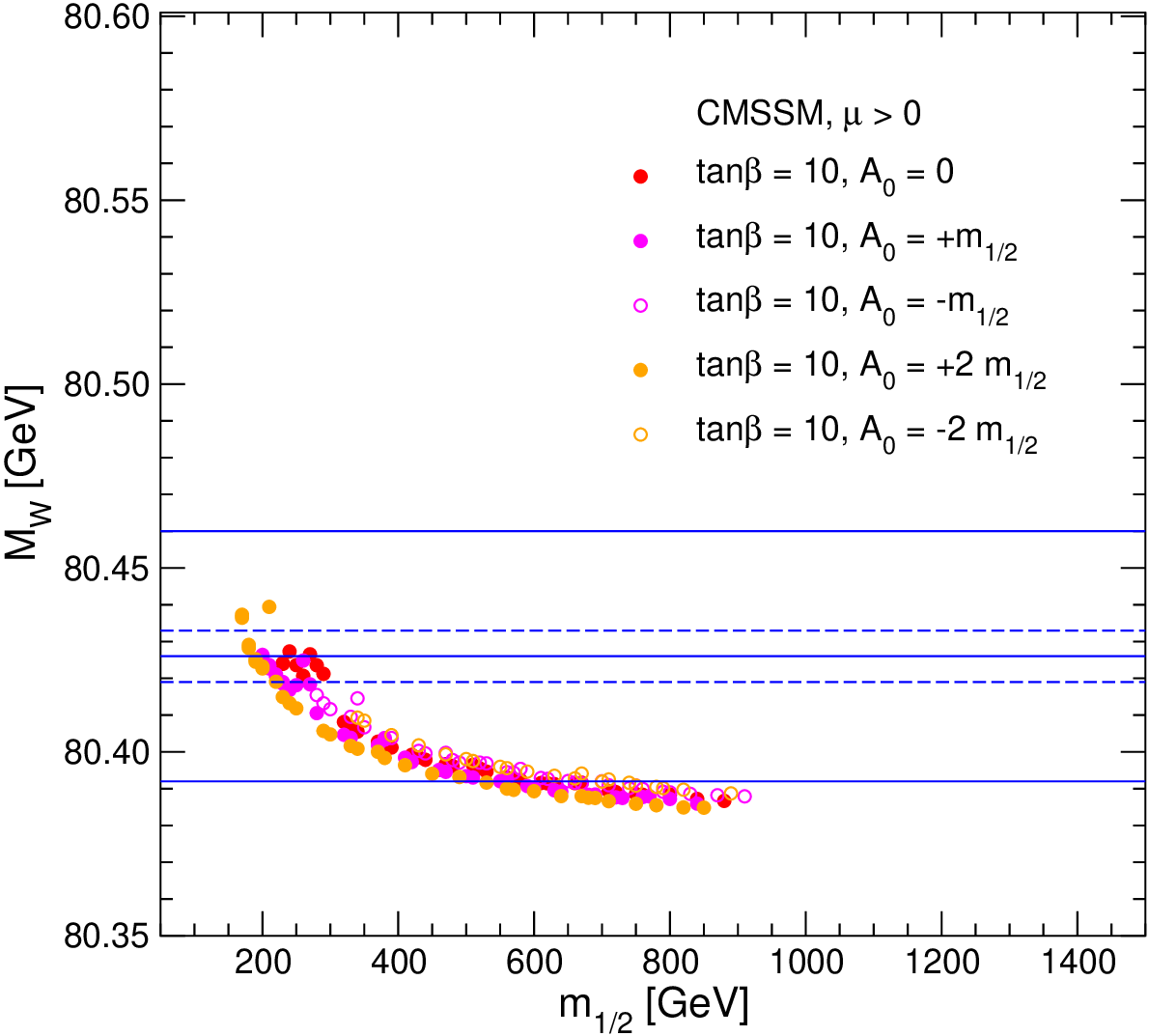}
\includegraphics[width=.48\textwidth]{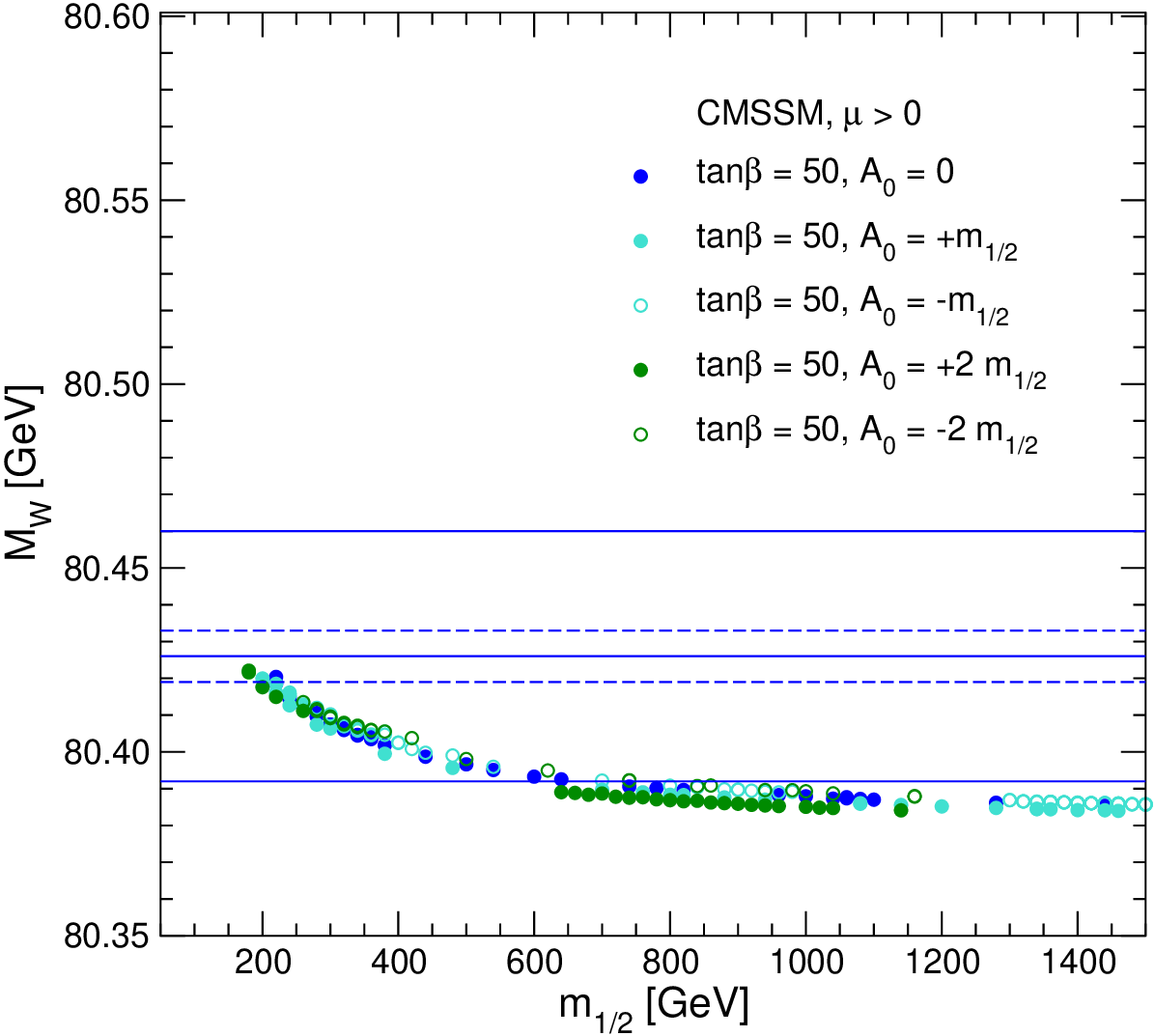}
\caption{%
The mSUGRA prediction for $\MW$ as a function of $m_{1/2}$ along the 
WMAP strips for (a) $\tb = 
10$ and (b) $\tb = 50$ for various $A_0$ values \cite{ehow3}. 
In each panel, the 
centre (solid) line is the present central
experimental value, and the (solid) outer lines show the current $\pm
1$-$\sigma$ range. The dashed lines correspond to the anticipated GigaZ
accuracy, assuming the same central value.
}
\label{fig:MW}
\end{center}
\end{figure}

\smallskip
We start with the prediction for $\MW$.
The evaluation is based on the corrections described in
\refse{sec:evalMW}. 
We display in \reffi{fig:MW} the mSUGRA prediction for $\MW$ and
compare it with the present measurement (solid lines) and a possible
future determination with GigaZ (dashed lines).
Panel (a) shows the values of $\MW$ obtained with $\tb = 10$ and 
$|A_0| \le 2$, and panel (b) shows the same for $\tb = 50$. It is striking 
that the present central value of $\MW$ (for both values of $\tb$)
favours low values of $m_{1/2} \sim 200$--$300$~GeV, though
values as large as 800~GeV are allowed at the 1-$\sigma$ level, and
essentially all values of $m_{1/2}$ are allowed at the $90 \%$ confidence
level. 
The GigaZ determination of $\MW$ might be able to determine
indirectly a low value of $m_{1/2}$ with
an accuracy of $\pm 50 \gev$, but even the GigaZ precision would still be
insufficient to determine $m_{1/2}$ accurately if $m_{1/2} \gsim 600 \gev$
(in accordance with the discussion in \refse{sec:delrhoasbs}).

\begin{figure}[htb!]
\vspace{1em}
\begin{center}
\includegraphics[width=.48\textwidth]{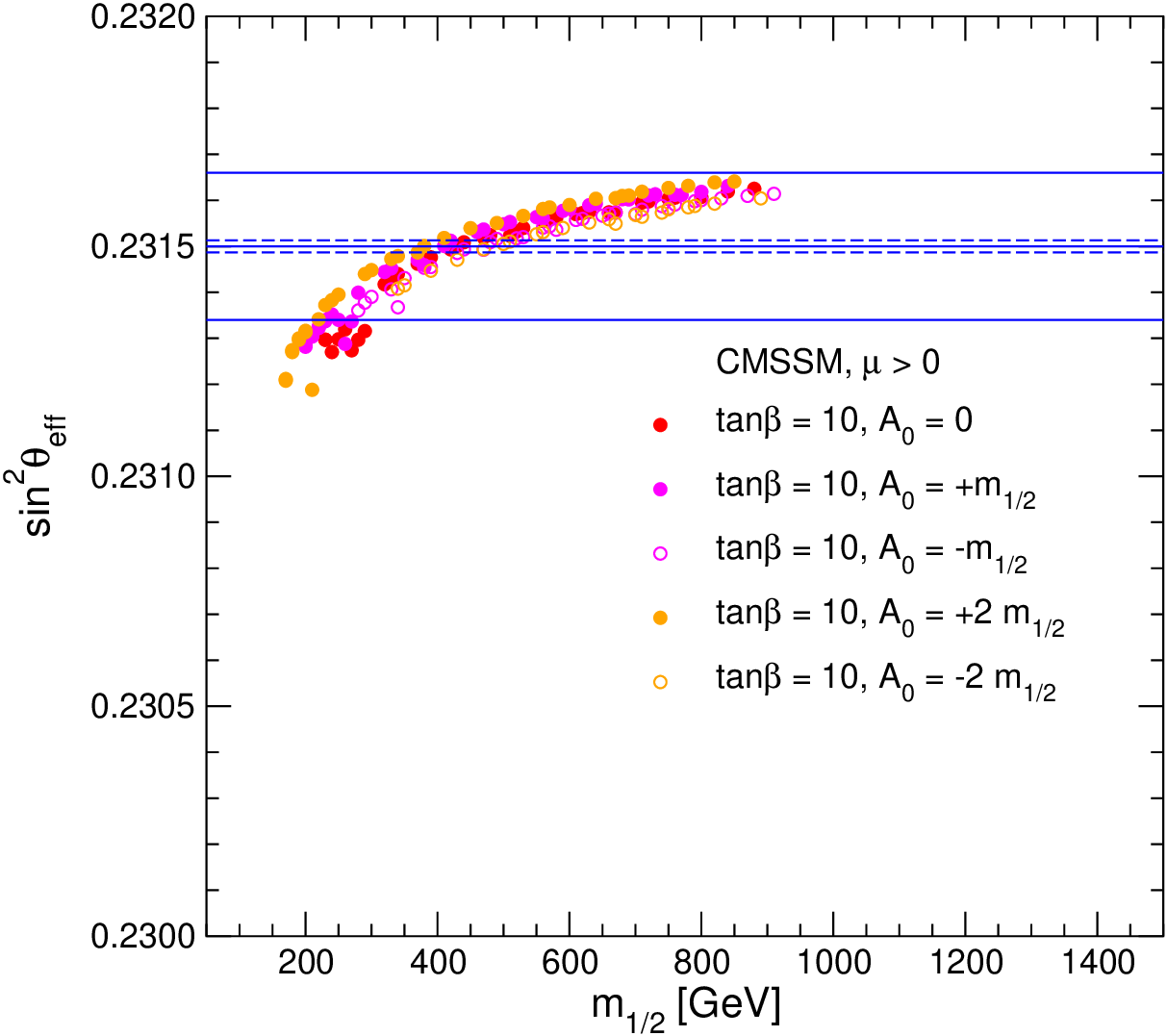}
\includegraphics[width=.48\textwidth]{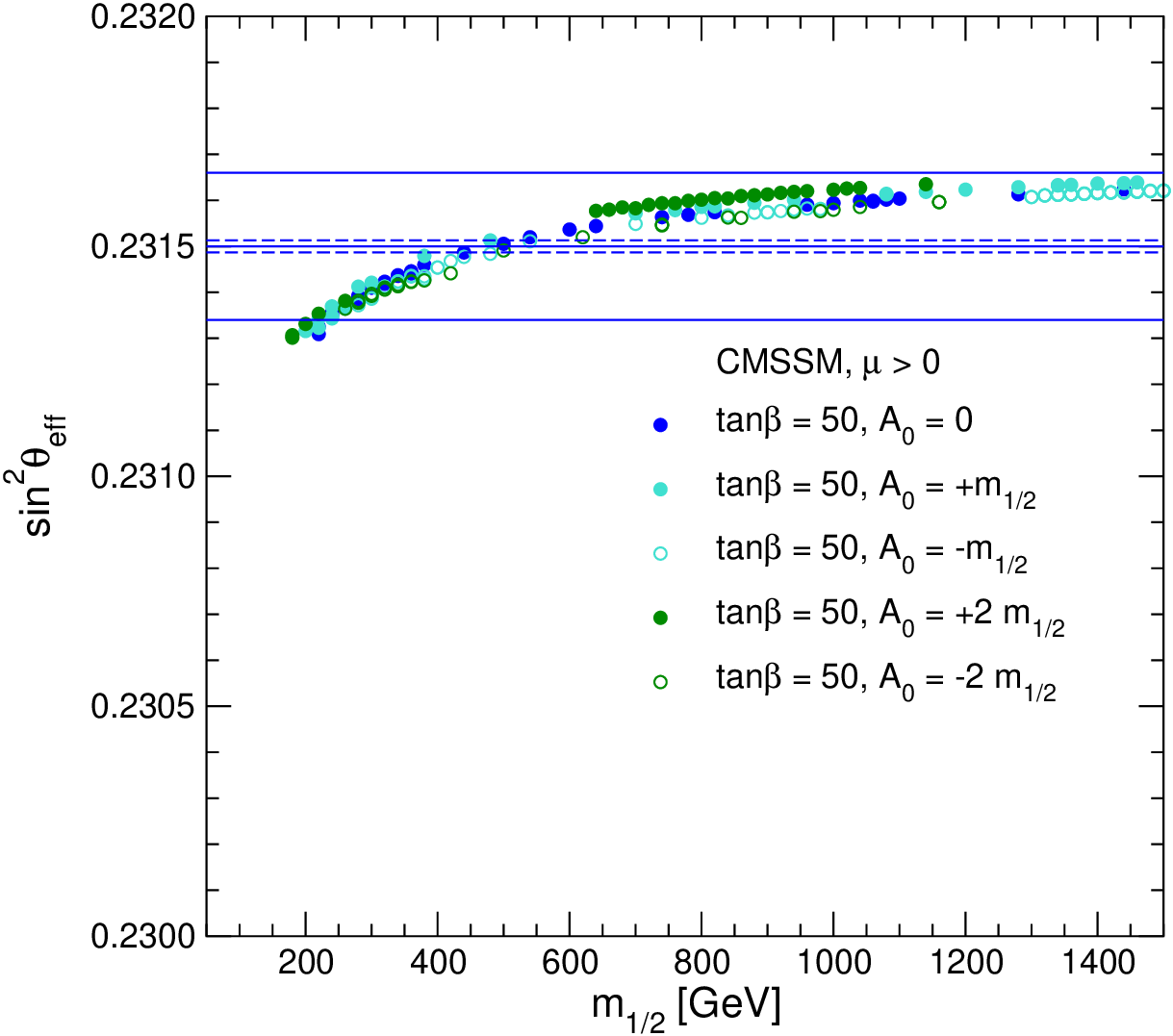}
\caption{%
The mSUGRA prediction for $\sweff$ as a function of $m_{1/2}$ along the 
WMAP strips for (a) 
$\tb = 10$ and (b) $\tb = 50$ for various $A_0$ values \cite{ehow3}.
In each panel, 
the centre (solid) line is the present central
experimental value, and the (solid) outer lines show the current $\pm
1$-$\sigma$ range. The dashed lines correspond to the anticipated GigaZ
accuracy, assuming the same central value.
}
\label{fig:SW}
\end{center}
\end{figure}

\smallskip
The situtation is similar for the prediction of $\sweff$ shown in
\reffi{fig:SW}. The results are based on the corrections
described in \refse{sec:Zobs} and are given for the same values of
$A_0$ and $\tb$ as in \reffi{fig:MW}. 
As in the case of $\MW$, low values of $m_{1/2}$ are also
favoured by $\sweff$. The present central value prefers 
$m_{1/2} = 300$--$500 \gev$, but the 1-$\sigma$ range extends beyond 1500~GeV
(depending on $A_0$), 
and all values of $m_{1/2}$ are allowed at the $90 \%$ confidence level. 
The GigaZ precision on $\sweff$ would 
be able to determine $m_{1/2}$ indirectly with even greater accuracy 
than $\MW$ at low $m_{1/2}$, but would also be insufficient if 
$m_{1/2} \gsim 700 \gev$.

\smallskip
Next the prediction of $\amu$ within mSUGRA is analyzed. 
The evaluation is based on the full one-loop result~\cite{g-2MSSMf1l},
the corresponding QED two-loop corrections~\cite{g-2MSSMlog2l} and the
two-loop corrections from the closed SM fermion/sfermion
loops~\cite{g-2FSf}. The very recent two-loop corrections of
\citere{g-2CNH} have been included via an approximation formula. 
For older evaluations of $\amu$
within mSUGRA (mostly based on the full one-loop result and the
corresponding QED corrections), see
\citeres{g-2appl1,g-2appl2,g-2scan1L,recentfit}.

\begin{figure}[bht!]
\begin{center}
\epsfig{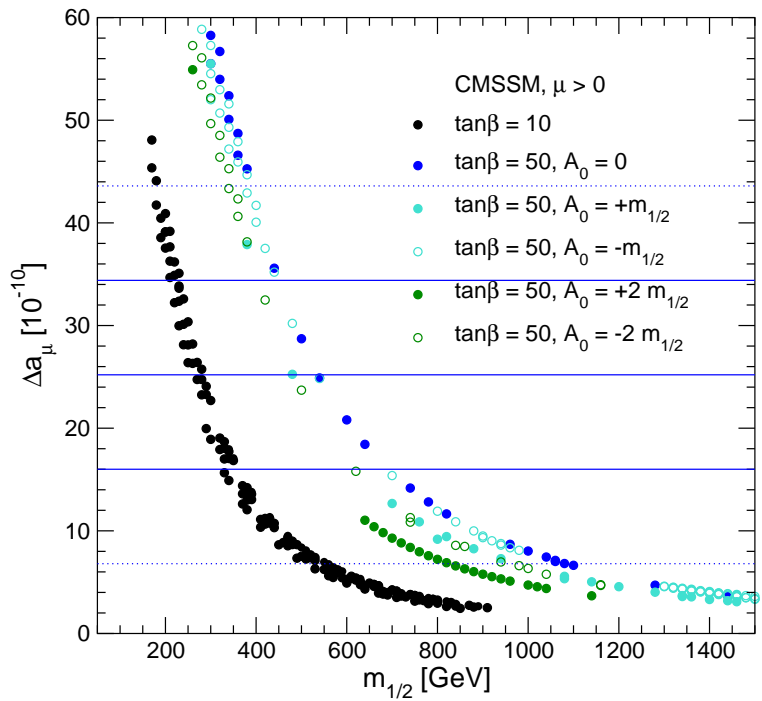}
\caption{%
The mSUGRA prediction for $\De \amu$ as a function of $m_{1/2}$ along 
the WMAP strips for $\tb = 
10, 50$ and different $A_0$ values \cite{ehow3}.
The central (solid) line is the 
central value of the present discrepancy between experiment and the SM 
value evaluated using $e^+ e^-$ data, and the other solid (dotted)
lines show the current $\pm 1(2)$-$\sigma$ ranges, see \refeq{deviationfinal}.
}
\label{fig:AMU}
\end{center}
\vspace{-2em}
\end{figure}

As seen in \reffi{fig:AMU}, the mSUGRA prediction for $\amu$ is almost 
independent of $A_0$ for $\tb = 10$, but substantial variations are possible 
for $\tb = 50$, except at very large $m_{1/2}$. In the case $\tb = 10$, 
$m_{1/2} \sim 200$--400~GeV is again favoured at the $\pm 1$-$\sigma$ 
level, but this preferred range shifts up to 400 to 800~GeV if $\tb = 50$,
depending on the value of $A_0$. 
For the two $\tb$ values the requirement of
agreement of the mSUGRA prediction with the experimental data at the
95\% C.L.\ restricts $m_{1/2}$ to 
\BEA
\tb = 10 ~:~~~ 200 \gev &\lsim~ m_{1/2}~ \lsim& 600 \gev~, \\
\tb = 50 ~:~~~ 350 \gev &\lsim~ m_{1/2}~ \lsim& 1100 \gev~.
\EEA

\smallskip
Now we turn to the decay $b \to s \ga$. 
Since this decay occurs at the loop level in the SM, the MSSM 
contribution might be of similar magnitude. The most up-to-date
theoretical estimate of the SM contribution to the branching ratio
is~\cite{bsgSM}
\BE
\br( b \to s \ga ) = (3.70 \pm 0.30) \times 10^{-4},
\label{bsga}
\end{equation}
where the calculations have been carried out completely to NLO in the 
\msbar\ renormalization scheme, and the error is dominated by 
higher-order QCD uncertainties. However, the error estimate for 
$\br(b \to s \ga)$ is still under debate, see e.g.~\citere{bsgneubert}.
The MSSM evaluation shown below is based on \citeres{bsgSM,bsgMSSM}.

\begin{figure}[htb!]
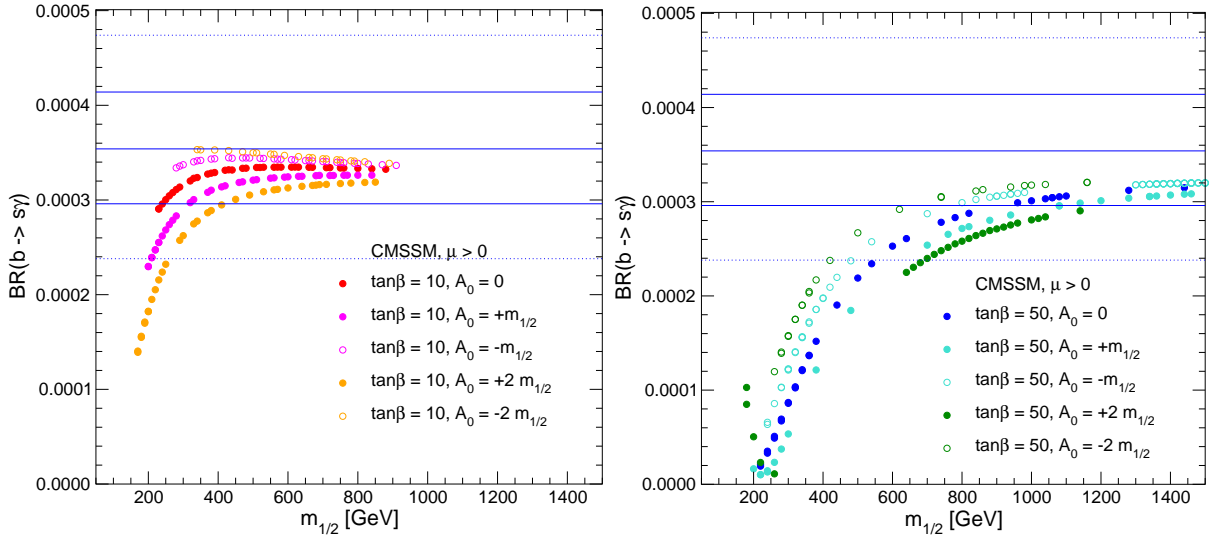

\begin{center}
\includegraphics[width=.48\textwidth]{figs2/ehow.BSG11a.cl.eps}
\includegraphics[width=.48\textwidth]{figs2/ehow.BSG11b.cl.eps}
\caption{%
The mSUGRA predictions for $\br(b \to s \ga)$ as a function of $m_{1/2}$ 
along the WMAP strips for 
(a) $\tb = 10$ and (b) $\tb = 50$ and various choices of $A_0$. The 
uncertainty shown combines linearly the current experimental 
error and the present theoretical uncertainty in the SM prediction.
The central (solid) line indicates the current experimental
central value, and the other solid (dotted)
lines show the current $\pm 1(2)$-$\sigma$ ranges
\cite{ehow3}.
}
\label{fig:BSG}
\end{center}
\end{figure}

For comparison, the present experimental 
value estimated by the Heavy Flavour Averaging Group (HFAG)
is~\cite{bsgexp}
\BE
\br( b \to s \ga ) = (3.54^{+ 0.30}_{- 0.28}) \times 10^{-4},
\label{bsgaexp}
\end{equation}
where the error includes an uncertainty due to the decay spectrum, as well 
as the statistical error. The very good agreement between \refeq{bsgaexp} 
and the SM prediction \refeq{bsga} imposes important constraints on the 
MSSM. The uncertainty range 
shown in \reffi{fig:BSG} combines linearly the current 
experimental error and the present theoretical uncertainty in the SM 
prediction. Since the mSUGRA corrections are
generally smaller for smaller $\tb$, even values of $m_{1/2}$ as low as
$\sim 200 \gev$ would be allowed at the $90 \%$ confidence level if
$\tb = 10$, whereas $m_{1/2} \gsim 400 \gev$ would be required if 
$\tb = 50$.
These limits are very sensitive to $A_0$, and, assuming that in the
future the experimental and theoretical uncertainty in 
$\br(b \to s \gamma)$ can be reduced by a factor $\sim 3$,
the combination of $\br(b \to s \gamma)$ with 
the other precision observables might be able, in principle,
to constrain $A_0$ significantly.

\begin{figure}[hbt!]
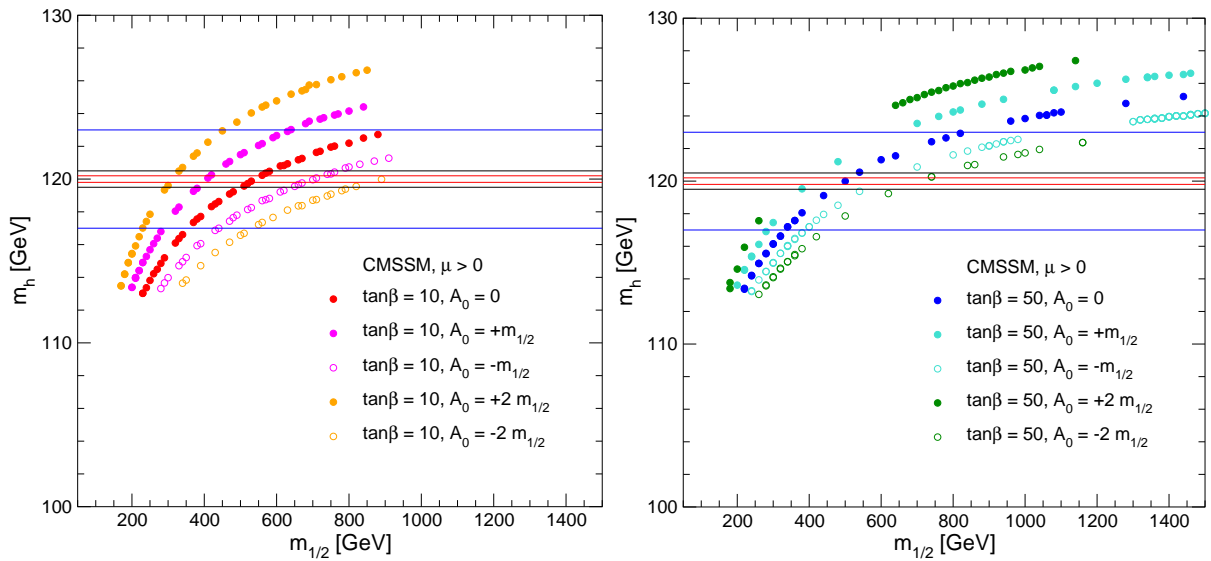

\begin{center}
\includegraphics[width=.48\textwidth]{figs2/ehow.Mh11a.cl.eps}
\includegraphics[width=.48\textwidth]{figs2/ehow.Mh11b.cl.eps}
\caption{%
The mSUGRA predictions for $\mh$ as functions of $m_{1/2}$ with 
(a) $\tb = 10$ and (b) $\tb = 50$ for various $A_0$ \cite{ehow3}. 
A hypothetical experimental value is shown, namely $\mh = 120 \gev$.
We display an optimistic anticipated theory uncertainty of $\pm 0.2 \gev$,
as well as a more realistic theory uncertainty of $\pm 0.5 \gev$ and 
the current theory uncertainty of $\pm 3 \gev$.
}
\label{fig:Mh}
\end{center}
\end{figure}

\smallskip
Finally we present results for the lightest Higgs boson mass in the CDM
allowed strips of the mSUGRA parameter space. 
In \reffi{fig:Mh} we show the results for $\mh$. A hypothetical
measurement at $\mh = 120 \gev$ is shown. Since the experimental error
at the ILC will be smaller than the prospective theory uncertainties
(see \refse{subsec:mhunc}), we
display the effect of the current and future intrinsic uncertainties. In
addition, a more optimistic value of $\De\mh = 200$~MeV is also shown. 
The figure clearly illustrates
the high sensitivity of this electroweak
precision observable to variations of the supersymmetric parameters 
(detailed results for Higgs boson phenomenology in
mSUGRA can be found in \citere{ehow,asbs2,asbs,ehow2}).
The comparison between the measured value of $\mh$ and a precise theory
prediction will allow to set tight constraints on the allowed parameter
space of $m_{1/2}$ and $A_0$.

\section{Fits in mSUGRA}

The results for EWPO presented in the last section have been used to
perform a fit for the mSUGRA parameter space with CDM 
constraints~\cite{ehow3}. We first
review the fit using the currently existing data on $\MW$, $\sweff$,
$\amu$ and $\br(b \to s \ga)$. Secondly, we show the precision that
can be obtained in the future, using improved measurements of the
EWPO and including also the $\mh$ measurement as well as the
measurement of Higgs boson branching ratios. More details can be found
in \citere{ehow3}


\subsection{Present situation}
\label{subsec:fitpresent}

We now investigate the combined sensitivity of the four low-energy
observables for which experimental measurements exist at present, namely
$\MW$, $\sweff$, $(g-2)_\mu$ and $\br(b \to s \ga)$. 
We begin with an analysis of the sensitivity to $m_{1/2}$ moving 
along the WMAP strips with fixed values of $A_0$ and $\tb$. 
The experimental uncertainties, the intrinsic errors from unknown
higher-order corrections and the parametric uncertainties have been
added quadratically, except for $\br(b \to s \ga)$, where they have
been added linearly. Assuming that the four observables are
uncorrelated, a $\chi^2$ fit has been performed with
\BE
\chi^2 \equiv \sum_{n=1}^{N} \KL \frac{R_n^{\rm exp} - R_n^{\rm theo}}
                                 {\si_n} \KR^2~.
\end{equation}
Here $R_n^{\rm exp}$ denotes the experimental central value of the
$n$th observable, so that $N = 4$ for the set of observables included in
this fit,
$R_n^{\rm theo}$ is the corresponding mSUGRA prediction and $\si_n$
denotes the combined error, as specified above.

\begin{figure}[thb!]
\begin{center}
\epsfig{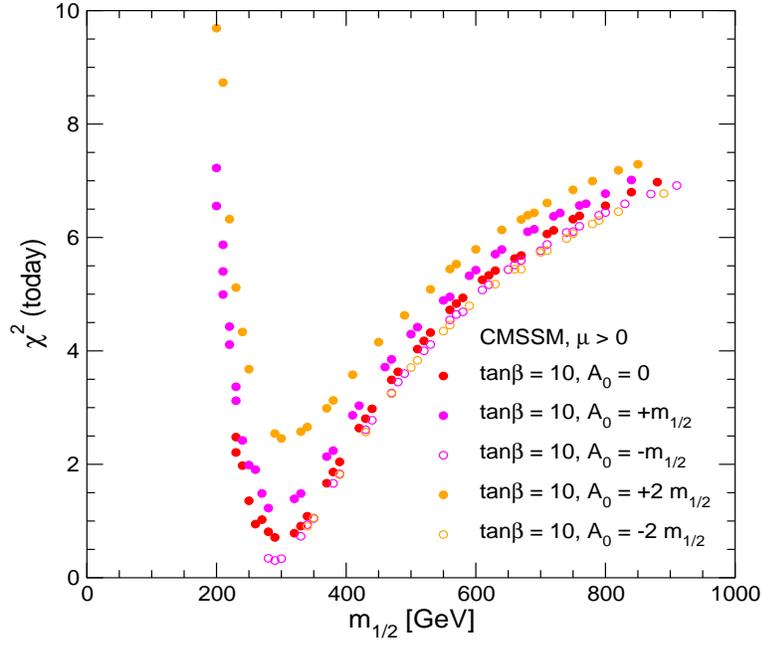}\\[4em]
\epsfig{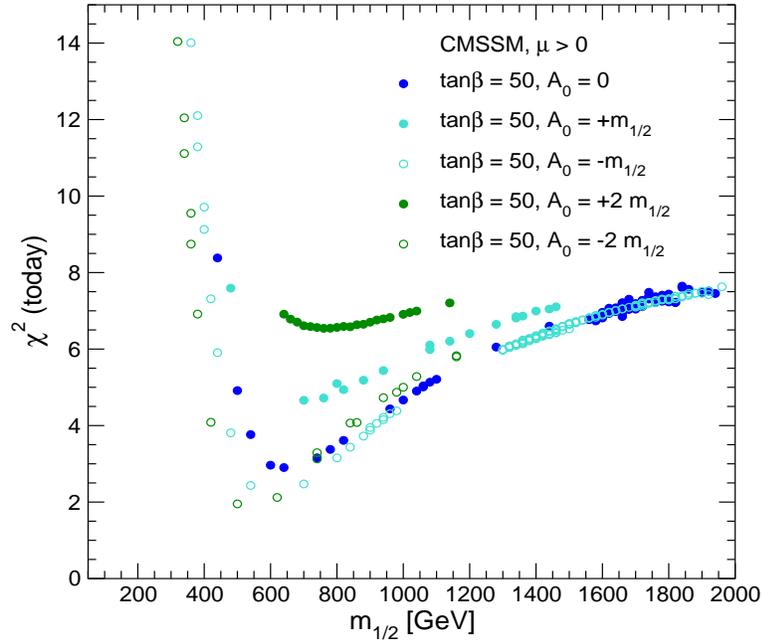}
\caption{%
The results of $\chi^2$ fits based on the current experimental
results for the precision observables $\MW$, $\sweff$, $(g-2)_\mu$ and
$\br(b \to s \ga)$ are shown as functions of $m_{1/2}$ in the mSUGRA
parameter space with CDM constraints for different values of
$A_0$ \cite{ehow3}.
The upper plot shows the results for $\tb = 10$, and the lower plot
shows the case $\tb = 50$.
}
\label{fig:CHI}
\end{center}
\end{figure}

The results are shown in \reffi{fig:CHI} for $\tb = 10$ and $\tb = 50$.  
They indicate that, already at the present level of experimental
accuracies, the electroweak precision observables combined with the WMAP
constraint provide a sensitive probe of the mSUGRA scenario, yielding
interesting 
information about its parameter space. For $\tb = 10$, mSUGRA provides a
very good description of the data, resulting in a remarkably small minimum
$\chi^2$ value. The fit shows a clear preference for relatively small
values of $m_{1/2}$, with a best-fit value of about $m_{1/2} = 300 \gev$.
The best fit is obtained for $A_0 \leq 0$, while positive values of $A_0$
result in a somewhat lower fit quality. 
The fit yields an upper bound on $m_{1/2}$ of about 600~GeV at the
90\%~C.L.\ (corresponding to $\Delta \chi^2 \le 4.61$).
The mass spectrum favored at the 90\%~C.L.\ contains many light states
that should be accessible at the LHC and the ILC, offering good
prospects of the direct detection of SUSY.

For $\tb = 50$ the overall fit quality is worse than for $\tb = 10$, and
the sensitivity to $m_{1/2}$ from the precision observables is lower. This
is related to the fact that, whereas $\MW$ and $\sweff$ prefer small values
of $m_{1/2}$ also for $\tb = 50$, as seen in \reffis{fig:MW} and
\ref{fig:SW}, the CMSSM predictions for $(g-2)_{\mu}$ and $\br(b \to s \ga)$ 
for high $\tb$ are in better agreement with the data for larger
$m_{1/2}$ values, as seen in \reffis{fig:AMU} and \ref{fig:BSG}.  Also in
this case the best fit is obtained for negative values of $A_0$, but the
preferred values for $m_{1/2}$ are 200--300~GeV higher than for $\tb = 10$.
The mass spectrum favored at the 90\%~C.L.\ is heavier than for $\tb = 10$.
However, still several SUSY particles should be accessible at the
ILC. Since colored SUSY particles should be within the kinematic 
reach of the LHC, also in this case there are good
prospects of the direct detection of SUSY.

\begin{figure}[th!]
\begin{center}
\epsfig{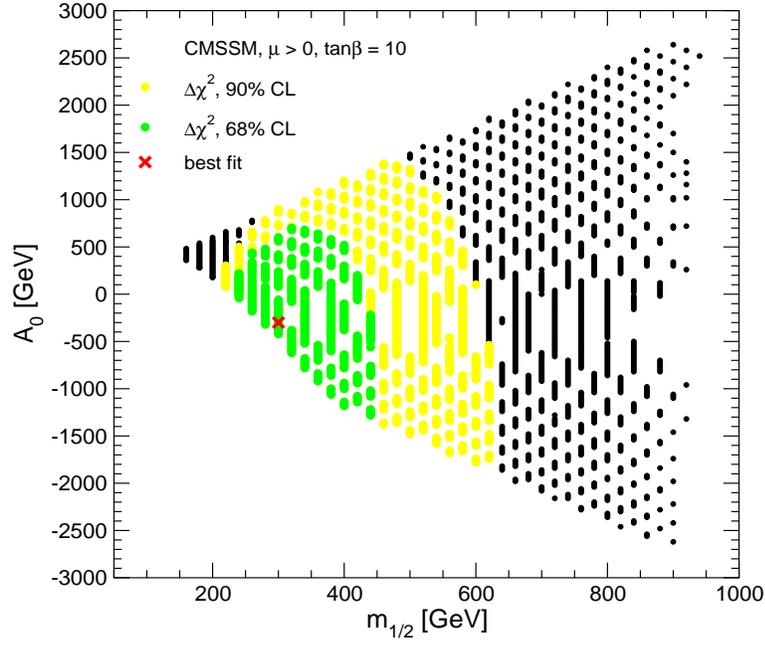}\\[4em]
\epsfig{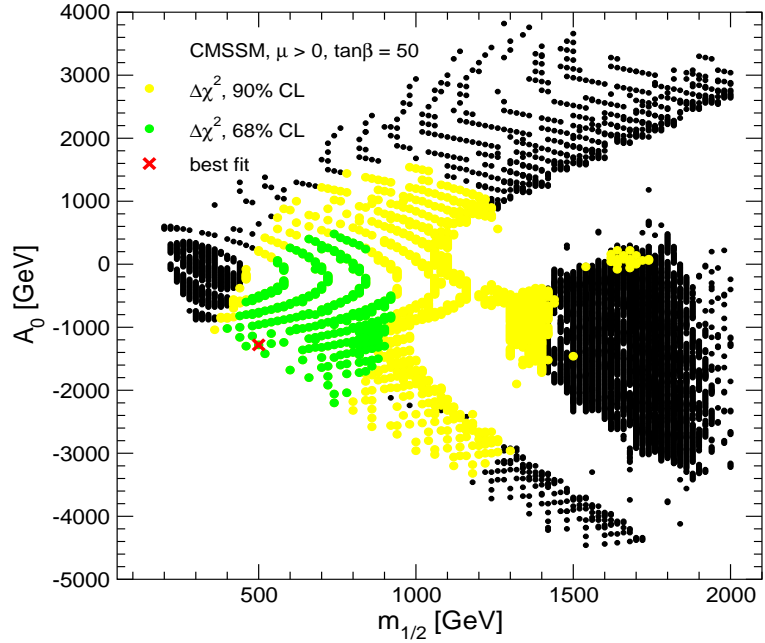}
\caption{%
The results of $\chi^2$ fits for $\tb = 10$ (upper plot) and $\tb = 50$
(lower plot) based on the current experimental
results for the precision observables $\MW$, $\sweff$, $(g-2)_\mu$ and
$\br(b \to s \ga)$ are shown in the \plane{m_{1/2}}{A_0}s of the mSUGRA
scenario with the WMAP constraint \cite{ehow3}.
The best-fit points are indicated, and 
the coloured regions correspond to the 68\% and 90\% C.L.\ regions,
respectively.
}
\label{fig:scancurrent}
\end{center}
\end{figure}

We now turn to the results obtained from a scan over the $m_{1/2}$--$A_0$
parameter plane. 
\reffi{fig:scancurrent} shows the CDM-allowed regions in the
\plane{m_{1/2}}{A_0} for $\tb = 10$ and $\tb = 50$. 
The current best-fit values
obtained via $\chi^2$ fits for $\tb = 10$ and $\tb = 50$ are indicated. The
coloured regions around the best-fit values correspond to the 68\% and 90\%
C.L.\ regions (corresponding to $\De \chi^2 \le 2.30, 4.61$,
respectively). 

For $\tb = 10$ (upper plot of \reffi{fig:scancurrent}),
the precision data yield sensitive constraints on the available
parameter space for $m_{1/2}$ within the WMAP-allowed region. The
precision data are less sensitive to $A_0$.
The 90\% C.L.\ region contains all the WMAP-allowed $A_0$ values in
this region of $m_{1/2}$ values. As expected from the discussion above,
the best fit is obtained for negative $A_0$ and relatively small values
of $m_{1/2}$. At the 68\% C.L., the fit
yields an upper bound on $m_{1/2}$ of about 450~GeV. This bound is
weakened to about 600~GeV at the 90\% C.L.%
\footnote{
A preference for relatively small values of $m_{1/2}$ within the mSUGRA
has also been noticed in \citere{deBoer:2003xm}, where only $(g-2)_\mu$ and
$\br(b \to s \ga)$ had been analyzed.}

As discussed above, the overall fit quality is worse for $\tb = 50$, and
the sensitivity to $m_{1/2}$ is less pronounced. This is demonstrated in
the lower plot of
\reffi{fig:scancurrent}, which shows the result of the fit in the 
\plane{m_{1/2}}{A_0} for $\tb = 50$. The best fit is obtained for
$m_{1/2} \approx 500 \gev$ and negative $A_0$. The upper bound on
$m_{1/2}$ increases to nearly 1~TeV at the 68\% C.L.


\subsection{Future expectations}
\label{subsec:fitfuture}

\begin{figure}[htb!]
\mbox{}\vspace{-1cm}
\begin{center}
\epsfig{figure=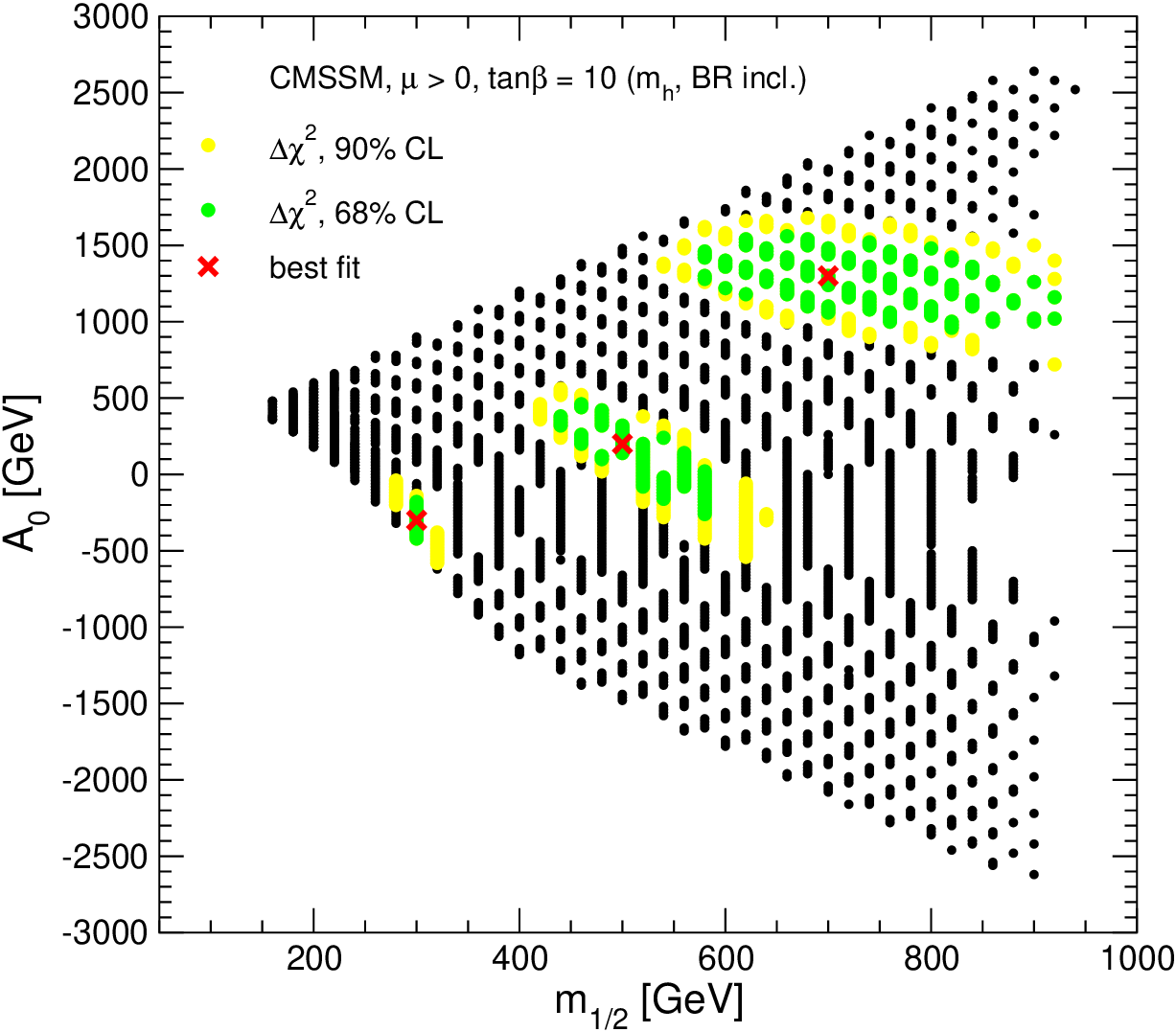,width=12cm,height=9.2cm}\\[2em]
\epsfig{figure=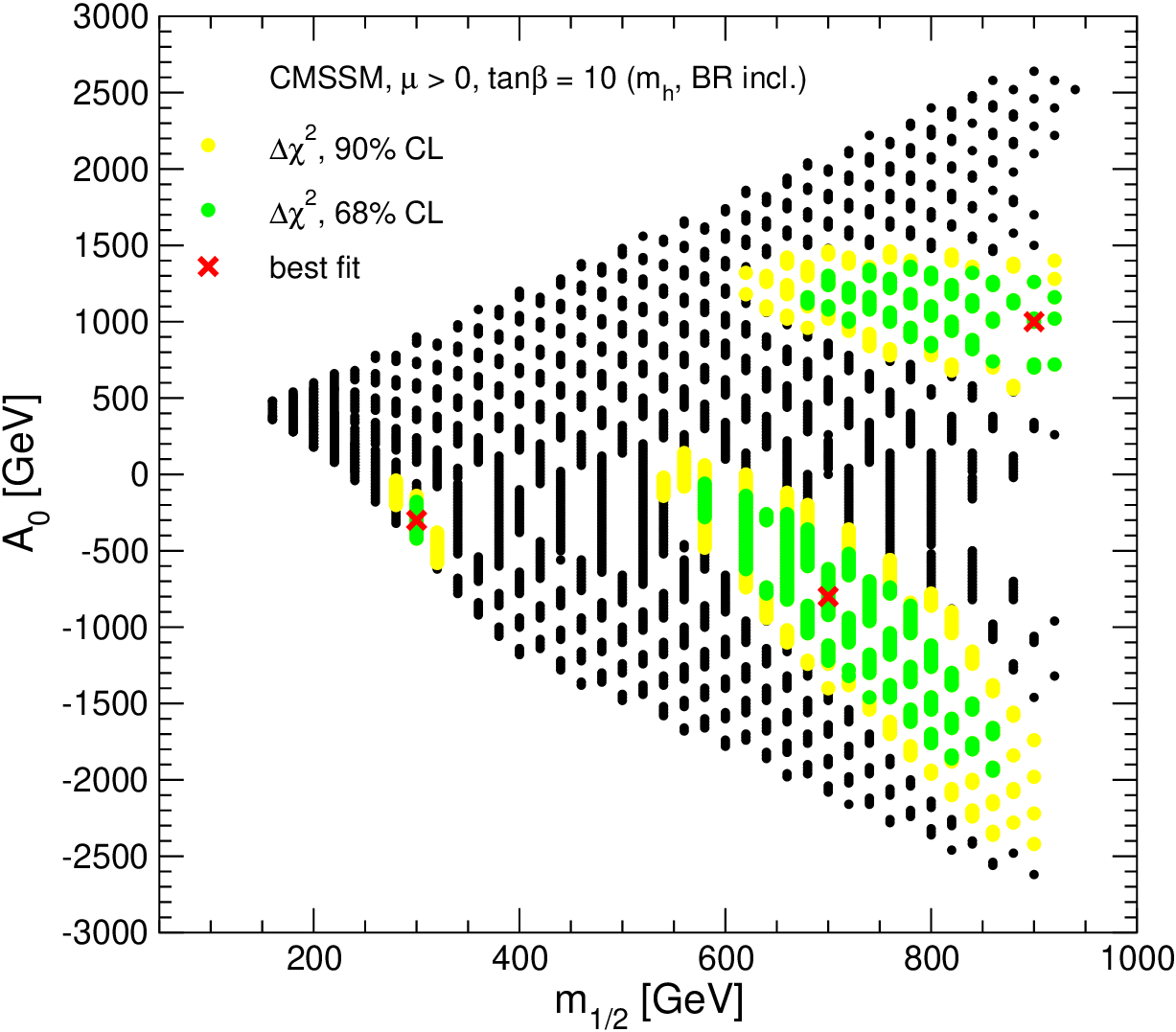,width=12cm,height=9.2cm}
\caption{%
The results of a $\chi^2$ fit based on the prospective experimental
accuracies for the precision observables $\MW$, $\sweff$, $(g-2)_\mu$,
$\br(b \to s \ga)$, $\mh$ and Higgs branching ratios at the ILC
are shown in the \plane{m_{1/2}}{A_0} of the mSUGRA
with WMAP constraints for $\tb = 10$ \cite{ehow3}.
In both plots the 
WMAP-allowed region and the best-fit point 
according to the current situation (see \reffi{fig:scancurrent}) are
indicated. In both plots two further hypothetical future `best-fit' values 
have been chosen for illustration.
The coloured regions correspond to the 68\% and 90\% C.L.\ regions
according to the ILC accuracies.
}
\label{fig:m12A0708}
\end{center}
\vspace{-0.8cm}
\end{figure}

\begin{figure}[htb!]
\mbox{}\vspace{-1cm}
\begin{center}
\epsfig{figure=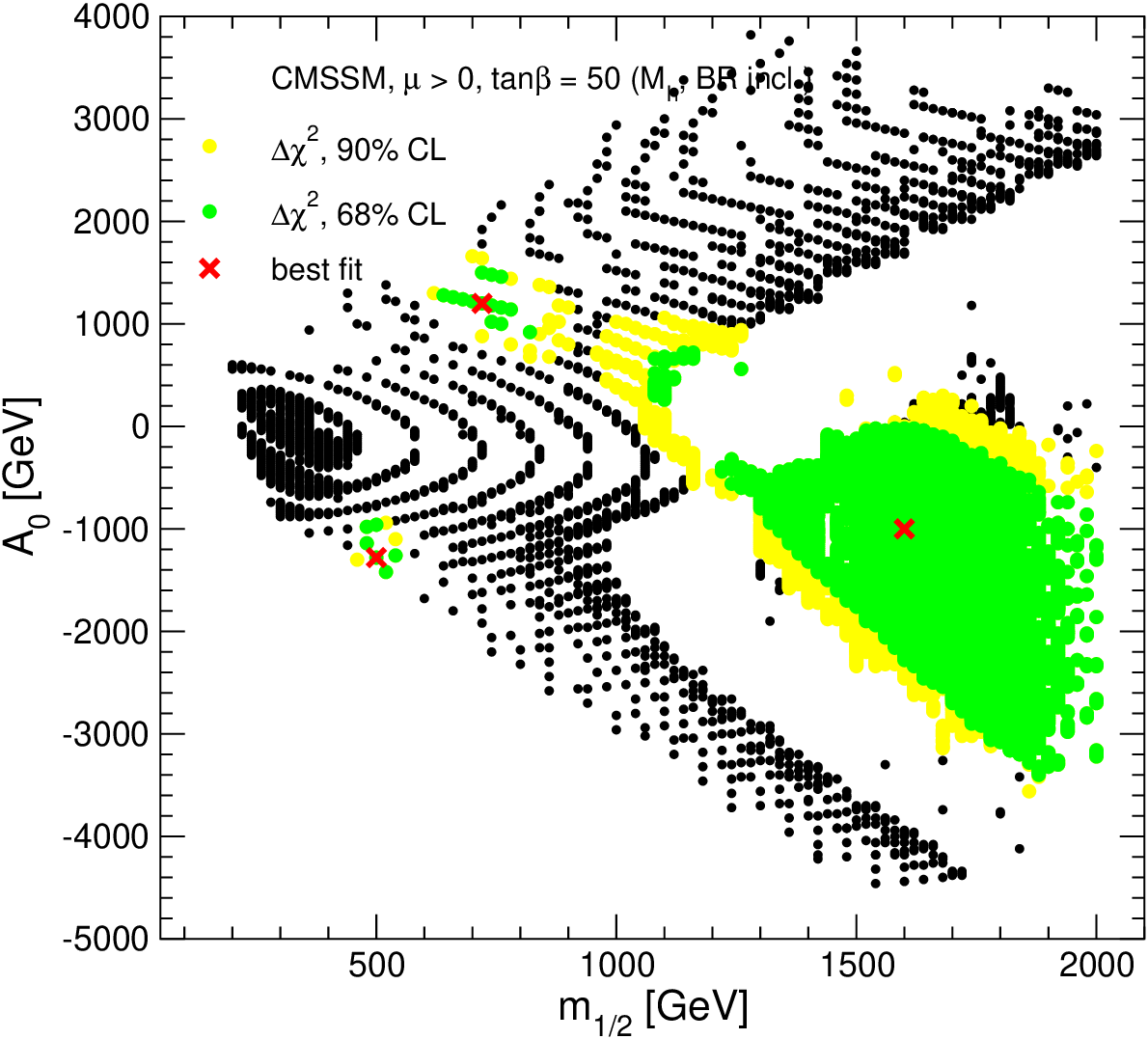,width=12cm,height=9.2cm}\\[2em]
\epsfig{figure=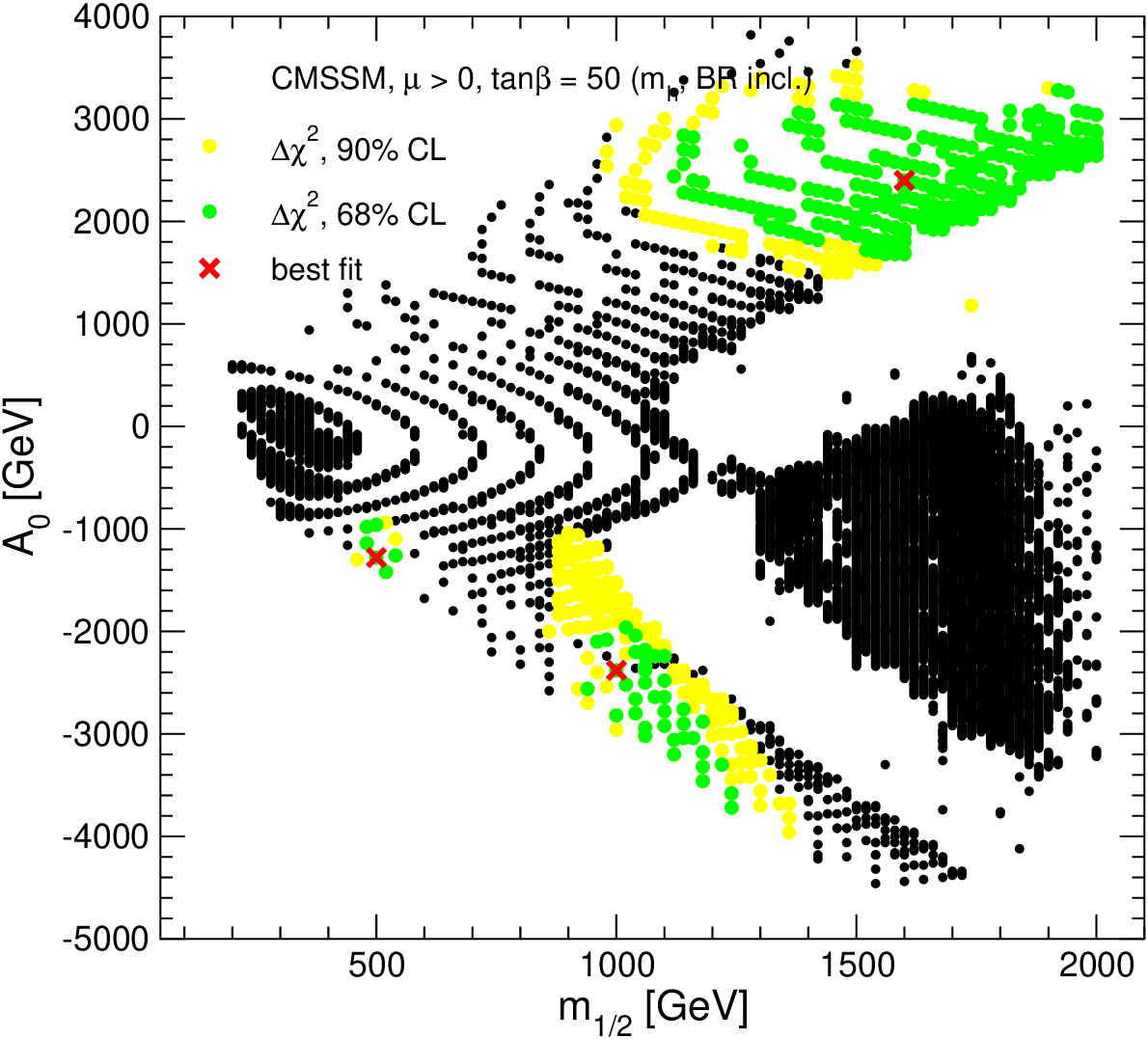,width=12cm,height=9.2cm}
\caption{%
The results of a $\chi^2$ fit based on the prospective experimental
accuracies for the precision observables $\MW$, $\sweff$, $(g-2)_\mu$,
$\br(b \to s \ga)$, $\mh$ and Higgs branching ratios at the ILC
are shown in the \plane{m_{1/2}}{A_0} of the mSUGRA scenario 
with WMAP constraints for $\tb = 50$ \cite{ehow3}. 
In both plots the 
WMAP-allowed region and the best-fit point for $\tb = 50$
according to the current situation (see \reffi{fig:scancurrent}) are
indicated. In both plots two further hypothetical future `best-fit' values 
have been chosen for illustration.
The coloured regions correspond to the 68\% and 90\% C.L.\ regions
according to the ILC accuracies.
}
\label{fig:m12A1718}
\end{center}
\vspace{-0.8cm}
\end{figure}

We now investigate the combined sensitivity of the precision
observables $\MW$, $\sweff$, \mbox{$(g-2)_\mu$,} $\br(b \to s \ga)$,
$\mh$ and the ratio $\br(\hbb) / \br(\hWW)$
in the \plane{m_{1/2}}{A_0} of the mSUGRA scenario using ILC (and GigaZ)
accuracies. For $(g-2)_\mu$ we assume a reduction of the error by two,
for $\br(b \to s \ga)$ by a factor of three. At the ILC with 
$\sqrt{s} = 1 \tev$ a measurement of 
$\br(\hbb) / \br(\hWW)$ with an accuracy of $\sim 1.5\%$ can be
envisaged~\cite{barklow}.
\reffi{fig:m12A0708} shows the fit results for $\tb = 10$, while 
\reffi{fig:m12A1718} shows the $\tb = 50$ case.

In each figure we show two plots, where the WMAP-allowed region and the
best-fit point according to the current situation (see
\reffi{fig:scancurrent}) are indicated. In both plots two further
hypothetical future `best-fit' points have been chosen for illustration.
For all the `best-fit' points, the assumed central experimental values of
the observables have been chosen such that they precisely coincide with the
`best-fit' points%
\footnote{
It was checked explicitly that assuming future experimental values of the 
observables with values distributed statistically around the present 
`best-fit' points with the estimated future errors does not degrade 
significantly the qualities of the fits.
}%
. The coloured regions correspond to the 68\% and 90\%
C.L.\ regions around each of the `best-fit' points according
to the ILC accuracies.

The comparison of \reffis{fig:m12A0708}, \ref{fig:m12A1718} with the result
of the current fit, \reffi{fig:scancurrent}, shows that the ILC
experimental precision will lead to a drastic improvement in the
sensitivity to $m_{1/2}$ and $A_0$ from comparing precision data with the
mSUGRA predictions. For the best-fit values of the current fits for $\tb =
10$ and $\tb = 50$, the ILC precision would allow one to narrow down the
allowed mSUGRA parameter space to very small regions in the 
\plane{m_{1/2}}{A_0}.
The comparison of these indirect predictions for $m_{1/2}$ and $A_0$
with the information from the direct detection of supersymmetric particles
would provide a stringent test of the model at the loop level. A
discrepancy could indicate that supersymmetry is realised in a more
complicated way than assumed in mSUGRA.

The additional hypothetical `best-fit' points shown in
\reffis{fig:m12A0708}, \ref{fig:m12A1718} illustrate the indirect
sensitivity to the mSUGRA parameters in scenarios where the precision
observables prefer larger values of $m_{1/2}$. Because of the decoupling
property of supersymmetric theories, the indirect constraints become
weaker for increasing $m_{1/2}$. 

For $\tb = 10$, we have investigated hypothetical `best-fit' values for
$m_{1/2}$ of 500~GeV, 700~GeV (for $A_0 > 0$ and $A_0 < 0$) and 900~GeV.
For $m_{1/2} = 500 \gev$, the 90\% C.L.\ region in the 
\plane{m_{1/2}}{A_0}
is significantly larger than for the current best-fit value of
$m_{1/2} \approx 300 \gev$, but interesting limits can still be set on both
$m_{1/2}$ and $A_0$. For $m_{1/2} = 700 \gev$ and $m_{1/2} = 900 \gev$, the
90\% C.L.\ region extends up to the boundary of the WMAP-allowed parameter
space for $m_{1/2}$. Even for these large values of $m_{1/2}$, however, the
precision observables (in particular the observables in the Higgs sector)
still allow one to constrain $A_0$.

\clearpage
For $\tb = 50$, where the WMAP-allowed region extends up to much higher
values of $m_{1/2}$, we find that for a `best-fit' value of $m_{1/2}$ as
large as 1~TeV the precision data still allow one to establish an upper
bound on $m_{1/2}$ within the CDM-allowed region. This indirect sensitivity
to $m_{1/2}$ could give important hints for supersymmetry searches at
high-energy colliders. For `best-fit' values of $m_{1/2}$ in excess of
1.5~TeV, on the other hand, the indirect effects of heavy sparticles become
so small that they are difficult to resolve even with ILC accuracies.
To conclude, the indirect sensitivity from the measurement of
precision observables at the ILC have a potential even to exceed the
direct search reach of both the LHC and ILC.


\newpage

\chapter{Conclusions}

An overview of the current status of precision tests of supersymmetry
has been given, and future prospects have been discussed. We have mainly 
focused on the $W$~boson mass, $\MW$, the effective leptonic
weak mixing angle, $\sweff$, the anomalous magnetic moment of the muon,
$(g-2)_\mu$, and the lightest $\cp$-even MSSM Higgs boson mass, $\mh$,
but constraints from $b$~physics, direct collider searches and
cosmological data have also been included in the discussion.

Precise experimental data are available for $\MW$, $\sweff$ and
$(g-2)_\mu$, while $\mh$ is expected to become a precision observable if 
a supersymmetric Higgs sector is realised in nature. Confronting the high
experimental precision with the theory predictions provides sensitivity
to quantum corrections of the theory, where the whole structure of the
model enters. This allows to set indirect constraints on the properties 
of particles even if they are
too heavy to be produced directly.
In order to exploit the experimental precision, 
the theoretical predictions for the electroweak precision
observables in supersymmetry (or in other models that are confronted
with the data) should be at least at the same level of accuracy. 
Ideally, the remaining theoretical uncertainties should be so small that
they are negligible compared to the experimental errors. 
Sophisticated higher-order calculations are necessary in order to match
this demand, and a considerable effort will be required for keeping up
with the prospective improvements of the experimental accuracies in
future experiments.

We have briefly discussed the necessary ingredients of higher-order
calculations in supersymmetry, focusing in particular on regularisation
and renormalization, and have pointed out important differences compared
to the case of the SM. The large number of parameters in the MSSM, most 
of which
are not directly related to any particular physical observable, and the 
relations imposed by the underlying symmetry make it quite involved to
formulate a coherent and easily applicable renormalization prescription
for the whole MSSM. Different prescriptions exist in the literature 
for various sectors of the MSSM, but no common standard has emerged yet. 

The current status of the theoretical predictions for the most important
precision observables has been revieved, and estimates of the remaining
theoretical uncertainties from unknown higher-order corrections and from
the experimental errors of the SM input parameters have been given.
The theoretical predictions have then been compared with the current
experimental results (in the case of $\mh$ the MSSM prediction has been
confronted with the exclusion bounds from the Higgs search at LEP). The
resulting constraints on the MSSM parameter space have been analyzed. We
have investigated how well the MSSM describes the data and whether the
data give some preference for the MSSM as compared to the SM. This has
been analyzed both for the unconstrained MSSM and for specific soft
SUSY-breaking scenarios. The mSUGRA scenario, characterised by four
parameters and a sign, can still simultaneously satisfy the constraints
from the electroweak precision data, direct collider searches and the
stringent bounds on cold dark matter in the universe from WMAP and other
cosmological data. It turns out that the mSUGRA scenario with
cosmological constraints in fact yields a very good fit to the data.
The fit results indicate a clear preference for a relatively
light mass scale of the SUSY particles, offering good prospects for
direct SUSY searches at the LHC and at the ILC.

We have investigated future prospects of electroweak precision tests of
supersymmetric models. Anticipated improvements in the experimental 
precision have been discussed in view of the LHC and the ILC, and the 
prospects for a further reduction
of the theoretical uncertainties have been analyzed. Based on these
estimates of future experimental and theoretical precisions, we find
that the sensitivity of the precision tests will improve very
significantly, leading to stringent constraints on the MSSM parameter
space (and on any other conceivable model of new physics).
If supersymmetric particles are discovered at the next generation of
colliders, the combination of information from the direct observation of 
SUSY particles and the indirect information from electroweak precision
observables will allow very powerful tests of the model. This can lead
to a discrimination between the minimal and non-minimal models, a
distinction between different SUSY-breaking scenarios, and indirect
predictions for parameters or particle masses that are not directly
experimentally accessible. These consistency tests at the quantum level
using all available experimental information will be crucial in the
quest to extrapolate the results of the next generation of colliders to
physics at high scales.

\section*{Acknowledgements}
We thank S.~Ambrosanio, A.~Dedes, G.~Degrassi, A.~Djouadi, J.~Ellis,
M.~Frank, P.~Gambino, C.~J\"unger, S.~Kraml, F.~Merz,
K.~Olive, S.~Pe\~naranda, W.~Porod,
H.~Rzehak, P.~Slavich, D.~St\"ockinger, S.~Su and A.~Weber for 
collaboration on results 
presented in this 
report and for useful discussions.
G.W.\ thanks the CERN Theory Division for kind hospitality during the final
stages of preparing this paper.
This work has been supported by the European Community's Human
Potential Programme under contract HPRN-CT-2000-00149 Physics at
Colliders.

\newpage
\addcontentsline{toc}{chapter}{Appendix}
\begin{appendix}

\chapter{Loop integrals}
\label{app:loopint}

In this appendix we present the loop integrals needed for the \twol\
evaluation of the SUSY contributions to the EWPO. $D$ denotes the
space-time dimension and $\de \equiv \frac{1}{2}(4 - D)$. 
%
%
In the following formulas we neglected the terms proportional to 
$\ga_E - \ln(4\pi\,\mu^2)$, which are connected to the divergent
parts. They always cancel for physical observables.

The analytical formulas for $A_0$ and $B_0$ are taken from
\citere{a0b0c0d0}, $T_{134}$ and $T_{234'}$ are taken from
\citere{t134}, the other integrals can be found in
\citere{twoloopint}. The notation for the integrals is as in 
\citere{twocalc}.

\section{$A_0(m)$}
\label{subsec:a0}
\BEA
A_0(m_1) &=& \frac{m_1^2}{\de} \non \\
 && + m_1^2 \KL 1 - \ln(m_1^2) \KR \non \\
 && + \de \; m_1^2 \KL \frac{\pi^2}{12} + \edz \ln^2(m_1^2) 
                     - \ln(m_1^2) + 1 \KR,
\EEA

\vspace{2mm}
\noindent
{\bf special cases:}
\BE
A_0(0) = 0 
\label{a00}
\end{equation}

\noindent{\bf derivatives:}
\BEA
\frac{\dd}{\dd m^2} A_0(m) &=& \frac{D/2 - 1}{m^2} A_0(m) \\
\frac{\dd^2}{\dd (m^2)^2} A_0(m) &=& \frac{D/2 - 1}{m^4} 
                                     \KL \frac{D}{2} - 2 \KR A_0(m)
\EEA


\section{$B_0(p^2, m_1, m_2)$}
\label{sec:b0}
\BEA
B_0(p^2, m_1, m_2) 
 &=& \frac{1}{\de} B_0^{1/\de} + B_0^{\rm fin}(p^2, m_1, m_2) 
                              + \de B_0^{\de}(p^2, m_1, m_2) \non \\
\mbox{with } 
  B_0^{1/\de} &=& 1 \non \\
  B_0^{\rm fin}(p^2, m_1, m_2) &=& 
  - \Big\{ \edz \KL \ln(m_1^2) + \ln(m_2^2) \KR
            - 2 + \frac{m_1^2/m_2^2 - 1}{2 p^2/m_2^2} 
                  \ln\KL \frac{m_1^2}{m_2^2}\KR \non \\
 && \;\;\;\;\;  - \edz \frac{r_1 - r_2}{p^2/m_2^2} 
                   \KL \ln(r_1) - \ln(r_2) \KR \Big\} \\
\EEA
$r_1$ and $r_2$ are the solutions of
\BE 
m_2^2\,r + \frac{m_1^2}{r} = m_1^2 + m_2^2 - p^2. 
\label{r12}
\end{equation}

\vspace{3mm}
\noindent
{\bf special cases:}
\BEA
B_0(0, m_1, m_2) &=& \frac{A_0(m_1) - A_0(m_2)}{m_1^2 - m_2^2} \\
B_0(0, m, m) &=& \frac{D/2 - 1}{m^2} A_0(m) \\
B_0(0, m, 0) &=& \frac{1}{m^2} A_0(m) 
\EEA

\vspace{3mm}
\noindent
{\bf derivatives:}
\BEA
B_0'(q^2,m_1,m_2) &=& \frac{1}{N} [(m_1^2+m_2^2)B_0(q^2,m_1,m_2) - m_1^2
    B_0(0,m_1,m_1) \\ \nonumber
    &-& m_2^2 B_0(0,m_2,m_2) - q^2 -\frac{(m_1^2-m_2^2)^2}{q^2}
    (B_0(q^2,m_1,m_2)-B_0(0,m_1,m_2))] \\ 
    N &=& [q^2-(m_1-m_2)^2][q^2-(m_1+m_2)^2]
\EEA

Special cases:
\BEA
B_0'(q^2,m,m) &=& \frac{1}{q^2-4m^2} \BL \frac{2m^2}{q^2}(B_0(q^2,m,m)-
                  B_0(0,m,m))-1 \BR \\ 
B_0'(q^2,0,m) &=& \frac{1}{(q^2-m^2)^2} \Big[ m^2B_0(q^2,0,m) -m^2B_0(0,m,m)
                  - q^2 \\ 
              & & \hspace{2cm} -\frac{m^4}{q^2}(B_0(q^2,0,m)-B_0(0,0,m)) \Big]
                                                         \\ 
B_0'(q^2,0,0) &=& - \frac{1}{q^2} 
\EEA


\section{$T_{134}$}
\label{sec:t134}

The masses have to fulfil the relation $m_3 > m_1, m_2$.
\BEA
T_{134} &=&
    \frac{1}{2\,\de^2} \KL m_1^2 + m_2^2 + m_3^2 \KR \non \\
 && + \frac{1}{\de} \KKKL \frac{3}{2} \KL m_1^2 + m_2^2 + m_3^2 \KR
                          - m_1^2 \ln(m_1^2) - m_2^2 \ln(m_2^2) 
                          - m_3^2 \ln(m_3^2) \KKKR \non \\
 && + \KL \frac{7}{2} + \frac{\pi^2}{12} \KR
      \KL m_1^2 + m_2^2 + m_3^2 \KR \non \\
 && + m_1^2 \KL \ln^2(m_1^2) - 3 \ln(m_1^2) \KR
    + m_2^2 \KL \ln^2(m_2^2) - 3 \ln(m_2^2) \KR
    + m_3^2 \KL \ln^2(m_3^2) - 3 \ln(m_3^2) \KR \non \\
 && + \frac{1}{4} \KL +m_1^2 - m_2^2 - m_3^2 \KR 
                  \ln^2 \KL \frac{m_2^2}{m_3^2} \KR
    + \frac{1}{4} \KL -m_1^2 + m_2^2 - m_3^2 \KR 
                  \ln^2 \KL \frac{m_1^2}{m_3^2} \KR \non \\
 && + \frac{1}{4} \KL -m_1^2 - m_2^2 + m_3^2 \KR 
                  \ln^2 \KL \frac{m_1^2}{m_2^2} \KR
    + \tilde{\Phi}(m_1^2, m_2^2, m_3^2) 
\EEA
with
\BEA
\label{phischlange}
\tilde{\Phi}(m_1^2, m_2^2, m_3^2) &=& \frac{1}{2} m_3^2 
            \lambda\KL \frac{m_1^2}{m_3^2}, \frac{m_2^2}{m_3^2} \KR \non\\
 & & \Bigg( 2 \ln(\alpha_1(m_1, m_2, m_3)) \ln(\alpha_2(m_1, m_2, m_3))
     - \ln\KL \frac{m_1^2}{m_3^2} \KR \ln\KL \frac{m_2^2}{m_3^2} \KR \non\\
 & & - 2 {\rm Li_2}(\alpha_1(m_1, m_2, m_3)) 
     - 2 {\rm Li_2}(\alpha_2(m_1, m_2, m_3))
     + \frac{\pi^2}{3} \Bigg) \\
 & & \non \\
 \lambda(x, y) &=& \sqrt{1 + x^2 + y^2 - 2x - 2y - 2xy} \\
 \alpha_i(m_1, m_2, m_3) &=& \frac{1}{2}
      \Big(1 - (-1)^i \frac{m_1^2}{m_3^2} + (-1)^i \frac{m_2^2}{m_3^2}
             - \lambda\KL \frac{m_1^2}{m_3^2}, \frac{m_2^2}{m_3^2} \KR \Big).
\EEA

\vspace{3mm}
\noindent
{\bf special cases:}

\BEA
T_{134'}(m_1^2, m_2^2, 0) &=&
    \frac{1}{2\,\de^2} \KL m_1^2 + m_2^2  \KR \non \\
 && + \frac{1}{\de} \KKKL \frac{3}{2} \KL m_1^2 + m_2^2 \KR
                          - m_1^2 \ln(m_1^2) - m_2^2 \ln(m_2^2) 
                          \KKKR \non \\
 && + \KL \frac{7}{2} + \frac{\pi^2}{12} \KR
      \KL m_1^2 + m_2^2 \KR \non \\
 && + m_1^2 \KL \ln^2(m_1^2) - 3 \ln(m_1^2) \KR
    + m_2^2 \KL \ln^2(m_2^2) - 3 \ln(m_2^2) \KR \\
 && + \frac{1}{4} \KL -m_1^2 - m_2^2 \KR 
                  \ln^2 \KL \frac{m_1^2}{m_2^2} \KR
    + \frac{1}{4} \KL +m_1^2 - m_2^2  \KR 
                  \KL \ln^2 (m_2^2) - \ln^2 (m_1^2) \KR \non \\
 && + \KL - \edz (m_2^2 - m_1^2) \ln(m_2^2) 
            \ln \KL \frac{m_1^2}{m_2^2} \KR 
          + (m_2^2 - m_1^2) {\rm Li_2}\KL 1-\frac{m_1^2}{m_2^2} \KR
      \hspace{-1mm}\KR \non \\
 && \non \\
T_{134'}(m^2, m^2, 0) &=&
    \frac{D/2 - 1}{(D - 3) m^2} (A_0(m))^2 \\
 && \non \\
T_{13'4'}(m^2, 0, 0) &=&
    \frac{1}{2\,\de^2} \KL m^2  \KR \non \\
 && + \frac{1}{\de} \KKKL \frac{3}{2} \KL m^2  \KR
                          - m^2 \ln(m^2) \KKKR \non \\
 && + m^2 \KL \frac{7}{2} + \frac{3 \pi^2}{12} 
                + \ln^2 (m^2) - 3 \ln (m^2) \KR
\EEA


\section{$T_{234'}$}
\label{sec:t234p}

Here $p^2$ contains a small imaginari part, 
$\ie, \epsilon > 0$.
\BEA
T_{234'} &=& 
    \frac{1}{2 \de^2} \KL m_1^2 + m_2^2 \KR \non \\
 && + \frac{1}{\de} \KKKL \frac{3}{2} \KL m_1^2 + m_2^2 \KR
                          - m_1^2 \ln(m_1^2) - m_2^2 \ln(m_2^2) 
                          - \frac{1}{4} p^2 \KKKR \non \\
 && + m_1^2 \KL \ln^2(m_1^2) - 3 \ln(m_1^2) \KR
    + m_2^2 \KL \ln^2(m_2^2) - 3 \ln(m_2^2) \KR
    + \edz p^2 \ln(-p^2) \non \\
 && + \frac{1}{4} p^2 \KKKL \ln\KL \frac{m_1^2}{-p^2} \KR
                           + \ln\KL \frac{m_2^2}{-p^2} \KR 
                           - \frac{13}{2} \KKKR \non \\
 && + (m_1^2 + m_2^2) \KKKL 3 + \frac{\pi^2}{12} 
      - \frac{1}{4} \ln^2 \KL \frac{m_1^2}{m_2^2} \KR \KKKR \non \\
 && + \edz (m_1^2 - m_2^2) 
      \KKKL {\rm Li_2} \KL \frac{m_1^2 - m_2^2}{m_1^2} \KR
           -{\rm Li_2} \KL \frac{m_2^2 - m_1^2}{m_2^2} \KR \KKKR \non\\
 && + \frac{p^2}{4} \KKKL \KL \frac{m_1^2}{p^2} \KR^2
                         -\KL \frac{m_2^2}{p^2} \KR^2 \KKKR
      \ln \KL \frac{m_1^2}{m_2^2} \KR \non \\
 && + \frac{1}{4} (p^2 + m_1^2 + m_2^2) \frac{m_2^2}{p^2}
      (r_1 - r_2) (-\ln(r_1) + \ln(r_2)) \non \\
 && + m_1^2 \KL 1 - \frac{m_2^2}{p^2} \KR
      \KKKL {\rm Li_2} \KL \frac{1 - r_1}{-r_1} \KR
           +{\rm Li_2} \KL \frac{1 - r_2}{-r_2} \KR
           -{\rm Li_2} \KL \frac{m_1^2 - m_2^2}{m_1^2} \KR \KKKR \non\\
 && + m_2^2 \KL 1 - \frac{m_1^2}{p^2} \KR
      \KKKL {\rm Li_2} \KL 1 - r_1 \KR
           +{\rm Li_2} \KL 1 - r_2 \KR
           -{\rm Li_2} \KL \frac{m_2^2 - m_1^2}{m_2^2} \KR \KKKR,
\EEA
where $r_1$ and $r_2$ are given by \refeq{r12}.


\section{$T_{123'4}$}
\label{sec:t123p4}

The following formula, looking at the series expansion in $1/\de$,
are correct up to \order{\de^0}.

\BEA
T_{123'4}(m_1^2, m_4^2, 0, m_1^2) &=& 
   T_{123'4}(m_1^2, m_4^2, m_1^2, 0) =  \\
 && \mbox{}\hspace{-2cm}\Bigg\{ \KL 1 - 2 (\ln(p^2) - i\pi) \de
                    + \edz \KL 2 \ln(p^2) - i\pi \KR^2 \de^2 \KR
                    \frac{1 + \de}{2}(\tilde{B}(p^2, m_1^2, m_4^2))^2
                    \non\\
 && + \edz \KKL 3 - \frac{x_{14} \ln^2(x_{14})}{(1 - x_{14})^2}
                  - \frac{x_{41} \ln^2(x_{41})}{(1 - x_{41})^2}
                  - G(x_{14}) + G(x_{41}) \KKR \Bigg\} \non
\EEA
with the following functions
\BEA
\tilde{B}(p^2, m_i^2, m_j^2) &=& 
   \frac{1}{\de} + \Bigg\{ \edz \KKL 2 + \KL \frac{m_i^2}{-p^2}
                                           -\frac{m_j^2}{-p^2}
           + \sqrt{\la\KL \frac{m_i^2}{-p^2}, \frac{m_j^2}{-p^2} \KR}
                                        \KR \ln(x_{ij}) 
                               - \ln\KL \frac{m_i^2}{-p^2} \KR \KKR \non\\
 &&                      + \edz \KKL 2 + \KL \frac{m_j^2}{-p^2}
                                           -\frac{m_i^2}{-p^2}
           + \sqrt{\la\KL \frac{m_i^2}{-p^2}, \frac{m_j^2}{-p^2} \KR}
                                        \KR \ln(x_{ji}) 
                               - \ln\KL \frac{m_j^2}{-p^2} \KR \KKR
                                                      \Bigg\}\non\\
 && \KL 1 + (\ln(p^2) - i\pi) \de \KR \non \\
 && +\de \KL B_0^{\de}(p^2, m_i, m_j) - \edz (\ln(p^2) - i\pi)^2 \KR \\
\tilde{B}'(p^2, m_i^2, m_j^2) &=&
   \frac{1}{2 \sqrt{\la\KL \frac{m_i^2}{-p^2}, \frac{m_j^2}{-p^2}
                      \KR}}
    \Big[ \KL 1 + \frac{m_i^2}{-p^2} - \frac{m_j^2}{-p^2} 
         + \sqrt{\la\KL \frac{m_i^2}{-p^2}, \frac{m_j^2}{-p^2} \KR} 
         \KR \ln(x_{ij}) \non \\
 &&     +\KL 1 + \frac{m_i^2}{-p^2} - \frac{m_j^2}{-p^2} 
         + \sqrt{\la\KL \frac{m_i^2}{-p^2}, \frac{m_j^2}{-p^2} \KR} 
         \KR \ln(x_{ji}) \Big]
         \KL 1 + (\ln(p^2) - i\pi) \de \KR \non \\
 && - \de \, p^2 \frac{\dd}{\dd (m_i)^2} B_0^{\prime\de}(p^2, m_i, m_j) \\
x_{ij} &=& \frac{2 \frac{m_i^2}{-p^2}}{1 + \frac{m_i^2}{-p^2} +
           \frac{m_j^2}{-p^2} 
           + \sqrt{\KKL \la \KL \frac{m_i^2}{-p^2},
                                \frac{m_j^2}{-p^2} \KR \KKR} } \\
\la(x, y) &=& 1 + 2 x + 2 y + (x - y)^2 \\
F(x) &=& 6 {\rm Li_3}(x) - 4 {\rm Li_2}(x) \ln(x) 
         - \ln^2(x) \ln(1-x) \\
G(x) &=& -2 {\rm Li_2}(1-x) + \frac{\pi^2}{3} 
         + \frac{x}{1-x} \ln^2(x) 
\EEA


\section{$T_{1123'4}$}
\label{sec:t1123p4}

The following formula, looking at the series expansion in $1/\de$,
correct up to \order{\de^0}.

\BEA
\lefteqn{
T_{1123'4}(m_1^2, m_4^2, 0, m_1^2) = 
  \Bigg\{ \KL 1 - 2 (\ln(p^2) - i\pi) \de
              + \edz \KL 2 \ln(p^2) - i\pi \KR^2 \de^2 \KR\non } \\
 &&  (-1 - \de) \tilde{B}(p^2, m_1^2, m_4^2)
                \tilde{B}'(p^2, m_1^2, m_2^2) \non\\
 && + \frac{1}{p^2} \Bigg\{
     \frac{1}{\sqrt{\la\KL \frac{m_i^2}{-p^2}, \frac{m_j^2}{-p^2} \KR}}
     \Bigg[ \frac{\ln(x_{14})}{1 - x_{14}}
            \frac{1 + \ln(x_{14})}{1 - x_{14}}
            - \frac{x_{41} \ln(x_{41})}{1 - x_{41}}
              \frac{1 + x_{41} \ln(x_{41})}{1 - x_{41}} \Bigg] \non\\
 && +\edz \frac{-p^2}{m_i^2} \Bigg[
     \sqrt{\la\KL \frac{m_i^2}{-p^2}, \frac{m_j^2}{-p^2} \KR}
     G(x_{14} x_{41}) \non \\
 && -\edz \KL 1 - \frac{m_i^2}{-p^2} + \frac{m_j^2}{-p^2}
       +\sqrt{\la\KL \frac{m_i^2}{-p^2}, \frac{m_j^2}{-p^2} \KR} \KR
          G(x_{14}) \non \\
 && +\edz \KL 1 - \frac{m_i^2}{-p^2} + \frac{m_j^2}{-p^2}
       -\sqrt{\la\KL \frac{m_i^2}{-p^2}, \frac{m_j^2}{-p^2} \KR} \KR
          G(x_{41}) \Bigg] \Bigg\}
\EEA


\section{$T_{123'45}$}
\label{sec:t123p45}

The following formula, looking at the series expansion in $1/\de$,
correct up to \order{\de^0}.

\BEA
T_{123'45}(m_1^2, m_1^2, 0, m_4^2, m_4^2) &=& 
  \KL 1 - 2 (\ln(p^2) - i\pi) \de
                    + \edz \KL 2 \ln(p^2) - i\pi \KR^2 \de^2 \KR\non\\
 && \frac{1}{p^2} \KL F(1) + F(x_{14} x_{41}) - F(x_{14}) - F(x_{41}) \KR
\EEA

This function is finite in the limit $\de \to 0$.

\chapter{Input parameters and benchmark scenarios}
\label{chap:benchmark}

For our numerical results, the following values of 
the SM parameters have been used (all other quark and lepton masses are
negligible):
\BE 
\begin{aligned}
G_F &= 1.16639\times 10^{-5}, &\quad
m_\tau &= 1.777 \gev, \\
\MW &= 80.450 \gev, &
\mt &= 174.3 \gev, \\
\MZ &= 91.1875 \gev, &
\mb &= 4.25 \gev, \\
\Gamma_Z &= 2.4952 \gev, &
m_c &= 1.5 \gev .
\end{aligned} 
\label{eq:inputpars}
\end{equation}
The predictions for the observables in this report are in some cases
expressed in terms of running bottom- and top-quark masses in order to
absorb QCD corrections. The numerical values of these running masses
differ from the pole masses given in \refeq{eq:inputpars}.

For our numerical evaluation we often refer to four benchmark
scenarios that have been defined in \citere{LHbenchmark} for Higgs boson
searches at hadron colliders and beyond.
The four benchmark scenarios are
(more details can be found in \citere{LHbenchmark}) 
\begin{itemize}

\item
the ``$\mhmax$'' scenario, which
yields a maximum value of $\Mh$ for given $\MA$ and $\tb$,
\BE
\begin{aligned}
{}& \mt = 174.3 \gev, \quad \msusy = 1 \tev, \quad
\mu = 200 \gev, \quad M_2 = 200 \gev, \\
{}& \Xt = 2\, \msusy, \quad
\Atau = \Ab = \At, \quad \mgl = 0.8\,\msusy\,,
\end{aligned}
\label{mhmax}
\end{equation}

\item
the ``no-mixing'' scenario, with no mixing in the $\Stop$~sector,
\BEA
{}&& \mt = 174.3 \gev, \quad \msusy = 2 \tev, \quad
\mu = 200 \gev, \quad M_2 = 200 \gev, \non \\
{}&& \Xt = 0, \quad
\Atau = \Ab = \At, \quad \mgl = 0.8\,\msusy\,,
\label{nomix}
\EEA

\item
the ``gluophobic-Higgs'' scenario, with a suppressed $ggh$ coupling,
\BEA
{}&& \mt = 174.3 \gev, \quad \msusy = 350 \gev, \quad
\mu = 300 \gev, \quad M_2 = 300 \gev, \non \\
{}&& \Xt = -750 \gev, \quad
\Atau = \Ab = \At, \quad \mgl = 500 \gev\,,
\label{gluophobicH}
\EEA

\item
the ``small-$\aeff$'' scenario, with possibly reduced decay rates for
$\hbb$ and $\htautau$,
\BEA
{}&& \mt = 174.3 \gev, \quad \msusy = 800 \gev, \quad
\mu = 2.5 \, \msusy, \quad M_2 = 500 \gev, \non \\
{}&& \Xt = -1100 \gev, \quad 
\Atau = \Ab = \At, \quad \mgl = 500 \gev\,.
\label{smallaeff}
\EEA

\end{itemize}
As explained above, for the sake of simplicity, $\msusy$ is chosen as a
common soft SUSY-breaking parameter for all three generations.

\end{appendix}

\newpage
\addcontentsline{toc}{chapter}{Bibliography}



\end{document}